%% file: manual.tex
\documentclass[11pt,BCOR12mm,DIV12,pdftex,a4paper,twoside,cleardoubleplain,
headsepline,chapterprefix,appendixprefix,fleqn,liststotoc,bibtotoc,pointlessnumbers,final]{scrbook}
\pdfoutput=1 

\usepackage{verbatim}
\usepackage[latin1]{inputenc}
\usepackage[T1]{fontenc}
\usepackage{ae}
\usepackage{amsmath, amssymb}
\usepackage[usenames,dvipsnames]{xcolor} 
\usepackage{parskip}
\usepackage{xspace}
\usepackage{noitemsep}
\usepackage[german, english]{babel}
\usepackage[level, nodayofweek]{datetime}
\usepackage{bigints}
\usepackage{hyperref}
\definecolor{linkcolor}{rgb}{.8,0,0}
\definecolor{urlcolor}{rgb}{0,0,.7}
\definecolor{citecolor}{rgb}{0,.5,0}
\definecolor{acrocolor}{rgb}{0,0,.7}
\hypersetup{colorlinks=true,linkcolor=linkcolor,citecolor=citecolor,urlcolor=urlcolor}
\usepackage{float}
\usepackage{tabularx}
\usepackage{mdframed}
\usepackage[pdftex]{graphicx}
\DeclareGraphicsExtensions{.pdf}  
\usepackage{epstopdf}

\usepackage{finesse}
\usepackage[numbers]{natbib}
\graphicspath{{pic/}}
\let \IG \includegraphics

\newcommand{\code}[1]{{\tt\small #1}}
\newcommand{\mFig}[1]{Figure~\ref{#1}}

\setcapindent{1em}

\deffootnote[1em]{1em}{1em}
{\textsuperscript{\normalfont\thefootnotemark\ }}

\usepackage{fancyvrb}

\DefineVerbatimEnvironment{finesse}{Verbatim}
{formatcom=\small}

\definecolor{katbg}{rgb}{0.99,0.99,0.99}
\newmdenv[linecolor=blue,backgroundcolor=katbg,innermargin =+0.5cm, outermargin =+0.5cm]{katbox}
\fvset{fontsize=\small}

\newcommand{\loadkat}[1]{
\begin{katbox}
\vspace{-18pt}
\VerbatimInput{files/#1}
\end{katbox}
}

\newcommand{\iprod}[1]{\left\langle #1 \right\rangle}

\newcounter{dummy}

\usepackage[new]{optional}

\begin{document}
\selectlanguage{english}
\frontmatter
\include{title}
\cleardoublepage
\include{syntax}
{
\setlength{\baselineskip}{12pt}
\tableofcontents
}
\mainmatter
\include{introduction} 
\include{program_files} 
\include{math_desc} 
\include{higher_order_modes} 

\include{advanced_usage} 
\appendix
\include{appendix_geo_sens}

\include{appendix_thermal}
\include{appendix_map_knm}

\include{numerical_math} 
\include{syntax_ref}   
\backmatter
\include{ack}
\include{bib}
\end{document}

%% file: title.tex

\begin{titlepage}
\pdfbookmark[1]{Title}{title}

\begin{center}
{\LARGE {\bf FINESSE 1.0} \\ \vspace{1cm}Frequency domain INterferomEter Simulation SoftwarE}

\vspace{.5cm}
{\Large \url{www.gwoptics.org/finesse}}

\vspace{5cm}
\IG [viewport=50 25 300 150,clip,angle=0,scale=1] {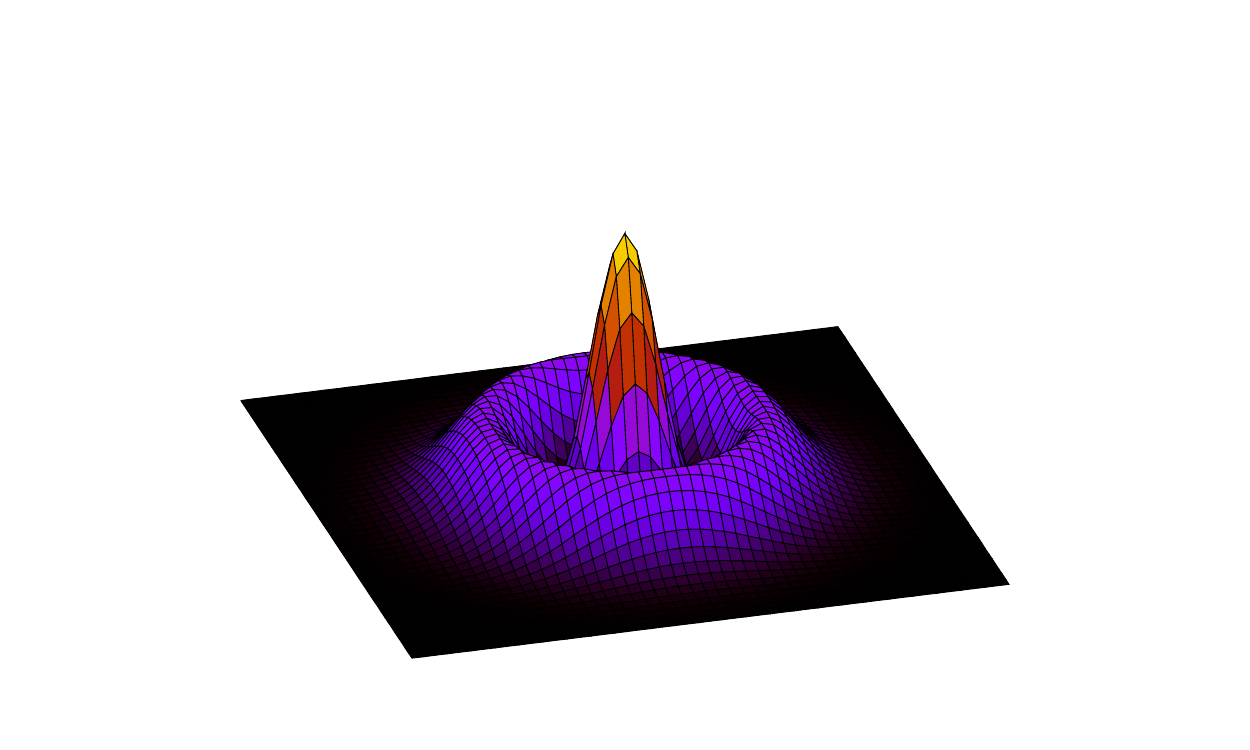} 
\end{center}
\vfill

\vspace{1cm}

\Finesse is a fast interferometer simulation program.  For a given optical
setup, it computes the light field amplitudes at every point in the
interferometer assuming a steady state.  To do so, the interferometer
description is translated into a set of linear equations that are solved
numerically. For convenience, a number of standard analyses can be performed
automatically by the program, namely computing modulation-demodulation error
signals, transfer functions, shot-noise-limited sensitivities, and beam
shapes.  \Finesse can perform the analysis using the plane-wave
approximation or \HG modes. The latter allows computation of the properties
of optical systems like telescopes and the effects of mode matching and
mirror angular positions. 

\vfill
\begin{center}
\newdate{date}{06}{06}{2013}
\displaydate{date}
\end{center}
\end{titlepage}
\thispagestyle{empty}

\vspace*{\fill}
\Finesse, the accompanying documentation, and the example files 
have been written by:

Andreas Freise\\
School of Physics and Astronomy\\
The University of Birmingham\\
Edgbaston, Birmingham, B15 2TT, UK\\
\href{mailto:andreas.freise@googlemail.com}{andreas.freise@googlemail.com}

\Finesse has been substantially developed further during the last year (2012 - 2013).
Daniel Brown provided significant contributions to the code,
the manual and the publication of \Finesse as open
source. Charlotte Bond has carefully tested the new code and provided
tutorials, examples and documentation. 

Parts of the original \Finesse source and `mkat' have been written by Gerhard
Heinzel, the document `sidebands.ps' by Keita Kawabe, the Octave
examples and its description by Gabriele Vajente, part of the
\Finesse source have been written by Paul Cochrane.

\textsc{The software and documentation is provided as is without any warranty 
of any kind.}
\textsc{Copyright \copyright\ by Andreas Freise 1999\,--\,\number\year.}

The source code for \Finesse is available as open source under the 
GNU General Public License version 3 as published by the Free Software Foundation.

This manual and all \Finesse documentation and examples available from
\url{www.gwoptics.org/finesse} and related pages are distributed under
a Creative Commons Attribution-Non\-com\-mercial-Share Alike License, 
see\\ \url{http://creativecommons.org/licenses/by-nc-sa/2.0/uk/}. 

This document has been assigned the LIGO DCC number: LIGO-T1300431.

\thispagestyle{empty}


%% file: syntax.tex

\vspace{-3cm}
{
\begin{center}
\small
\pdfbookmark[1]{Short Syntax Reference}{syntax}
\begin{verbatim}
------------------------------------------------------------------------
  FINESSE 1.0     - Help Screen -                      05.06.2013
------------------------------------------------------------------------
** Usage (1) kat [options] infile [outfile [gnufile]] 
   or    (2) kat [options] basename
   in (2) e.g. basename 'test' means input filename : 'test.kat', 
   output filename : 'test.out' and Gnuplot file name : 'test.gnu'.
** Support :
   User support forums:     http://www.gwoptics.org/finesse/forums/
   Online syntax reference: http://www.gwoptics.org/finesse/reference/
** Available options:
 -v : prints version number and build date
 -h : prints this help (-hh prints second help screen)
 -c : check consistency of interferometer matrix
 -max : prints max/min
 -sparse, -klu : switch to SPARSE or KLU library respectively
 --server : starts Finesse in server mode
 --noheader : suppresses header information in output data files
 --perl1 : suppresses printing of banner
 --quiet : suppresses almost all screen outputs
 --convert : convert knm files between text and binary formats
** Available interferometer components:
 l name P f [phase] node                          - laser
 m name R T phi node1 node2                       - mirror
 (or: m1 name T Loss phi ...          
      m2 name R Loss phi ... )        
 s name L [n] node1 node2                         - space
 bs name R T phi alpha node1 node2 node3 node4    - beamsplitter
 (or: bs1 name T Loss phi ... 
      bs2 name R Loss phi ... )             
 gr[n] name d node1 node2 [node3 [node4]]         - grating
 isol name S node1 node2                          - isolator
 mod name f midx order am/pm [phase] node1 node2  - modulator
 lens f node1 node2                               - thin lens
** Detectors:
 pd[n] name [f1 [phase1 [f2... ]]] node[*]        - photodetector [mixer]
 pdS[n] name [f1 phase1 [f2... ]] node[*]         - sensitivity
 pdN[n] name [f1 phase1 [f2... ]] node[*]         - norm. photodetector
 ad name [n m] f node[*]                          - amplitude detector
 shot name node[*]                                - shot noise
 bp name x/y parameter node[*]                    - plots beam parameters
 cp cavity_name x/y parameter                     - plots cavity parameters
 gouy name x/y space-list                         - plots gouy phase
 beam name [f] node[*]                            - plots beam shape
 qshot name num_demod f [phase] node[*]           - quantum shotnoise detector
 qshotS name num_demod f [phase] node[*]          - quantum shotnoise sens.
** Available commands:
 fsig name component [type] f phase [amp]         - signal
 tem[*] input n m factor phase                    - input power in HG/LG modes
 mask detector n m factor                         - mode mask for outputs
 pdtype detector type-name                        - set detector type
 attr component M value Rcx/y value x/ybeta value - attributes of m/bs
 (alignment angles beta in [rad])
 map component filename                           - read mirror map file
 knm component_name filename_prefix [flag]        - save coefficients to file
 maxtem order                                     - TEM order: n+m<=order
 gauss name component node w0 z [wy0 zy]          - set q parameter
 gauss* name component node q [qy] (q as 'z z_R') - set q parameter
 gauss** name component node w(z) Rc [wy(z) Rcy]  - set q parameter
 cav name component1 node component2 node         - trace beam in cavity
 startnode node                                   - startnode of trace
 retrace [off]                                    - re-trace beam on/off
 phase 0-7  (default: 3)                          - change Gouy phases
 (1: phi(00)=0, 2: gouy(00)=0, 4: switch ad phase)
 conf component_name setting value                - configures component
** Plot and Output related commands :
 xaxis[*] component param. lin/log min max steps  - parameter to tune
 x2axis[*] component param. lin/log min max steps - second axis for 3D plot
 noxaxis                                          - ignore xaxis commands
 const name value                                 - constant $name
 variable name value                              - variable $name
 set name component parameter                     - variable $name
 func name = function-string                      - function $name
 lock[*] name $var gain accuracy                  - lock: make $var to 0
 put[*] component parameter $var/$x1/$x2          - updates parameter
 noplot output                                    - no plot for 'output'
 trace verbosity                                  - verbose tracing
 yaxis [lin/log] abs:deg/db:deg/re:im/abs/db/deg  - y-axis definition
 scale factor [output]                            - y-axis rescaling
 diff component parameter                         - differentiation
 deriv_h value                                    - step size for diff
** Auxiliary plot commands :
 gnuterm terminal [filename]                      - Gnuplot terminal
 pause                                            - pauses after plotting
 multi                                            - plots all surfaces
                                                    save/load knm file
 GNUPLOT \ ... \ END                              - set of extra commands
                                                    for plotting.
\end{verbatim}
\newpage
\enlargethispage*{4.\baselineskip}
\vspace{-4cm}
\begin{verbatim}
------------------------------------------------------------------------
  FINESSE 1.0   - Help Screen (2) -                    05.06.2013
------------------------------------------------------------------------
 ** Alternative calls:
  kat --server <portnumber (11000 to 11010)> [options] inputfile 
   starts Finesse in server mode listening on a TCP/IP port
  kat --convert knm_file_prefix 1/2 [new_knm_file_prefix]
   converts binary knm file to ASCII and vice versa
 ** Some conventions:
  names (for components and nodes) must be less than 15 characters long
  angles of incidence, phases and tunings are given in [deg]
  (a tuning of 360 deg corresponds to a position change of lambda)
  misalignment angles are given in [rad]
 ** Geometrical conventions:
  tangential plane: x, z (index n), saggital plane: y, z (index m)
  xbeta refers to a rotation in the x, z plane, i.e. around the y-axis
  R<0 when the center of the respective sphere is down beam
  (the beam direction is defined locally through the node order:
  i.e. mirror: node1 -> node2, beam splitter: node1 -> node3)
  beam parameter z<0 when waist position is down beam
 ** trace n: `n' bit coded word, produces the following output:
     trace 1:   list of TEM modes used
     trace 2:   cavity eigenvalues and cavity parameters like FWHM, 
                FSR optical length and Finesse
     trace 4:   mode mismatch parameters for the initial setup
     trace 8:   beam parameters for every node, nodes are listed in
                the order found by the tracing algorithm
     trace 16:  Gouy phases for all spaces
     trace 32:  coupling coefficients for all components
     trace 64:  mode matching parameters during calculation, if
                they change due to a parameter change, for example
                by changing a radius of curvature.
     trace 128: nodes found during the cavity tracing
 ** phase 0-7: also bit coded, i.e. 3 means `1 and 2'
     phase 1:   phase of coupling coefficients k_00 set 0
     phase 2:   Gouy phase of TEM_00 set to 0
     phase 4:   `ad name n m f' yields amplitude without Gouy phase
     (default: phase 3)
 ** bp, possible parameters for this detector:
     w  : beam radius
     w0 : waist radius
     z  : distance to waist
     zr : Rayleigh range
     g  : Gouy phase
     r  : Radius of curvature (phase front) in meters
     q  : Gaussian beam parameter
 ** isol S, suppression given in dB:
     amplitude coefficient computed as 10^-(S/20)
 ** maxtem O/off : O=n+m (TEM_nm) the order of TEM modes, 
     `off' switches the TEM modus off explicitly
 ** conf:
    This command is used to configure settings of a component
    that are not physical parameters.
    - conf component integration_method n
      sets the numerical integration method. Cubature refers to a
      self-adapting routine which is faster but less robust
      1 - Riemann Sum
      2 - Cubature - Serial
      3 - Cubature - Parallel (default)
    - conf component interpolation_method n
      sets the interpolation method for the numerical integration of
      surface maps (use NN for maps with sharp edges):
      1 - Nearest Neighbour
      2 - Linear (default)
      3 - Spline
    - conf component interpolation_size n
      sets the size of the interpolation kernel, must be odd and > 0
      (default=3)
    - conf component knm_flags n
      Sets the knm computation flags which define if coeffs are calculated
      numerically or analytically if possible, see below for values of 'n'.
    - conf component show_knm_neval 0/1
      Shows the number of integrand evaluations used for the map integration.
    - conf component save_knm_matrices 0/1
      If true the knm matrices are saved to .mat files
      for distortion, surface map and the final result
    - conf component save_knm_binary 0/1
      If true the knm and merged map data is stored in a
      binary format rather than ASCII, see --convert option
      for converting between the 2 formats
    - conf component save_interp_file 0/1
      If true a file is written for each knm
      to output each interpolated point. The file
      has 4 columns: x, y, amplitude, phase.
    - conf component save_integration_points 0/1
      If true all points used for integration are
      saved to a file (use this only with the Riemann
      integrator, Cubature can use millions of points!)
    - conf component knm_order 12/21
      changes the order in which the coupling coefficient matrices
      are computed. 1 = Map, 2 = Bayer-Helms
    - conf component knm_change_q 1/2
      specifies the expansion beam parameter q_L.
      If 1 then q_L = q'_1 and if 2 then q_L = q_2.
 ** knm flags: `n' bit coded word, set computation of coupling coeffs
               use with the conf knm_flag command option:
     0  : analytic solution of all effects used
     1  : verbose, i.e. print coupling coefficients
     2  : numerical integration if x and y misalignment is set
     4  : numerical integration if x or y misalignment is set
     8  : calculates aperture knm by integration
     16 : calculates curvature knm by integration
     32 : calculates bayer-helms knm by integration
     (default: knm 8)
\end{verbatim}
\end{center}
}


%% file: introduction.tex

\chapter{Introduction} 

\Finesse is a simulation program for interferometers.  The user can build
any kind of virtual interferometer using the following components:

\begin{itemize}
\item[-] lasers, with user-defined power, wavelength and shape of the output 
beam;
\item[-] free spaces with arbitrary index of refraction;
\item[-] mirrors and beam splitters, with flat or spherical surfaces;
\item[-] modulators to change amplitude and phase of the laser light;
\item[-] amplitude or power detectors with the possibility of demodulating
the detected signal with one or more given demodulation frequencies;
\item[-] lenses and \FI s.
\end{itemize}

For a given optical setup, the program computes the light field amplitudes
at every point in the interferometer assuming a steady state.  To do so, the
interferometer description is translated into a set of linear equations that
are solved numerically. For convenience, a number of standard analyses can
be performed automatically by the program, namely computing
modulation-demodulation error signals and transfer functions.  \Finesse can
perform the analysis using plane waves or \HG modes. The latter allows
computation of the effects of mode matching and misalignments. In addition,
error signals for automatic alignment systems can be simulated.

\begin{figure}[h]
\centering
\IG [scale=1, angle=0] {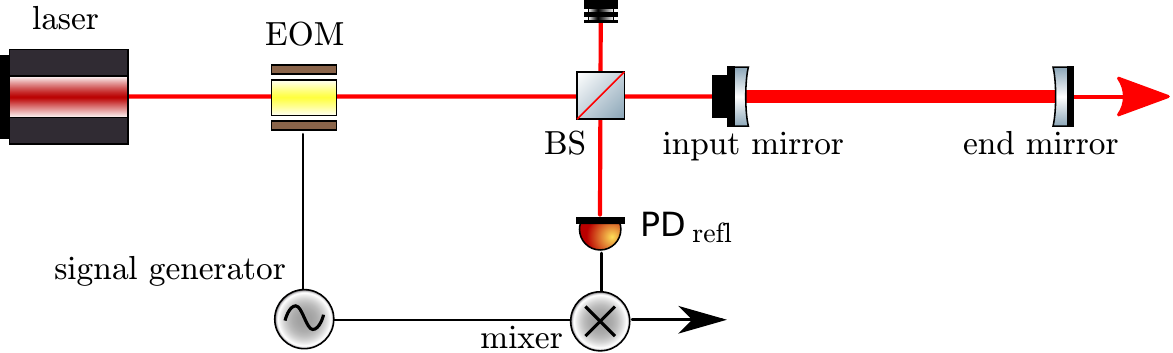} 
\caption{A schematic diagram of a laser interferometer which can be
  modelled using \Finesse (in this case a Fabry-Perot cavity with a
  Pound-Drever-Hall control scheme).}
\label{fig:schematic01}
\end{figure}

Literally every parameter of the interferometer description can be tuned
during the simulation. The typical output is a plot of a photodetector
signal as a function of one or two  parameters of the interferometer
(e.g.~arm length, mirror reflectivity, modulation frequency, mirror
alignment).  \Finesse automatically calls Gnuplot (a free graphics
program~\cite{gnuplot}) to create 2D or 3D plots of the output data
(alternatively plotting of the output data can be performed with
Python or Matlab, see sections~\ref{sec:mfiles} and \ref{sec:pyfiles}.
Optional text output provides information about the optical setup like, for
example, mode mismatch coefficients, eigenmodes of cavities and beam sizes.

\Finesse provides a fast and versatile tool that has proven to be very
useful during design and commissioning of interferometric \gw
detectors.  However, the program has been designed to allow the analysis of
arbitrary, user-defined optical setups. In addition, it is easy to install
and easy to use.  Therefore \Finesse is very well suited to study basic
optical properties, like, for example, the power enhancement in a resonating
cavity or modulation-demodulation methods.

\section{Motivation}

The search for \gws with interferometric detectors has led to a new type of
laser interferometer: new topologies are formed combining known
interferometer types.  In addition, the search for \gws requires optical
systems with a very long baseline, large circulating power and an enormous
stability.  The properties of this new class of laser interferometers have
been the subject of extensive research.

\begin{figure}[h]
\centering
\IG [width=0.9\textwidth, angle=0] {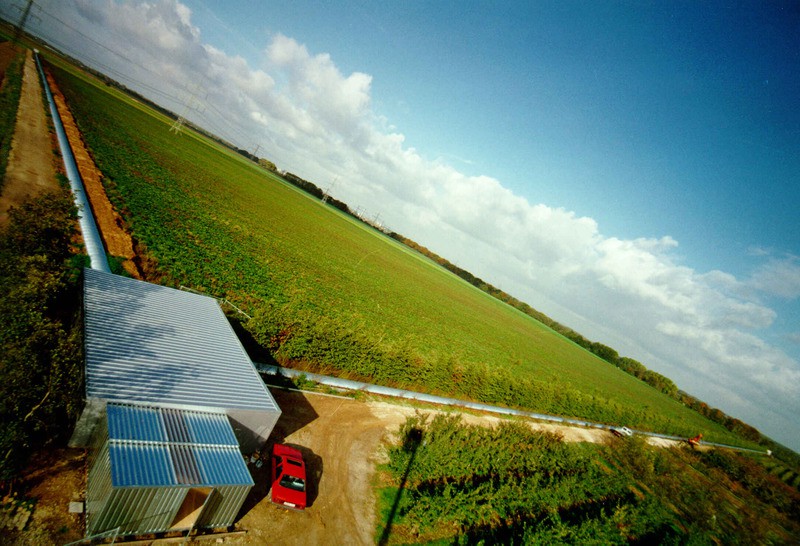} 
\caption{Bird's eye view of the GEO\,600 gravitational wave detector
  near Hannover, Germany. Image courtesy of Harald L\"uck, Albert Einstein Institute Hannover.}
\label{fig:geo600}
\end{figure}

Several prototype interferometers have been built during the last few
decades to investigate their performance in detecting \gws.
The optical systems, \FP cavities, a \Mi and combinations thereof are in
principle simple and have been used in many fields of science for many
decades.  The sensitivity required for the detection of the expected small
signal amplitudes of \gws, however, has put new constraints on the design of
laser interferometers.  The work of the \gw research groups has led to a new
exploration of the theoretical analysis of laser interferometers.
Especially, the clever combination of known interferometers has produced new
types of interferometric detectors that offer an optimised sensitivity for
detecting \gws.  Work on prototype interferometers has shown that the models describing the
optical system become very complex even though they are based on simple
principles.  Consequently, computer programs have been developed to automate
the computational part of the analysis.  To date, several programs for
analysing optical systems are available to the \gw community~\cite{gwic}. 

The idea for \Finesse was first raised in 1997, when I was visiting
the Max-Planck-Institute for Quantum Optics
in Garching, to assist Gerhard Heinzel with his work on Dual Recycling at
the 30\,m prototype interferometer.  We were using optical simulations which
were rather slow and not very flexible.  At the same time Gerhard Heinzel
had developed a linear circuit simulation~\cite{Liso} that used a numerical
algorithm to solve the set of linear equations representing an electronic
circuit. The similarities of the two computational tasks and the outstanding
performance of \Liso lead to the idea to use the same methods for an optical
simulation. Gerhard Heinzel kindly allowed me to copy the \Liso source code
which saved me much time and trouble in the beginning; and even today many
of the \Liso routines are still used in their original form inside \Finesse.

In the following years \Finesse was continually developed during my work at
the university in Hannover within the \GEO project~\cite{geo,GEO600:prop}.
\Finesse has been most frequently utilised during the commissioning of
\GEO, some of these simulation results have been published
in~\cite{adf-amaldi5,hal,icm,hrg} and in~\cite{adf}.  
\Finesse is now actively developed as an open source project and it is
now widely used in many other projects; the \Finesse
home page lists more than 60 documents citing it~\cite{impact}.

\newpage 

\section{How does it work?}
When the program is run, \Finesse performs the following steps:

{\bf Reading a text input file}:\hspace{0.3cm}  One has to write an input
text file\footnote{There is also a graphical user interface~\cite{luxor}
that can be used to generate the input file.} that describes the
interferometer in the form of components and connecting nodes (see
Section~\ref{sec:comp+nodes}).  Several commands specify the computational
task and the output format.  The command \cmd{xaxis}, for example, defines
the parameter to be tuned during the simulation. 

{\bf Generating the set of linear equations}:\hspace{0.3cm} The mutual
coupling of all light amplitudes inside the interferometer can be described
by linear equations.  \Finesse converts the list of components and nodes
given in the input file into a matrix and a `right hand side' vector (see
Section~\ref{sec:matrix}) which together represent a set of linear
equations.  The calculation is initialised by the commands in the input
file.

{\bf Solving the linear equation system numerically}:\hspace{0.3cm} For each
data point, the linear set of equations is updated, the `right hand side'
vector is generated and the system is solved numerically (using a sparse
matrix solver such as~\cite{sparse}).  This step is repeated for each data point and each
possible light frequency (i.e.~modulation sidebands).

{\bf Writing the data to an output file}:\hspace{0.3cm} After solving the
system of equations, all light amplitudes inside the interferometer are
known. \Finesse then computes the specified output signals (amplitudes,
powers, demodulated signals, etc.) and writes them to a text file (extension
`.out'). The screen output is also stored in a file with the extension
`.log').

{\bf Plotting the data}:\hspace{0.3cm} \Finesse uses
external programs to generate and
display plots of the output data, it creates a so called batch file
with plotting instructions and then calls the external program, by
default Gnuplot~\cite{gnuplot}. Gnuplot can display graphs on screen or
write the data to a file of a specified graphics format (postscript, gif, etc.) \Finesse creates
an additional file (extension `.gnu') that serves as a batch file for
Gnuplot and calls Gnuplot to automatically produce the plot.
Gnuplot is a free program available for different operating systems. If you
do not have Gnuplot installed yet, you should do so
(\url{http://www.gnuplot.info}). 

Alternatively you can use Python or Matlab to plot the data. \Finesse saves
a Python file (extension `.py') and a Matlab script file (extension
`.m'). Python can be called automatically to plot the data similar to
the Gnuplot scenario described above. However, plotting the
results with Matlab has to be performed by the user:
Inside Matlab, change into
the working directory containing the `katfilename.out' and `katfilename.m' file and 
call the latter with the command `katfilename' (replace `katfilename' by the
actual name of the file). By using the command
`gnuterm no' in the input file the automatic call to Gnuplot is suppressed.

\section{Quick start}

This section presents some example simulations. 
\Finesse it is very easy to use,
despite this rather voluminous manual. The manual contains much basic
information about optical systems and typical tasks in interferometer
analysis. Anyone familiar with the analysis of optical systems should be
able to install \Finesse and do a first calculation in roughly half an hour.
If you do not have much experience with interferometers, you can use
\Finesse to learn more about them. 

You can find more information online starting at
\url{www.gwoptics.org/finesse}, such
as simple and advanced examples for \Finesse simulations, an online
reference page for the \Finesse syntax and advanced installation
instructions. Further \Finesse examples in the context of advanced
interferometry can also be found in the free online article
`Interferometer Techniques for Gravitational-Wave Detection'~\cite{interferometry}.

\subsection{Installation}

The installation is simple:
\begin{itemize}
\item[-] You only have to copy all files into your working directory. The 
required files are: {\bf kat} or {\bf kat.exe} (the executable) and 
{\bf kat.ini}, the initialisation file.  In addition, you may want to try 
the program using my example input files.  These have the extension 
{\bf *.kat}.
\item[-] \Finesse is a text-based application that you have {\bf to run from 
within a terminal or command window}. You can start the program by typing 
`kat', which will print a small banner; `kat -h' will give you a short 
syntax reference for input files; `kat -hh' prints further help.
\item[-] If you want \Finesse to automatically generate graphical output, 
you have to have Gnuplot or Python installed 
and \Finesse must know the correct command to start it. You can add the
command for calling Gnuplot and Python on your system to the file `kat.ini' (see 
Section \ref{sec:ini} for more details).
\item[-] You should test the program with one of the example files:
e.g.~`kat bessel.kat' will cause the program to calculate light field
amplitudes behind a phase modulator as a function of the modulation strength
(modulation index).  The calculated data will be written to `bessel.out'; also a
batch file (`bessel.gnu') for plotting the data with Gnuplot will be created
and Gnuplot will be started.
\item[-] Please let me know if the above did not work for you!
\end{itemize}

Jan Harms has created a graphical user interface for \Finesse~\cite{luxor}\footnote{The
functionality of \Luxor does not necessarily include the latest features of
the current version of \Finesse. Nevertheless, it should be able
to handle the majority of simulation tasks.}. This manual does not 
consider the use of \Luxor but describes the original use of \Finesse, 
i.e. with ASCII text files and the terminal window. Personally 
I would only recommend \Luxor for \Finesse beginners. In my opinion
the careful use of text files is preferable when the
interferometers or the simulation tasks become more complex.

\subsection{How to perform a simulation}

In order to do a simulation, you have to create a text (ASCII) input file
for \Finesse that specifies the interferometer and the simulation task.  In
the following sections we discuss two example files.  The first example is
meant for people without much experience in interferometry. It shows some
basic syntax without any complicated optics. The second example is aimed at
people who know what a `transfer function' or a `Pound-Drever-Hall scheme'
is and only need to understand the input syntax of \Finesse.  

\begin{figure}[h]
\centering
\IG [scale=.6, angle=0] {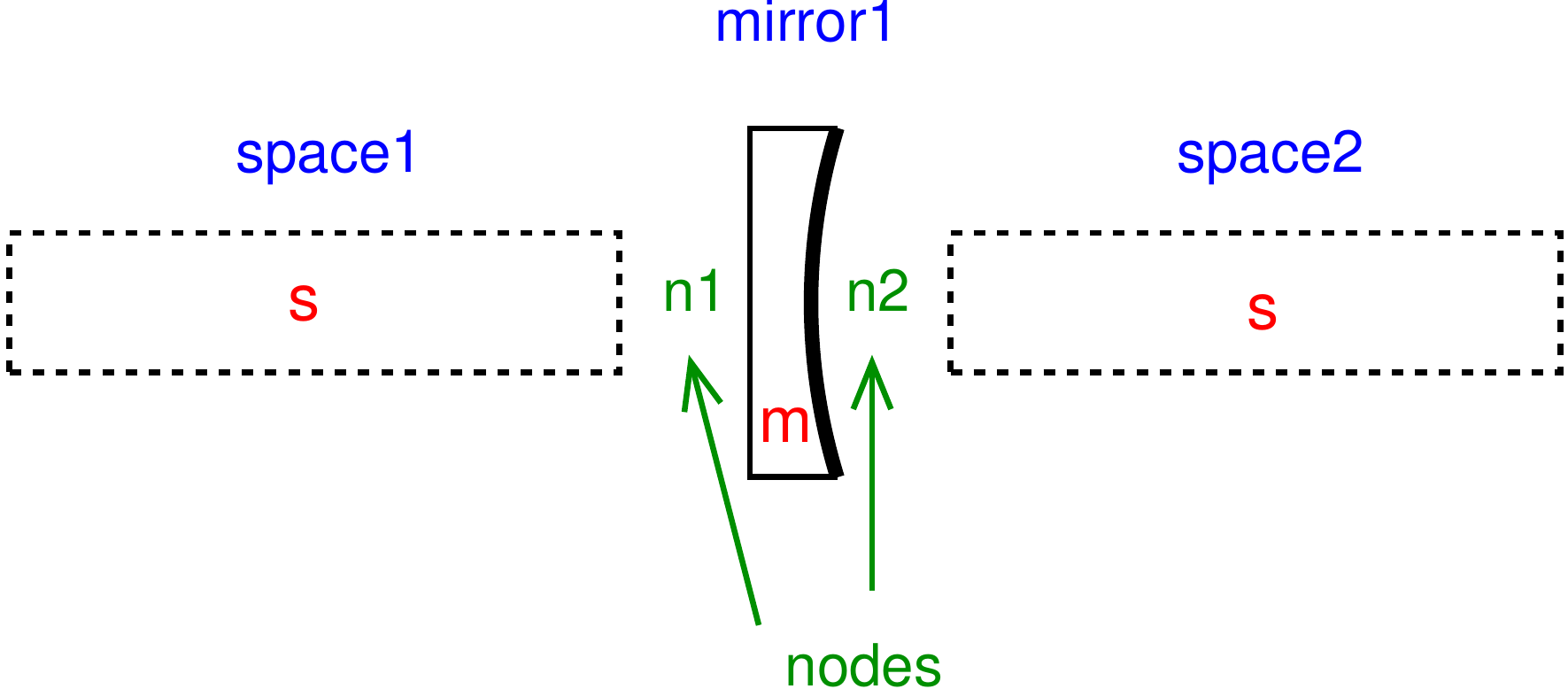} 
\caption{A mirror (m) and two spaces (s) connected via nodes n1 and n2.}
\label{fig:node1}
\end{figure}

I believe that, before you start writing an input file, it is essential {\bf
to first draw the optical setup on a piece of paper}.
Next, you have to break
down the optical system into its components and nodes (see
\Sec{sec:comp+nodes} for details on `nodes').  The components include
mirrors, lasers or `free spaces', separated by nodes.  When, for
example, a mirror is next to a `free space' there is a node between them
(see \fig{fig:node1}).  All components and nodes have to be given a name.
The name will be used to refer to the component, either with a command in
the input file, or by \Finesse in warning or error messages. 

An input file can consist of a series of component descriptions, commands
and some comments (text after `\texttt{\#}').  Component descriptions are
typically entered (one per line) as `\texttt{keyword name parameter-list
node-list}'.  The mirror of \fig{fig:node1}, for example, can be described
by

\begin{finesse}
m mirror1 0.9 0.1 0 n1 n2
\end{finesse}

The keyword for a mirror is \cmd{m}. The name of the component, in this
example \cmd{mirror1}, is followed by three numerical values.  These are the
values for the parameters of the component mirror, namely: power
reflectivity (R), power transmission (T) and tuning (phi) (see
Section~\ref{sec:tuning} for the definition of `tuning').  You do not have
to memorise all the parameters: calling \Finesse with the command `kat -h'
prints a help screen that includes a short description of the input file
syntax with all component parameters. In addition, we provide a handy
online reference at \url{http://www.gwoptics.org/finesse/reference/}.

The last two entries in the component description of the mirror above are
the `node list'. Most components are connected to exactly two nodes. Beam
splitters are connected to four nodes, a laser to one. Every node is
connected to at least one component and never to more than two, otherwise
the description is inconsistent.

In the component description the keyword specifies the type of component
(\cmd{m} for mirror, \cmd{bs} for beam splitter, etc.), you can choose an
arbitrary name but \textbf{it must be less than 15 characters long}.

Detectors are special components; they can be located anywhere in the
interferometer. Every detector defines one output variable that will be
computed by \Finesse. By specifying several detectors, you may compute
several signals at the same time.

In addition to component description, commands can be given (one per line).
The commands are used to initialise the simulation, they specify what output
is to be computed and which parameters are to be changed. For example, the
command

\begin{finesse}
xaxis space1 L lin 1 10 100
\end{finesse}

defines the $x$-axis of the output data. In this case it is the length (L) of
the component \cmd{space1}. The length will be tuned linearly \cmd{lin} from
1 to 10 metres in 100 steps. The \cmd{xaxis} command has to be given for
every simulation. Other commands are optional (like \cmd{scale}, which can
scale outputs by a user-defined factor). In addition, some commands can be
used to customise the graphical output.

The description of the following example input files does not include a
detailed explanation of the syntax for commands or component description.
Please refer to the syntax reference while studying the following examples.
\textbf{The help screen (type `kat -h') gives a short syntax reference. The
full syntax reference is given in \app{sec:syntax_reference}.}

Note: Text from the input files (like commands, keywords, etc.) is printed
in a \cmd{fixed-width font} throughout this manual.

\subsubsection{A simple example: Bessel functions}
\label{sec:examples1}

This example features a laser, a `phase modulator' (usually an electro-optic
device that can modulate the phase of a passing light beam) and `amplitude
detectors'. Amplitude detectors can measure the amplitude and phase of a
light field. Such a device does not exist in reality but is a very useful
tool in simulations.

The phase modulation of a light field (at one defined frequency, the
\emph{modulation frequency}) can be described in the frequency domain as the
generation of `modulation sidebands'. These sidebands are new light fields
with a frequency offset to the initial light. In general, a symmetric pair
of such sidebands with a frequency offset of plus or minus the modulation
frequency is always generated. For stronger modulations, symmetric pairs at
multiples of the modulation frequency are also generated.

The amplitude of these sidebands can be mathematically described using
Bessel functions (see Section \ref{sec:phasemod} for details on phase
modulation and Bessel functions).  In this example, the amplitudes of three
modulation sidebands are detected.  The result can be used to check whether
the implementation of Bessel functions in \Finesse is correct.

The input file `bessel.kat' for this simulation looks as follows:
\loadkat{bessel.kat}
 
The only two components of the setup in this example are the laser
defined by :

\begin{finesse}
l i1 1 0 n0 
\end{finesse}
 
and the modulator:
\begin{finesse}
mod eo1 40k .05 5 pm n0 n1
\end{finesse}

\begin{figure}[htb]
\centering
\IG [scale=0.9, angle=0] {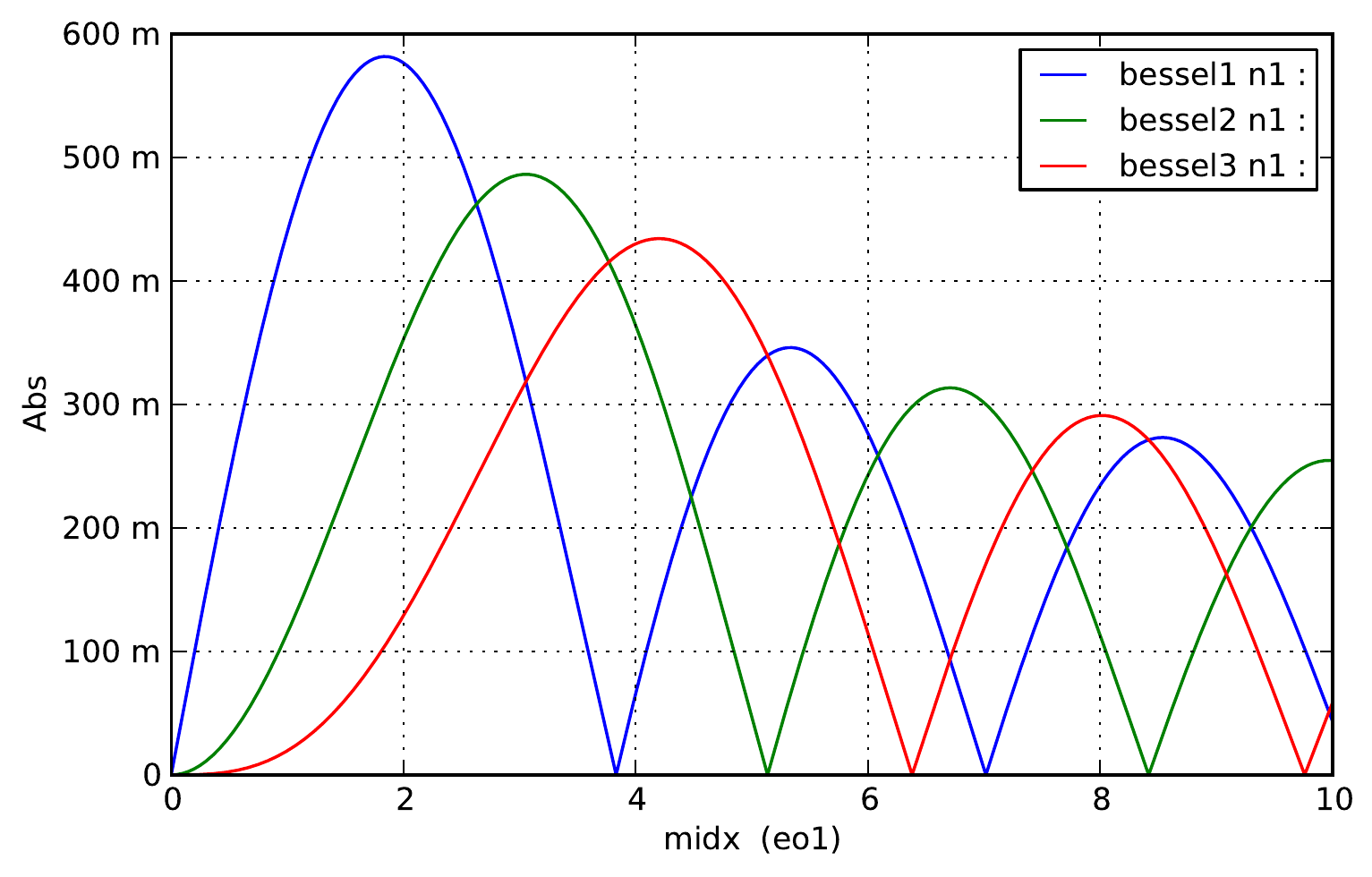} 
\caption{Simple example: testing the Bessel functions. (The plots
  shown in this manual have been automatically created by \Finesse using the Python batch file.)}
\label{fig:bessel1}
\end{figure}
These two components are connected via node \cmd{n0}, and the exit of the
modulator is node \cmd{n1}. The laser provides the input field at frequency
$0$\,Hz and a power of $1$\,W. All frequency values have to be understood as
offset frequencies to a default frequency, which can be set by specifying a
default wavelength in the init file `kat.ini'.  When the laser beam passes
the modulator, sidebands are added at multiples of the modulation frequency.
In this case, the modulation frequency is 40\,kHz. The strength (or depth)
of the modulation can be specified by the modulation index (\cmd{midx});
here \cmd{midx}=0.5. In general, the sidebands generated by a phase
modulation with modulation frequency $\w_m/2\pi$ and $\cmd{midx}=m$ can be
described by the following sum (see Section~\ref{sec:phasemod}):
\begin{equation}
\sum_{k=-\infty}^{\infty}i^{\,k}~J_k(m)~e^{\I k \w_m\T},
\end{equation}

with $J_k(x)$ as the Bessel function of order $k$. For small modulation
indices, the Bessel function becomes very small with increasing $k$.
Therefore, usually only a finite part of the above sum has to be taken into
account: 
\begin{equation}
\sum_{k=-order}^{order}i^{\,k}~J_k(m)~e^{\I k \w_m\T}.
\end{equation}

The maximum value for $k$ (`order') is set as a parameter in the component
description of the modulator (\cmd{order}). In this example, \cmd{order} is
set to 5 which will result in 11 light fields leaving the modulator: 1 laser
field, 5 sidebands with positive frequency offsets (40\,kHz, 80\,kHz,
120\,kHz, 160\,kHz, 200\,kHz) and 5 sidebands with negative frequency
offsets (-40\,kHz, -80\,kHz, -120\,kHz, -160\,kHz, -200\,kHz).  In order to
detect some of the light fields after the modulator, we connect `amplitude
detectors' (\cmd{ad}) to node \cmd{n1}:

\begin{finesse}
ad bessel1 40k n1       
ad bessel2 80k n1       
ad bessel3 120k n1     
\end{finesse} 
  
For each detector, a different frequency is specified (40\,kHz, 80\,kHz and
120\,kHz).  This means that we will compute field amplitudes for three
different sidebands.  The setup is now completely described. Next, we
have to define the simulation task:

\begin{finesse}
xaxis eo1 midx lin 0 10 1000
yaxis abs                   
\end{finesse}
 
The compulsory command \cmd{xaxis} specifies the parameter we want to tune
during the simulation. In this example, the parameter \cmd{midx} of the
modulator (\cmd{eo1}) will be changed linearly (\cmd{lin}) from 0 to 10 in
1000 steps. The command \cmd{yaxis abs} specifies that the absolute values
of the computed complex field amplitudes will be plotted.  The command
\cmd{gnuterm x11} specifies a screen output for Gnuplot in a typical Unix
environment. Windows users should use  \cmd{gnuterm windows}. In addition,
there are a number of predefined graphics formats (Gnuplot terminals) for file output (like ps, eps,
gif). If no Gnuplot terminal is given \Finesse uses `x11' on Unix and `windows' on
Windows systems.

In summary, we have set up a very simple optical system which generates
phase-modulated sidebands. By running the simulation, we can now compute
amplitudes of the sidebands as a function of the modulation index. The
resulting plot is shown in Figure \ref{fig:bessel1}.

\subsubsection{A more complex example: A \FP\ cavity in a Pound-Drever-Hall setup}
\label{sec:examples2}

This example consists of two sequential simulations: a) the generation of a
Pound-Drever-Hall error signal for a simple \FP\ cavity, and b) the transfer
function with respect to this error signal (mirror motion to error signal).

\paragraph{The error signal: pdh-signal.kat}

The input file is:
\loadkat{pdh-signal.kat}

\begin{figure}[h]
\centering
\IG [scale=0.9, angle=0] {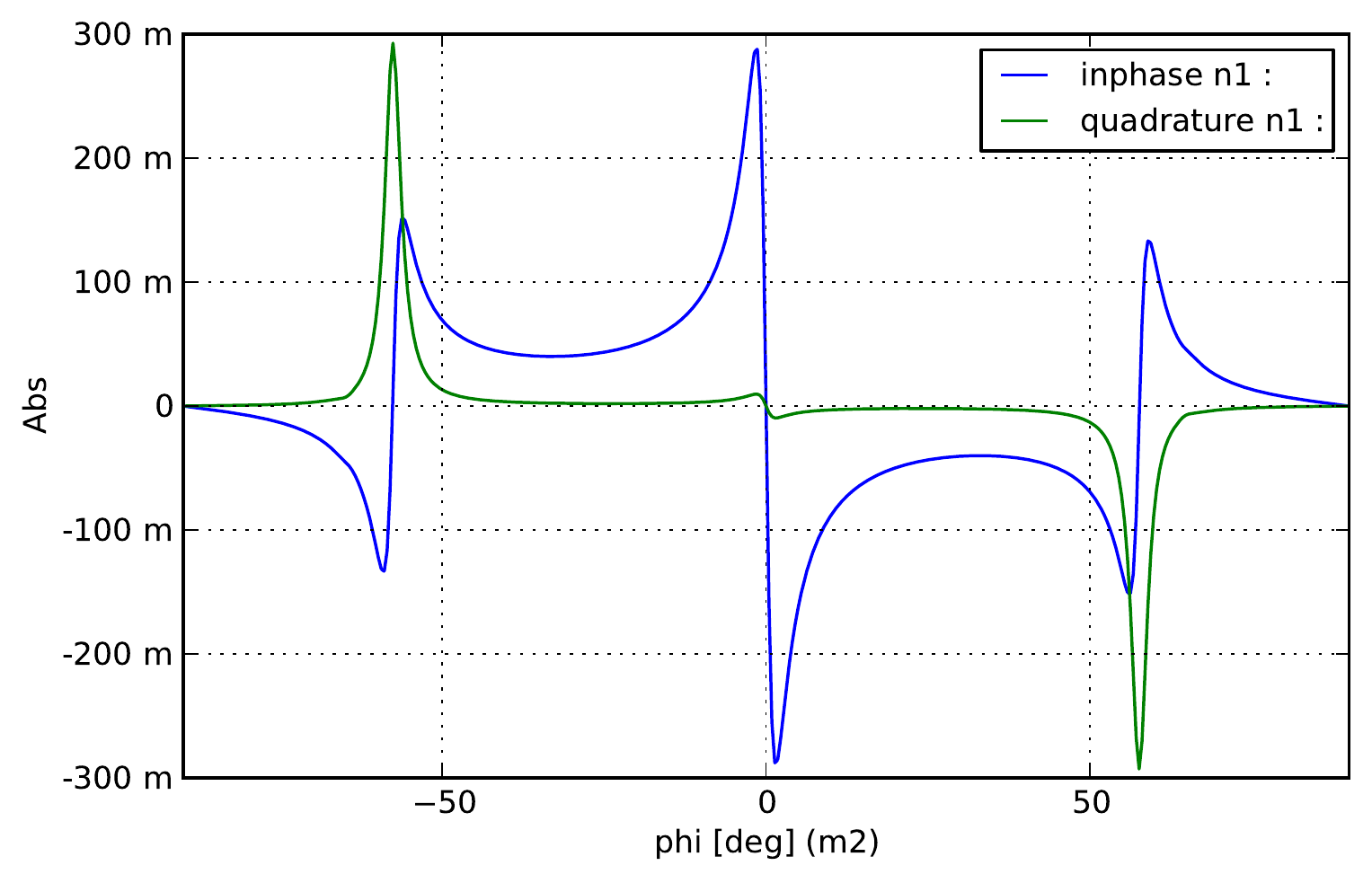} 
\caption{Second example: a Pound-Drever-Hall error signal.}
\label{fig:pdh1}
\end{figure}

The interferometer is a simple \FP\ cavity that consists of two mirrors
(\cmd{m1}) and (\cmd{m2}) and a `free space' (\cmd{s1}) in between. The laser
(\cmd{i1}) provides an input field with a power of 1~Watt. This beam is
passed through a modulator which applies a phase modulation.  The connecting
nodes are \cmd{n0, n1, n2, n3}. 

The first components define a 1200\,m long over-coupled cavity: 
\enlargethispage*{2\baselineskip}

\begin{finesse}
m m1 0.9 0.0001 0 n1 n2         
s s1 1200 n2 n3                 
m m2 1 0 0 n3 dump              
\end{finesse}

The cavity is resonant for the default laser frequency because the two
mirror tunings (\cmd{phi}) are set to zero (see Section~\ref{sec:tuning}).
The default frequency is set by the value `lambda' in the init file
`kat.ini', the default value for `lambda' is 1064\,nm. The laser:
\begin{finesse}
l i1 1 0 n0
\end{finesse}
has an offset frequency of 0\,Hz, which means the laser light has the
default laser frequency and is thus resonant in the cavity. Next, we need to
phase modulate the light (the Pound-Drever-Hall scheme is a
modulation-demodulation method):
\begin{finesse}
mod eo1 40k 0.3 3 pm n0 n1
\end{finesse}
The Pound-Drever-Hall signal can now be generated with a photodetector and
one `mixer'.  A mixer is an electronic device that can perform a
demodulation of a signal by multiplying it with a reference signal at the modulation
frequency (local oscillator). For demodulation one has to specify a
\emph{demodulation phase}. In general, when the optimum phase is not yet
known, two components of the signal, `in-phase' and `quadrature', are
usually computed where the demodulation phase of the `quadrature' signal has
a 90-degree offset to the `in-phase' demodulation:
\begin{finesse}
pd1 inphase 40k 0 n1
pd1 quadrature 40k 90 n1
\end{finesse}

These two detectors detect the light power reflected by the mirror
(\cmd{m1}) and demodulate the signal at 40\,kHz. The typical
Pound-Drever-Hall error signal is plotted as a function of the mismatch of
laser frequency to cavity resonance. In this example, we choose the
\cmd{xaxis} to be the microscopic position of the second mirror:
\begin{finesse}
xaxis m2 phi lin -90 90 400
\end{finesse}
The command \cmd{xaxis} defines the parameter that is varied during the
simulation.  At the same time, it defines the $x$-axis of the output (plot).
Here, the previously defined mirror (\cmd{m2}) is moved by changing the
tuning (\cmd{phi}) linearly (\cmd{lin}) from -90 degrees to 90 degrees in
400 steps.  This starts the simulation with the cavity being anti-resonant
for the laser light and  sweeps the cavity through the resonance until the
next anti-resonance is reached.  The resulting plot is shown in Figure
\ref{fig:pdh1}.

\paragraph{The transfer function: `pdh.kat'}
The second part of this example is very similar to the first part; it also
employs a \FP\ cavity with a Pound-Drever-Hall setup. This time, however, we
are not interested in the error signal as a function of a mirror position,
but in the transfer function of the optical system in a potential feedback
loop.  We assume that an actuator can move the input mirror (\cmd{m1}) and
we want to know the transfer function from a displacement of \cmd{m1} to the
output of the photodetector plus mixer (from the previous example). Please
note that the cavity parameters are now different. The input file is:
\loadkat{pdh.kat}

Most of the above is similar to the previous example. Only one new component
has been added:
\begin{finesse}
fsig sig1 m1 10 0
\end{finesse}
This is the \emph{signal frequency}. It can be understood as connecting the
source from a network analyser to an actuator which can move mirror
\cmd{m1}. This means a periodic signal called \cmd{sig1} now `shakes' the
mirror at 10 Hz (phase=0).  The periodic movement of the mirror can be
described as a phase modulation of the light that is reflected by the
mirror, i.e.~phase modulation sidebands are generated.

The photodetector used in the previous example to compute the
Pound-Drever-Hall error signal has to be extended by another mixer to detect
the field amplitude \emph{at the signal frequency}:

\begin{finesse}
pd2 inphase 40k 0 10 n1
\end{finesse}

The first demodulation at 40\,kHz (demodulation phase 0 degrees) is still
the same as before. The second demodulation is at 10 Hz, the signal
frequency. You will note that for the second demodulation no demodulation
phase is given. If the demodulation phase is not given, the output is
(mathematically) simply a complex number representing the amplitude and
relative phase of the error signal at the signal frequency\footnote{In an
experiment, this is slightly more complex: a network analyser would perform
the demodulation at the signal frequency \emph{twice} with two different
demodulation phases and then calculate the amplitude and phase of the
signal}.  If we now sweep the signal frequency (simultaneously at the source
and the second mixer), we will get a transfer function. This can be done by
the following commands:

\begin{finesse}
xaxis sig1 f log .01 100 400
put inphase f2 $x1             
\end{finesse}

\begin{figure}[bt]
\centering
\IG [scale=0.9] {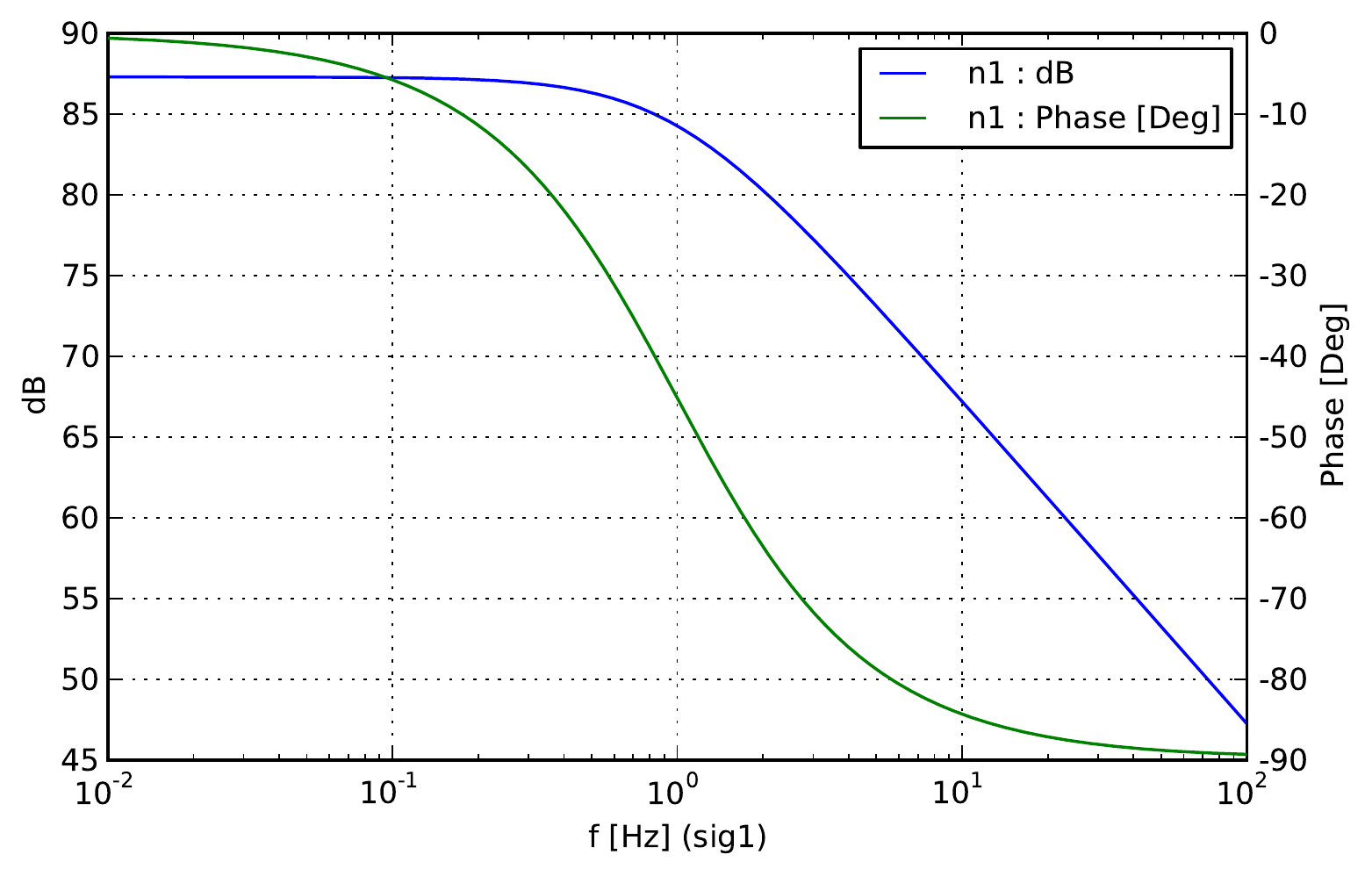} 
\caption{Second example: the transfer function of the a Pound-Drever-Hall 
error signal (mirror motion to error signal).}
\label{fig:pdh2}
\end{figure}

The parameter to be swept is \cmd{sig1}, the signal frequency.  In order to
always compute the transfer function, the frequency of the second
demodulation at the photodetector must also be changed accordingly. 
This is assured by the \cmd{put} command. \cmd{put} sets an interferometer
parameter to the value of a variable. In this case it puts the $x$-axis value
$x1$ to the second frequency of the photodetector (\cmd{f2}).  The resulting
plot is shown in Figure~\ref{fig:pdh2}.

%% file: program_files.tex

\chapter{The program files}

\section{kat---the main program}

The name of the binary, i.e.~the command to start \Finesse is `kat'.
The syntax is:\\
\cmd{kat [options] infile [outfile [gnufile]] }\\
or\\
\cmd{kat [options] basename}\\
where e.g.~basename `test' means input filename : `test.kat', output
filename : `test.out' and Gnuplot batch filename : `test.gnu' (parameters in
square brackets are optional).  The input file has to be provided by the
user, output and Gnuplot files are created by \Finesse. Available options :

{\renewcommand{\labelitemi}{ }
\begin{itemize}
\item -h : prints first help screen with short syntax reference
\item -hh : prints second help screen with some conventions
\item -v : prints version and exits
\item -c : forces consistency check of interferometer matrix (slow)
\item -max : prints maximum and minimum of every plot
\item -klu : forces KLU sparse matrix solver
\item -sparse : forces SPARSE sparse matrix solver
\item -\/-noheader : suppresses the printing header information in output data files
\item -\/-server : starts \Finesse in server mode (see \ref{sec:servermode})
\item -\/-perl1 : suppresses the printing of the banner and Gnuplot
command
\item -\/-convert : converts {\tt knm} files between text and binary formats,
  see appendix~\ref{sec:mapknm}.
\item -\/-quiet : suppresses almost all screen outputs
\end{itemize}}

\section{kat.ini---the init file for kat}
\label{sec:ini}

The file `kat.ini' is read when the program is started. 
It is a text file in which some program parameters can be defined.

For \Finesse to find the `kat.ini' file you have to either place one
copy into the current working directory (i.e. the directory containing the
\Finesse input files you are working with), or you specify a global
variable `KATINI' on you computer which contains the full path to a 
`kat.ini' file. Please refer to help service of your operating system in order
find out how to set such a variable. On many Unix-like system this can be
done by adding some command like:

\begin{finesse}
export KATINI=$HOME/work/kat/kat.ini
\end{finesse}

to a configuration file.

The following parameters can be set within the `kat.ini' file:
{\renewcommand{\labelitemi}{ }
\begin{itemize}
\item clight : speed of light
\item lambda : main wavelength of the input laser light (`lambda' sets
$\lambda_0$ and thereby defines $\omega_0$)
\item deriv\_h : step size for numerical differentiation (this parameter can be 
overwritten in the input file with the same syntax, i.e. {\tt deriv\_h value} see also page \pageref{sec:derivh})
\item qeff : quantum efficiency of photodetectors
\item epsilon\_c : $\epsilon_c=\epsilon_0 \cdot c$ used for relating light
power to field amplitudes, $P=\epsilon_c |EE^*|$
\item n0 : default refractive index for spaces
\item gnuversion : version number of your Gnuplot binary (a two digit number, for example, 4.2)
\item PDTYPE : photodetector definition (see also page \pageref{sec:pdtype})
\item GNUCOMMAND : system command to start Gnuplot
\item GNUTERM : Gnuplot plotting terminal description
\item PYTHONCOMMAND : system command to start Python
\item PYTERM : Python plotting terminal description
\item PLOTTING : selecting default plotting program (Gnuplot/Python)
\end{itemize}}

The following parameters can be used for customising the locking algorithm :
{\renewcommand{\labelitemi}{ }
\begin{itemize}
\item locksteps : total number of steps for trying to achieve the locking
condition
\item autostop : if set to 1, stop locking after first failure
\item sequential : bit coded, 0/1 sequential off/on, 5 first lock
sequential. A sequential lock includes a lock hierarchy based on the order
of the {\Co lock} commands in the input file. The first lock is kept at zero
while the second is changing, the first two are kept locked while the third
is changing etc. The sequential lock has proven to be slower but more
successful in finding the operating point.  Often it is convenient to use
sequential locking only for the first data point and then switch to the
faster parallel locking in which all loops are iterating together.
\item autogain : switch for the automatic gain control, 0/1/2 =
off/on/verbose
\item lockthresholdlow : threshold for `gain too low' check
\item lockthresholdhigh : threshold for `gain too high' check
\item locktest1 : number of steps to wait until loop gain is checked
\item locktest2 : number of checks to be invalid before the gain is changed
\item gainfactor : in case of action, change gain by this factor
\end{itemize}}

The `\texttt{\#}' sign is used for comment lines, and parameters are
specified as `name value', e.g.~`clight 300000000.0'.  If the program
cannot read or find the `kat.ini' file it uses the following default
values:
{\renewcommand{\labelitemi}{ }
\begin{itemize}
\item clight : 299792458.0
\item lambda : 1.064e-6
\item deriv\_h : 1e-3\footnote{Note that you have to use a smaller value for
`deriv\_h' if you use alignment angles because the angles are typically of
the order of 1e-6 and `deriv\_h' must be smaller.}
\item qeff : 1.0
\item epsilon\_c : 1.0
\item n0 : 1.0
\item gnuversion : 4.2
\item GNUCOMMAND : 
`c:$\backslash$programs$\backslash$gnuplot$\backslash$wgnuplot.exe' for Windows systems, 
`gnuplot -persist' for Linux systems and 
`/sw/bin/gnuplot -persist' on OS X.
\item locksteps : 10000
\item autostop : 1
\item sequential : 5
\item autogain : 2
\item lockthresholdlow : 0.01
\item lockthresholdhigh : 1.5
\item locktest1 : 5
\item locktest2 : 40
\item gainfactor : 3
\end{itemize}}
The Gnuplot terminal `x11' or `windows' is selected with respect to the
operating system. 
In order to plot the data with Gnuplot you may have to adjust GNUCOMMAND which is the
system command used by \Finesse to start Gnuplot.  The command must include
the full pathname and all options.  

Several Gnuplot terminals are predefined in `kat.ini'.  The syntax is as
follows:

\begin{finesse}
GNUTERM name
(some Gnuplot commands like e.g.: 
set term postscript eps
set title
...)
END
\end{finesse}
Please read the Gnuplot manual for information about the Gnuplot commands.

Similarly, if you prefer to create the graphical output with Python
you need to adjust the PYTHONCOMMAND and can add details
using the PYTERM commands. Finally, the command PLOTTING
may be used to define the default plotting, i.e. Gnuplot or Python.
Note that in all cases the script files fro plotting with Gnuplot, Python and
Matlab will be generated, so that you can conveniently plot the result with any
of these program at a later time.

Furthermore, the file `kat.ini' hosts definitions for photodetector types
that have some special features with respect to the detection of \HG modes.
Please also see \Sec{sec:split_pd} and page \pageref{sec:pdtype} for an explanation of these
detector definitions.  Many different types of real detectors (like split
detectors) or (spatially) imperfect detection can be simulated using this
feature.  The syntax for the type definitions: 

\begin{finesse}
PDTYPE name
...
END
\end{finesse}

Between {\Co PDTYPE} and {\Co END} several lines of the following format may
be given:
\begin{itemize}
\item[1.]{`\texttt{0 1 0 2 1.0}',
       the beat between \M{01} and \M{02} is scaled by a factor of $1.0$}
\item[2.]{`\texttt{0 0 * 0 1.0}',
       `*' means `any': the beats of \M{00} with \M{00}, \M{10},
       \M{20}, \M{30}, etc.~are scaled by a factor of $1.0$}
\item[3.]{`\texttt{x y x y 1.0}',
       `\texttt{x}' or `\texttt{y}' also means `any' but here all instances
       of `\texttt{x}' are always the same number (likewise for `\texttt{y}'). 
       So, in this example, all beats of a mode with itself are scaled by
       $1.0$}
\end{itemize}
All beat signals not explicitly given are scaled by 0.0.  Please take care
when entering a definition, because the parser is very simple and cannot
handle extra or missing spaces or extra characters.  The file `kat.ini' in
the \Finesse package includes the definitions for split photodetectors.

\section{*.kat---the input files (how to do a calculation)}
\label{sec:howto}

The program does not work interactively, i.e.~all the information about the
optical setup and the calculation task has to be stored in one input text
file before the program is called. This section describes the syntax of the
input files. For a better understanding please also look at the online
examples.  In addition, \app{sec:syntax_reference} gives an
extensive syntax description.  Together with the given examples this should
allow one to understand the input file syntax for all possible simulation
tasks.

A line of the input file can be empty, specify {\bf one} component, or
specify {\bf one} command. Text after a `\cmd{\#}' sign is treated as a
comment.  A component entry has the following syntax:

\cmd{component\_type name parameter\_list node\_list}

{\bf Component names and node names must be less than 15 characters long}.
For example a mirror can be specified by

\cmd{m mirror1 0.9 0.1 0 n1 nout3}

where `\cmd{m}' is the keyword for the component mirror, `\cmd{mirror1}' is
the name of the component. The parameter list of a mirror is
`power-reflectivity power-transmittance tuning' or in short `R T phi'. The
above example therefore specifies a mirror with R=0.9, T=0.1 and phi=0
connected to nodes `\cmd{n1}' and `\cmd{nout3}'.

{\bf Node names can be chosen by the user and must not be longer than 15
characters}. If a special node has only one connection and will not be used
for detection either, the special name `\cmd{dump}' can be used to indicate
a beam dump. This does not affect the results but reduces the set of linear
equations by one and thus speeds up the calculation.

Note that {\bf even if you want to tune (or sweep) a certain parameter you
have to enter a fixed value at the proper place first}. Imagine that you
have to build the full interferometer before you start moving or shaking
things. The commands then follow the interferometer description. 

\section{*.out---the output files}

The output files (*.out) are the main output of \Finesse, i.e.~they contain
the result of the simulation run. These files contain the calculated pure
data in text format. The first three lines are a header containing information
about the simulation and the output data, a typical header might look like:

\begin{finesse}
\end{finesse}
The first line contains the version, build number and build date of the
\Finesse binary. The second line defines whether the data refers to a
2D or 3D plot and what y-axes have been specified. The third line
then gives the labels of the data columns, i.e. the name of the 
x-axis (or x-axes), in this case 'phi [deg] (m1)',
and the names of the detectors, here 'tr1' and 'tr2'.

The '\%' sign is used as a comment char in this case because Matlab and 
Gnuplot can recognise this as a comment char. If your Gnuplot
version complains about the header you can try to add the following 
to your Gnuplot configuration file:

\begin{finesse}
set datafile commentschars "#!
\end{finesse}

Alternatively you can of course run \Finesse with the \cmd{-\/-noheader} option,
which suppresses the header in the output files.

The header is followed by the data, stored in rows and columns:

\cmd{x y1 [y2 y3 y4 ...]}\qquad    for 2D plots\\

\cmd{x1 x2 y1 [y2 y3 y4 ...]}\qquad    for 3D plots\\ 

where \cmd{x1} is the first $x$-axis, in 3D plots \cmd{x2} is used for the
second $x$-axis. The $y$ values correspond to various graphs, for example:

\cmd{x amplitude1 phase1 amplitude2 phase2}\\

\section{*.gnu---the Gnuplot batch files}

These files are batch files for Gnuplot. They are text files with a few
simple commands that tell Gnuplot which file it should read and how it
should plot it. You can easily change the file yourself to vary the look
of the plot or to do some calculations within Gnuplot with the data. Be
aware that if you don't rename the file, other runs with the same input
file will overwrite the Gnuplot batch file.

\section{*.m---the Matlab script files}\label{sec:mfiles}
These files are Matlab input files containing the
necessary commands to plot the data in the `.out' files with Matlab.
To do so, start Matlab, then  inside Matlab, change into
the working directory containing the `katfilename.out' and `katfilename.m' file and 
call the latter with the command `katfilename' (replace `katfilename' by the
actual name of the file). Please note that Matlab does not recognise all
filenames, for example, you must not use minus or plus signs in Matlab
script names. Therefore \Finesse will replace any '-' in the 
basename by '\_' for creating the corresponding name of the Matlab file.
Again, please be aware that if you don't rename the file, 
other runs with the same input file will overwrite the Matlab file.

The Matlab files actually are not scripts but contain function. In order
to get more information about using these you can get help by
typing \cmd{help katfilename} (again replace `katfilename' with the actual name
of the file). This should print something like:

\begin{finesse}
 ----------------------------------------------------------------
  function [x,y,z] = katfilename(noplot)
  Matlab function to plot Finesse output data
  Usage: 
  [x,y,z] = katfilename    : plots and returns the data
  [x,y,z] = katfilename(1) : just returns the data
            katfilename    : just plots the data
  Created automatically Wed Jun 11 11:34:17 2008
  by Finesse 0.99.8 (3200), 08.06.2008
 ----------------------------------------------------------------
\end{finesse}

This explains the three different possibilities to call the function
and either load the data, plot the data or do both.

\section{*.py---the Python script files}\label{sec:pyfiles}
Similarly, these files are Python scripts that include the necessary
commands to plot the \Finesse output with Python, using matplotlib.

This top comment block shows the usage:
\begin{finesse}
"""-----------------------------------------------------------------
  Python file for plotting Finesse ouput ttt.out
  created automatically Thu Apr 25 00:40:52 2013

  Run from command line as: python ttt.py
  Load from python script as: import ttt
  And then use:
  ttt.run() for plotting only
  x,y=ttt.run() for plotting and loading the data
  x,y=ttt.run(1) for only loading the data
-----------------------------------------------------------------"""
\end{finesse}


%% file: math_desc.tex

\chapter{Mathematical description of light beams and optical components}

\section{Introduction}

The following sections provide information about how the various aspects of
an interferometer simulation are coded within the \Finesse source code.  The
analysis of optical systems described here is based on the principle of
superposition of light fields: a laser beam can be described as the sum of
different light fields.  The possible degrees of freedom are:
\begin{itemize}
\item[-] frequency,
\item[-] geometrical shape and position,
\item[-] polarisation.
\end{itemize}

In the analysis of interferometric \gw\ detectors, the amplitudes and
frequencies of light fields are of principal interest. The polarisation is
neglected in the analysis given here, but the formalism can in principle be
easily extended to include polarisation also.

This chapter describes the mathematical formalism based on plane waves only.
In \chap{sec:HGmodes} the formalism with respect to \HG modes
will be given; it is a straightforward extension of the plane
wave analysis and makes use of the methods described here.

\subsection{Static response and frequency response}

The optical system shall be modelled by a set of linear equations that
describes the light field amplitudes in a steady state. When a vector of
input fields is provided, the set of linear equations can be mathematically
solved by computing a \emph{solution vector} that holds the field amplitudes
at every component in the optical system. 

The analysis provides information about the light field amplitudes as a
function of the parameters of the optical system.  Two classes of
calculations can be performed:

\begin{itemize} 
\item[a)] \textbf{Static response}: Computing the light field amplitudes as
a function of a quasi-static change of one or more parameters of the optical
components. For example, the amplitude of a light field leaving an
interferometer as a function of a change in an optical path length.  The
settling time of the optical system can usually be estimated using the
optical parameters. Parameter changes that are negligible during the
settling time can be assumed to be quasi-static. In a well-designed optical
system many parameter changes can be treated as quasi-static so that the
static response can be used to compute, for example, the (open-loop) error
signal of the optical system's control loop.

\item[b)] \textbf{Frequency response}: In general, the frequency response
describes the behaviour of an output signal as a function of the frequency
of a fixed input signal.  In other words, it represents a transfer function;
in this context, a transfer function of an optical system.  The input signal
is commonly the modulation of light fields at some point in the
interferometer.  The frequency response allows computation of the optical
transfer functions as, for example, required for designing control loops.
\end{itemize} 

\subsection{Transfer functions and error signals}
\label{sec:trans+err}

Two common tasks for interferometer analysis are the computations of error
signals and optical transfer functions. Both are important for the design of
servo loops to control the interferometer.  In an interferometer, several
degrees of freedom for the optical components exist (e.g.~positions,
alignment angles) and active stabilisation is necessary to enhance the
sensitivity.

An error signal is the output of a sensor (or in general a measurable
signal) as a function of one degree of freedom (of the interferometer).  The
transfer function now gives the frequency-dependent coupling of a signal
that is present in that particular `degree of freedom' into the error
signal.
 
Transfer functions can be used to compute the coupling of noise in the
interferometer and thus to estimate the sensitivity.  The following sections
give an introduction into the computation of error signals and transfer
functions with \Finesse.

\subsubsection{Modulation-demodulation methods}

Several standard techniques exist to generate error signals for controlling
an interferometer. Many of them use modulation-demodulation schemes in which
at some point inside the optical setup a light field is modulated (in phase
or amplitude) at a fixed frequency. To derive error signals, the output of a
photodetector is then demodulated (using a mixer) at that frequency.
Modulation-demodulation is a well known technique which is commonly used for
the transmission of low frequency signals (e.g. radio transmission).  It has
the advantage of shifting low frequency signals to higher frequencies.
Typically, many noise contributions are frequency dependent such that the
noise decreases at higher frequencies. Therefore, the signal-to-noise ratio
can be enhanced in many cases using modulation-demodulation.

\subsubsection{Error signals}

In general, an error signal is an output of any kind of detector that is
suitable for stabilising a certain parameter $p$ with a servo loop.
Therefore, the error signal must be a function of the parameter $p$. In most
cases it is preferable to have a bipolar signal with a zero crossing at the
operating point $p_0$.  The slope of the error signal at the operating point
is a measure of the `gain' of the sensor (which in general is a combination
of optics and electronics).
\begin{figure}[tb]
\centering
\IG [viewport=0 0 360 220,scale=.93] {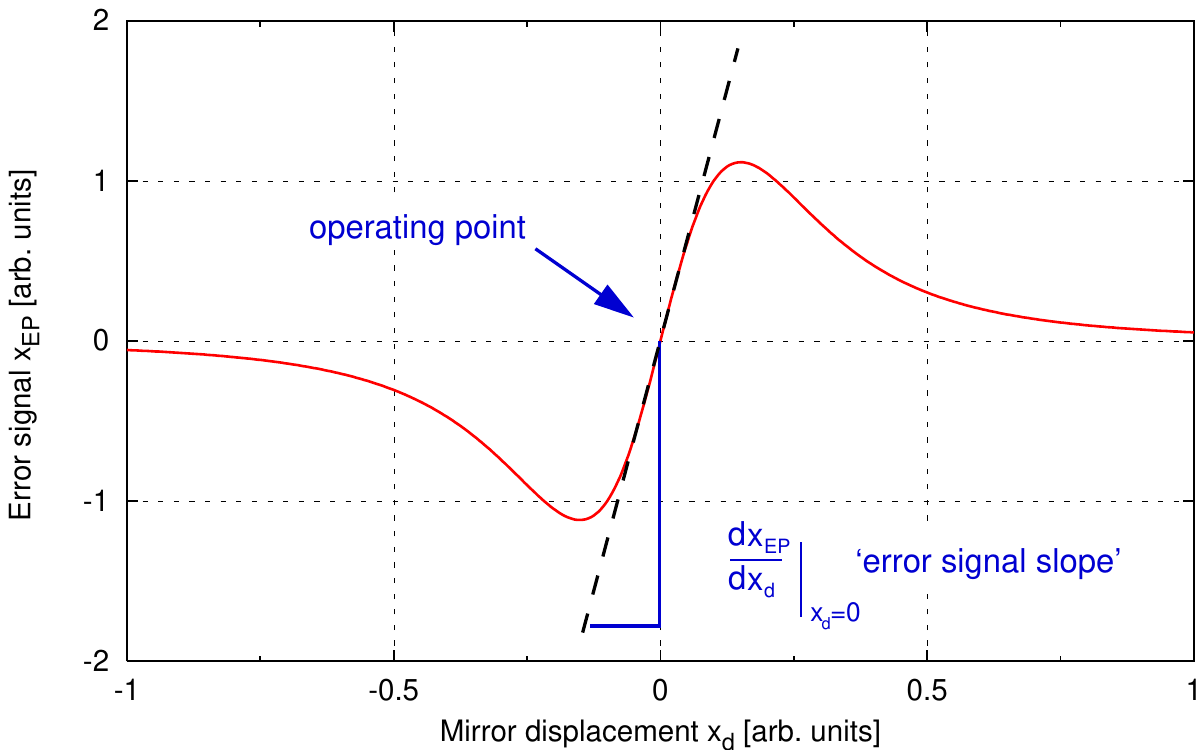}\\ 
\IG [viewport=0 0 360 230,scale=.93] {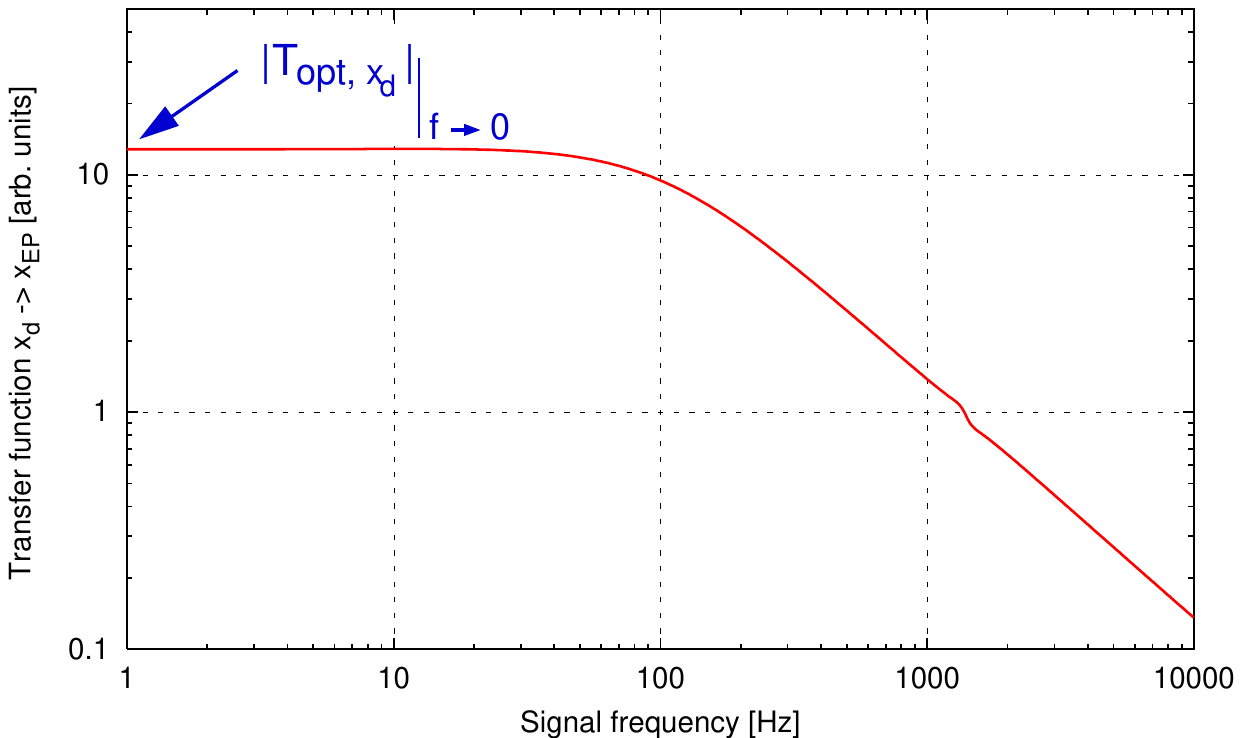} 
\caption[Example error-signal slope]
{Example of an error signal: the top graph shows the electronic
interferometer output signal as a function of mirror displacement.  The
operating point is given as the zero crossing, and the \emph{error-signal
slope} is defined as the slope at the operating point.  The bottom graph
shows the magnitude of the transfer function \emph{mirror displacement
$\rightarrow$ error signal}. The slope of the error signal (upper graph) is
equal to the low frequency limit of the transfer function magnitude (see
\eq{eq:slope_tf}).}
\label{fig:slope1}
\end{figure}

\subsubsection{Transfer functions}

Transfer functions describe the propagation of a periodic signal through a
\emph{plant} and are usually given as frequency plots of amplitude and
phase.  A transfer function describes the \emph{linear} coupling of
signals inside a system. This means a transfer function is independent of
the actual signal size. For small signals or small deviations, most systems
can be linearised and correctly described by transfer functions.

Experimentally, network analysers are commonly used to measure a transfer
function: One connects a periodic signal (the \emph{source}) to an actuator
of the plant (which is to be analysed) and to an input of the analyser.  A
signal from a sensor that monitors a certain parameter of the plant is
connected to the second analyser input. By mixing the source with the sensor
signal the analyser can determine the amplitude and phase of the input
signal with respect to the source (amplitude equals one and the phase equals
zero when both signals are identical).

In \Finesse the transfer function is calculated in a similar manner with the
limitation that all {\it plants} are---of course---optical systems.  The
command:
\begin{finesse}
fsig name component [type] f phase [amp] 
\end{finesse}
specifies the source signal. One has to set a frequency and a phase and can
optionally set an amplitude and the type of the signal. Giving an amplitude
or phase makes sense only if the frequency is applied to more than one
component and the {\it relative} driving phases and amplitudes are of
interest.  {\bf A signal can be added to the following components: mirror,
beam splitter, space, input, and modulator}.  All of these can add a
modulation to a light field, i.e.~they can act dynamically on the light field amplitudes.
In all cases the modulation is only applied to laser fields or modulation
sidebands (i.e.~those which are generated by a modulator).  The practical
reason for this restriction is the difficulty of avoiding endless loops when
signal sidebands are generated around signal sidebands. From the physics
point of view the restriction also makes sense since transfer functions can
be calculated with infinitesimally small signals (i.e.~perturbations of order
$\epsilon$) so that all terms of the order $\epsilon^2$ can be omitted. 

In \Finesse, modulation at the signal frequency is realised by adding two
signal sidebands to the light field. Applying the signal to the various
components results in different amplitudes and phases of these sidebands.
The exact numbers for each possible component are given in Section
\ref{sec:rhs}.  The detection of the signal for creating a transfer function
is included in the photodetector components \cmd{pd} (see below).

\fig{fig:slope1} shows an example of an error signal and its corresponding
transfer function. The operating point will be at:
\begin{equation}
x_{\rm d} = 0 \qquad \mbox{and} \qquad x_{\SSm EP}(x_{\rm d}=0) = 0
\end{equation} 

The optical transfer function $T_{\rm opt, x_{\rm d}}$ with respect to this
error signal is defined by:
\begin{equation}
\WT{x}_{\SSm EP}(f) = T_{\rm opt, x_{\rm d}} T_{\rm det} \WT{x}_d(f), 
\end{equation}

with $T_{\rm det}$ as the transfer function of the sensor. In the following,
$T_{\rm det}$ is assumed to be unity.  At the zero crossing the slope of the
error signal represents the magnitude of the transfer function for low
frequencies:
\begin{equation}\label{eq:slope_tf}
\left|\frac{d x_{\SSm EP}}{d x_{\rm d}}\right|_{~\bigl|x_{\rm
d}=0\bigr.}~=~|T_{\rm opt, x_{\rm d}}|_{~\bigl|f\rightarrow 0\bigr.}
\end{equation}
The quantity above will be called the \emph{error-signal slope} in the
following text. It is proportional to the \emph{optical gain} $|T_{\rm opt,
x_{\rm d}}|$, which describes the amplification of the \gw\ signal by the
optical instrument.

\subsection{The interferometer matrix}
\label{sec:matrix}

The task for \Finesse is to compute the coupling of light field amplitudes
inside a given interferometer.  \Finesse assumes the following
simplifications:
\begin{itemize}
\item the interferometer can be described via linear coupling 
of the light field amplitudes,
\item there is no polarisation of the light, nor polarising components,
\item the frequency of a given light field is never changed, in particular
frequency shifting is not possible.
\end{itemize}
 
With these simplifications all interactions at optical components can be
described by a simple set of linear equations. For a given number of input
fields this set of equations can be `solved' (either numerically or
analytically) and the output fields can be computed. \Finesse first creates
local matrices with the local coupling coefficients for every optical
component. Next, the full interferometer matrix is compiled from these
`local' coupling matrices. The full interferometer matrix then transforms a
vector with all local fields (the `solution' vector) into a vector that
contains non-zero entries for the input light fields in all interferometer
inputs.  The latter is called the `right hand side' (RHS) vector. 
\begin{equation}
\left(\begin{array}{c} \\ \mbox{interferometer}\\ \mbox{matrix}\\
\\\end{array}\right)\times
\left(\begin{array}{c} \\\vec{x}_{\rm sol}  \\ \\ \\\end{array}\right)
=\left(\begin{array}{c} \\  \vec{x}_{\rm RHS}\\ \\ \\\end{array}\right)
\end{equation}
The number of rows (the matrix is of the type $n\times n$) is determined by
the number of distinct light field amplitudes inside the interferometer.
If, for example,  we consider only one frequency component and one geometrical mode,
exactly two light fields are present at every node and the number of rows is two
times the number of nodes.

Usually one interferometer matrix is sufficient to compute all light
fields in one step. However, in order to decrease the size of the matrix
and thus to increase the speed of the computation, \Finesse creates independent
matrices for each frequency components and solves these sequentially.
This has the disadvantage that couplings between different frequency
components need to be described in a simplified form. 

The RHS vector consists mostly of zeros since usually there are only a few
distinct sources of light in an interferometer. These sources are `lasers',
`modulators' and `signal frequencies'. The `modulators' and `signal
frequencies' shift light power from a light field at a specified frequency
to one or more field components with a frequency offset.  Therefore, when 
independent matrices are constructed for each frequency, these components
must be treated as light sources (in general as
devices that can create or destroy light power at a given frequency).

Naturally, the entries in the matrix and in the RHS vector change during a
simulation. In fact, the coefficients of the matrix are updated every time a
parameter has been changed. Then, for each frequency that may be present in
the interferometer, an RHS vector is set up and the system of linear
equations is solved numerically.  The solution vector is computed and thus
the field amplitudes at all frequencies inside the interferometer.

\section{Conventions and concepts}

This section presents an overview of the conventions and definitions that
are used in \Finesse. Several methods for describing the same physics are
commonly used.  Therefore, the knowledge of the definitions used by \Finesse
is essential for understanding the syntax of the input files, the results of
the computation and, in some cases, the descriptions in this manual. 

\subsection{Nodes and components}\label{sec:comp+nodes}

\begin{figure}[h]
\centering
\IG [scale=0.6] {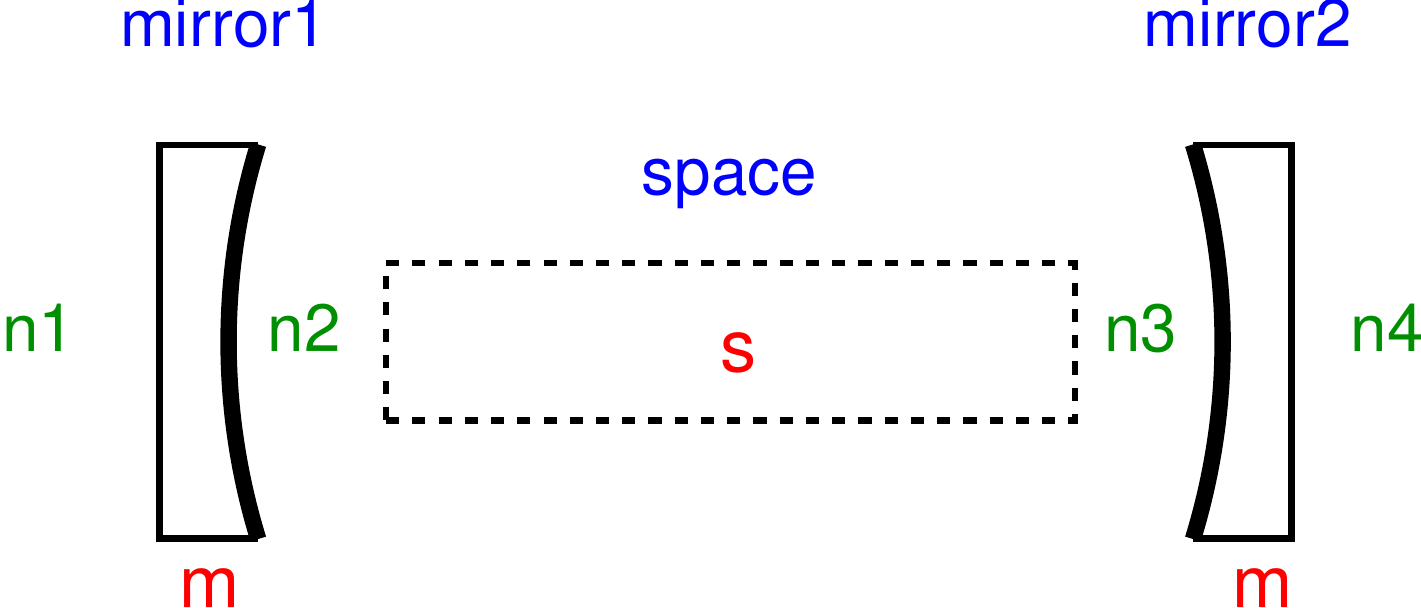}
\caption{A simple \FP\ cavity shown as components and nodes.}
\label{fig:cavity1}
\end{figure}

The interferometer has to be specified as a group of components connected by
`nodes'. For example, a two mirror \FP\ cavity as in
Figure~\ref{fig:cavity1} could be: 
\begin{itemize}
\item{mirror one (\cmd{m1}) with nodes \cmd{n1} and \cmd{n2}}
\item{free space (\cmd{s}) with nodes \cmd{n2} and \cmd{n3}}
\item{mirror two (\cmd{m2}) with nodes \cmd{n3} and \cmd{n4}}
\end{itemize}

Node \cmd{n2} connects mirror \cmd{mirror1} to space \cmd{s} and node \cmd{n3}
connects space \cmd{s} to mirror \cmd{mirror2}. The nodes \cmd{n1} and \cmd{n4}
are now the input and output nodes of the cavity.  The program calculates
the light fields at all nodes. As opposed to reality, you can put a detector
at every node without disturbing the interferometer. {\bf Two light fields
are present at every node: one in each direction of propagation}, e.g.~in
the example above at node \cmd{n2}, one light field approaching the mirror
\cmd{mirror1} from the right and one leaving the mirror \cmd{mirror1} to the right. If
a detector is located at node \cmd{n2}, only {\bf one} of these two fields
is detected; you have to know (or specify) which. If the node where you
put a detector has only one connection (like \cmd{n4} above), the non-empty
light field (i.e.~coming from the mirror \cmd{mirror2}) is chosen automatically.
If the node is inside the system (e.g.~\cmd{n2})  {\bf the beam going into a
space coming from a different component is usually detected}. If there are
no spaces in between, the following rules apply in the given order:
\begin{itemize} 
\item if a mirror is connected to the node, the beam coming {\it from} the
mirror is detected,
\item if a beam splitter is connected to the node, the beam coming {\it
from} the beam splitter is detected,
\item if a modulator is connected to the node, the beam coming {\it from}
the modulator is detected,
\item if two components of the same type (i.e.~two mirrors) are connected to
the node, the beam coming from the first specified component (in the input
file) is detected.
\end{itemize}
The rules above state which beam is detected by default. Of course, you can
specify that the respective other beam should be detected (see syntax
reference for `\cmd{ad}' or `\cmd{pd}' in \app{sec:syntax_reference}).

\subsection{Mirrors and beam splitters}\label{sec:m+bs}

The main components of interferometers in \Finesse are mirrors
and beam splitters. Following the designs of the numerical methods 
used in \Finesse I have defined mirrors and beam splitters in a 
slightly counter-intuitive way:
\begin{itemize}
\item mirrors (\cmd{m}) are defined as single optical surfaces with two nodes
\item beam splitters (\cmd{bs}) are defined as single optical surfaces with four nodes
\end{itemize}

In other words, a mirror always retro-reflects a beam into the
incoming node while a beam splitter separates the reflected beam
from the incoming beam.

The counter-intuitive part in this is that when a real beam splitter
(i.e. a partial reflective surface) is used under normal incidence
we would still call it a beam splitter while in \Finesse it has
to be modeled as mirror. On the other hand if we employ a mirror
as a turning mirror, a mirror of $R=1$ with an angle of incidence of 45 degrees,
this must be modelled as a beam splitter in \Finesse.

In short the terms mirror and beam splitter do \emph{not} refer to the
reflectance or transmission of the optical surface nor it's use
in reality but solely on the angle of incidence of the incoming beam. 

Please note also that even though many diagrams in this
manual depict mirrors and beam splitters as components with two
surfaces the actual components \cmd{m} and \cmd{bs} represent
\emph{single} optical surfaces. Such surfaces of course do not resemble
physical objects as such. However, they can be used to model simplified
interferometer layouts. In more detailed optical layouts it is often
wise to use more realistic models for the optical components:
a 'real' mirror or beam splitter would consist of two optical surfaces
with a substrate in between. To model this one needs to employ a number
of basic \Finesse components. A mirror then can be modeled as:\\
\cmd{m Mfront \dots nM1 nMi1}\\
\cmd{s Msubstrate \dots nMi1 nMi2}\\
\cmd{m Mback \dots nMi2 nM2}\\

A beam splitter is more complex because the outgoing beams would pass 
the back surface at different location. \Finesse cannot
handle this directly, instead one has
to consider the two locations at the back surface as two independent
components (this also makes sense in practice since the surface properties for the two locations
are not necessarily equal). Thus a beam splitter can be modeled as:\\
\cmd{bs BSfront \dots nBS1 nBS2 nBSi1 nBSi3}\\
\cmd{s BSsubstrate1 \dots nBSi1 nBSi2}\\
\cmd{s BSsubstrate2 \dots nBSi3 nBSi4}\\
\cmd{bs BSback1 \dots nBSi2 nBS3}\\
\cmd{bs BSback2 \dots nBSi4 nBS4}\\

\section{Frequencies and wavelengths}

\Finesse distinguishes between three types of light fields:
\begin{itemize}
\item laser light (input light),
\item modulation sidebands (generated by the component `modulator':
\cmd{mod}), and
\item signal sidebands (generated by the command `signal frequency':
\cmd{fsig}).
\end{itemize}

Throughout this manual, different representations for the frequencies of
light fields are used: wavelength ($\lambda$), frequency ($f$), or angular
frequency ($\w$). In \Finesse it is simpler: in the file `kat.ini' the
default laser frequency is defined via a wavelength $\lambda_0$.  All other
frequencies must be given as frequency offsets ({\bf not angular frequency})
to that reference.  The terms `carrier' frequency or `carrier' light are
used in this manual to refer to a light field that is subject to some kind
of modulation, either by a modulator or a signal.  The modulation will
create `sidebands' around the `carrier'.  Laser light fields can
be carriers. But also modulation sidebands created by the component
\cmd{mod} can serve as carrier fields but \emph{only} for signal modulation.

Please note that modulators cannot create sidebands of sidebands.  This
directly affects some simulation tasks. It is quite easy to make an error if
this limitation is not considered.  Two typical examples for analysis tasks
that require some care are the transfer of frequency noise through an
optical system or error signals for any kind of control scheme that needs
double modulation (2 sets of RF sidebands, or higher harmonics of the RF
sidebands).  Please have a look at Keita Kawabe's note on `Sidebands of
Sidebands' \cite{keita}, which is part of the \Finesse package.

A modulator can be used as a phase or amplitude modulator.  In both cases,
modulators add symmetric sidebands to {\em input fields}, i.e.~laser light
and {\bf not} to other modulation sidebands or signal sidebands.  In
addition, a modulator can be used in the {\em single sideband mode} so that
only one modulation sideband is added to the laser field.  Signal
frequencies perform a modulation on {\em input fields} and  {\em modulation
sidebands} from a modulator but {\bf not} on other signal sidebands.

\subsection{Phase change on reflection and transmission}
\label{sec:phase}

When a light field passes a \bs, a phase jump in either the reflected,
transmitted, or both fields is required for energy conservation; the actual
phase change for the different fields depends on the type of \bs (see
\cite{atr1} and \cite{ghh_phd}).  In practice, the absolute phase of the
light field at a \bs is of little interest so to calculate interferometer
signals one can choose a convenient implementation for the relative phase.
Throughout this work, the following convention is used: mirrors and \bss
are assumed to be symmetric (not in how they split the light power but with
respect to the phase change) and the phase is not changed upon reflection;
instead, the phase changes by $\pi/2$ at every transmission.

Please be aware that this is directly connected to the resonance condition
in the simulation: if, for example, a single surface with power
transmittance $T=1$ is inserted into a simple cavity, the extra phase change
by the transmission will change the resonance condition to its opposite.
Inserting a `real' component with two surfaces, however, does not show this
effect.

\subsection{Lengths and tunings}
\label{sec:tuning}

The interferometric \gw\ detectors typically use three different types of
light fields: the laser with a frequency of $f\approx2.8\cdot10^{14}\,{\rm
Hz}$, modulation sidebands used for interferometer control with frequencies
(offsets to the laser frequency) of $f\approx30\cdot10^{6}\,{\rm Hz}$, and
the signal sidebands at frequencies of 10\,Hz to 1000\,Hz\footnote{The
signal sidebands are sometimes also called \emph{audio sidebands} because of
their frequency range.}.

The resonance condition inside the cavities and the operating point of the
interferometer depend on the optical path lengths modulo the laser
wavelength, i.e.~for the light of a Nd:YAG laser length differences of less
than $1\,\mu{\rm m}$ are of interest, not the absolute length. The
propagation of the sideband fields depends on the much larger wavelength of
the (offset) frequencies of these fields and thus often on absolute lengths.
Therefore, it is convenient to split distances $D$ between optical
components into two parameters~\cite{ghh_phd}: one is the macroscopic
`length' $L$ defined as that multiple of the default wavelength $\lambda_0$
yielding the smallest difference to $D$. The second parameter is the
microscopic \emph{tuning} that is defined as the remaining difference
between $L$ and $D$. This tuning is usually given as a phase $\Tun$ (in
radian) with $2\pi$ referring to one wavelength\footnote{Note that in other
publications the tuning or equivalent microscopic displacements are
sometimes defined via an optical path length difference and then often
$2\pi$ is used to refer to the change of the optical path length of one
wavelength which, for example, if the reflection at a mirror is described,
corresponds to a change of the mirror's position of $\lambda_0/2$.}.  In
\Finesse tunings are entered and printed in degrees, so that a tuning of
$\phi=360$ degrees refers to a change in the position of the component by
one wavelength ($\lambda_0$).

This convention provides two parameters that can describe distances with a
markedly improved numerical accuracy. In addition, this definition often
allows simplification of the algebraic notation of interferometer signals.

In the following, the propagation through free space is defined as a
propagation over a macroscopic length $L$, i.e.~a free space is always
`resonant', i.e.~a multiple of $\lambda_0$.  The microscopic tuning appears
as a parameter of mirrors and beam splitters. It refers to a microscopic
displacement perpendicular to the surface of the component. If, for example,
a cavity is to be resonant to the laser light, the tunings of the mirrors
have to be the same whereas the length of the space in between can be
arbitrary.

Note that if you change the frequency of the input lasers the spaces are
still resonant to the default wavelength $\lambda_0$ (as given in `kat.ini')
and not to the wavelength of the input light.


\section{The \pwa}
\label{sec:code}
In many simulations the shape of the light beams or, in general, the 
geometric properties of a beam transverse to the optical axis are
not of interest. In that case one can discard this information and 
restrict the model to the field on the optical axis. This is equivalent
to a model where all light fields are plane waves traveling along 
one optical axis. This is the standard mode of \Finesse and is called
\emph{\pwa} in the following.

In the \pwa all light fields are described in one dimension. All beams 
and optical components are assumed to be centered on the optical axis and of infinite size.
Using plane waves, it is very simple to compute interferometer signals
depending on the phase and frequency of the light, and of the longitudinal
degrees of freedom. Furthermore it can be easily extended to include other
degrees of freedom, such as polarisation or transverse beam shapes. 

This section introduces the \pwa as used in \Finesse by default. It also
presents the basis for the \HG extension given in \Sec{sec:HGmodes}.

\subsection{Description of light fields}
\label{sec:fields}

A laser beam is usually described by the electric component of its
electromagnetic field:
\begin{equation}
\vec{E}(t,\vec{x})\,=\,\vec{E}_0 \mCos{\w\T - \vec{k}\vec{x}}.
\end{equation}
In the following calculations, only the scalar expression for a fixed point
in space is used. The calculations can be simplified by using the full
complex expression instead of the cosine:
\begin{equation}
E(t)\,=\,E_0\, \mExB{\I\left(\w\T + \OPh\right)}\,=a\,\mEx{\I \w\T},
\end{equation}
where $a=E_0\, \mEx{\I\OPh}$.  The real field at that point in space can
then be calculated as:
\begin{equation}
\vec{E}(t) = \myRe{E(t)}\cdot \vec{e}_{\rm pol},
\end{equation}
with $\vec{e}_{\rm pol}$ as the unit vector in the direction of
polarisation.

Each light field is then described by the complex amplitude $a$ and the
angular frequency $\w$. Instead of $\w$, also the frequency $f=\w/2\pi$ or
the wavelength $\lambda=2\pi c/\w$ can be used to specify the light field.
It is often convenient to define one \emph{default frequency} (also called
the \emph{default laser frequency}) $f_0$ as a reference and describe all other
light fields by the offset  $\Delta f$ to that frequency. In the following,
some functions and coefficients are defined using $f_0$, $\w_0$, or
$\lambda_0$ referring to a previously defined default frequency. The setting
of the default frequency is arbitrary, it merely defines a reference for
frequency offsets and does not influence the results. 

The electric component of electromagnetic radiation is given in Volt per
meter. The light power computes as:
\begin{equation}
P = \epsilon_0 c ~E E^*,
\end{equation}
with $\epsilon_0$ the electric permeability of vacuum and $c$ the speed
of light.  However, for more intuitive results the light fields can be given
in converted units, so that the light power can be computed as the square of
the light field amplitudes.  Unless otherwise noted, throughout this work
the unit of light field amplitudes is the square root of Watt. Thus, the
power computes simply as:
\begin{equation}
P=E E^*.
\end{equation}
The parameter `epsilon\_c' in the init file `kat.ini' can be used to set the
value of $\epsilon_0\cdot c$ (the default is $\epsilon_0\cdot c=1$).

\subsection{Photodetectors and mixers}
\label{sec:pd+mix}

In plane-wave mode \Finesse offers two methods for detecting light in an interferometer,
amplitude detectors and photodetectors. An amplitude detector (\cmd{ad})
detects {\bf only} the light amplitude at the given frequency even if other
light fields are present. An amplitude detector is a virtual device.

A photodetector (\cmd{pd}) does not only detect light at the given
frequency, but also beat signals at that frequency. For example, at DC the
photodetector detects the full DC power of all present light fields. This
`photodetector' refers to real photodetectors except for the fact that it
does not destroy (or change in any sense) the light field.

The photodetectors can perform a demodulation of the detected signal (light
power).  In reality this would be done by a mixer.  In \Finesse\
photodetectors can be specified with up to 5 mixer frequencies and phases:
when a mixer frequency (and phase) is given, the signal is demodulated at
this frequency. When more frequencies are specified, the signal is
demodulated at these frequencies sequentially.  Please note that \Finesse
does not simulate a mixer: in the frequency domain the demodulation can be
achieved by simply selecting only amplitudes at the modulation frequency
when computing the output of a photodetector (see Section~\ref{sec:mod}). 

A real mixer always demodulates the signal with a certain demodulation
phase. The output is then a real number which represents the amplitude of
the signal at the specified frequency and phase. More information can be
obtained if two mixers are used with different demodulation phases.  Using
two mixers that demodulate the same signal at the same frequency but with a
difference in the demodulation phase of $\pi/2$ the amplitude and
phase of the signal at the specified frequency can be reconstructed. This
is used in network analysers to measure transfer functions.

In \Finesse the demodulation automatically preserves the phase of the signal
anyway. If a demodulation phase is specified, the complex amplitude is
projected onto that phase and thus converted to a real number. On the other
hand, a network analyser can be simulated by simply leaving out the last
step: if the demodulation phase for the last specified frequency is omitted,
\Finesse keeps the full complex amplitude.  This feature is (as in
network analysers) commonly used for computing transfer functions.

\subsection{Modulation of light fields}
\label{sec:mod}

In principle, all parameters of a light field can be modulated. This section
describes the modulation of the amplitude, phase and frequency of the light.

Any sinusoidal modulation of amplitude or phase generates new field
components that are shifted in frequency with respect to the initial field.
Basically, light power is shifted from one frequency component, the
\emph{carrier}, to several others, the \emph{sidebands}. The relative
amplitudes and phases of these sidebands differ for different types of
modulation and different modulation strengths.

\subsubsection{Phase modulation}
\label{sec:phasemod}

Phase modulation can create a large number of sidebands. The amount of
sidebands with noticeable power depends on the modulation strength (or
depths) given by the \emph{modulation index} $m$.

Assuming an input field:
\begin{equation}
E_{\rm in}~=~E_0~\mEx{\I\w_0 \T},
\end{equation}
a sinusoidal phase modulation of the field can be described as:
\begin{equation}\label{eq:mod1}
E~=~E_0~\mExB{\I(\w_0 \T + m \mCos{\w_{\rm m} \T})}.
\end{equation}
This equation can be expanded using the Bessel functions $J_k(m)$ to:
\begin{equation}\label{eq:bessel0}
E~=~E_0~\mEx{\I\w_0 \T}~\sum_{k=-\infty}^{\infty}\I^{\,k}~J_k(m)~\mEx{\I k
\w_{\rm m}\T}.
\end{equation}

The field for $k=0$, oscillating with the frequency of the input field
$\w_0$, represents the carrier. The sidebands can be divided into
\emph{upper} ($k>0$) and \emph{lower} ($k<0$) sidebands. These sidebands are
light fields that have been shifted in frequency by $k\, \w_{\rm m}$. The
upper and lower sidebands with the same absolute value of $k$ are called a
pair of sidebands of order $k$. 

\eq{eq:bessel0} shows that the carrier is surrounded by an infinite number
of sidebands.  However, the Bessel functions decrease for large $k$, so for
small modulation indices ($m<1$), the Bessel functions can be approximated
by:
\begin{equation}\label{eq:besselapprox}
J_k(m)~=~\frac{1}{k!}\left(\frac{m}{2}\right)^k+O\left(m^{k+2}\right).
\end{equation}
In which case, only a few sidebands have to be taken into account. For
$m\ll1$ we can write:
\begin{equation}
{\renewcommand{\arraystretch}{1.5}
\begin{array}{lcl}
E&=&E_0~\mEx{\I\w_0 \T}\\
& & \times\Bigl(J_0(m)-\I J_{-1}(m)~\mEx{-\I \w_{\rm m}\T}+\I J_{1}(m)~\mEx{\I
\w_{\rm m}\T}\Bigr),
\end{array}}
\end{equation}
and with
\begin{equation}
J_{-k}(m)=(-1)^kJ_k(m),
\end{equation}
we obtain:
\begin{equation}
E~=~E_0~\mEx{\I\w_0 \T}~\left(1+\I\frac{m}{2}\Bigl(\mEx{-\I \w_{\rm
m}\T}+\mEx{\I \w_{\rm m}\T}\Bigr)\right),
\end{equation}
as the first-order approximation in $m$.

When the modulator functions as a phase modulator, then the {\em order} of
sidebands can be given. For example: 
\begin{finesse}
mod eom1 10M 0.6 2 pm node1 node2
\end{finesse}
applies a cosine phase modulation at 10\,MHz with a modulation index of
$m=0.6$ and order 2, i.e.~4 sidebands are added to the laser field.

The given number for {\em order} in the modulator command simply specifies
the highest order of Bessel function which is to be used in the sum in
\eq{eq:bessel0}, i.e.~the program code uses the equation:
\begin{equation}
E~=~E_0~\mEx{\I\w_0 \T}~\sum_{k=-order}^{order}i^{\,k}~J_k(m)~\mEx{\I k
\w_m\T},
\label{eq:bessel1}
\end{equation}


\subsubsection{Frequency modulation}

For small modulation indices phase modulation and frequency modulation can
be understood as different descriptions of the same effect \cite{ghh_phd}.
With the frequency defined as $f= d\OPh/dt$ a sinusoidal frequency
modulation can be written as:
\begin{equation}
E~=~E_0~\mEx{\I\left(\w_0\T + \frac{\Delta\w}{\w_{\rm m}} \mCos{\w_{\rm m}
\T}\right)},
\end{equation}
with $\Delta\w$ as the frequency swing (how \emph{far} the frequency is
shifted by the modulation) and $\w_{\rm m}$ the modulation frequency (how
\emph{fast} the frequency is shifted).  The modulation index is defined as:
\begin{equation}
m=\frac{\Delta\w}{\w_{\rm m}}.
\end{equation}

\subsubsection{Amplitude modulation}

In contrast to phase modulation, (sinusoidal) amplitude modulation always
generates exactly two sidebands.  Furthermore, a natural maximum modulation
index exists: the modulation index is defined to be one ($m=1$) when the
amplitude is modulated between zero and the amplitude of the unmodulated
field.

If the amplitude modulation is performed by an active element, for example
by modulating the current of a laser diode, the following equation can be
used to describe the output field:
\begin{equation}
{\renewcommand{\arraystretch}{1.5}
\begin{array}{lcl}
E&=&E_0~\mEx{\I\w_0 \T}~\Bigl(1+m\mCos{\w_{\rm m} \T}\Bigr)\\
&=&E_0~\mEx{\I\w_0 \T}~\Bigl(1+\frac{m}{2}~\mEx{\I \w_{\rm m}
\T}+\frac{m}{2}~\mEx{-\I \w_{\rm m} \T}\Bigr).
\end{array}}
\end{equation}
However, passive amplitude modulators (like acousto-optic modulators or
electro-optic modulators with polarisers) can only reduce the amplitude. 
In these cases, the following equation is more useful:
\begin{equation}
{\renewcommand{\arraystretch}{1.5}
\begin{array}{lcl}
E&=&E_0~\mEx{\I\w_0 \T}~\left(1-\frac{m}{2}\Bigl(1-\mCos{\w_{\rm m}
\T}\Bigr)\right)\\
&=&E_0~\mEx{\I\w_0 \T}~\Bigl(1-\frac{m}{2}+\frac{m}{4}~\mEx{\I \w_{\rm m}
\T}+\frac{m}{4}~\mEx{-\I \w_{\rm m} \T}\Bigr).
\end{array}}
\end{equation}


\subsubsection{Single sideband}

The modulator components in \Finesse can be switched to a \emph{single
sideband mode} where only one sideband is added to the input light.  This
sideband can be either identical to one phase modulation sideband or one
amplitude modulation sideband (see above). 
The modulation index is used as usual, so that a single
sideband created with modulation index $m$ has the same amplitude as, for
example, the upper sideband in an ordinary phase modulation (order=1) with
modulation index $m$. Therefore, the amplitude of the input field remains
larger in the single sideband case.  If $A_0$ is the amplitude of the input
light before modulation and $A_m$ is the amplitude of the carrier light
after a normal modulation, the amplitude of the carrier after a single
sideband modulation is:
\begin{equation}
A_{ssb}~=~A_0-\frac{A_0-A_m}{2}.
\end{equation}

\subsubsection{Oscillator phase noise}
\label{sec:phasenoise}

The oscillator phase noise (or modulator phase noise) can give some
information about the performance of a modulation scheme in connection with
a certain interferometer configuration.  The term \emph{phase noise}
describes the change of the phase of the modulation frequency. In
Equation~\ref{eq:mod1} the phase of the modulation frequency was supposed to
be zero and is not given explicitly. In general, the modulated light has to
be written with a phase term:
\begin{equation}
E~=~E_0~\exp\left(\I\left(\w_0 \T + m \cos{\left(\w_m \T
+\varphi_m(t)\right)}\right))\right).
\end{equation}
Using Equation~\ref{eq:bessel1} phase noise can be expressed like this:
\begin{equation}
E~=~E_0~e^{\I\w_0 \T}~\sum_{k=-order}^{order}i^{\,k}~J_k(m)~e^{\I k (\w_m\T
+ \varphi_m(t))}.
\end{equation}
To investigate the coupling of $\varphi_m(t)$ into the output signal, we
apply a cosine modulation at the signal frequency ($\omega_{\rm noise}$):
\begin{equation}
\varphi_m(t)=m_2~\cos(\w_{\rm noise}\T),
\end{equation}
which results in the following field:
\begin{equation}
E~=~E_0~e^{\I\w_0 \T}~\sum_{k=-order}^{order}i^{\,k}~J_k(m)~e^{\I k
\w_m\T}~\sum_{l=-\infty}^{\infty}i^{\,l}~J_l(k~m_2)~e^{\I l \w_{\rm
noise}\T}.
\label{eq:osci1}
\end{equation}
The extra modulation of $\varphi_m$ thus adds extra sidebands to the light
(which will be called `audio sidebands' in the following since in most cases
the interesting signal frequencies are from DC to some kHz whereas the phase
modulation frequencies are very often in the MHz regime).  The audio
sidebands are generated around each phase modulation sideband.  We are
interested in the coupling of the audio sidebands into the interferometer
output because these sidebands will generate a false signal and therefore
limit the sensitivity of the interferometer. For a computation of a transfer
function, the amplitude of the signal sidebands  (here: modulation index of
audio sidebands) is assumed to be very small so that only the terms for
$l=-1,0,1$ in the second sum in Equation~\ref{eq:osci1} have to be taken
into account and the Bessel functions can be simplified to:
\begin{equation}
\begin{array}{rl}
E~=~E_0&e^{\I\w_0 \T}~\sum_{k=-order}^{order}i^{\,k}~J_k(m)~e^{\I k
\w_m\T}\\
&\times\left( 1+\I\frac{k~m_2}{2}~e^{-i \w_{\rm
noise}\T}~+\I\frac{k~m_2}{2}~e^{\I \w_{\rm noise}\T}+O((km_2)^2)\right).
\end{array}
\end{equation}
 
\Finesse automatically generates the above signal sidebands for oscillator
phase noise when the command \cmd{fsig} (see Section~\ref{sec:trans+err}) is
used with a modulator as component.  For example,
\begin{finesse}
fsig signal1 eom1 10k 0
\end{finesse}
adds audio sidebands ($\w_{\rm noise}=2\pi\, 10\,{\rm kHz}$) to the
modulation sidebands (which are generated by \cmd{eom1}).

\subsubsection[Oscillator amplitude noise]{Oscillator amplitude noise\footnote{This section has been contributed by Joshua Smith} }
\label{sec:oscamppnoise}
Oscillator amplitude noise has not yet been implemented in \Finesse. This section describes
preparatory work towards a future implementation.

In order to derive the coupling between sidebands we start with a phase modulated light field,
\begin{equation}
E = E_0 \exp(i (\w_0 t + m \cos(\w_m t)) )
\end{equation}
and replace the modulation index of the phase modulation with one that
is amplitude modulated with amplitude $m_2$ and frequency $\w_{m_2}$,
\begin{equation}
m = m_1 (1+m_2 \cos(\w_{m_2} t))
\end{equation}
The light field now has form,
\begin{equation}
E = E_0 \exp(i(\w_0 t + m_1 \cos(\w_{m_1} t) + m_1 m_2 \cos(\w_{m_1} t) \cos(\w_{m_2} t)))
\end{equation}
Using the identity,
\begin{equation}
\cos(A)\cos(B) = \frac{1}{2} (\cos(A+B)+\cos(A-B))
\end{equation}
we can obtain,
\begin{eqnarray}
E &=& E_0 \exp(i\w_0 t) \nonumber\\
&&\cdot\exp(i m_1 \cos(\w_{m_1} t)) \nonumber\\
&&\cdot\exp(i \frac{m_1 m_2}{2} \cos((\w_{m_1}+\w_{m_2})t) \\ 
&&\cdot\exp(i \frac{m_1 m_2}{2} \cos((\w_{m_1}-\w_{m_2})t) \nonumber
\end{eqnarray}
This can be expanded into three sums of Bessel functions following Equation~\ref{eq:bessel0} and 
Equation~\ref{eq:besselapprox}
\begin{eqnarray}
E &=& E_0 \exp(i\w_0 t) \nonumber\\
&&\cdot\sum_{k=-\infty}^\infty i^k J_k(m_1) \exp(ik\w_{m_1} t) \nonumber\\
&&\cdot\sum_{l=-\infty}^\infty i^l J_l(m_{12}) \exp(il(\w_{m_1}+\w_{m_2}) t) \\
&&\cdot\sum_{n=-\infty}^\infty i^n J_n(m_{12}) \exp(in(\w_{m_1}-\w_{m_2}) t) \nonumber\\
 &= &  E_0 \exp(i\w_0 t) \left(\sum_{k=-\infty}^\infty i^k J_k(m_1) \exp(ik\w_{m_1} t)\right)\cdot C \nonumber
\end{eqnarray}
with 
\begin{eqnarray}
C &=& 
\sum_{l=-\infty}^\infty i^l J_l(m_{12}) \exp(il(\w_{m_1}+\w_{m_2}) t) \\
&&\cdot\sum_{n=-\infty}^\infty i^n J_n(m_{12}) \exp(in(\w_{m_1}-\w_{m_2}) t) \nonumber
\end{eqnarray}
where $m_{12} = m_1 m_2/2$. As before we restrict this analysis to small modulation
indices $m_2$ and only consider sidebands with $l=0,\pm1$ and $n=0,\pm1$.
This yields
\begin{eqnarray}
C &=& 
\left(1+i\frac{m_{12}}{2}\exp(-i(\w_{m_1}+\w_{m_2}) t) +i\frac{m_{12}}{2}\exp(i(\w_{m_1}+\w_{m_2}) t) 
\right) \nonumber\\
&& \cdot
\left(1+i\frac{m_{12}}{2}\exp(-i(\w_{m_1}-\w_{m_2}) t) +i\frac{m_{12}}{2}\exp(i(\w_{m_1}-\w_{m_2}) t) 
\right)\\
&=&1+i\frac{m_{12}}{2} \left(\exp(-i(\w_{m_1}+\w_{m_2}) t) + \exp(i(\w_{m_1}+\w_{m_2}) t)\right.  \nonumber\\
&&\left.+\exp(-i(\w_{m_1}-\w_{m_2}) t) + \exp(i(\w_{m_1}-\w_{m_2}) t)  \right) + O(m_2^2) \nonumber\\
&=&1+i\frac{m_{12}}{2} \left(\exp(i\w_{m_1} t) + \exp(-i\w_{m_1} t)  \right)\left(\exp(i\w_{m_2} t) 
+ \exp(-i\w_{m_2} t)  \right) + O(m_2^2) \nonumber
\end{eqnarray}

\subsection{Coupling of light field amplitudes}
\label{sec:coupling}

Many optical systems can be described mathematically using linear coupling
of light field amplitudes. Passive components, such as mirrors, \bss and
lenses, can be described well by linear coupling coefficients. Active
components, such as electro-optical modulators cannot be described so
easily. Nevertheless, simplified versions of active components can often be
included in a linear analysis. 

The coupling of light field amplitudes at a simple (flat, symmetric, etc.)
mirror under normal incidence can be described as follows: there are two
input fields, $\rm In1$ impinging on the mirror on the front surface and
$\rm In2$ on the back surface.  Two output fields leave the mirror,
$\rm Out1$ and $\rm Out2$. With the amplitude coefficients for reflectance
and transmittance ($r$, $t$) the following equations can be composed:
\begin{equation}
{\renewcommand{\arraystretch}{1.5}
\begin{array}{l}
{\rm Out1}=r~ {\rm In1} + \I t ~{\rm In2},\\
{\rm Out2}=r~ {\rm In2} + \I t~ {\rm In1}.
\end{array}}
\end{equation}
Possible loss is included in this description because the sum $r^2+t^2$ may be  
less than one; see \Sec{sec:phase} about the convention for the phase change.

The above equations completely define this simplified optical component.
Optical systems that consist of similar components can be described by a set
of linear equations. Such a set of linear equations can easily be solved
mathematically, and the solution describes the equilibrium of the optical
system: given a set of input fields (as the `right hand side' of the set of
linear equations), the solution provides the resulting field amplitudes
everywhere in the optical system. This method has proven to be very powerful
for analysing  optical systems. It can equally well be adapted to an
algebraic analysis as to a numeric approach. 

In the case of the plane-wave approximation, the light fields can be described
by their complex amplitude and their angular frequency.  The linear
equations for each component can be written in the form of local coupling
matrices in the format:
\begin{equation}
\rule{0pt}{1.3cm}
\left( 
\begin{array}{c}
\rm Out1 \\
\rm Out2 
\end{array}
\right)=\left(
\begin{array}{cc}
a_{11} &a_{21}  \\
 a_{12} & a_{22}
\end{array}\right)
\left(
\begin{array}{c}
\rm In1 \\
\rm In2 
\end{array}\right)
\end{equation}
\hspace{9cm}\IG [viewport=0 -30 4 -10, scale=0.4] {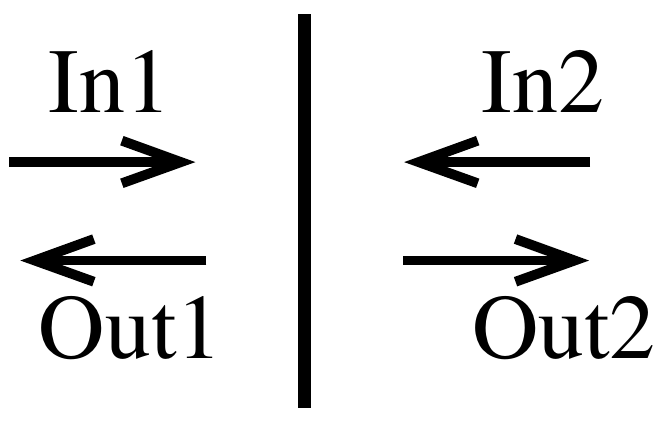}\\
with the complex coefficients $a_{ij}$.  These matrices serve as a compact
and intuitive notation of the coupling coefficients.  For solving a set of
linear equations, a different notation is more sensible: a linear set of
equations can be written in the form of a matrix that represents the
interferometer: the \emph{interferometer matrix} times the vector of field
amplitudes (solution vector). Together with the right hand side vector that
gives numeric values for the input field, the set of linear equations is
complete:
{\mathindent 1.5cm
\begin{equation}
\left(\begin{array}{c} \\ \mbox{interferometer}\\ \mbox{matrix}\\
\\\end{array}\right)\times
\left(\begin{array}{c} \\\vec{x}_{\rm sol}  \\ \\ \\\end{array}\right)
=\left(\begin{array}{c} \\  \vec{x}_{\rm RHS}\\ \\ \\\end{array}\right).
\end{equation}}

For the above example of a simple mirror the linear set of equations in
matrix form looks as follows:
{\mathindent 1.5cm
\begin{equation}
\left(\begin{array}{cccc}
 1 & 0 & 0 & 0 \\
 -a_{11} & 1 & -a_{21} & 0 \\
 0 & 0 & 1 & 0 \\
 -a_{12} & 0 &-a_{22} & 1 
\end{array}\right)\times
\left(\begin{array}{c} {\rm In1} \\ {\rm Out1} \\ {\rm In2}\\ {\rm
Out2}\\\end{array}\right)=
\left(\begin{array}{c} {\rm In1} \\ 0 \\ {\rm In2}\\ 0\\\end{array}\right).
\end{equation}
}

\subsubsection{Space}

The component `space' is propagating a light field through free space over a
given length $L$ (index of refraction $n$).  The length times index of
refraction is by definition (in \Finesse) always a multiple of the default
laser wavelength $\lambda_0$.  This defines a \emph{macroscopic length}
(see Section~\ref{sec:tuning}).  If the actual length between two other
components is not a multiple of the default wavelength, the necessary extra
propagation is treated as a feature of one or both end components. For
example, a space between two mirrors is always resonant for laser light at
the default wavelength. To tune the cavity away from it, one or both mirrors
have to be {\it tuned} (see Section~\ref{sec:tuning}) accordingly while the
length of the component `space' is not changed.

\vspace{.5cm}{
 \hspace{1cm}$
\left( 
\begin{array}{c}
\rm Out1 \\
\rm Out2 
\end{array}
\right)=\left(
\begin{array}{cc}
0 &s_1  \\
 s_2 & 0
\end{array}\right)
\left(
\begin{array}{c}
\rm In1 \\
\rm In2 
\end{array}\right)
$}

\nopagebreak
\vspace{-.2cm}
\hspace{10cm}\IG [viewport=0 0 4 2, scale=0.4] {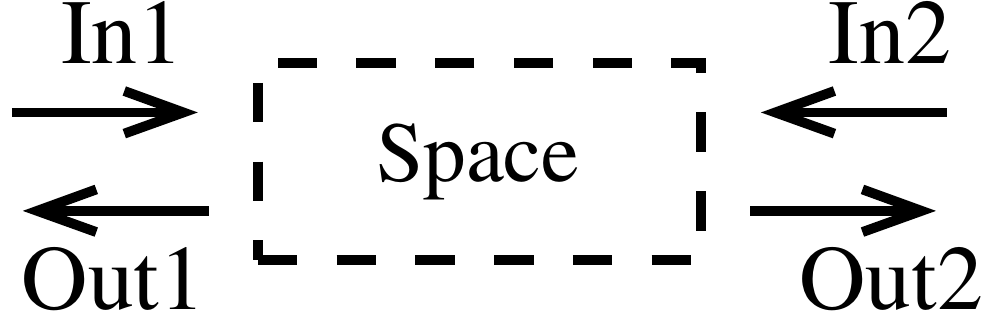} 
\vspace{.2cm}

The propagation only affects the phase of the field:
\begin{equation}
s_1~=~s_2~=~\mEx{-\I\w n L/c}~=~\mEx{-\I(\w_0+\Delta\omega) n
L/c}~=~\mEx{-\I\Delta\w n L/c},
\label{eq:pwspace}
\end{equation}
where $\mEx{-\I\omega_0 n L/c}=1$ following from the definition of
macroscopic lengths (see above).  The used parameters are the length $L$,
the index of refraction $n$, the angular frequency of the light field $\w$,
and the offset to the default frequency $\Delta \w$.

\subsubsection{Mirror}

From the definition of the component `space' that always represents a
macroscopic length, follows the necessity to perform microscopic
propagations inside the mathematical representation of the components
\emph{mirror} and \emph{\bs}.  In this description the component mirror is
always hit at normal incidence.  Arbitrary angles of incidence are discussed
for the component \bs, see below.

A light field $E_{\rm in}$ reflected by a mirror is in general changed 
in phase and amplitude:
\begin{equation}
E_{\rm refl}~=~r~\mEx{\I\OPh}~E_{\rm in},
\end{equation}
where $r$ is the amplitude reflectance of the mirror and $\OPh~=~2kx$ the
phase shift acquired by the propagation towards and back from the mirror if
the mirror is not located at the reference plane ($x=0$).

The \emph{tuning} $\Tun$ gives the displacement of the mirror expressed in
radian (with respect to the reference plane).  A tuning of $\Tun=2\pi$
represents a displacement of the mirror by one carrier wavelength:
$x=\lambda_0$. The direction of the displacement is arbitrarily defined to
be in the direction of the normal vector on the front surface, i.e.~a
positive tuning moves the mirror from node2 towards node1 (for a mirror
given by `\cmd{m \dots node1 node2}').

If the displacement $x_{\rm m}$ of the mirror is given in meters, then
the corresponding tuning $\Tun$ computes as follows:
\begin{equation}
\Tun~=~kx_{\rm m}~=~x_{\rm m}\frac{2\pi}{\lambda_0}~=~x_{\rm m}\frac{\omega_0}{c}.
\end{equation}

A certain displacement results in different phase shifts for light fields
with different frequencies. The phase shift a general field acquires at the
reflection on the front surface of the mirror can be written as:
\begin{equation}
\OPh~=~2 \Tun \frac{\w}{\w_0}.
\end{equation}
If a second light beam hits the mirror from the other direction the phase
change $\OPh_2$ with respect to the same tuning would be: 
\begin{equation}
\OPh_2=-\OPh.
\end{equation}
The tuning of a mirror or \bs does not represent a change in the path
length but a change in the position of component. The transmitted light is
thus not affected by the tuning of the mirror (the optical path for the
transmitted light always has the same length for all tunings). Only the
phase shift of $\pi/2$ for every transmission (as defined in
Section~\ref{sec:phase}) has to be taken into account:
\begin{equation}
E_{\rm trans}~=\I~t~E_{\rm in},
\end{equation}
with $t$ as the amplitude transmittance of the mirror.

The coupling matrix for a mirror is:
\begin{equation}
\rule{0pt}{1.3cm}
\left( 
\begin{array}{c}
\rm Out1 \\
\rm Out2 
\end{array}
\right)=\left(
\begin{array}{cc}
m_{11}  & m_{21} \\
m_{12}  & m_{22} \\
\end{array}\right)
\left(
\begin{array}{c}
\rm In1 \\
\rm In2 
\end{array}\right)
\end{equation}
\hspace{9cm}\IG [viewport=0 -30 4 -10, scale=0.4] {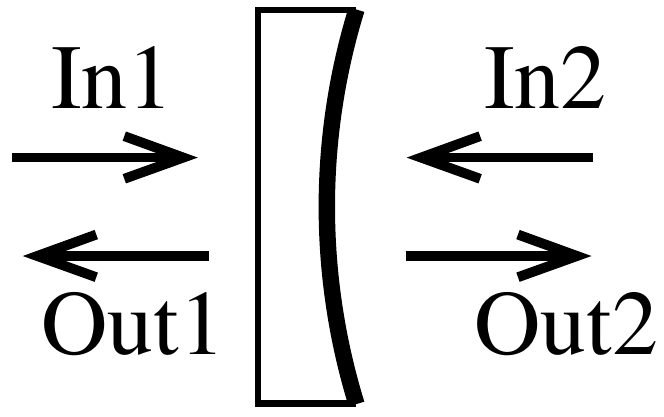} \\
with the coefficients given as:
\begin{eqnarray}
m_{12}&=&m_{21}~=~\I t,\nonumber\\
m_{11}&=&r~\mEx{\I 2\Tun~\omega/\omega_0},\nonumber\\
m_{22}&=&r~\mEx{-\I 2\Tun~\omega/\omega_0},\nonumber
\end{eqnarray}
with $\Tun=2\pi~\mbox{\tt phi}/360$, {\tt phi} as the tuning of the mirror
given in the input file, and $\omega$ the angular frequency of the reflected
light.

\subsubsection{Beam splitter}

\begin{figure}[ht]
\begin{center}
\IG [viewport= 0 0 360 260,clip,scale=0.6] {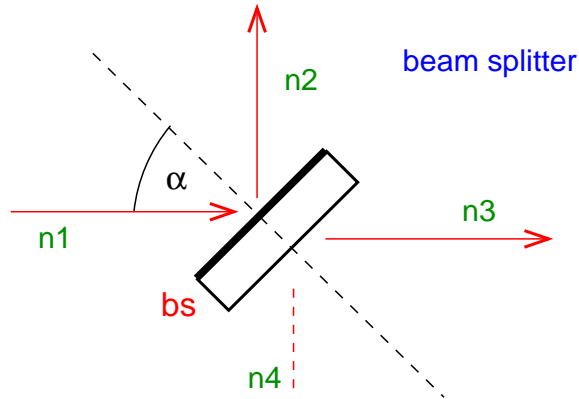}
\caption{Beam splitter.}
\end{center}
\label{fig:bs1}
\end{figure}

A \bs is similar to a mirror except for the extra parameter $\alpha$ which
indicates the angle of incidence of the incoming beams and that it can be
connected to four nodes.  The order in which these nodes have to be entered
is shown in Figure~\ref{fig:bs1}.

Since, in this work, a displacement of the \bs is assumed to be
perpendicular to its optical surface, the angle of incidence affects the
phase change of the reflected light. Simple geometric calculations lead to
the following equation for the optical phase change $\OPh$:
\begin{equation}
\OPh~=~2\,\Tun\,\frac{\omega}{\omega_0}\cos(\alpha).
\end{equation}
The coupling matrix has the following form:\\
\begin{minipage}{\textwidth}
\begin{equation}
\rule{0pt}{1.3cm}
\left( 
\begin{array}{c}
\rm Out1 \\
\rm Out2 \\
\rm Out3 \\
\rm Out4 
\end{array}
\right)=\left(
\begin{array}{cccc}
0& bs_{21}  & bs_{31} & 0\\
bs_{12}&0&0  & bs_{42}\\
bs_{13}&0&0  & bs_{43}\\
0& bs_{24}  & bs_{34} & 0
\end{array}\right)
\left(
\begin{array}{c}
\rm In1 \\
\rm In2 \\
\rm In3 \\
\rm In4
\end{array}\right)
\end{equation}
\hspace{10.5cm}\IG [viewport=0 -30 4 -10, scale=0.3] {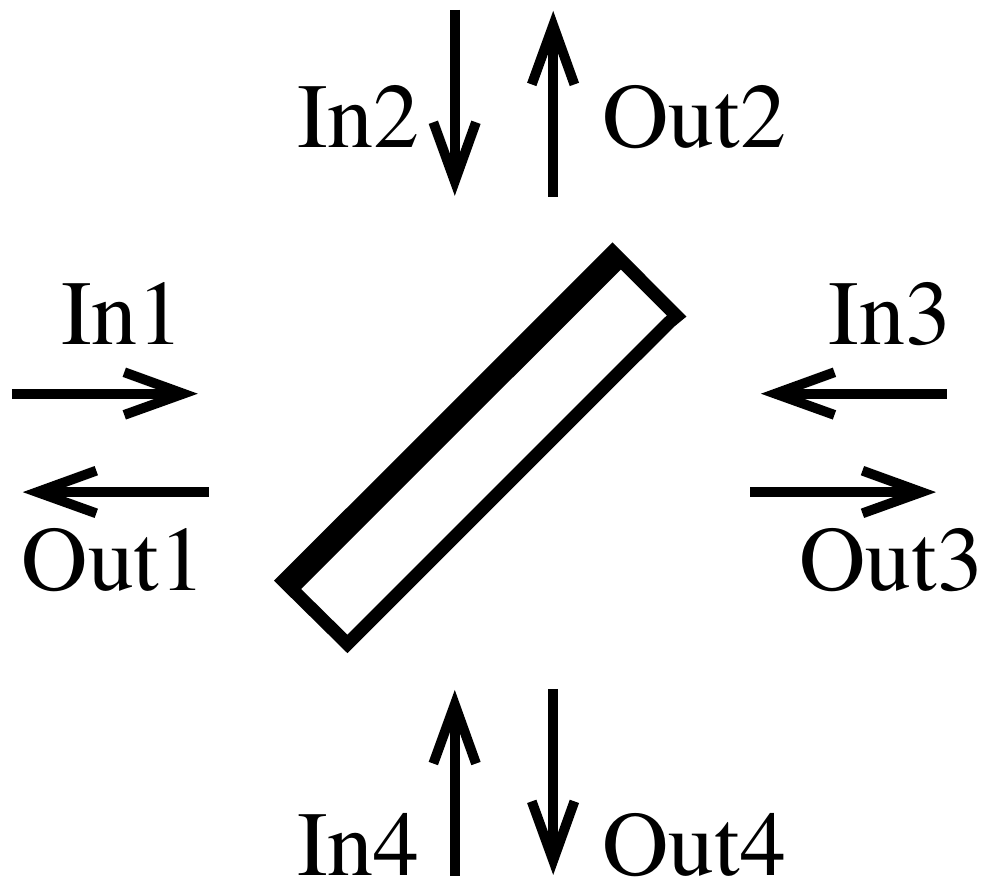}
\end{minipage}
with the coefficients:
\begin{eqnarray}
bs_{12}&=&bs_{21}~=~r~\mEx{\I 2\Tun\omega/\omega_0\cos{\alpha}},\nonumber\\
bs_{13}&=&bs_{31}~=~\I t,\nonumber\\
bs_{24}&=&bs_{42}~=~\I t,\nonumber\\
bs_{34}&=&bs_{43}~=~r~\mEx{-\I 2\Tun\omega/\omega_0\cos{\alpha}},\nonumber
\end{eqnarray}
and $\Tun=2\pi~\mbox{\tt phi}/360$.

\subsubsection{Modulator}

The modulation of light fields is described in \Sec{sec:mod}. A small
modulation of a light field in amplitude or phase can be described as
follows: a certain amount of light power is shifted from the carrier into
new frequency components (sidebands). In general, a modulator can create a
very large number of sidebands if, for example, the modulator is located
inside a cavity: on every round trip the modulator would create new
sidebands around the previously generated sidebands.  This effect
\emph{cannot} be modelled by the formalism described here.

Instead, a simplified modulator scheme is used. The modulator only acts on
specially selected light fields and generates a well-defined number of
sidebands. With these simplifications the modulator can be described as:
\begin{itemize}
\item[-] an attenuator for the light field that experiences the modulation
(at the carrier frequency);
\item[-] a source of light at a new frequency (the sideband frequencies), see
\Sec{sec:rhs}.
\end{itemize}
All other frequency components of the light field are attenuated by the
modulator in accordance with the $J_0$ Bessel coefficient.  The coupling
matrix for the modulator is:
\vspace{.8cm}{
 \hspace{1cm}$
\left( 
\begin{array}{c}
\rm Out1 \\
\rm Out2 
\end{array}
\right)=\left(
\begin{array}{cc}
0  & eo_{21} \\
eo_{12}  & 0 \\
\end{array}\right)
\left(
\begin{array}{c}
\rm In1 \\
\rm In2 
\end{array}\right)
$}

\nopagebreak
\vspace{-.2cm}
\hspace{10cm}\IG [viewport=0 0 4 2, scale=0.4] {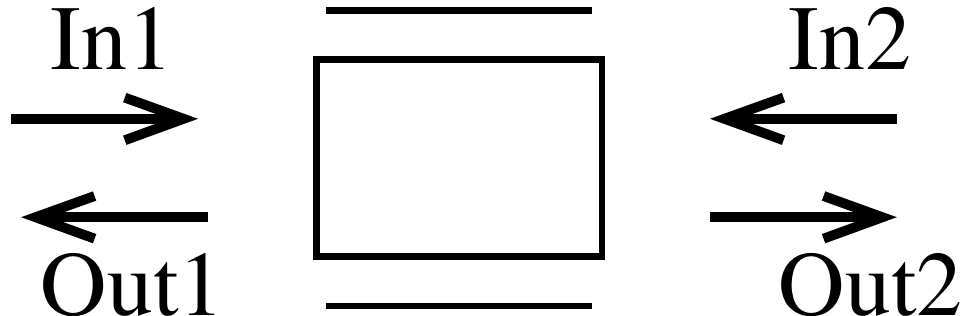} 
\vspace{.2cm}

The modulation and signal sidebands are not affected by the modulator. The
coupling coefficients for field amplitudes at these frequencies are simply:
\begin{eqnarray}
eo_{12}&=&eo_{21}~=~1.
\end{eqnarray}
When the input field is a laser field, the modulator shifts power from the
main beam to generate the modulation sidebands. Therefore a modulator
reduces the amplitude of the initial field. The phase is not changed:
\begin{eqnarray}
eo_{12}&=&eo_{21}~=~C,
\end{eqnarray}
with 
\begin{equation}
C~=~1 - \frac{m}{2},
\end{equation}
($m$ is the modulation index \cmd{midx}) for amplitude modulation and
\begin{equation}
C~=~J_0(m),
\end{equation}
for phase modulation. If the `single sideband' mode is used, then $C$ is
replaced by $C'$:
\begin{equation}
C'~=~1-\frac{1-C}{2}.
\end{equation}

\subsubsection{Isolator (diode)}

The isolator represents a simplified \FI : light passing in one direction is
not changed, whereas the power of the beam passing in the other direction is
reduced by a specified amount:

\begin{minipage}{\textwidth}
\begin{equation}
\rule{0pt}{1.3cm}
\left( 
\begin{array}{c}
\rm Out1 \\
\rm Out2 
\end{array}
\right)=\left(
\begin{array}{cc}
0  & d_{21} \\
    d_{12}  & 0 \\
\end{array}\right)
\left(
\begin{array}{c}
\rm In1 \\
\rm In2 
\end{array}\right)
\end{equation}
\hspace{9cm}\IG [viewport=0 -50 4 -30, scale=0.4] {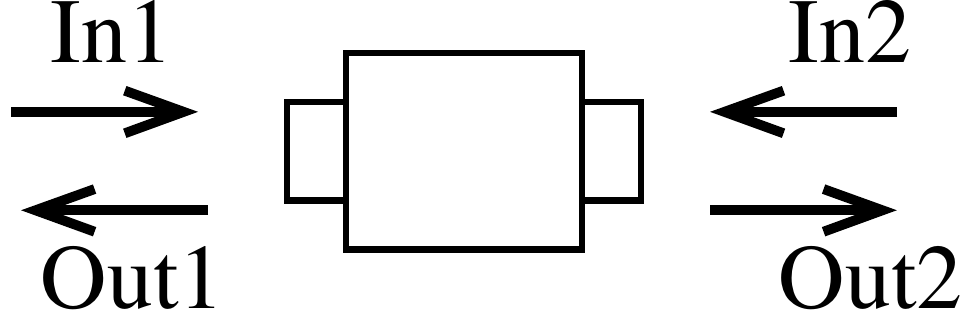} \\
\end{minipage}
The amplitude coupling coefficients are:
\begin{eqnarray}
d_{12}&=&~1,\nonumber\\
d_{21}&=&~10^{-S/20},\nonumber
\end{eqnarray}
with $S$ the specified suppression given in dB.

\subsubsection{Lens}

The thin lens does not change the amplitude or phase of the light fields.

\vspace{.5cm}{
 \hspace{1cm}$
\left( 
\begin{array}{c}
\rm Out1 \\
\rm Out2 
\end{array}
\right)=\left(
\begin{array}{cc}
 0  & 1 \\
1  & 0 \\
\end{array}\right)
\left(
\begin{array}{c}
\rm In1 \\
\rm In2 
\end{array}\right)
$}

\nopagebreak
\vspace{-.2cm}
\hspace{10.6cm}\IG [viewport=0 0 4 20, scale=0.4] {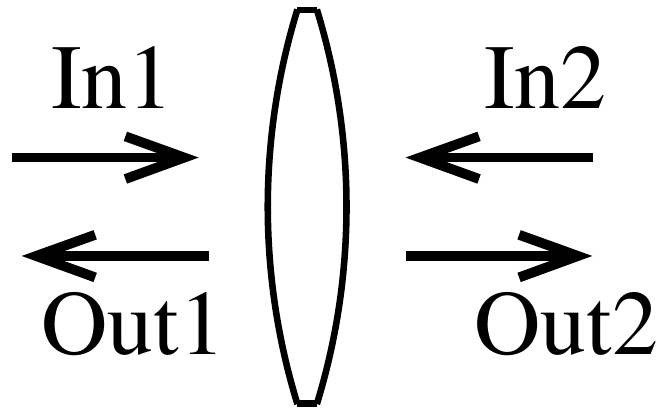} 
\vspace{.2cm}

\subsubsection{Gratings}

Gratings are optical components which require a more complex treatment than
the components above. The context of \Finesse allows simulation of certain
aspects of a grating in a well defined configuration.  This section gives a
short introduction to the implementation of gratings in \Finesse. This
work has been done with help by Alexander Bunkowski and the notation is
based on his paper~\cite{gitter1}.

The name grating is used for various very different types of optical
components.  The following description is restricted to phase gratings used
in reflection.  However, the implemented formalism can also be used to
simulate some properties of optical setups with other grating types.

This phase grating in reflection has been chosen because it can possibly be
manufactured with similar optical and mechanical qualities as the high
quality mirrors used in gravitational wave detectors today. Thus low-loss
laser interferometers with an all-reflective topology can be envisaged.

\begin{figure}[ht]
\begin{center}
\IG [scale=0.4] {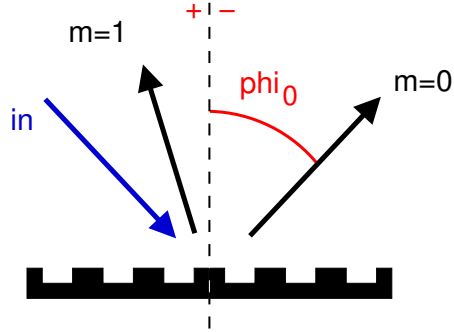}
\caption{A grating illuminated by a beam (in). The number of outgoing beams
is given by the grating equation \eq{eq:grating-eq}. The beams are
numbered by an integer ($m$) and the angles with respect to the grating
normal are given as $\phi_m$ (angles left of the grating normal are
positive, right of the normal negative). The geometry is chosen such that
the angle of incidence is always positive. $m=0$ marks the zeroth order
reflection which corresponds to a normal reflection. Thus
$\phi_0=-\alpha$.  The beams with negative orders ($m=-1,-2,$\dots) are with
angles $-\pi/2<\phi<-\alpha$ whereas positive orders have angles of
$\pi/2>\phi>-\alpha$. In the example shown here only two orders exist.}
\label{fig:gr1}
\end{center}
\end{figure}

The gratings in \Finesse are characterised by the number of ports. In
general a grating is defined by its grating period, given in [nm]. With the
wavelength and the angle of incidence $\alpha$ all possible outgoing beams
can be computed with the grating equation:
\begin{equation}\label{eq:grating-eq}
\sin{(\phi_m)}+\sin(\alpha)=\frac{m\,\lambda}{d},
\end{equation}
with $m$ an integer to label the order of the outgoing beam.  An example is
shown in \fig{fig:gr1}.  The geometry is chosen so that always
$\alpha\geq0$.  This is possible since the setup is (so far) symmetric.  All
orders with angles $\phi_m$ between $90^\circ$ and $-90^\circ$ exist and
will contain some amount of light power. The zeroth order represents the
reflection as on a mirror surface with $\phi_0=-\alpha$. Orders with
negative number leave the grating with an angle $-90<\phi<-\alpha$, positive
orders have angles with $90>\phi>-\alpha$.

In order to construct a device with a small number of ports the grating
period has to be chosen such that $d\approx\lambda$.  \eq{eq:grating-eq}
can be used to compute limits for the grating parameters with respect to the
configuration used.  We can write \eq{eq:grating-eq} as:
\begin{equation}
\frac{m\,\lambda}{d}=\sin{(\phi_m)}+\sin(\alpha)=[-1,2].
\end{equation}

In all following cases more than just the zeroth order ($m=0$) should exist,
$n$ shall be a positive integer.  For the positive order $m=n$ to be allowed
we get:
\begin{equation}
\frac{\lambda}{d}\leq\frac{2}{n}.
\end{equation} 
And the negative order $m=-n$ can exist only if:
\begin{equation}
\frac{\lambda}{d}\leq\frac{1}{n}.
\end{equation} 
\tab{tab:orders} gives on overview of the modes that the grating equation
allows to exist in certain intervals. Note that we have not yet specified
$\alpha$. Whether an order exists or not can be completely determined only
for a given angle of incidence.

\begin{table}[h]
\begin{center}
\begin{tabular}{|cl|c|c|c|c|c|c|c|c|c|}
\hline
                             & $m$ &   0 &   1 &   2 &   3 &    4  &    -1 &   -2 &   -3 & number \\  
$\lambda/d$                  &     &     &     &     &     &       &       &      &      & of orders\\  
\hline                                                                                    
$]2,\inf]$                   &     &   x &     &     &     &       &       &      &      & 1 \\  
$]1,2]$                      &     &   x &   x &     &     &       &       &      &      & 2 \\  
$]\frac{2}{3},1]$            &     &   x &   x &   x &     &       &     x &      &      & 4 \\  
$]\frac{1}{2},\frac{2}{3}]$  &     &   x &   x &   x &  x  &       &     x &      &      & 5 \\  
$]\frac{1}{3},\frac{1}{2}]$  &     &   x &   x &   x &  x  &    x  &     x &   x  &      & 7 \\  
$]\frac{2}{5},\frac{1}{3}]$  &     &   x &   x &   x &  x  &    x  &     x &   x  &   x  & 8 \\  
\hline
\end{tabular}
\caption{Possible existing orders with (for a well-chosen angle of incidence
$\alpha$) in dependence of $\lambda/d$.}
\label{tab:orders}
\end{center}
\end{table}

\paragraph{Littrow configuration}

one special setup is the \emph{Littrow} configuration in which the angle of
incidence coincides with one mode angle. The $n$th order Littrow
configuration is given by:
\begin{equation}
\phi_n=\alpha,
\end{equation}
which yields:
\begin{equation}
\sin(\alpha)=\frac{n \lambda}{2d}.
\end{equation}
 
\paragraph{Grating components in \Finesse}

\Finesse offers the following three grating types:
\begin{itemize}
\item[gr2] : a 2 port grating in first order Littrow configuration
\item[gr3] : a 3 port grating in second order Littrow configuration
\item[gr4] : a 4 port device, only the first order exists and is used {\bf
not} in Littrow configuration
\end{itemize}
Each configuration corresponds to a set of limits, for example, the angle
of incidence.

In the following, these limits and the coupling matrices for these grating
configurations are given.  The matrix is given in the form:
\begin{equation}
b_i=A_{ij} a_j,
\end{equation}
with $a_j$ being the vector of incoming fields and $b_i$ the vector of
outgoing fields.

\paragraph{gr2 component}

a grating in first order Littrow configuration is defined by the fact
that only the orders $m=0,1$ exist and that $\phi_1=\alpha$. The grating
equation therefore reduces to
\begin{equation}
\sin(\phi_m)=(m-1/2)\lambda/d.
\end{equation}
The existence of $m=1$ gives $\lambda/d<2$, the non-existence of $m=-1$ or
$m=2$ yields $\lambda/d>2/3$. Written together, we get:
\begin{equation}
\lambda/2<d<3/2 ~\lambda.
\end{equation}

The angle of incidence $\alpha$ and the grating period are related as:
\begin{equation}
\alpha=\arcsin\left(\frac{\lambda}{2d}\right).
\end{equation}
The angle of incidence is set automatically by \Finesse (the positive value
is chosen by default).

The two coupling efficiencies $\eta_0$, $\eta_1$ are constrained by energy
conservation\footnote{The gratings are defined as lossless components in \Finesse. This
corresponds to the employed phase relations between different orders. Currently, losses can be added
only be inserting extra mirrors with $R=0, T\neq1$.} (as for the \bs) as:
\begin{equation}
\eta_0^2+\eta_1^2=1.
\end{equation}
The coupling matrix is given by:

\vspace{.8cm}{
 \hspace{1cm}$
\left( 
\begin{array}{c}
\rm b_1 \\
\rm b_2 
\end{array}
\right)=\left(
\begin{array}{cc}
\I \eta_1 & \eta_0  \\
\eta_0 & \I \eta_1
\end{array}\right)
\left(
\begin{array}{c}
\rm a_1 \\
\rm a_2 
\end{array}\right)
$}

\nopagebreak
\vspace{-.2cm}
\hspace{10cm}\IG [viewport=0 0 4 2, scale=0.2] {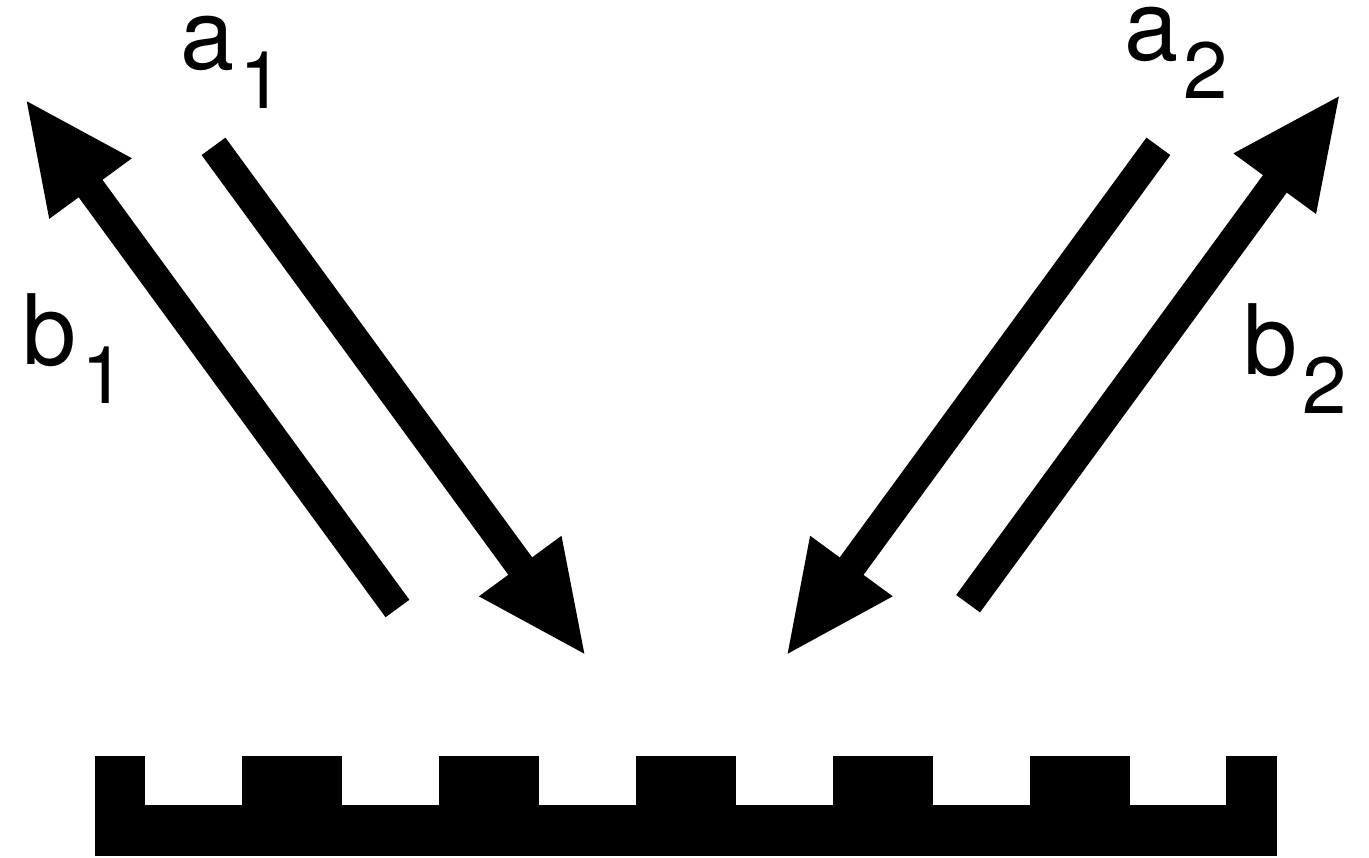} 
\vspace{.2cm}

\paragraph{gr3 component}

The second order Littrow configuration. Only the orders $m=0,1,2$ 
exist. This gives:
\begin{equation}
\lambda < d < 2\lambda.
\end{equation}
And $\phi_2$ must be equal to $\alpha$. This yields:
\begin{equation}
\alpha=\arcsin\left(\frac{\lambda}{d}\right).
\end{equation}

The coupling matrix for this grating configuration is rather
complex~\cite{gitter1}. It can be written as:

\vspace{.8cm}{
 \hspace{1cm}$
\left( 
\begin{array}{c}
\rm b_1 \\
\rm b_2 \\
\rm b_3
\end{array}
\right)=\left(
\begin{array}{ccc}
\eta_2 e^{\I \phi_2} & \eta_1 e^{\I \phi_1} & \eta_0 e^{\I \phi_0} \\
\eta_1 e^{\I \phi_1} & \rho_0 e^{\I \phi_0} & \eta_1 e^{\I \phi_1} \\
\eta_0 e^{\I \phi_0} & \eta_1 e^{\I \phi_1} & \eta_2 e^{\I \phi_2} 
\end{array}\right)
\left(
\begin{array}{c}
\rm a_1 \\
\rm a_2 \\
\rm a_3
\end{array}\right)
$}

\nopagebreak
\vspace{-.2cm}
\hspace{10cm}\IG [viewport=0 0 4 2, scale=0.2] {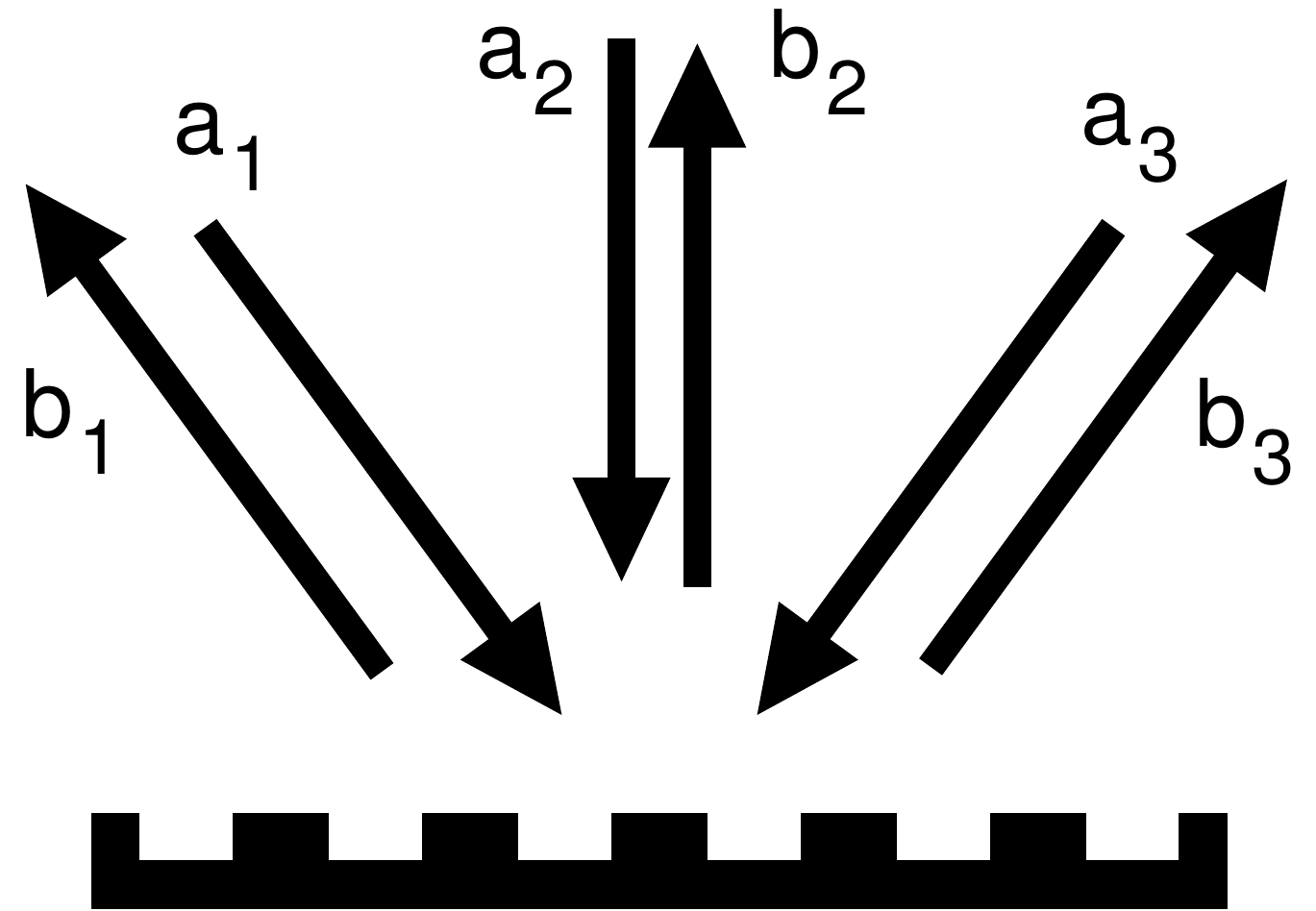} 
\vspace{.2cm}

with:
\begin{equation}
\begin{array}{lcl}
\phi_0 & = & 0,\\
\phi_1 & = &
-\frac{1}{2}\arccos\left(\frac{\eta_1^2-2\eta_0^2}{2\rho_0\eta_0}\right),\\
\phi_2 & = & \arccos\left(\frac{-\eta_1^2}{2\eta_2\eta_0}\right).
\end{array}
\end{equation}

The coupling efficiencies are limited by energy conservation to:
\begin{equation}
\begin{array}{lcl}
\rho_0^2+2\eta_1^2=1,\\
\eta_0^2 +\eta_1^2 +\eta_2^2=1.
\end{array}
\end{equation}

Further limits for the coupling efficiencies follow from the coupling
phases:
\begin{equation}
\frac{1-\rho_0}{2}\leq \eta_0, \eta_2 \leq\frac{1+\rho_0}{2}.
\end{equation}

\paragraph{gr4 component}

The grating is \emph{not} used in any Littrow configuration. Only two orders
are allowed to exist.  From the grating equation one can see that these can
only be $m=0,1$.  To compute the limits for $\lambda/d$ and $\alpha$ we
rewrite the grating equation as:
\begin{equation}
a+b=mc,
\end{equation}
with $a=\sin(\alpha)\in[0,1]$, $b=\sin(\phi)\in[-1,1]$ and $c=\lambda/d>0$.
From the fact that the first order $m=1$ should exist we get:
\begin{equation}
a+b=c.
\end{equation}
This can only be true if 
\begin{equation}
c<2,
\end{equation}
and
\begin{equation}
a>c-1.
\end{equation}

We can derive the next limit from the fact that $m=-1$ must not exist, i.e.:
\begin{equation}
a+b\neq -c,
\end{equation}
this is true if
\begin{equation}
a+c < -1 \qquad \lor  \qquad a+c > 1.
\end{equation}
The first condition is never fulfilled so we get the remaining limit as:
\begin{equation}\label{eq:gr4lim1}
a>1-c.
\end{equation}

Now, we must make sure that $m=2$ does not exist:
\begin{equation}
a+b\neq 2c.
\end{equation}
As before we can write this as
\begin{equation}\label{eq:gr4lim2}
a< 2c-1 \qquad \lor  \qquad a> 2c+1.
\end{equation}
The second condition can never be fulfilled. Also the first limit immediately
gives $c>\frac{1}{2}$ but together with \eq{eq:gr4lim1} we can restrict
possibles values for $c$ even further.  Combining \eq{eq:gr4lim1} and
\eq{eq:gr4lim2} we get:
\begin{equation}
a>1-c\qquad \land  \qquad a< 2c-1.
\end{equation}
This is only possible if
\begin{equation}
1-c< 2c-1,
\end{equation}
and thus $c>2/3$.

In summary we get:
\begin{equation}
\begin{array}{l}
\lambda/2~~<~~d~~<~~3/2 ~\lambda, \\
1-\lambda/d~~ <~~ \sin(\alpha)~~ <~~ 2\lambda/d -1 \qquad\mbox{for}\qquad
\lambda/d<1,\\
\lambda/d-1~~ <~~ \sin(\alpha)\qquad\mbox{for}\qquad \lambda/d>1.
\end{array}
\end{equation}

The coupling of the field amplitude then corresponds to that of a \bs.  The two
coupling efficiencies $\eta_0$, $\eta_1$ are constrained by energy
conservation as:
\begin{equation}
\eta_0^2+\eta_1^2=1.
\end{equation}
The coupling matrix is given by:

\vspace{.8cm}{
\hspace{1cm}
$\rule{0pt}{1.3cm}
\left( 
\begin{array}{c}
\rm b1 \\
\rm b2 \\
\rm b3 \\
\rm b4 
\end{array}
\right)=\left(
\begin{array}{cccc}
0& A_{21}  & A_{31} & 0\\
A_{12}&0&0  & A_{42}\\
A_{13}&0&0  & A_{43}\\
0& A_{24}  & A_{34} & 0
\end{array}\right)
\left(
\begin{array}{c}
\rm a1 \\
\rm a2 \\
\rm a3 \\
\rm a4
\end{array}\right)
$}

\nopagebreak
\vspace{-.6cm}
\hspace{10cm}\IG [viewport=0 0 4 2, scale=0.2] {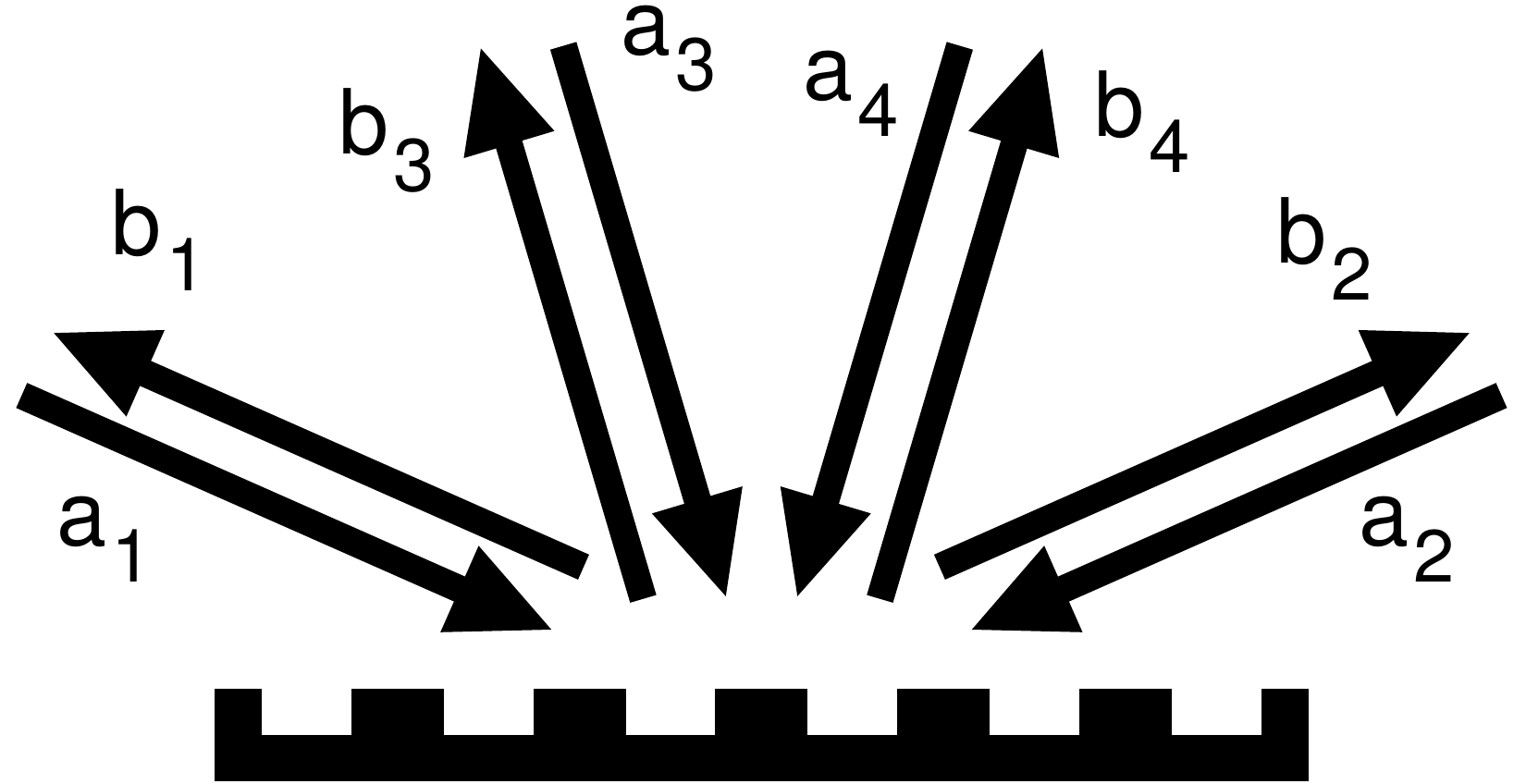} 
\vspace{.6cm}

with the coefficients:
\begin{eqnarray}
A_{12}&=&A_{21}~=~\eta_0,\\
A_{13}&=&A_{31}~=~\I \eta_1,\\
A_{24}&=&A_{42}~=~\I \eta_1,\\
A_{34}&=&A_{43}~=~\eta_0.
\end{eqnarray}

\subsection{Input fields or the `right hand side' vector}
\label{sec:rhs}

After the set of linear equations for an optical system has been determined,
the input light fields have to be given by the user.  The respective fields
are entered into the `right hand side' (RHS) vector of the set of linear
equations. The RHS vector consists of complex numbers that specify the
amplitude and phase of every input field. Input fields are initially set to
zero, and every non-zero entry describes a light source. The possible
sources are lasers, modulators and `signal sidebands'. 

\subsubsection{Laser}

The principal light sources are, of course, the lasers. They are connected
to one node only. The input power is specified by the user in the input
file.  For every laser the field amplitude is set as:
\begin{equation}
a_{in}=\sqrt{(P/\epsilon_c)}~e^{\I \varphi},
\end{equation}
with 
\begin{equation}
P~=~\epsilon_0 c |a|^2,
\end{equation}
as the laser power and $\varphi$ the specified phase.  The conversion factor
$\epsilon_c=\epsilon_0 \cdot c$ can be set in the init file `kat.ini'. The
default value is $\epsilon_c=1$. This setting does not yield correct
absolute values for light field amplitudes, i.e. when amplitude detectors
are used. Instead, one obtains more intuitive numbers from which the
respective light power can be easily computed. For the correct units of
field amplitudes, the value for $\epsilon_c$ can be set to
$\epsilon_c=\epsilon_0 \cdot c\approx0.0026544$.

\subsubsection{Modulators}

Modulators produce non-zero entries in the RHS vector for every modulation
sideband generated.  Depending on the order ($k\ge 0$) and the modulation
index ($m$), the input field amplitude for amplitude modulation is:
\begin{equation}
a_{\rm in}=\frac{m}{4},
\end{equation}
and for phase modulation:
\begin{equation}
a_{ \rm in}=(-1)^k~J_k(m)~\mEx{\I \OPh},
\end{equation}
with $\OPh$ given as (Equation~\ref{eq:bessel0}):
\begin{equation}
\OPh=\pm k\cdot(\frac{\pi}{2}+\OPh_{\rm s}),
\end{equation}
where $\OPh_{\rm s}$ is the user-specified phase from the modulator
description.  The sign of $\OPh$ is the same as the sign of the frequency
offset of the sideband.  For `lower' sidebands ($f_{\rm mod}<0$) we get
$\OPh=-\dots$, for `upper' sidebands ($f_{\rm mod}>0$) it is $\OPh=+\dots$.

\subsubsection{Signal frequencies}

The most complex input light fields are the signal sidebands. They can be
generated by many different types of modulation inside the interferometer
(signal modulation in the following). The components mirror, \bs, space,
laser and modulator can be used as a source of signal sidebands.  Primarily,
artificial signal sidebands are used as the input signal for  computing
transfer functions of the optical system. The amplitude, in fact the
modulation index, of the signal is assumed to be much smaller than unity so
that the effects of the modulation can be described by a linear analysis. If
linearity is assumed, however, the computed transfer functions are
independent of the signal amplitude; thus, only the relative amplitudes of
output and input are important, and the modulation index of the signal
modulation can be arbitrarily set to unity in the simulation.
 
Signal frequencies can be `applied' to a number of different components
using the command \cmd{fsig}. The connection of the signal frequency causes
the component to---in some way---modulate the light fields at the component.
The frequency, amplitude and phase of the modulation can be specified by
\cmd{fsig}. 

\Finesse always assumes a numerical amplitude of 1. The numerical value of 1
has a different meaning for applying signals to different components (see
below). The amplitude that can be specified with \cmd{fsig} can be used to
define the relative amplitudes of the source when the signal is applied to
several components at once. Please note that \Finesse does not
correct the transfer functions for strange amplitude settings.  An amplitude
setting of two, for example, will scale the output (a transfer function) by
a factor of two. 

In order to have a determined number of light fields, the signal modulation
of a signal sideband has to be neglected. This approximation is sensible
because in the steady state the signal modulations are expected to be tiny
so that second-order effects (signal modulation of the signal modulation
fields) can be omitted.


In general, the carrier field at the `signal component' can be written as:
\begin{equation}
E_{in}~=~E_0~\mEx{\I\w_c \T +\varphi_c},
\end{equation}
with $\w_c$ the carrier frequency, and $\varphi_c$ the phase of the carrier.
In most cases the modulation of the light will be a
phase modulation.  Then the field after the modulation can be expressed
in general as:
\begin{equation}
E_{out}~=~A~E_0~\mEx{\I\w_c \T +\varphi_c +\varphi(t)+B},
\end{equation}
with $A$ as a real amplitude factor, $B$ a constant phase term and
\begin{equation}
\varphi(t)~=~m\cos{(\w_s\T +\varphi_s)},
\end{equation}
with $m$ the modulation index, $\w_s$ the signal frequency and $\varphi_s$
the phase as defined by \cmd{fsig}. The modulation index will in general
depend on the signal amplitude $a_s$ as given by \cmd{fsig} and also other
parameters (see below).  As mentioned in Section~\ref{sec:trans+err}, the
simple form for very small modulation indices ($m \ll 1$) can be used: only
the two sidebands of the first order are taken into account and the Bessel
functions can be approximated by:
\begin{equation}
\begin{array}{l}
J_0(m) \approx 1,\\
J_{\pm1}(m) \approx \pm\frac{m}{2}.
\end{array}
\end{equation}
Thus, the modulation results in two sidebands (`upper' and `lower' with a
frequency offset of $\pm\w_s$ to the carrier) which can be written as:
\begin{equation}
\begin{array}{rcl}
E_{sb}&=&\I\frac{m}{2}~A~E_0~\mEx{\I((\w_c\pm\w_s) \T +\varphi_c+B\pm\varphi_s)}\\
&=&\frac{m}{2}~A~E_0~\mEx{\I (\w_{sb} \T +\pi/2 +\varphi_c+B\pm\varphi_s)}.
\end{array}
\end{equation}

\paragraph{Mirror}

\begin{figure}[h]
\centering
\IG [scale=.4, angle=0] {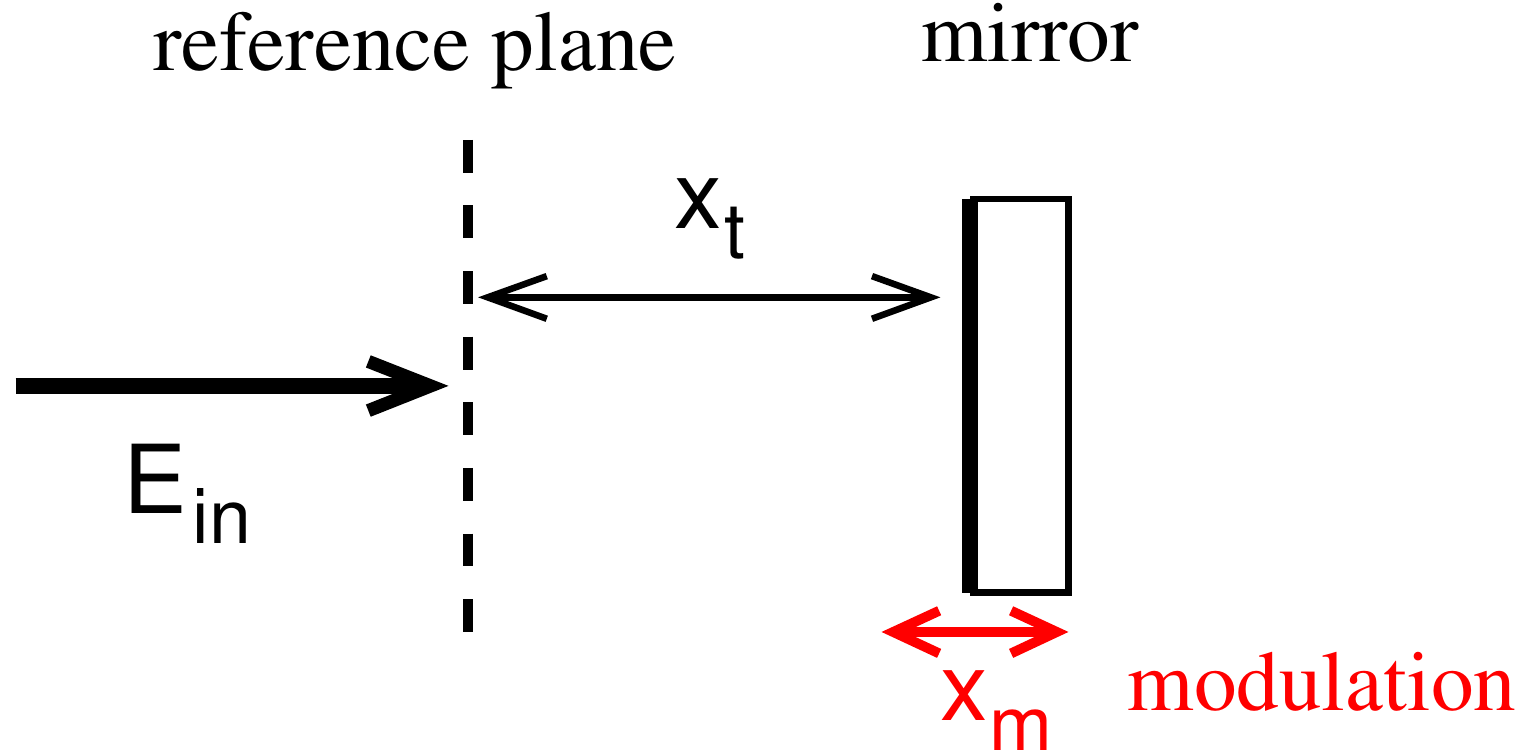} 
\caption{Signal applied to a mirror: modulation of the mirror position.}
\label{fig:mirror_mod}
\end{figure}
 
Mirror motion does not change transmitted light. The reflected light will be
modulated in phase by any mirror movement.  The relevant parameters are
shown in Figure~\ref{fig:mirror_mod}. At a reference plane (at the nominal
mirror position when the tuning is zero), the field impinging on the mirror
is:
\begin{equation}
E_{\rm in}~=~E_0~\mExB{\I(\w_{\rm c} t +\OPh_{\rm c} -k_{\rm c}
x)}~=~E_0~\mEx{\I\w_{\rm c} t+ \OPh_{\rm c}}.
\end{equation}
If the mirror is detuned by $x_{\rm t}$ (here given in meters) then the
electric field at the mirror is:
\begin{equation}
E_{\rm mir}~=~E_{\rm in}~\mEx{-\I k_{\rm c} x_{\rm t}}.
\end{equation}

With the given parameters for the signal frequency, the position modulation
can be written as $x_{\rm m}=a_{\rm s}\cos(\w_{\rm s}t+\OPh_{\rm s})$ and
thus the reflected field at the mirror is:
\begin{equation}
E_{\rm refl}~=~r~E_{\rm mir}~\mEx{\I2k_{\rm c} x_{\rm m}}~=~r~E_{\rm
mir}~\mExB{\I2k_{\rm c} a_{\rm s}\cos(\w_{\rm s}t+\OPh_{\rm s})},
\end{equation}
setting $m=2k_{\rm c} a_{\rm s}$, this can be expressed as:
\begin{equation}
{\renewcommand{\arraystretch}{1.5}
\begin{array}{rcl}
E_{\rm refl}&=&r~E_{\rm mir}~\Bigl(1+\I\frac{m}{2}\mExB{-\I(\w_{\rm
s}t+\OPh_{\rm s})}+\I\frac{m}{2}\mExB{\I(\w_{\rm s}t+\OPh_{\rm s})}\Bigr)\\
&=&r~E_{\rm mir}~\Bigl(1+\frac{m}{2}\mExB{-\I(\w_{\rm s}t+\OPh_{\rm
s}-\pi/2)}~\Bigr.\\
& &+\Bigl.\frac{m}{2}\mExB{\I(\w_{\rm s}t+\OPh_{\rm s}+\pi/2)}\Bigr).
\end{array}}
\end{equation}
This gives an amplitude for both sidebands of:
\begin{equation}
a_{\rm sb}~=r~m/2~E_0~=~r~k_{\rm c} a_{\rm s}~E_0.
\end{equation}
The phase back at the reference plane is:
\begin{equation}\label{eq:plane-phisb}
\OPh_{\rm sb}~=~\OPh_{\rm c}+\frac{\pi}{2}\pm\OPh_{\rm s}-(k_{\rm c}+k_{\rm
sb})~x_{\rm t},
\end{equation}
where the plus sign refers to the `upper' sideband and the minus sign to
the `lower' sideband.  As in \Finesse the tuning is given in degrees, 
i.e.~the conversion from $x_{\rm t}$ to $\Tun$ has to be taken into
account:
\begin{equation}
{\renewcommand{\arraystretch}{1.5}
\begin{array}{rcl}
\OPh_{\rm sb}&=&\OPh_{\rm c}+\frac{\pi}{2}\pm\OPh_{\rm s}-(\w_{\rm
c}+\w_{\rm sb})/c~ x_{\rm t}\\
&=&\OPh_{\rm c}+\frac{\pi}{2}\pm\OPh_{\rm s}-(\w_{\rm c}+\w_{\rm sb})/c
~\lambda_0/360~  \Tun\\
&=&\OPh_{\rm c}+\frac{\pi}{2}\pm\OPh_{\rm s}-(\w_{\rm c}+\w_{\rm sb})/\w_0
~2\pi/360~  \Tun.
\end{array}}
\end{equation}
With a nominal signal amplitude of $a_{\rm s}=1$, the sideband amplitudes
become very large.  For an input light field at the default wavelength one
typically obtains:
\begin{equation}
a_{\rm sb}=r~k_{\rm c}~E_0=r~\w_{\rm
c}/c~E_0=r~2\pi/\lambda_0~E_0\approx6\cdot10^6.
\end{equation}
Numerical algorithms have the best accuracy when the various input numbers
are of the same order of magnitude, usually set to a number close to one.
Therefore, the signal amplitudes for mirrors (and \bss) should be scaled: a
natural scale is to define the modulation in radians instead of meters. The
scaling factor is then $\w_0/c$, and setting $a=\w_0/c~a'$ the reflected field
at the mirror becomes:
\begin{equation}
{\renewcommand{\arraystretch}{1.5}
\begin{array}{rl}
E_{\rm refl}&=~r~E_{\rm mir}~\mEx{\I2 \w_{\rm c}/\w_0~x_{\rm m}}\\
&=~r~E_{\rm mir}~\mExB{\I2 \w_{\rm c}/\w_0~ a'_{\rm s}\cos(\w_{\rm
s}t+\OPh_{\rm s})},
\end{array}}
\end{equation}
and thus the sideband amplitudes are:
\begin{equation}
a_{\rm sb}~=r~~\w_{\rm c}/\w_0~ a'_{\rm s}~E_0,
\end{equation}
with the factor $\w_{\rm c}/\w_0$ typically being close to one.  The units
of the computed transfer functions are `output unit per radian'; which are
neither common nor intuitive.  The command \cmd{scale meter} converts the
units into the more common `Watts per meter' by applying the inverse scaling
factor  $c/\w_0$.

When a light field is reflected at the back surface of the mirror, the
sideband amplitudes are computed accordingly. The same formulae as above can
be applied with $x_{\rm m} \rightarrow -x_{\rm m}$ and $x_{\rm t}
\rightarrow -x_{\rm t}$, yielding the same amplitude as for the reflection
at the front surface, but with a slightly different phase:
\begin{equation}
{\renewcommand{\arraystretch}{1.5}
\begin{array}{rcl}
\OPh_{\rm sb,back}&=&\OPh_{\rm c}+\frac{\pi}{2}\pm(\OPh_{\rm s}+\pi)+(k_{\rm
c}+k_{\rm sb})~ x_{\rm t}\\
&=&\OPh_{\rm c}+\frac{\pi}{2}\pm(\OPh_{\rm s}+\pi)+(\w_{\rm c}+\w_{\rm
sb})/\w_0 ~2\pi/360~  \Tun.
\end{array}}
\end{equation}

\paragraph{Beam splitter}

When the signal frequency is applied to the beam splitter, the reflected
light is modulated in phase. In fact, the same computations as for mirrors
can be used for beam splitters.  However, all distances have to be scaled by
$\cos(\alpha)$ (see Section~\ref{sec:coupling}).  Again, only the reflected
fields are changed by the modulation and the front side and back side
modulation have different phases. The amplitude and phases compute to:
\begin{align}
a_{\rm sb}~&=r~~\w_c/\w_0~ a_{\rm s}\cos(\alpha)~E_0,\\
\phi_{\rm sb,front}~&=~\varphi_c+\frac{\pi}{2}\pm\varphi_{\rm
s}-(k_{c}+k_{sb}) x_t\cos(\alpha),\\
\phi_{\rm sb,back}~&=~\varphi_c+\frac{\pi}{2}\pm(\varphi_{\rm
s}+\pi)+(k_{c}+k_{sb}) x_t\cos(\alpha).
\end{align}

\paragraph{Space}

For interferometric gravitational wave detectors, the `free space' is an
important source for a signal frequency: a passing gravitational wave
modulates the length of the space (i.e.~the distance between two test
masses). A light field passing this length will thus be modulated in phase.
The phase change $\phi(t)$ which is accumulated over the full length is
(see, for example, \cite{jun1}):
\begin{equation}
\phi(t)~=~\frac{\w_c~n~L}{c}+\frac{a_s}{2}\frac{\w_c}{\w_s}
\sin{\left(\w_s\frac{n~L}{c}\right)}
\cos{\left(\w_s\left(t-\frac{n~L}{c}\right)\right)},
\end{equation}
with $L$ the length of the space, $n$ the index of refraction and $a_s$ the
signal amplitude given in strain (h).  This results in a  signal sideband
amplitude of:
\begin{align}
a_{\rm
sb}~&=~\frac{1}{4}~\frac{\w_c}{\w_s}\sin{\left(\w_s\frac{n~L}{c}\right)}
a_{\rm s}~E_0,\\
\phi_{\rm sb}~&=~\varphi_c+\frac{\pi}{2}\pm\varphi_s-(\w_c+\w_s)\frac{nL}{c}.
\end{align}

\paragraph{Laser}

Applying a signal to a laser is treated as a frequency modulation of the
laser:
\begin{equation}
E~=~E_0~e^{\I (\w_c\,t +a_s/\w_s\cos(\w_s\,t+\varphi_s) +\varphi_c)}.
\end{equation}
Therefore the amplitude of the sidebands is scaled with frequency as:
\begin{equation}
\begin{array}{l}
a_{sb}=\frac{a_{\rm s}}{2 \w_{\rm s}}~E_0,\\
\phi_{\rm sb}=\varphi_c+\frac{\pi}{2}\pm \varphi_s.
\end{array}
\end{equation}

\paragraph{Modulator}

Signal frequencies at a modulator are treated as `phase noise'.  See
Section~\ref{sec:phasenoise} for details. The electric field that leaves the
modulator can be written as:
\begin{equation}
\begin{array}{rl}
E~=~E_0&e^{\I(\w_0
\T+\varphi_0)}~\sum_{k=-order}^{order}i^{\,k}~J_k(m)~e^{\I k(
\w_m\T+\varphi_m)}\qquad\\
&\times\left( 1+\I\frac{k~m_2~a_s}{2}~e^{-i
(\w_s\T+\varphi_s)}~+\I\frac{k~m_2~a_s}{2}~e^{\I(
\w_s\T+\varphi_s)}+O((km_2)^2)\right),
\end{array}
\end{equation}
with
\begin{equation}
\begin{array}{l}
E_{\rm mod}~=~E_0~J_k(m),\\
\varphi_{\rm mod}=\varphi_0 + k\frac{\pi}{2} + k \varphi_m.
\end{array}
\end{equation}
The sideband amplitudes are:
\begin{equation}
\begin{array}{l}
a_{sb}=\frac{a_{\rm s}k m_2}{2}~E_{\rm mod},\\
\phi_{\rm sb}=\varphi_{\rm mod}+\frac{\pi}{2}\pm \varphi_s.
\end{array}
\end{equation}

\subsection{Photodetectors and demodulation}

With all the different ways of plotting signals in \Finesse it is important
to understand that every photodiode output represents only \emph{one}
frequency component of a signal.  The `signal' can be a light field, a light
power, or a mixer output.  Of course, you can calculate more than one
frequency component at a time, but only by specifying different outputs for
each of them. 

When the program has calculated the light field amplitudes for every
frequency and every output port, the interferometer is completely `solved'.
With the field amplitudes and phases you can now calculate every error
signal or frequency response. The different possible detectors you can use
with \Finesse are meant to simplify this task for the most common purposes.
Each photodetector type calculates a special output from the available
field amplitudes. The following paragraphs show how this is done. For
simplicity the calculations will be given for one output port only. 

Common to several detector types are some scaling factors which can be
applied: the command \cmd{scale} can be used to scale the output by a
given factor. Several preset factors can be used:
\begin{itemize}
\item{{\tt\small scale ampere output} in the input file scales light power
to photocurrent for the specified detector. The scaling factor (from Watts
to Amperes) is:
\begin{equation}\label{eq:watts}
C_{\rm ampere}= \frac{e~q_{\rm eff}~\lambda_0 }{hc}~[\frac{\rm A}{\rm W}],
\end{equation} 
with $e$ the electron charge, $q_{\rm eff}$ the quantum efficiency of the
detector, $h$ Planck's constant, and $c$ the speed of light.  $\lambda_0$ is
the default laser wavelength; if several light fields with different
wavelengths are present or sidebands are concerned, still only one
wavelength is used in this calculation. The differences in $\lambda$ should
be very small in most cases so that the resulting error is negligible.}
\item{{\tt\small scale meter output} can be useful when a transfer function
has been computed and {\tt\small fsig} was applied to a mirror or beam
splitter.  The output is scaled by $2\pi/\lambda_0$ (or $\lambda_0/2\pi$ for
{\tt\small pdS}).  Because the microscopic movement of these components is
always set via the tuning, a computed transfer function with the signal
inserted at a mirror or beam splitter will have the units Watt/radian.  With
{\tt\small scale meter} the result will be rescaled to Watt/meter. In case
of a sensitivity ({\tt\small pdS}), the output will be scaled to $\rm
m/\sqrt{\rm Hz}$.} 
\item{{\tt\small scale deg output} will scale the output by $180/\pi$. This
may be useful in some cases. For example, if the DC value of a transfer
function is to be compared to the slope of an error signal (at the operating
point). The latter is usually given in Watt/degree whereas the transfer
function is typically Watt/radian.}
\end{itemize}

In general, several light fields with different amplitudes, phases and
frequencies will be present on a detector. The resulting light field in an
interferometer output (i.e. on a detector) can be written as
\begin{equation}
\begin{array}{rcl}
E&=&e^{\I\w_0 \T}\sum\limits_{n=0}^{N}a_n~e^{\I\w_n \T},
\end{array}
\label{eq:field}
\end{equation} 
where the $a_n$ are complex amplitudes.

The frequency $\w_0$ is the default laser frequency, and $\w_n$ are offset
frequencies to $\w_0$ (either positive, negative or zero).  Note that very
often a slightly different representation is chosen
\begin{equation}
\begin{array}{rcl}
E&=&e^{\I\w_0\T} \left(b_0+b_1~e^{\I\w_1t}+b_{-1}~e^{-\I\w_1t}+
\dots+b_M~e^{\I\w_Mt}+b_{-M}~e^{-\I\w_Mt}\right),
\end{array}
\end{equation} 
where $\w_0$ is the carrier frequency and $\w_1$, $\w_2, \dots, \w_M>0$ are
the (symmetric) sidebands.  However, in a general approach there might be
more than one carrier field and the sidebands are not necessarily symmetric,
so Equation~\ref{eq:field} is used here. 

\subsubsection{Amplitude detector}

The amplitude detector simply plots the already calculated amplitudes of the
light field at the specified frequency.  The amplitude at frequency $\w_m$
is a complex number ($z$), and is calculated as follows:
\begin{equation}
z=\sum_n a_n \quad \mbox{with}\quad
\{n~|~n\in\{0,\dots,N\}~\wedge~\w_n=\w_m\}.
\end{equation}
Note that the amplitude detector distinguishes between positive and negative
frequencies.

\subsubsection{Photodetectors}\label{sec:pd}

For real detectors we have to look at the intensity at the output port:
\begin{equation}
\begin{array}{rcl}
|E|^2&=&E\cdot E^*~=~\sum\limits_{i=0}^N\sum\limits_{j=0}^N
a_ia_j^*~e^{\I(\w_i-\w_j)\T}\\
&=& A_0+A_1e^{\I\wb_1\T}+A_2e^{\I\wb_2\T}+\dots,
\end{array}
\label{eq:intensity}
\end{equation}
with the $A_i$ being the amplitudes of the light power sorted by the beat
frequencies $\bar{\omega}_i$.  In fact, \Finesse never calculates the light
power as above, instead it only calculates parts of it depending on the
photodetector you specify.  There are basically two ways of using
photodetectors in \Finesse: a simple photodetector for DC power and a
detector with up to 5 demodulations. These detectors see different parts of
the sum in Equation~\ref{eq:intensity}. 

\paragraph{The DC detector}

looks for all components without frequency dependence.  The
frequency dependence vanishes when the frequency becomes zero, i.e.~in all
addends of Equation~\ref{eq:intensity} with $\w_i = \w_j$.  The output is a
real number, calculated like this:
\begin{equation}\label{eq:dc_det}
x=\sum\limits_i\sum\limits_j a_ia_j^*\quad\mbox{with}\quad
\{i,j~|~i,j\in\{0,\dots,N\}~\wedge~\w_i=\w_j\}.
\end{equation}

\paragraph{A single demodulation}

can be described by a multiplication of the output with a cosine:
$\cos(\w_x+\varphi_x)$ ($\w_x$ is the demodulation frequency and $\varphi_x$
the demodulation phase) which is also called the `local oscillator'. In
\Finesse, the term `demodulation' also implies a low pass filtering of the
signal after multiplying it with a local oscillator.  After whatever
demodulation was performed only the {\bf DC} part of the result is taken
into account.  The signal is 
\begin{equation}
S_0~=~|E|^2~=~E\cdot E^*~=~\sum\limits_{i=0}^N\sum\limits_{j=0}^N
a_ia_j^*~e^{\I(\w_i-\w_j)\T},
\end{equation}
multiplied with the local oscillator it becomes
\begin{equation}
\begin{array}{rcl}
S_{1}&=&S_0\cdot
\cos(\w_xt+\varphi_x)~=~S_0\frac12\left(e^{\I(\w_xt+\varphi_x)} +
e^{-\I(\w_xt+\varphi_x)}\right)\\
&=&\frac12\sum\limits_{i=0}^N\sum\limits_{j=0}^N
a_ia_j^*~e^{\I(\w_i-\w_j)\T}\cdot\left(e^{\I(\w_xt+\varphi_x)} +
e^{-\I(\w_xt+\varphi_x)}\right).
\end{array}
\end{equation}
With $A_{ij}=a_ia_j^*$ and $e^{\I\w_{ij}\T}=e^{\I(\w_i-\w_j)\T}$ we can
write
\begin{equation}
S_{1}~=~\frac12\left(\sum\limits_{i=0}^NA_{ii}+\sum\limits_{i=0}^N 
\sum\limits_{j=i+1}^N (A_{ij}~e^{\I\w_{ij}\T}+A_{ij}^*~e^{-\I\w_{ij}\T})\right)\cdot
\left(e^{\I(\w_xt+\varphi_x)}+e^{-\I(\w_xt+\varphi_x)}\right).
\end{equation}
When looking for the DC components of $S_1$ we get the following
\begin{equation}
\begin{array}{lll}
S_{\rm
1,DC}&=&\sum\limits_{ij}
\frac12(A_{ij}~e^{-\I\varphi_x}+A_{ij}^*~e^{\I\varphi_x})\quad\mbox{with}\quad
\{i,j~|~i,j\in\{0,\dots,N\}~\wedge~\w_{ij}=\w_x\}\\
&=&\sum\limits_{ij}\Re\left\{A_{ij}~e^{-\I\varphi_x}\right\}.
\label{eq:single_demod}
\end{array}
\end{equation}
This would be the output of a mixer. The results for $\varphi_x=0$ and
$\varphi_x=\pi/2$ are called \emph{in-phase} and \emph{in-quadrature}
respectively (or also \emph{first} and \emph{second quadrature}).  They are
given by:
\begin{equation}
\begin{array}{rcl}
S_{\rm 1,DC,phase}&=&\sum\limits_{ij}\Re\left\{A_{ij}\right\},\\
S_{\rm 1,DC,quad}&=&\sum\limits_{ij}\Im\left\{A_{ij}\right\}.
\end{array}
\end{equation}
When the user has specified a demodulation phase the output given by
\Finesse is real
\begin{equation}
x=S_{\rm 1,DC}.
\end{equation}
If no phase is given the output is a complex number 
\begin{equation}
z=\sum\limits_{ij}A_{ij}\quad\mbox{with}\quad
\{i,j~|~i,j\in\{0,\dots,N\}~\wedge~\w_{ij}=\w_x\}.
\end{equation}

\paragraph{A double demodulation}

is a multiplication with two local oscillators and taking the DC component
of the result. First looking at the whole signal we can write:
\begin{equation}
S_{2}=S_0\cdot \cos(\w_x+\varphi_x)\cos(\w_y+\varphi_y).
\end{equation}
This can be written as 
\begin{equation}
\begin{array}{rcl}
S_{2}&=&S_0\frac12(
\cos(\w_y+\w_x+\varphi_y+\varphi_x)+\cos(\w_y-\w_x+\varphi_y-\varphi_x))\\
&=&S_0\frac12( \cos(\w_++\varphi_+)+\cos(\w_-+\varphi_-)),
\end{array}
\end{equation}
and thus reduced to two single demodulations. Since we now only care for the
DC component we can use the expression from above
(Equation~\ref{eq:single_demod}).  These two demodulations give two complex
numbers:
\begin{equation}
\begin{array}{rcl}
z1&=&\sum\limits_{ij}A_{ij}\quad\mbox{with}\quad
\{i,j~|~i,j\in\{0,\dots,N\}~\wedge~\w_i-\w_j=\w_+\},\\
z2&=&\sum\limits_{ij}A_{kl}\quad\mbox{with}\quad
\{k,l~|~k,l\in\{0,\dots,N\}~\wedge~\w_k-\w_l=\w_-\}.\\
\end{array}
\end{equation}
The demodulation phases are applied as follows to get a real output (two
sequential mixers):
\begin{equation}
x=\Re\left\{(z_1~e^{-\I\varphi_x}+z_2~e^{\I\varphi_x})~e^{-\I\varphi_y}\right\}.
\end{equation}
A demodulation phase for the first frequency (here $\varphi_x$) must be
given in any case. To get a complex output the second phase can be omitted:
\begin{equation}
z=z_1~e^{-\I\varphi_x}+z_2~e^{\I\varphi_x}.
\end{equation}

\paragraph{More demodulations}

can also be reduced to single demodulations as above. In fact the same code
computes output signals for up to 5 demodulations.

\newpage

\begin{figure}[htb]
\centering
\IG [scale=.7] {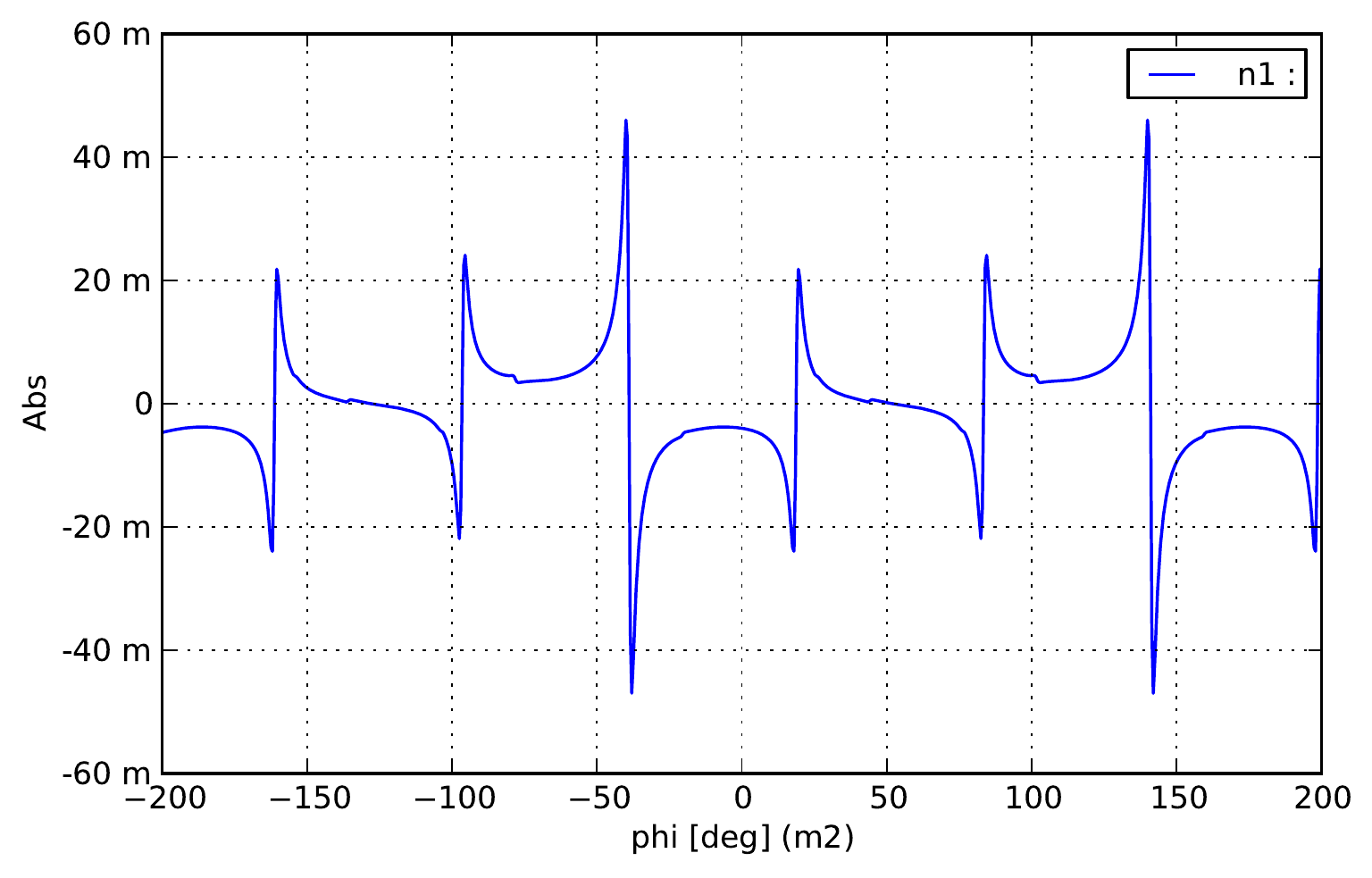}\\
\IG [scale=.7] {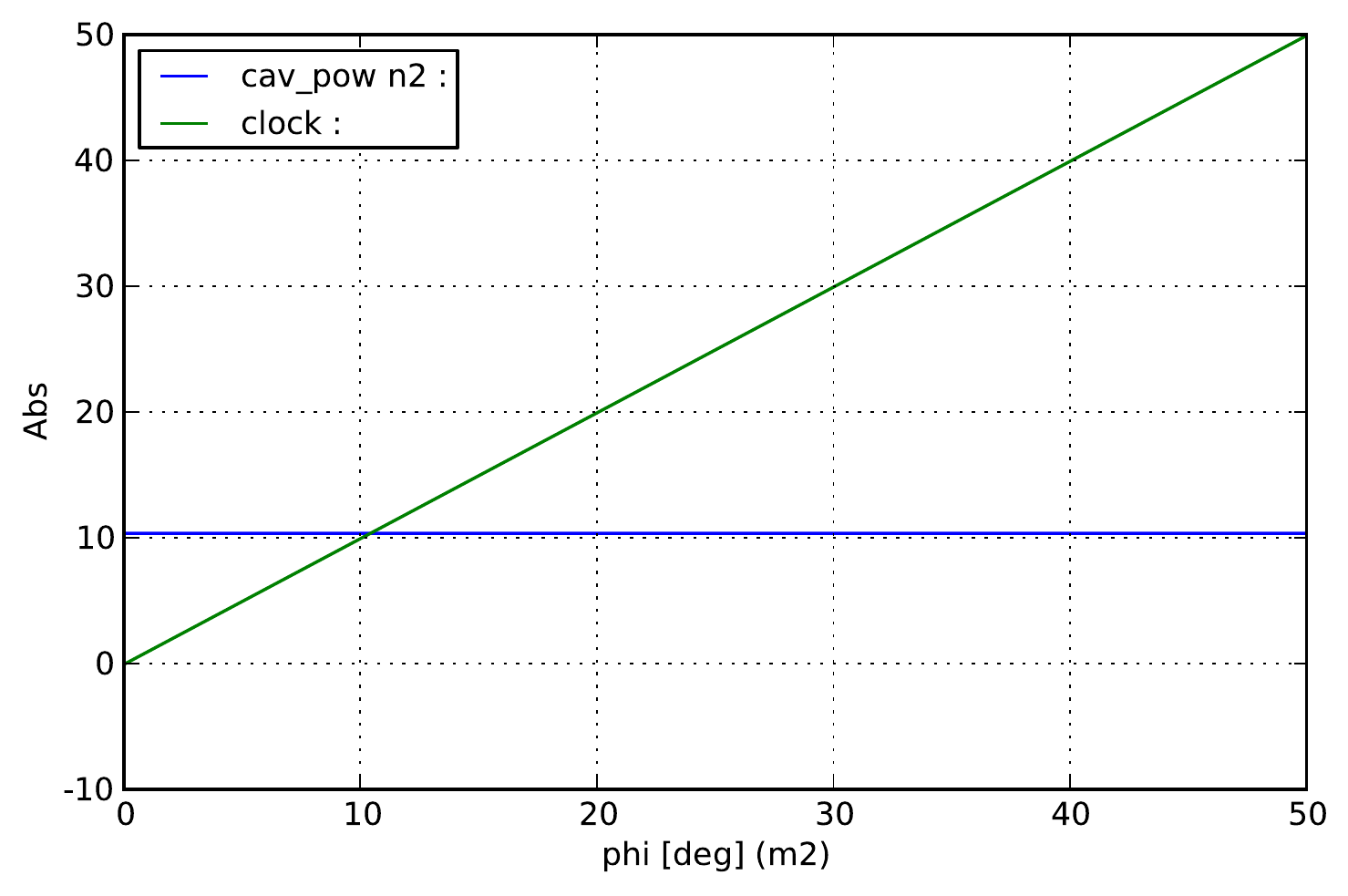} 
\caption[Pound-Drever Hall lock]{Using a Pound-Drever signal for the \code{lock} command: the top plot
  shows the error signal, the lower plot shows the correction signal
  when the lock is active and the cavity circulating power as a test
  signal. The correction signal `clock' increases proportional to the
  detuning as expected and the cavity power is maintained throughout.}
\label{fig:lock1}
\end{figure}
\section{The {\Co lock} command}\label{sec:lock}
The {\Co lock} command as described in the
\nameref{sec:syntax_reference} 
can be used to implement a feedback system using
an internal iterative process. 
For each data point as specified by the {\Co xaxis} commands
the {\Co lock} command solves the interferometer matrix several 
times in order to perform the iteration. Used correctly the
{\Co lock} command can be an effective tool. However, it should 
be noted that even a simple
{\Co lock} often increases the computation time by factors of five.

One of the main uses of the \code{lock} command is to keep a cavity
on resonance in the presence of beam shape distortions, either through
mirror misalignments, mode mismatch, or more complex surface
distortions introduced via mirror maps, see \Sec{sec:mirrormaps}.
In all these cases higher-order modes play a significant role in the
description of the optical fields, and as a result the 
simple trick of making all spaces to be multiples of the wavelength does
not guarantee resonance condition for mirror tunings of $\Tun=0$.
Instead the correct mirror tunings need to be set explicitly, for
example, by first scanning the cavity and finding the resonance 
condition. Alternatively a \code{lock} command can be used to
iteratively adjust the tuning of a cavity for each data point.

\subsection{Using a real error signal for a lock}
One of the best ways to ensure that a lock is performing as expected
is to use a real error signal, for example to lock a cavity with a
Pound-Drever Hall (PDH) error signal, as shown in \mFig{fig:lock1}.
In the presence of higher-order modes, the cavity resonance does
not necessarily coincide with the maximum circulating power.
If  a PDH signal is used in the experiment, you should use the
same error signal in \Finesse to ensure that the cavity in the
model is always tuned to the same operating point as the 
experiment.

This example is on purpose using a mode-mismatched and
misaligned cavity to show the impact of these distortions on the
error signal. The top plot in \mFig{fig:lock1} clearly shows the
zero crossings of the carrier and sideband fields. In order to
guarantee a successful lock the simulation should be started on
resonance, for example we can tune m1 so that the cavity is 
on resonance and if we chose to tune m2, as shown in the bottom
plot, this should start at a tuning of zero.
The \Finesse file for using a Pound-Drever Hall signal with the
\code{lock} command is:
\loadkat{pdh-lock.kat}

\subsection{Tuning the {\Co lock}}
If you use a {\Co lock} you should carefully tuned the following parameters:
\begin{itemize}
\item gain: the iteration routines have an `autogain' feature which tries to 
correct for wrongly set gain values. However, it can only check for
large deviations from optimal gains. For example, setting the gain wrong by a factor
of two typically reduces the speed of the simulation by a factor of two or worse.
\item locking accuracy: the lock accuracy defines how large the residual deviation
between the set locking point and the iterative results should be. Make sure
that you know what accuracy you require and do try not to set more stringent
values than those required.
\item {\Co xaxis} step size: One must make sure that the starting point of
the xaxis command is actually at or close to the set point of any
active {\Co lock}. Otherwise the {\Co lock} iteration might fail to find the
set point. 
\enlargethispage*{2\baselineskip}

Furthermore, in many cases the lock uses on an error signal
that does not change linearly with the parameter with tuned by the {\Co xaxis} 
command. In that case the step size given in the {\Co xaxis} command should be
set small enough so that for each step the change in the error signal can
be still approximated as an almost linear change. The table below shows the
computing times for a Michelson {\Co lock} in the GEO\,600 input file.
This very rough comparison of computation times
shows that various step sizes work quite well
(in this example the accuracy was set to 1\,pW):
\vspace{3mm}
\begin{center}
\begin{tabular}{|c|c|c|c|c|c|c|c|}
\hline
number of steps & 700 & 800 & 900 & 1000 & 2000 & 3000 & 10000\\
\hline
computation time & 61s & 6s & 5s & 5s & 6s & 9s & 28s\\
\hline
\end{tabular}
\end{center}
\vspace{3mm}
Please note that in case of the 700 steps \Finesse's autogain function has changed the 
loop gain automatically which also re-calibrates the step size. Even though
in this case the lock was still succesful, it demonstrated that too large step 
sizes should be avoided. In the case of the 2000 steps, each step required 
about 22 internal iterations to reach the locking accuracy of 1\,pW. 
An accuracy of 0.1\,mW could be achieved with 5 iterations.
\end{itemize}


\begin{figure}[htb]
\centering
\IG [scale=.7] {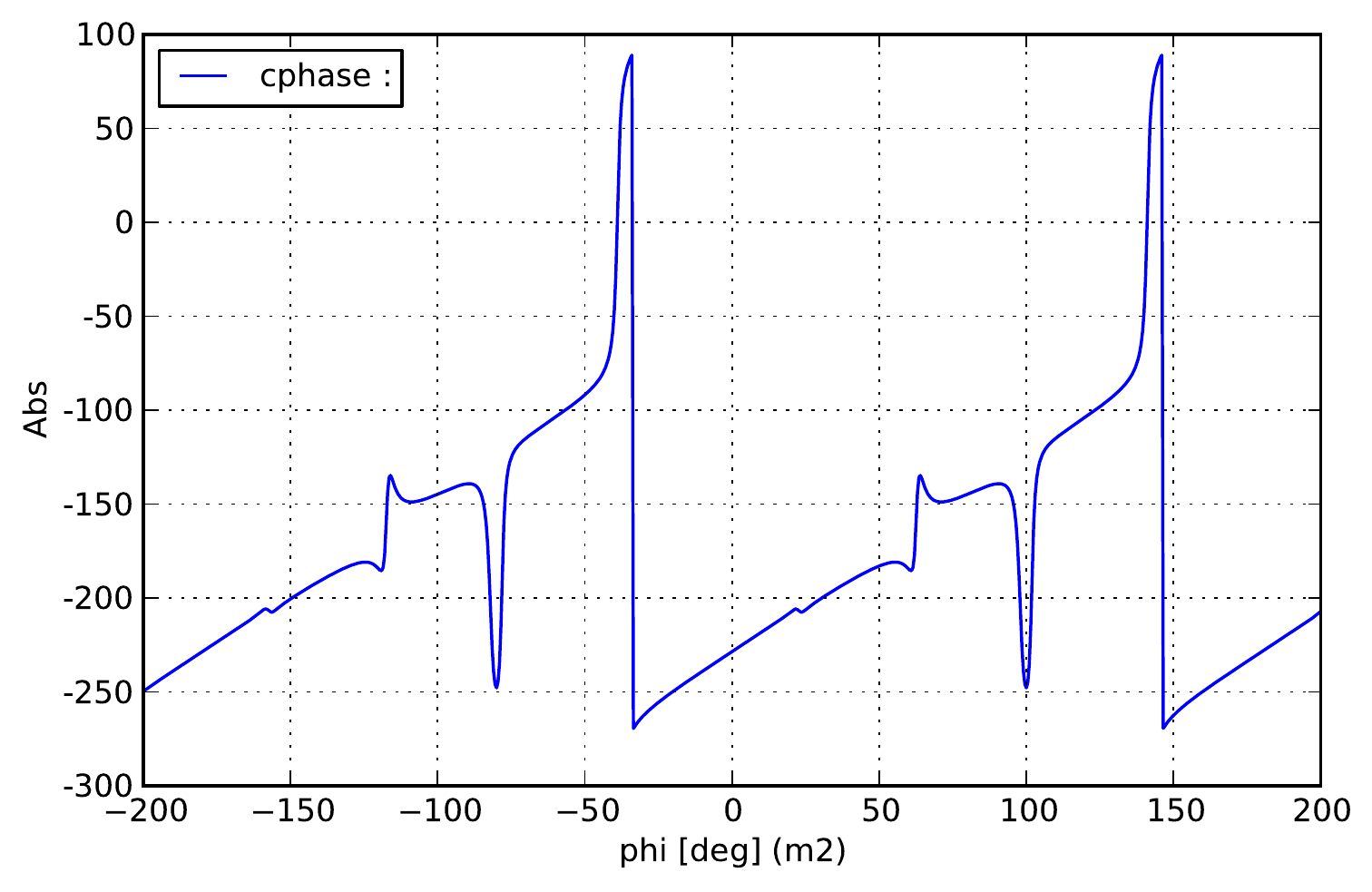}\\
\IG [scale=.7] {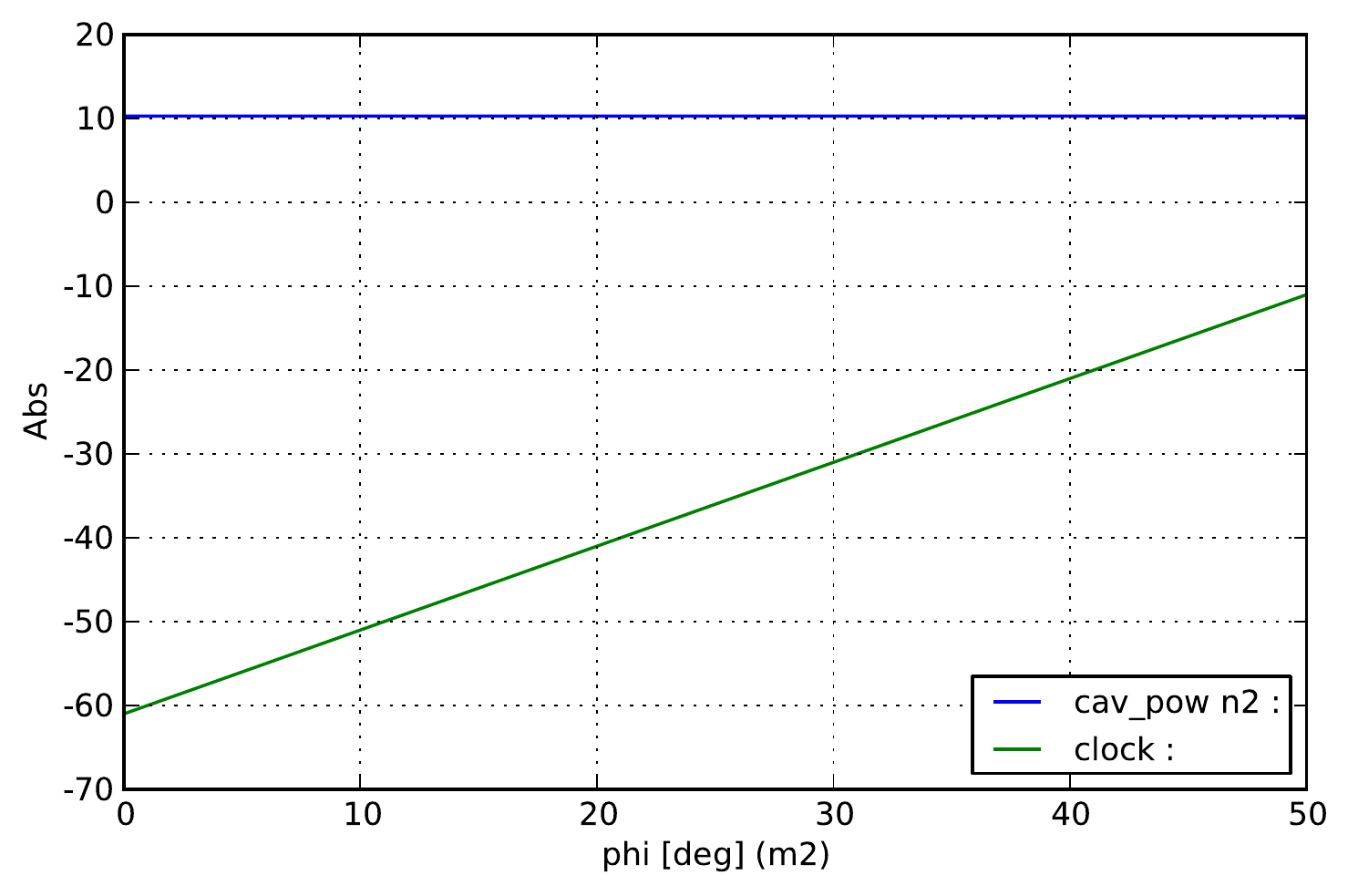} 
\caption[Pound-Drever Hall lock]
{Using a pseudo-lock signal for the \code{lock} command: the top plot
  shows the error signal, the lower plot shows the correction signal
  when the lock is active and the cavity circulating power as a test
  signal.}
\label{fig:lock2}
\end{figure}\subsection{A pseudo-lock}
Another possibility to create a signal for the \code{lock} command is
to use a so-called `pseudo-lock' system in which the simulation
signals, such as the phase of a light field is used to generate an
error signal. Since such signal are not available in the experiment,
this cannot represent a real control system. However, it has the
advantage of generating an error signal without the need of extra
optical components, suc as the electro-optical modulator for the
PDH scheme shown above. An example for a pseudo-lock in shown 
in \mFig{fig:lock2} and the \Finesse file is given below.

It should be noted that pseudo-locks can be problematic, if they are
not set up very carefully, their operating point might not be
identical to that of the real control loop, which would cause the
model to behave differently than the experiment and thus produce
misleading and wrong results. This has been the reason behind a great
many wasted modelling efforts in the GW community!

\loadkat{pseudo-lock.kat}

\section{Shot-noise-limited sensitivity}\label{sec:shotnoise}

\Finesse allows one to compute the shot-noise-limited sensitivity using the
command \cmd{pdS}. However, to date this \shot computation must be
considered as a approximation; as it does not represent the true
quantum noise but only the shot-noise. This still provides 
correct results for interferometers in which radiation pressure effects can be neglected.
\Finesse is currently being extended to correctly compute the quantum noise
of the light upon detection with a photodiode.

This section gives a detailed description on how \Finesse computes shot-noise
using the {\tt shot} command and the respective 
linear spectral density that is often called sensitivity. In has been
shown to be difficult to remember all factors of 2 correctly, and thus
a step by step explanation for a simple example system is provided here.

Appendix~\ref{sec:geo_sens} further provides a comparison between
different methods of computing the shot noise with \Finesse:
analytically from the DC power, using the {\tt shot} command and
using the new {\tt qshot} command, which can correctly compute the
shotnoise for heterodyne detection schemes. The appendix 
is based on a dedicated effort by the GEO collaboration in
2007 to understand and verify the shot noise level limiting the GEO
sensitivity at high frequencies.

The \shot computation is based on the Schottky formula for the
(single-sided) power spectral density of the fluctuation of the photocurrent 
for a given mean current $\bar{I}$:
\begin{equation}
S_I(f) =2\,e\,\bar{I},
\end{equation}
with $e$ the electron charge.  Here $S_X (f)$ denotes the single-sided power
spectral density of $X$ over the Fourier frequency $f$. 

The link between (mean) photocurrent $\bar{I}$ and (mean) light power
$\bar{P}$ is given by the relation:
\begin{equation}
\bar{I}= e N= \frac{e~\eta~\lambda }{\hbar 2 \pi c} \bar{P},
\end{equation}
with $N$ as the number of photons and $\eta$ the quantum efficency of the
diode.  Intead of Planck's constant we write $\hbar\,2\,\pi$ to avoid
confusion with the typical use of $h(t)$ for the strain of a gravitational
wave.

Thus we can now give a power spectral density for the fluctuations of the
photocurrent. The conversion between Watts in Ampere is defined
by the constant $C_2$ in the relation $\bar{P}=C_2\,\bar{I}$.
Thus the power spectral densities must be converted as
$S_P(f)=C_2^2S_P(f)$. With the relation $S_I(f)=C_1\,\bar{I}$ we can 
then write:
\begin{equation}\label{eq:shotnoise}
S_P(f)=C_2^2S_I(f)=C_2^2~C_1\,\bar{I}=C_2^2~C_1~\bar{P}/C_2=2\,\frac{2\pi\,\hbar\,c~}{\lambda}\bar{P}.
\end{equation}


\subsection{Simple Michelson interferometer on a half fringe}\label{sec:half_fringe}

A simple example for computing a shot-noise-limited sensitivity is a Michelson
interfeormeter held on a half fringe. In this case no modulation sidebands
are required to obtain the output signal. In the following examples we will
use:
\begin{itemize}
\item an input laser power of $P_0=1$\,W,
\item the quantum efficiency is \cmd{qeff}$=\eta=1$,
\item a symmetric beam splitter with $R=T=0.5$ (angle of incidence in
\Finesse is set to $0$\,deg),
\item we denote the interferometer outputs (arbitrarily following the \GEO\
layout) as `north', `east', `south' and `west', with the input port being in
the west and the light reflected by the beam splitter entering the north
arm,
\item the Michelson interferometer arm lengths are chosen as $L_{\rm
north}=1201$\,m and
$L_{\rm east}=1200$\,m.
\end{itemize}

\subsubsection{The noise amplitude}

The Michelson interferometer is set to a half fringe by detuning the beam
splitter by $22.5$\,degrees so that the power in both the west and south
output ports is $0.5$\,W. From \eq{eq:shotnoise} we expect a value of:
\begin{equation}
S_P(f)=\frac{6.6262\,10^{-34}~ 299792458}{1064\,10^{-9}}\,{\rm W}^2/{\rm
Hz}=1.867\,10^{-19}\,{\rm W}^2/{\rm Hz},
\end{equation}
or as a linear spectral density $\sqrt{S_P}=4.321\,10^{-10}$\,W/$\sqrt{\rm
Hz}$. \Finesse returns the same value if the \cmd{shot} detector is used.

\subsubsection{The signal amplitude}

The `signal' in this example will be a differential motion of the end
mirrors.  In order to compute the signals' amplitude in the photodiode we
can compute the transfer function of the mirror motion to the photodiode. A
motion of the end mirrors will modulate the reflected light in phase.

In the following we set all macroscopic lengths to be multiples of the
wavelength; the signal frequencies are assumed to be very small so that the
phase evolution for the carrier and the signal sidebands due to the free
propagation through the interferometer arms can be considered equal.

The light fields entering the arms are given by:
\begin{align}
E_{\rm N\,in}=\frac{1}{\sqrt{2}}E_{\rm in}\exp{\left(\I\frac{\pi}{4}\right)},\\
E_{\rm E\,in}=\frac{1}{\sqrt{2}}E_{\rm in}\exp{\left(\I\frac{\pi}{2}\right)}.
\end{align}

Sidebands are created upon reflection on the end mirrors. The phase of the
modulation is set to be $0\degrees$ at the north mirror and $180\degrees$ at
the east mirror. The phase of the sidebands is given by
\eq{eq:plane-phisb}:
\begin{equation}
\OPh_{\rm sb}~=~\OPh_{\rm c}+\frac{\pi}{2}\pm\OPh_{\rm s}-(k_{\rm c}+k_{\rm
sb})~x_{\rm t}.
\end{equation}

The light reflected by the end mirrors (with $r=1$) can then be written as:
\begin{equation}
\begin{array}{lcl}
E_{\rm N}&=&\frac{1}{\sqrt{2}}E_{\rm in} \exp{(\I\frac{\pi}{4})}
\left(1+ a_s \exp\left(\I \omega_s t + \frac{\pi}{2}\right) + a_s
\exp\left(-\I \omega_s t - \frac{\pi}{2}\right)\right)\\
&=&\frac{1}{\sqrt{2}}E_{\rm in} \exp{(\I \frac{\pi}{4})}
\left(1+ a_s \I \left(\exp\left(\I \omega_s t\right) + \exp\left(-\I
\omega_s t\right)\right)\right)\\
&=& \frac{1}{\sqrt{2}}E_{\rm in} \exp{(\I \frac{\pi}{4})} \left(1+2 a_s \I
\cos\left(\omega_s t\right)\right),\\
 & & \\
E_{\rm E}&=&\frac{1}{\sqrt{2}}E_{\rm in} \exp{(\I \frac{\pi}{2})}
\left(1+ a_s \exp\left(\I \omega_s t + \frac{3\pi}{2}\right) + a_s
\exp\left(-\left(\I \omega_s t + \frac{\pi}{2}\right)\right)\right)\\
&=&\frac{1}{\sqrt{2}}E_{\rm in} \exp{(\I \frac{\pi}{2})}
\left(1-2 a_s \I \cos\left(\omega_s t\right)\right),
\end{array}
\end{equation}
with $\omega_s$ as the signal frequency and $a_s$ the amplitude of the
mirror motion (given in radians, with $2\pi$ referring to a position change
of the mirror of $\lambda$).  For a computation of the transfer function in
our example we later set $a_s=1$ while at the same time we use the
approximations for $a_s \ll 1$.

At the beam splitter the reflected fields will be superimposed, in the south
port we get:
\begin{equation}
{\renewcommand{\arraystretch}{1.5}
\begin{array}{lcl}
E_{\rm S}&=&\frac{1}{\sqrt{2}}\exp{(\I\frac{\pi}{2})}E_{\rm
N}+\frac{1}{\sqrt{2}}\exp{(-\I\frac{\pi}{4})}E_{\rm E}\\
&=&\frac{1}{2}E_{\rm in}\exp{(\I\frac{\pi}{4})}\left(1+\I-2a_s \I
\cos\left(\omega_s t\right) - 2 a_s \cos\left(\omega_s t\right)\right)\\
&=&\frac{1}{2}E_{\rm in}\exp{(\I\frac{\pi}{4})}\left(1+i\right)\left(1-2 a_s
\cos\left(\omega_s t\right)\right)\\
&=&\frac{\I}{\sqrt{2}}E_{\rm in}\left(1-2 a_s \cos\left(\omega_s
t\right)\right).\\
\end{array}}
\end{equation}

With $|E_{\rm in}|^2=P_0=1$\,W the power detected by the diode in each
output port is thus:
\begin{equation}
{\renewcommand{\arraystretch}{1.5}
\begin{array}{lcl}
I_{\rm out}&=&\frac{1}{2}P_0\left(1+4 a_s^2 \cos^2\left(\omega_s t\right) -
4 a_s \cos\left(\omega_s t\right)\right)\\
&=&\frac{1}{2}\left(1+4 a_s^2 \cos^2\left(\omega_s t\right) - 4 a_s
\cos\left(\omega_s t\right)\right)\,{\rm [W]}.\\
\end{array}}
\end{equation}

The power in the signal sidebands can be neglected so that the DC power in
both output ports is given as $P_0/2=0.5$\,W.

The signal amplitude is given by $2 a_s P_0$.  In \Finesse to get the signal
amplitude we demodulate at the signal frequency, i.e.~we multiply the output
by $\cos{\omega_s t}$ and take the DC part of the resulting sum:
\begin{equation}
\begin{array}{lcl}
I_{\rm demod}&=&2 a_s P_0 \cos^2{(\omega_s t)}+O(\omega)+O(3\omega)\\
&=&a_s P_0 + a_s P_0\cos{(2\omega_s t)}+O(\omega)+O(3\omega).
\end{array}
\end{equation}
The signal amplitude is thus given by $a_s P_0$. By default a demodulation in 
\Finesse is
understood as a multiplication with a cosine and thus reduces
the signal size by a factor of 2, with the exception that in the case of the 
transfer function this is not wanted for the demodulation at the signal frequency.  

With \Finesse it is simple to compute a transfer function for a differential
displacement of the end mirrors into the detector output. For example, with
the commands
\begin{finesse}
fsig sig1 mE 1 0
fsig sig2 mN 1 180
pd1 south1 1 max n10
\end{finesse}
we compute the signal transfer function at 1\,Hz. The \Finesse output is:
$2$\,W/rad.  To obtain the more useful units W/m we have to multiply by
$2\pi/\lambda$.

It is important to understand which lengths we refer to with this transfer
function.  Due to the fact that \Finesse can compute more general optical
configurations than a Michelson interferometer, the amplitude of the
transfer function amplitude refers to the motion of each single mirror. For
example, a transfer function amplitude of 1 \,W/m means that the output
power changes by one 1\,nW when the east mirror is moved by 1\,nm and the
north mirror by -1\,nm. If we want to compute the transfer function
referring to the differential displacement $\Delta L=L_{\rm north}-L_{\rm
east}$ we have to divide the transfer function by a factor of two. Thus we
get:
\begin{equation}
T_{\Delta L \rightarrow P}=\frac{2\pi
P_0}{\lambda}=\frac{2\pi}{1064\,10^{-9}}\,{\rm W}/{\rm m}.
\end{equation}
If, in fact, the transfer function is to be given with respect to an optical
path length difference $\Delta L'$ one has to divide by another factor of
two:
\begin{equation}
T_{\Delta L' \rightarrow P}=\frac{\pi
P_0}{\lambda}=\frac{\pi}{1064\,10^{-9}}\,{\rm W}/{\rm m}.
\end{equation}

\subsubsection{Apparent length noise}

Now we can compute the apparent end-mirror motion (measured as an optical
path length difference) due to \shot dividing the noise spectral density
from \eq{eq:shotnoise} by the transfer function:
\begin{equation}
\sqrt{S_{\Delta
L'}(f)}=\frac{\sqrt{\frac{2\pi\,\hbar\,c~}{\lambda}P_0}}{\frac{\pi
P_0}{\lambda}}=\sqrt{2 \frac{\hbar c \lambda}{\pi P_0}}=1.463 10^{-16}\,{\rm
m}/\sqrt{\rm Hz}.
\end{equation}
As expected we get exactly $0.25$ times this value (i.e.~$3.658504314e-17$)
using \Finesse with:
\begin{finesse}
fsig sig1 mE 1 0
fsig sig2 mN 1 180
pdS1 south1 1 max n10
scale meter
\end{finesse}

With two detectors, one in the west and one in the south port we can expect
to have a better sensitivity by a factor of $\sqrt{2}$.  The detected signal
is twice the signal detected in a single port.  Also, the detected total DC
power increases by a factor of 2. Thus we expect the signals-to-\shot to
increase by a factor of $\sqrt{2}$. For our example we get $\sqrt{S_{\Delta
L}(f)}=\sqrt{\frac{\hbar c \lambda}{\pi P_0}}=1.034
10^{-16}$\,m/$\sqrt{\mathrm{Hz}}$.

\subsection{Simple Michelson interferometer on a dark fringe}

The following section demonstrates how to compute the shot-noise-limited sensitivity
for a Michelson interferometer on the dark fringe. However, since phase
modulation is employed the results based on the Schottky formula alone
are not correct \cite{meers91, niebauer91}. The following calculation is meant as an exercise 
to show that -- when the effects of the modulation are neglected -- the shot-noise-linited
sensitivity at the dark fringe is exactly as for a Michelson interferometer on a 
half fringe with two detectors.

Again we start with the light fields entering the arms which are now given by:
\begin{equation}
\begin{array}{l}
E_{\rm N\,in}=\frac{1}{\sqrt{2}}E_{\rm in}\exp{(\I\frac{\pi}{2})}, 
\quad\mathrm{and}\\
E_{\rm E\,in}=\frac{1}{\sqrt{2}}E_{\rm in}\exp{(\I\frac{\pi}{2})}.
\end{array}
\end{equation}
The light reflected by the end mirrors (with $r=1$) can then be written as:
\begin{equation}
\begin{array}{lcl}
E_{\rm N}&=&\frac{1}{\sqrt{2}}E_{\rm in} \exp{(\I \frac{\pi}{2})}
\left(1+ a_s \I \left(\exp\left(\I \omega_s t\right) + a_s \exp\left(-\I
\omega_s t\right)\right)\right)\\
&=&\frac{1}{\sqrt{2}}E_{\rm in} \left(\I - 2 a_s \cos\left(\omega_s
t\right)\right),\\
 & & \\
E_{\rm E}&=&\frac{1}{\sqrt{2}}E_{\rm in} \exp\left(\I \frac{\pi}{2}\right)
\left(1- a_s \I \left(\exp\left(\I \omega_s t\right) + a_s \exp\left(-\I
\omega_s t\right)\right)\right)\\
&=&\frac{1}{\sqrt{2}}E_{\rm in} \left(\I + 2 a_s \cos\left(\omega_s
t\right)\right).
\end{array}
\end{equation}
And hence in the south port we get:
\begin{equation}
{\renewcommand{\arraystretch}{1.5}
\begin{array}{lcl}
E_{\rm S}&=&\frac{1}{\sqrt{2}}\exp{(\I\frac{\pi}{2})}E_{\rm
N}+\frac{1}{\sqrt{2}}\exp{(-\I\frac{\pi}{2})}E_{\rm E}\\
&=&-2 \I E_{\rm in}a_s \cos{(\omega_s t)}.
\end{array}}
\end{equation}

The photocurrent generated by this output field does not contain a signal
at the frequency $\omega_s$. Such a signal component can be produced with a
modulation-demodulation technique. In this example we can simply add another
pair of phase modulation sidebands and ignore how these are exactly
generated.  Let us assume we have symmetric modulation sidebands at
$\omega_m$ reaching the photodiode in the south port. We get:
\begin{align}
E_{\rm S} &= \I b\left(\exp\left(\I\omega_m t\right)+ \exp\left(-\I\omega_m
t\right)\right)-2 \I E_{\rm in}a_s \cos\left(\omega_s t\right)\\
&= 2 \I b \cos{(\omega_m t)} -2 \I E_{\rm in}a_s \cos{(\omega_s t)}.
\end{align}
The power detected by the diode in each output port is thus:
\begin{align}
I_{\rm out}&=4 b^2 \cos^2\left(\omega_m t\right)+ 4a_s^2 E^2_{\rm in}
\cos^2\left(\omega_s t\right)- 8 b a_s E_{\rm in}\cos\left(\omega_m
t\right)\cos\left(\omega_s t\right)\\
&=2 b^2 + 2a_s^2 P_0 - 8 b a_s \sqrt{P_0}\cos\left(\omega_m
t\right)\cos\left(\omega_s t\right) + O(2 \omega_m) +O(2 \omega_s).\\
\end{align}
A demodulation at $\omega_m$ gives a signal amplitude proportional to $4 b
a_s \sqrt{P_0}$.

An example calculation can be done with a modulation at $10$\,MHz:
\begin{finesse}
mod eom1 10M .1 1 pm n2 n3
\end{finesse}

Then we first check the field amplitudes and the DC power in the output port
(at node \cmd{n10}) with:
\begin{finesse}
pd dc n10
ad c 0 n10
ad b 10M n10
ad as 1 n10
\end{finesse}

We get:
\begin{finesse}
dc=0.0002158906915
c=2.929386532e-17
b=0.01038967496
as=0.9975015621
\end{finesse}
Thus the carrier power (\cmd{c}) is approximately zero, the signal sideband
amplitude around $1$ and the DC power is given by $2b^2$.

Using the Schottky formula, we expect to have a shot noise spectral density \emph{before the
demodulation} of
\begin{equation}
\begin{array}{lcl}
\sqrt{S_{P}(f)}&=&\sqrt{2 \frac{2\pi\hbar c}{\lambda} 2 b^2} /\sqrt{\rm
Hz}\\
&=&8.962\,10^{-12}~{\rm W}/\sqrt{\rm Hz}.
\end{array}
\end{equation}
The \cmd{shot} detector in \Finesse gives \cmd{8.978e-12}.  The transfer
function then computes as $0.041454868$ which is exactly $4 b a_s$.
Please note that in the presence of modulation sidebands this is not quite correct. 
However, in many cases it can be used as a good approximation.

\subsubsection{Signal-to-noise}

The transfer function for an optical path length difference is once more
obtained by multiplying the above result by $2\pi/\lambda$ and dividing by a
factor of 4:
\begin{equation}
T_{\Delta L' \rightarrow P}=\frac{2\pi}{\lambda} b \sqrt{P_0}.
\end{equation}

Now we have to propagate the shot-noise through the demodulator as well.
Please consider that this is \emph{not} done automatically by \Finesse even
if you use a detector like, for example, \cmd{pdS2}. We consider the
amplitude spectral density $\sqrt{S_{P}(f)}$: because we are only interested in the
DC signal after the demodulation we can approximate the white spectrum 
by two uncorrelated noise amplitudes at $\pm \omega_m$ with $\omega_m$ being the 
demodulation frequency. Thus the amplitude noise spectral density after demodulation can
be approximated as:
\begin{equation}
\left[\sqrt{S_{P}(f)}\cos(\omega_m t)\right]_{\vert_{DC}}\approx \left(A_1 \cos\left(-\omega_m t\right) + A_2 \cos\left(\omega_m t\right) \right)\cos(\omega_m t)
\end{equation}
with $A_1$ and $A_2$ as two uncorrelated amplitudes\footnote{It should be clear that in a realistic scenario the noise amplitudes cannot
be assumed to be always uncorrelated.} of the magnitude $\sqrt{S_{P}(f)}$.
The right side of above equation yields:
\begin{equation}
\frac{1}{2}(A_1+A_2)=\frac{1}{\sqrt{2}}A_1=\frac{1}{\sqrt{2}}\sqrt{S_{P}(f)}=\sqrt{\frac{2\pi\hbar c}{\lambda} 2 b^2} /\sqrt{\rm Hz}
\end{equation}

This yields an apparent mirror motion of:
\begin{equation}
\begin{array}{lcl}
\sqrt{S_{\Delta L}(f)}&=&\frac{\sqrt{\frac{2\pi \hbar c}{\lambda} 2
b^2}}{\frac{2\pi}{\lambda}b\sqrt{P_0}}=\sqrt{\frac{\hbar c \lambda}{\pi
P_0}}\\
&=&1.03410 10^{-16}\,{\rm m}/\sqrt{\rm Hz},
\end{array}
\end{equation}
which is the same as in the case of the Michelson interferometer at a half
fringe with two detectors.



%% file: higher_order_modes.tex
\chapter{Higher-order spatial modes, the paraxial approximation}
\label{sec:HGmodes}

The analysis using a \pwa as described in the previous chapter allows one to
perform a large variety of simulations. Some analysis tasks, however,
include the beam shape and position, i.e.~the properties of the field
transverse to the optical axis.  The effects of misaligned components, for
example, can only be computed if beam shape and position are taken into
account.

The following sections describe a straightforward extension of the previous
chapter's analysis using transverse electromagnetic modes (TEM).  The
expression \emph{mode} in connection with laser light usually refers to the
eigenmodes of a cavity. Here, one distinguishes between longitudinal modes
(along the optical axis) and transverse modes, the spatial distribution of
the light beam perpendicular to the optical axis.  In the following, we are
looking at the spatial properties of a laser beam. A \emph{beam} in this
sense is a light field for which the power is confined to a small volume
around one axis (the optical axis, always denoted by $z$).

\section{\Finesse with Hermite-Gaussian beams}

By default \Finesse performs simulations using the plane-wave approximation.
If the input file contains commands which refer explicitly to Gaussian
beams, like, for example, \cmd{gauss}, \Finesse will use Hermite-Gaussian
beams instead. This is henceforth called the 'Hermite-Gauss extension'.

The command \cmd{maxtem} is used to switch manually between plane-waves and 
Gaussian beams and to set the maximum order for higher order TEM modes:\\
\cmd{maxtem off} switches to plane waves,\\
\cmd{maxtem order} with \cmd{order} a integer between 0 and 100 switches to Hermite-Gauss beams.
The simulation includes higher order modes \M{nm} with $n+m\leq$\cmd{order}.

The Hermite-Gauss extension of \Finesse is a powerful tool. However, it requires
some expert knowledge about the physics and its numerical representation.
The following sections provide the mathematical description of the 
Gaussian beams as it is used in \Finesse. Please see the extra
section \ref{sec:hg_commands} in the syntax reference for a description of commands relevant
to Hermite-Gauss beams.

\section{Gaussian beams}


Imagine an electric field that can be described as a sum of the
different frequency components and of the different spatial modes:
\begin{equation}\label{eq:HG_intro1}
E(t,x,y,z)~=~\sum_{j}~\sum_{n,m}~a_{jnm}~u_{nm}(x,y,z)~\mEx{\I(\w_j \T -k_j
z)},
\end{equation}
with $u_{nm}$ describing the spatial properties of the beam and $a_{jnm}$ as
complex amplitude factors ($\w_j$ is the angular frequency of the light field
and $k_j=\w_j/c$). 

Please note that in this case the amplitude coefficients $a_{jnm}$ are not
equivalent to the field amplitudes as computed by \Finesse. There is a
difference in phase as a consequence of the chosen implementation of the
Gouy phase; see \Sec{sec:gouy} for details. In the following
the amplitudes $a_{jnm}$ refer to the coefficients as defined in
\eq{eq:HG_intro1}. The amplitude coefficients computed and stored by
\Finesse will be denoted $b_{jnm}$. 

For simplicity we restrict the following description to a single frequency
component at one moment in time ($t=0$):
\begin{equation}
\label{eq:HG_intro2} 
E(x,y,z)~=~\mEx{-\I k z}~\sum_{n,m}~a_{nm}~u_{nm}(x,y,z).
\end{equation}

A useful mathematical model for describing spatial properties of light
fields in laser interferometers are the Hermite-Gauss modes, which are the
eigenmodes of a general spherical cavity (an optical cavity with spherical
mirrors) and represent an exact solution of the \emph{paraxial wave
equation}; see \app{sec:pax}.

The following section provides an introduction to \HG modes, including some
useful formulae and the description of the implementation of \HG modes in
\Finesse. Please note that the paraxial approximation requires a well
aligned and well mode-matched interferometer; \Sec{sec:limits_paraxial}
gives a short overview of the limits of this approximation.

\begin{figure}[htb]
\begin{center}
\IG [viewport=127 418 474 605,scale=1.1] {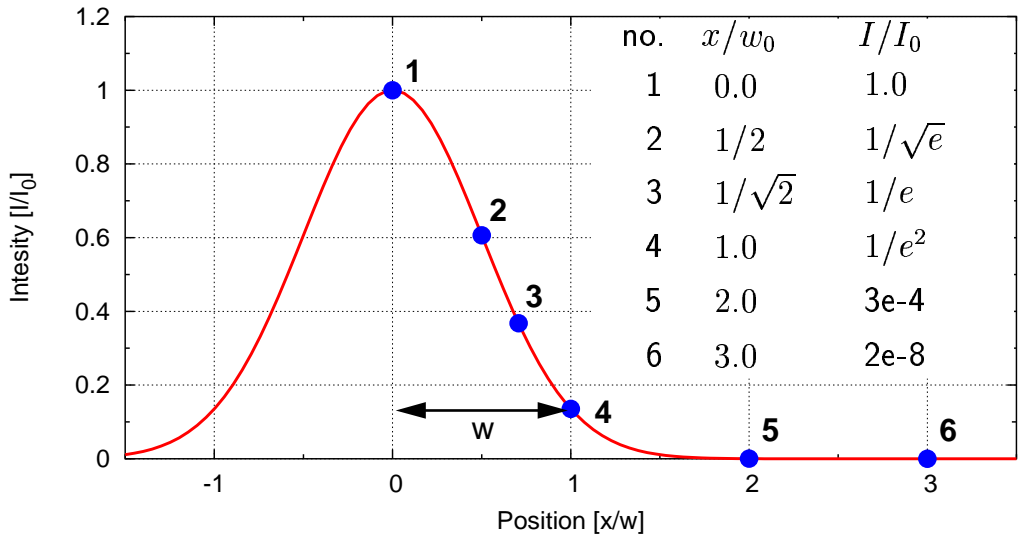} \\
\end{center}
\caption[Gaussian beam profile]
{One dimensional cross-section of a Gaussian beam. The width of the beam
is given by the radius $w$ at which the intensity is $1/e^2$ of the maximum
intensity.}
\label{fig:gauss_profile}
\end{figure}

The \emph{Gaussian beam} often describes a simple laser beam to a good
approximation. The Gaussian beam as such is the lowest-order Hermite-Gauss
mode $u_{00}$ which will be discussed later.  The electric field (again
assuming a single frequency and $t=0$) is given as:
\begin{equation}\label{eq:a_hg_base}
\begin{array}{lll}
E(x,y,z)&=&E_0 u_{00} \mEx{-\I kz}\\
&=&E_0\left(\frac{1}{R_C(z)}-\I\frac{\lambda}{\pi
w^2(z)}\right)\cdot\mEx{-\I
k\frac{x^2+y^2}{2R_C(z)}-\frac{x^2+y^2}{w^2(z)}-\I kz}.
\end{array}
\end{equation}
The shape of a Gaussian beam is quite simple: the beam has a circular
cross-section, and the radial intensity profile of a beam with total power
$P$ is given by:
\begin{equation}\label{eq:gauss_profile}
I(r)=\frac{2P}{\pi w^2(z)}\mEx{-2r^2/w^2},
\end{equation}
with $w$ the \emph{spot size}, defined as the \emph{radius} at which the
intensity is $1/e^2$ times the maximum intensity $I(0)$. This is a Gaussian
distribution, hence the name \emph{Gaussian beam}.  \fig{fig:gauss_profile}
shows a cross-section through a Gaussian beam and the radial intensity for
different positions with respect to the beam position and beam size.

Such a beam profile (for a beam with a given wavelength $\lambda$) can be
completely determined by two parameters: the size of the minimum spot size
$w_0$ (called \emph{beam waist}) and the position $z_0$ of the beam waist
along the $z$-axis. 

To characterise a Gaussian beam, some useful parameters can be derived from
$w_0$ and $z_0$.  A Gaussian beam can be divided into two different sections
along the $z$-axis: a \emph{near field} (a region around the beam waist) and
a \emph{far field} (far away from the waist). The length of the near-field
region is approximately given by the \emph{Rayleigh range} $\zr$.  The
Rayleigh range and the spot size are related by the following expression:
\begin{equation}
\zr=\frac{\pi w_0^2}{\lambda}.
\end{equation}
With the Rayleigh range and the location of the beam waist, we can write 
the following useful expression:
\begin{equation}
w(z)=w_0\sqrt{1+\left(\frac{z-z_0}{\zr}\right)^2}.
\end{equation}
This equation gives the size of the beam along the $z$-axis. In the
far-field regime ($z\gg \zr,z_0$), it can be approximated by a linear
equation:
\begin{equation}
w(z)\approx w_0\frac{z}{\zr}=\frac{z \lambda}{\pi \omega_0}.
\end{equation}

The angle $\Theta$ between the $z$-axis and $w(z)$ in the far field is
called the \emph{diffraction angle}\footnote{Also known as the
\emph{far-field angle} or the \emph{divergence} of the beam.} and is defined
as:
\begin{equation}
\Theta=\arctan\left(\frac{w_0}{\zr}\right)=\arctan\left(\frac{\lambda}{\pi
w_0}\right)\,\,\approx\frac{w_0}{\zr}.
\end{equation}

Another useful parameter is the \emph{radius of curvature} of the wavefront
at a given point $z$. The radius of curvature describes the curvature of the
`phase front' of the electromagnetic wave (a surface across the beam with
equal phase) at the position $z$. We obtain for the radius of curvature as a
fuction of $z$:
\begin{equation}
R_C(z)=z-z_0+\frac{\zr^2}{z-z_0}.
\end{equation}
For the radius of curvature we also find:
\begin{equation}
{\renewcommand{\arraystretch}{1.5}
\begin{array}{lll}
R_C\approx \infty,&  z-z_0\ll\zr\ & \qquad \mbox{\rm (beam waist)}\\
R_C\approx z, & z\gg\zr,~ z_0 & \qquad \mbox{\rm (far field)}\\
R_C=2\zr, & z-z_0=\zr & \qquad \mbox{\rm (maximum curvature)}\\
\end{array}}
\end{equation}

\section{Higher order \HG modes} 

The \HG modes are usually given in their orthonormal form as:
\begin{equation}\label{eq:HG_mode2}
{\renewcommand{\arraystretch}{1.5}
\begin{array}{lcl}
u_{\rm
nm}(x,y,z)&=&\left(2^{n+m-1}n!m!\pi\right)^{-1/2}
\frac{1}{w(z)}~\mEx{\I(n+m+1)\Psi(z)}\\
&&\times\qquad
H_n\left(\frac{\sqrt{2}x}{w(z)}\right)H_m\left(\frac{\sqrt{2}y}{w(z)}\right)
\mEx{-\I\frac{k(x^2+y^2)}{2R_C(z)}-\frac{x^2+y^2}{w^2(z)}},
\end{array}}
\end{equation}
with $n$, $m$ being the \emph{mode numbers} or \emph{mode indices}. In this
case $n$ refers to the modes in the $y$-$z$ plane (saggital) and $m$ to the
$x$-$z$ plane (tangential). The following functions are used in the equation
above:
\begin{itemize}
\item{$H_n(x)$: Hermite polynomial of the order $n$ (unnormalised), see
\app{sec:h_poly}},
\item{$w(z)$: beam radius or spot size},
\item{$\roc(z)$: radius of curvature of the phase front},
\item{$\Psi(z)$: Gouy phase}.
\end{itemize}
The definition of $\Psi(z)$ and some explanation is given in
\Sec{sec:gouy}.  The \HG modes can also be given in a very compact form
using the Gaussian beam parameter $q$; see below.

The \HG modes as given above are orthonormal and thus:
\begin{equation}\label{eq:HG_mode3}
\int\!\!\!\int\!dxdy~u_{n m}u^*_{n' m'}~=~\delta_{n n'}\delta_{m m'}.
\end{equation}
Therefore the power of a beam, as given by \eq{eq:HG_intro2}, being detected
on a single-element photodetector (provided that the area of the detector
is large with respect to the beam) can be computed as
\begin{equation}
P=\sum_{n,m} a_{nm}a_{nm}^*.
\end{equation}
Or for a beam with several frequency components (compare with
\eq{eq:dc_det}):
\begin{equation}\label{eq:hg_dc_det}
P=\sum_{n,m}\sum\limits_i\sum\limits_j a_{inm}a_{jnm}^*\quad\mbox{with}\quad
\{i,j~|~i,j\in\{0,\dots,N\}~\wedge~\w_i=\w_j\}.
\end{equation}

The $x$ and $y$ dependencies can be separated so that:
\begin{equation}\label{eq:HG_mode4}
u_{n m}(x,y,z)~=~u_{n}(x,z)u_{m}(y,z).
\end{equation}


\subsection{Gaussian beam parameter}
\label{sec:beam_param}

A set of \HG modes $u_{nm}$ can be described by one constant beam parameter
$q_0$: the \emph{Gaussian beam parameter}.  It is defined as:
\begin{equation}
\frac{1}{q(z)}=\frac{1}{R_C(z)}-\I\frac{\lambda}{\pi w^2(z)},
\end{equation}
and can also be written as:
\begin{equation}
q(z)=\I\zr +z-z_0=q_0+z-z_0,\qquad\mbox{where}\qquad q_0=\I\zr.
\end{equation}
The beam parameter $q_0$ is in general changed when the beam interacts with
a spherical surface.

Using this parameter \eq{eq:a_hg_base} can be rewritten as:
\begin{equation}
u(x,y,z)=\frac{1}{q(z)}\mEx{-\I k\frac{x^2+y^2}{2q(z)}}.
\label{eq:a_hg_qbase}
\end{equation}

The complete set of solutions as given in \eq{eq:HG_mode2} can now be
written as\footnote{Please note that this formula from~\cite{siegman} is
very compact. Since the parameter $q$ is a complex number, the expression
contains at least two complex square roots. The complex square root requires
a different algebra than the standard square root for real numbers.
Especially the third and fourth factors \emph{cannot} be simplified in any
obvious way i.e.: $\left(\frac{q_0}{q(z)}\right)^{1/2}
\left(\frac{q_0q^*(z)}{q_0^*q(z)}\right)^{n/2}
\neq\left(\frac{q_0^{n+1}{q^*}^n(z)}{q^{n+1}(z){q_0^*}^n}\right)^{1/2}$ !}:
\begin{equation}
u_{\rm nm}(x,y,z)=u_{\rm n}(x,z)u_{\rm m}(y,z),
\label{eq:a_hg_qmode}
\end{equation}
with
\begin{equation}
u_{\rm
n}(x,z)=\left(\frac{2}{\pi}\right)^{1/4}\left(\frac{1}{2^nn!w_0}\right)^{1/2}
\left(\frac{q_0}{q(z)}\right)^{1/2}\left(\frac{q_0~q^*(z)}{q_0^*~q(z)}\right)^{n/2}
H_n\left(\frac{\sqrt{2}x}{w(z)}\right)
\mEx{-\I\frac{kx^2}{2q(z)}}
\end{equation}
again, $H_n(x)$ represents a Hermite polynomial of order n.

The beam size and radius of curvature can also be written in terms of the
beam parameter $q$:
\begin{equation}
w^2(z)=\frac{\lambda}{\pi}\frac{|q|^2}{\myIm{q}},
\end{equation}
and
\begin{equation}
R_C(z)=\frac{|q|^2}{\myRe{q}}.
\end{equation}

It is clear that when using \HG modes one has to choose a base system of
beam parameters for describing the spatial properties. In \Finesse this
means a beam parameter has to be set for every node. Much of this task
is automated; see \Sec{sec:trace}.  In my experience the quality of the
simulations and the correctness of the results depend critically on the
choice of these beam parameters. One might argue that the choice of the base
system should not alter the result.  This is correct but there is a
practical limitation: the number of modes having non-neglible power might
become very large if the beam parameters are not optimised, so that in
practise to achieve a sensible computation time a good set of beam
parameters must be used. \Sec{sec:limits_paraxial} gives some advice how to
ensure that the simulation does not fail because of this limitation.

\subsection{Tangential and sagittal plane}

If the interferometer is confined to a plane as in \Finesse, it is
convenient to use projections of the three-dimensional description into two
planes: the tangential plane, defined as the $x$-$z$ plane and the sagittal
plane as given by $y$ and $z$. 

The beam parameter can then be split into two beam parameters: $q_{\rm s}$
for the sagittal plane and $q_{\rm t}$ for the tangential plane so that the
\HG modes can be written as:
\begin{equation}
u_{nm}(x,y)=u_n(x,q_{\rm t})~u_m(y,q_{\rm s}).
\end{equation}
Remember that these \HG modes form a base system. This means one can use
the separation in sagittal and tangential planes even if the analysed
optical system does not show this special type of asymmetry.  This
separation is very useful in simplifying the mathematics.

In the following, the term \emph{beam parameter} generally refers to a
simple $q_0$ but all the results can also be applied directly to a pair
of parameters $q_s, q_t$.

\subsection{Gouy phase shift}
\label{sec:gouy}

The introduction of spatial beam properties using \HG modes gives rise to
an extra longitudinal phase lag, this is the \emph{Gouy phase}. Compared
to a plane wave, the \HG modes have a slightly slower phase velocity,
especially close to the waist. The Gouy phase can be written as:
\begin{equation}
\Psi(z)=\arctan\left(\frac{z-z_0}{\zr}\right),
\end{equation}
or, using the Gaussian beam parameter:
\begin{equation}
\Psi(z)=\arctan\left(\frac{\myRe{q}}{\myIm{q}}\right).
\end{equation}

Compared to a plane wave, the phase lag $\OPh$ of a \HG mode is:
\begin{equation}
\OPh=(n+m+1)\Psi(z).
\end{equation}
With an astigmatic beam, i.e.~different beam parameters in the tangential
and saggital planes this becomes:
\begin{equation}
\OPh=\left(n+\frac{1}{2}\right)\Psi_t(z)+\left(m+\frac{1}{2}\right)\Psi_s(z),
\end{equation}
with 
\begin{equation}
\Psi_t(z)=\arctan\left(\frac{\myRe{q_t}}{\myIm{q_t}}\right),
\end{equation}
as the Gouy phase in the tangential plane (and $\Psi_s$ similarly the Gouy phase in the sagittal plane).

The command {\Co phase} can be used to specify how the Gouy phase is used
within the \Finesse simulation, see \Sec{sec:hg_commands}.

\subsection{ABCD matrices}
\label{sec:abcd}
The transformation of the beam parameter can be performed by the 
ABCD matrix-formalism \cite{siegman}. When a beam passes a mirror, \bs, lens
or free space, a beam parameter $q_1$ is transformed to $q_2$. This
transformation can be described by four real coefficients like so:
\begin{equation}
\frac{q_2}{n_2}=\frac{A\frac{q_1}{n_1}+B}{C\frac{q_1}{n_1}+D},
\end{equation}
with the coefficient matrix,
\begin{equation}
M=\left( 
\begin{array}{cc}
 A& B \\
 C& D 
\end{array}
\right),
\end{equation}
and $n_1$ being the index of refraction at the beam segment defined by
$q_1$ and $n_2$ the index of refraction at the beam segment described by
$q_2$.

The ABCD matrices for the optical components used by \Finesse are given
below, for the sagittal and tangential plane respectively.

\paragraph{Transmission through a mirror:}

A mirror in this context is a single, partly reflecting surface with an
angle of incidence of $90^\circ$.  The transmission is described by:\\
\begin{minipage}{\textwidth}
\begin{equation}\label{eq:abcd-mi1}
\rule{0pt}{1.3cm}M=\left( 
\begin{array}{cc}
1 & 0\\
\frac{n_2-n_1}{R_{\rm C}}& 1 
\end{array}
\right)\end{equation}
\hspace{7.75cm}\IG [viewport=0 -20 4 0, scale=0.5] {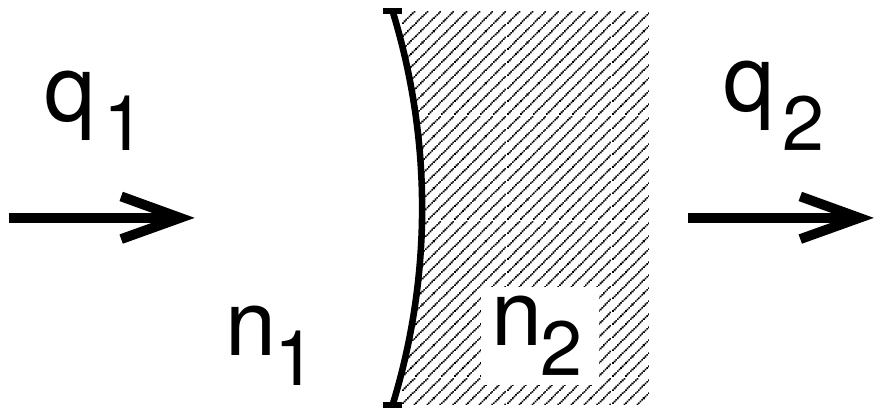}\end{minipage}\\
with $R_{\rm C}$ being the radius of curvature of the spherical surface. The
sign of the radius is defined such that $R_{\rm C}$ is negative if the
centre of the sphere is located in the direction of propagation. The
curvature shown above (in \eq{eq:abcd-mi1}), for example, is described by a
positive radius.

The matrix for the transmission in the opposite direction of propagation is
identical.

\paragraph{Reflection at a mirror:}

The matrix for reflection is given by:\\
\begin{minipage}{\textwidth}
\begin{equation}\label{eq:abcd_mr}
\rule{0pt}{1.3cm}
M=\left( 
\begin{array}{cc}
1 & 0\\
-\frac{2 n_1}{R_{\rm C}}& 1 
\end{array}
\right)
\end{equation}
\hspace{7.5cm}\IG [viewport=0 -15 4 5, scale=0.5] {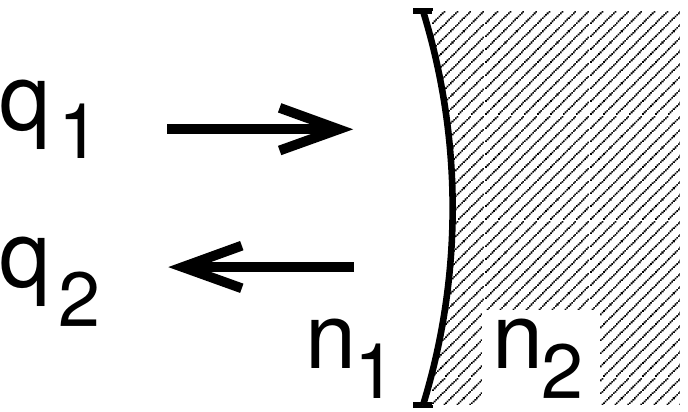}\end{minipage}\\
The reflection at the back surface can be described by the same type of
matrix by setting $C=2n_2/\roc$.

\paragraph{Transmission through a \bs:}

A \bs is understood as a single surface with an arbitrary angle of
incidence $\alpha_1$.  The matrices for transmission and reflection are
different for the sagittal and tangential planes ($M_{\rm s}$ and $M_{\rm
t}$):\\
\begin{minipage}{\textwidth}
\begin{equation}
\rule{0pt}{1.3cm}
\begin{array}{l}
M_{\rm t}=\left( 
\begin{array}{cc}
\frac{\mCos{\alpha_2}}{\mCos{\alpha_1}} & 0\\
\frac{\Delta n}{R_{\rm C}} & \frac{\mCos{\alpha_1}}{\mCos{\alpha_2}}
\end{array}
\right)\\
\ \\
M_{\rm s}=\left( 
\begin{array}{cc}
1 & 0\\
\frac{\Delta n}{R_{\rm C}} &1
\end{array}
\right)
\end{array}
\end{equation}
\hspace{7cm}\IG [viewport=0 -50 4 -30, scale=0.5] {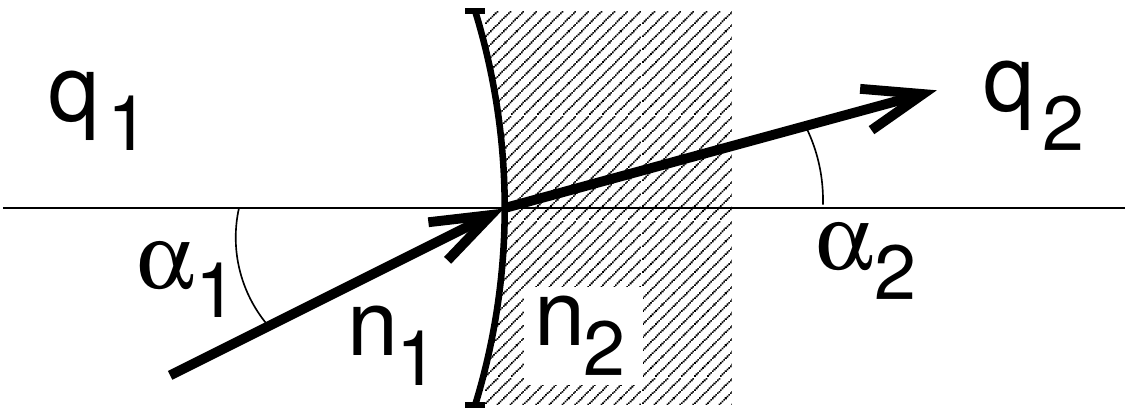} \end{minipage}\\
with $\alpha_2$ given by Snell's law:
\begin{equation}
n_1\mSin{\alpha_1}=n_2\mSin{\alpha_2},
\end{equation}
and $\Delta n$ by:
\begin{equation}
\Delta
n=\frac{n_2\mCos{\alpha_2}-n_1\mCos{\alpha_1}}{\mCos{\alpha_1}\mCos{\alpha_2}}.
\end{equation}

If the direction of propagation is reversed, the matrix for the sagittal
plane is identical and the matrix for the tangential plane can be obtained
by changing the coefficients A and D as follows:
\begin{equation}
\begin{array}{l}
A\longrightarrow1/A,\\
D\longrightarrow1/D.
\end{array}
\end{equation}

\paragraph{Reflection at a \bs:}

The reflection at the front surface of a \bs is given by:\\
\begin{minipage}{\textwidth}
\begin{equation}
\rule{0pt}{1.3cm}
\begin{array}{l}
M_{\rm t}=\left( 
\begin{array}{cc}
1 & 0\\
-\frac{2 n_1}{R_{\rm C} \mCos{\alpha_1}} & 1
\end{array}
\right)\\
\ \\
M_{\rm s}=\left( 
\begin{array}{cc}
1 & 0\\
-\frac{2 n_1 \mCos{\alpha_1}}{R_{\rm C} } & 1
\end{array}
\right)
\end{array}
\end{equation}
\hspace{7cm}\IG [viewport=0 -50 4 -30, scale=0.5] {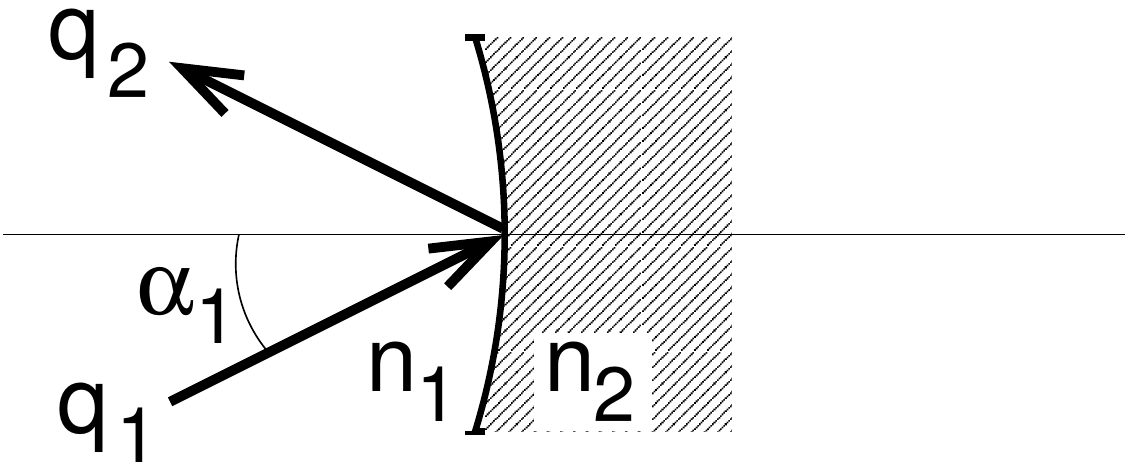}
\end{minipage} \\

To describe a reflection at the back surface the matrices have to be changed
as follows:
\begin{equation}
\begin{array}{l}
R_{\rm C}\longrightarrow-R_{\rm C},\\
n_1\longrightarrow n_2,\\
\alpha_1\longrightarrow-\alpha_2.
\end{array}
\end{equation}

\paragraph{Transmission through a thin lens:}

A thin lens transforms the beam parameter as follows:\\
\begin{minipage}{\textwidth}
\begin{equation}
\rule{0pt}{1.3cm}
M=\left( 
\begin{array}{cc}
1 & 0\\
-\frac{1}{f}& 1
\end{array}
\right)
\end{equation}
\hspace{7.85cm}\IG [viewport=0 -20 4 0, scale=0.5] {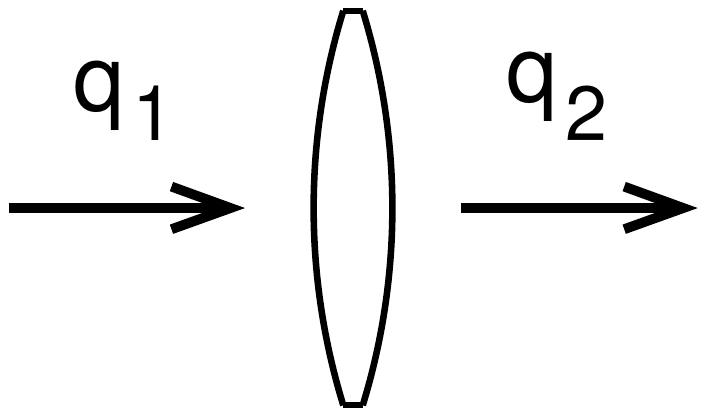} \end{minipage}\\
where $f$ is the focal length.  The matrix for the opposite direction of
propagation is identical.  Please note that a thin lens has to be surrounded
by `spaces' with index of refraction $n=1$.

\paragraph{Transmission through a free space:}

As mentioned above, the beam in free space can be described by one base
parameter $q_0$. In some cases it is convenient to use a similar matrix as
for the other components to describe the $z$-dependency of $q(z)=q_0+z$.  On
propagation through a free space of the length $L$ and index of refraction
$n$ the beam parameter is transformed as follows:

\begin{minipage}{\textwidth}
\begin{equation}
\rule{0pt}{1.3cm}
M=\left( 
\begin{array}{cc}
1 & \frac{L}{n}\\
0& 1
\end{array}
\right)
\end{equation}
\hspace{7cm}\IG [viewport=0 -40 4 -20, scale=0.5] {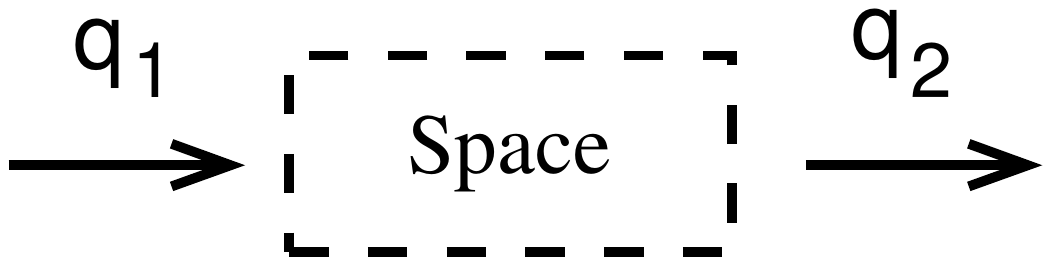} 
\end{minipage}\\
The matrix for the opposite direction of propagation is identical.

\paragraph{Reflection at a grating:}

Consider a curved diffraction grating with radius of curvature $R$, ruling
in the $y$-direction and grating spacing $d$ in the $x$-direction. An
incident beam striking the grating at an angle $\alpha$ from the normal in
the $x-z$ plane will be diffracted in $m$th order into angle $\phi_m$ in the
same plane given by the grating equation \eq{eq:grating-eq}.

The matrix for reflection from the grating in the tangential or $x-z$ plane
is given by:

\begin{minipage}{\textwidth}
\begin{equation}
\rule{0pt}{1.3cm}
\begin{array}{l}
M_{\rm t}=\left( 
\begin{array}{cc}
M & 0\\
-2/R_t &1/M\\
\end{array}
\right)\\
\ \\
M_{\rm s}=\left( 
\begin{array}{cc}
1 & 0\\
-2/R_s &1\\
\end{array}
\right)
\end{array}
\end{equation}
\hspace{9cm}\IG [viewport=0 -50 4 -30, scale=0.2] {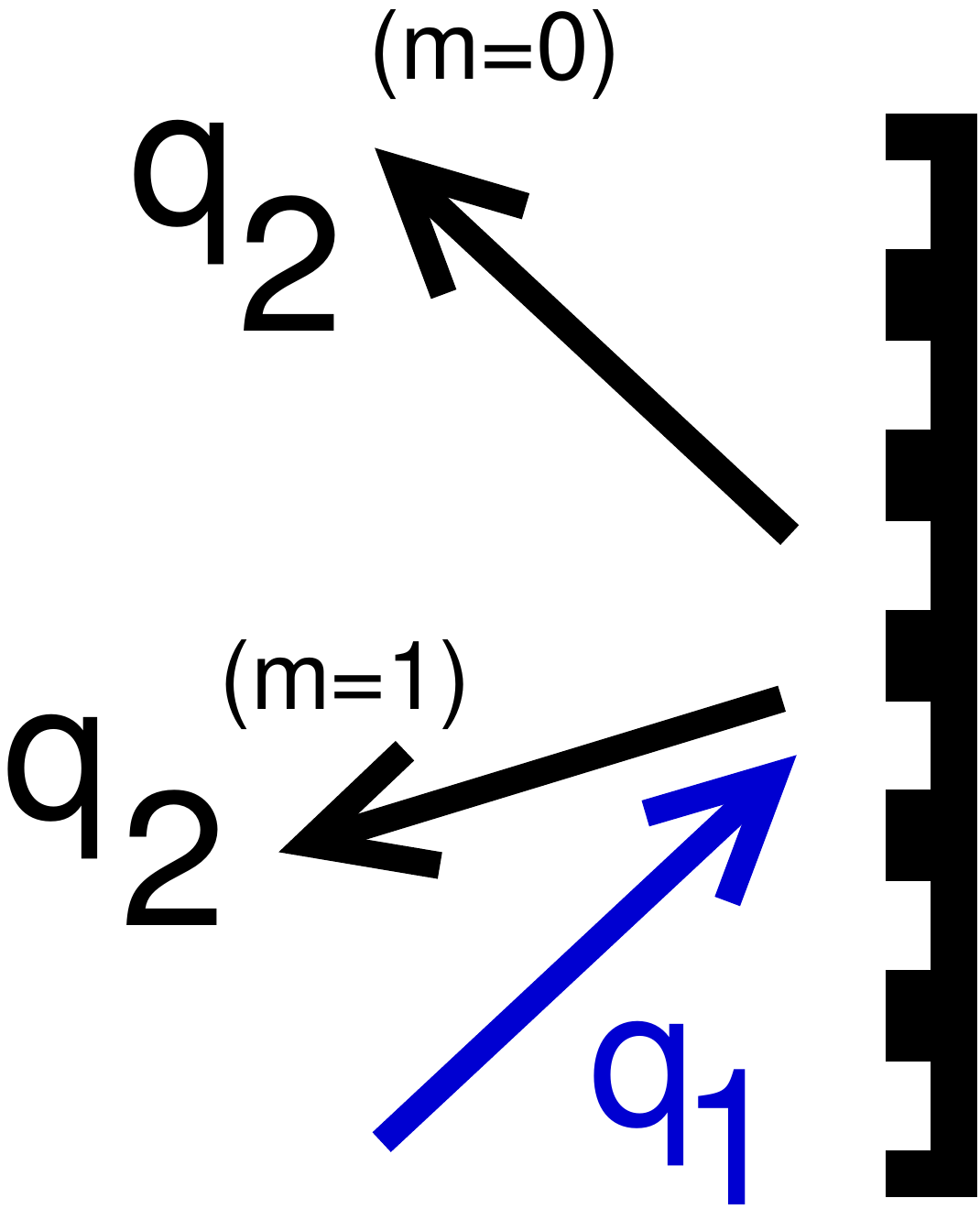} \end{minipage}\\
with $M$ given by:
\begin{equation}
M~=~\cos(\phi_m)/\cos(\alpha),
\end{equation}�
and effective radii as:
\begin{equation}
R_t~=~R~\frac{\cos(\alpha)\cos(\phi_m)}{\cos(\alpha)+\cos(\phi_m)},
\end{equation}
\begin{equation}
R_s~=~\frac{2\,R}{\cos(\alpha)+\cos(\phi_m)}.
\end{equation}

\section{Tracing the beam}\label{sec:trace}

As described in \Sec{sec:beam_param} one important step in the simulation
is the choice of beam parameters. In general the Gaussian beam parameter of
a \M{00} mode is changed at every optical surface (see \Sec{sec:abcd}).  In
other words, for each location inside the interferometer where field
amplitudes are to be computed a certain beam parameter has to be set for the
simulation.

A possible method to find reasonable beam parameters for every location in
the interferometer (every node in \Finesse) is to first set only some
specific beam parameters and then derive the remaining beam parameters from
these initial ones: usually it is sensible to assume that the beam at the
input can be properly described by the (hopefully known) beam parameter of
the laser's output mode. In addition, in most cavities the light fields can
be described safely by using cavity eigenmodes.  \Finesse provides two
commands, {\Co gauss} and {\Co cav}, that can be used for setting beam
parameters. The command {\Co cav} computes the eigenmodes of a (stable)
cavity and sets the respective beam parameters on every node that is part of
the cavity. Whereas the command {\Co gauss} is used to set a beam parameter
at one specific node, for example, at the input. The two types of commands are
executed as follows:
\begin{itemize}
\item Each {\Co cav} command checks if the respective cavity is stable and, if so, it
computes the eigenmode, and sets the respective beam parameters at every node that 
is inside the cavity.
\item The {\Co cav} commands are executed in the order they appear in the input file. If two cavities
include the same node, the later command overwrites the beam parameter set for this node
by previous {\Co cav} commands.
\item After the cavities have been all set, the {\Co gauss} commands are executed. If a {\Co
gauss} command is set to a node inside a cavity the beam parameter set by the {\Co cav}
command will be overwritten by the one given in the {\Co gauss} command.
\end{itemize}

After some beam parameters have been set by {\Co gauss} or {\Co cav} commands 
\Finesse uses a beam tracing algorithm to set beam parameters
for the remaining nodes.  `Trace' in this context means that a beam starting
at a node with an already known beam parameter is propagated through the
optical system and the beam parameter is transformed according to the
optical elements encountered.

The tracing algorithm in \Finesse works as follows:
\begin{itemize}
\item the starting point of the tracing can be set explicitly by the user with
the command {\Co startnode nodename} (the beam parameter of the respective
node has to be set with either {\Co gauss} or \cmd{cav}). If this command is not
used the starting point is automatically set:
\begin{itemize}
\item to the input node of the (first) cavity if the user has specified at
least one stable cavity;
\item to the node given in the first \cmd{gauss} command, if the user
specified at least one Gaussian parameter but no cavity;
\item to the node of the first laser if the user did not specify any beam
parameter. In addition, \Finesse sets the beam parameter of that input node
to a default beam parameter: $q=\I (2\,{\rm mm})^2 \pi/\lambda\, $ 
(i.e.~$w_0=2$\,mm, $z_0=0$\,mm).
\end{itemize}
\item from the starting point the beam is traced through the full
interferometer simply by following all possible paths successively. This is
done by moving from the start node to a connecting component then to the
next node, to the next component and so on.  At every optical element along
the path the beam parameter is transformed according to the ABCD matrix of
the element (see below). If more than one possibility exists (for example
at a beam splitter) the various paths are followed one after the other. Each
path ends, i.e.~is considered to be traced completely, when an already
encountered node or a `dump' node is found.

\item Every time a new node is found for which no beam parameter has been
yet set the current beam parameter is set for that node.

\item If a new node is found that already has a beam parameter (for example,
by the user with a {\Co gauss} command) the current beam parameter is
dropped and the parameter of the node is kept and used for further tracing
along that path instead. In this case the tracing algorithm makes sure that
no such beam parameter change occurs inside space components\footnote{It is
important for the chosen implementation of the Gouy phase (see
\Sec{HGmatrix}) that the beam parameter for both nodes of a space component
refer to the same beam waist.}.

\item When all paths have been traced completely, the number of nodes found
is compared to the total number of nodes of the setup and an error is
generated if these numbers do not match.
\end{itemize}

During the simulation, if a length, radius of curvature, or focal length is
changed, the optimum set of base parameters changes. 
When \Finesse detects a change in one of these parameters is automatically recompute 
the best beam parameters for each data
point. This will slow down the simulation a little but in almost all cases
it yields much better results. 
You can use the command
{\Co retrace} to force \Finesse to recompute beam parameters for each data
point.  Or you can force it to switch retracing off in all cases, using the command 
\cmd{retrace off}.

\Finesse can provide plenty of information about the tracing and the
resulting beam parameters. The command {\Co trace} can be used to set the
verbosity of \Finesse's tracing algorithm or, more generally, the verbosity
of the \HG mode; see the table on page \pageref{tab:trace} in the syntax
reference.

\section{Interferometer matrix with \HG modes}
\label{HGmatrix}

In the plane-wave analysis, a laser beam was described in general by the sum
of various frequency components of its electric field:
\begin{equation}
E(t,z)~=~\sum_{j}a_j~\mExB{\I(\w_j \T - k_j z)}.
\end{equation} 
Now, the geometric shape of the beam is included by describing each
frequency component by a sum of \HG modes:
\begin{equation}\label{eq:HG_intro}
E(t,x,y,z)~=~\sum_{j}~\sum_{n,m}~a_{jnm}~u_{nm}(x,y)~\mEx{\I(\w_j \T -k_j
z)}.
\end{equation}
The shape of such a beam does not change along the $z$-axis (in the
paraxial approximation). More precisely, the spot size and the position of
the maximum intensity with respect to the $z$-axis may change, but the
relative intensity distribution across the beam does not change its shape. 

Each part of the sum may be treated as an independent field that can be
described using the equation for the \pwa with only two exceptions:
\begin{itemize}
\item the propagation through free space has to include the Gouy phase
shift, and
\item upon reflection or transmission at a mirror or \bs the different
\HG modes may be coupled (see below).
\end{itemize}

The Gouy phase shift can be included into the simulation in several ways.
For reasons of flexibility is has been included in \Finesse as a phase shift
of the component space. The beam trace algorithm has been designed to set
the beam parameters of a space component so that at both nodes the beam
parameter gives the same Gouy phases. Therefore it is possible to associate
the component with a known phase delay. The amplitude of a field propagating
through a space is thus given by:
\begin{equation}
b_{\rm out}~=b_{\rm in} \mEx{\I\Delta\w n_r L/c-
\left(\frac{1}{2}+n\right)\Psi_x+\left(\frac{1}{2}+m\right)\Psi_y},
\end{equation}
(compare to \eq{eq:pwspace}).


This means the Gouy phases are stored explicitly in the amplitude
coefficients.  Therefore, the amplitudes $b_{\rm in/out}$ are not equivalent
to these $a_{jnm}$ in \eq{eq:HG_intro} or \eq{eq:HG_intro1}.  In fact, the
field amplitude is given in \Finesse (for one point in space, and $t=0$) as:
\begin{equation}\label{eq:myfield}
E(t,x,y,z)~=~\sum_{j}~\sum_{n,m}~b_{jnm}~u_{n}(x)
u_{m}(y)~\mEx{-\I\left(\frac{1}{2}+n\right)\Psi_t)} 
~\mEx{-\I\left(\frac{1}{2}+m\right)\Psi_s)},
\end{equation}
with 
\begin{equation}\label{eq:myfield2}
\Psi_t=\arctan\left(\frac{\myRe{q_t}}{\myIm{q_t}}\right), \qquad
\Psi_s=\arctan\left(\frac{\myRe{q_s}}{\myIm{q_s}}\right).
\end{equation}
This formula is used, for example, with the {\Co beam} detector.

Also, changing from one TEM base system to another it is necessary to turn
back the Gouy phase with respect to the old beam parameter and add the Gouy
phase with respect to the new beam parameter.  This is required because the
coupling coefficients used in the computation in \Sec{sec:knmnm} were
derived from the field description given by \eq{eq:HG_intro1} for which the
Gouy phase is not stored in the amplitude coefficients but implicitly given
by the spatial distribution. 

\section{Coupling of \HG modes}

The following is based on the work of F.~Bayer-Helms~\cite{bayer}. I later
discovered that there exists a very good description of coupling
coefficients by J.~Y.~Vinet~\cite{vpb}.

Let us assume two different cavities with different sets of eigenmodes. The
first set is characterised by the beam parameter $q_1$ and the second by
the parameter $q_2$. A beam with all power in the fundamental mode
\M{00}$(q_1)$ leaves the first cavity and is injected into the second.
Here, two `mis-configurations' are possible:
\begin{itemize}
\item if the optical axes of the beam and the second cavity do not overlap
perfectly, the setup is called \emph{misaligned},
\item if the beam size or shape at the second cavity does not
match the beam shape and size of the (resonant) fundamental eigenmode 
($q_1(z_{\rm cav})\neq q_2(z_{\rm cav})$), the beam is then not
\emph{mode-matched} to the second cavity, i.e.~there is a \emph{mode
mismatch}.
\end{itemize}

The above mis-configurations can be used in the context of simple beam
segments.  In the simulation, the beam parameter for the input light is
specified by the user.  Ideally, the ABCD matrices allow one to trace a beam
through the optical system by computing the proper beam parameter for each
beam segment.  In this case, the basis system of \HG modes is transformed
in the same way as the beam so that the modes are \emph{not coupled}.

For example, an input beam described by the beam parameter $q_1$ is passed
through several optical components, and at each component the beam parameter
is transformed according to the respective ABCD matrix. Thus, the electric
field in each beam segment is described by \HG modes based on different
beam parameters, but the relative power between the \HG modes with
different mode numbers remains constant, i.e.~a beam in a \M{00} mode is
described as a pure \M{00} mode throughout the full system. 

In practice, it is usually impossible to compute proper beam parameters for
each beam segment as above, especially when the beam passes a certain
segment more than once. The most simple example is the reflection at a
spherical mirror. Let the input beam be described by $q_1$. From
\eq{eq:abcd_mr} we know that the proper beam parameter of the reflected
beam is: 
\begin{equation}
q_2=\frac{q_1}{-2q_1/R_{\rm C}+1},
\end{equation}
with $R_{\rm C}$ being the radius of curvature of the mirror. In general, we
get $q_1\neq q_2$ and thus two different `proper' beam parameters for the
same beam segment.  Only one special radius of curvature would result in
matched beam parameters ($q_1=q_2$).

\subsection{Coupling coefficients for TEM modes}
\label{sec:knmnm}

The \HG modes are coupled whenever a beam is not matched to a cavity or to
a beam segment or if the beam and the segment are misaligned. In this case,
the beam has to be described using the parameters of the beam segment (beam
parameter and optical axis).  This is always possible (provided that the
paraxial approximation holds) because each set of \HG modes (defined by
the beam parameter at a position $z$) forms a complete set. Such a change of
the basis system results in a different distribution of light power in the
(new) \HG modes and can be expressed by coupling coefficients that yield
the change in the light amplitude and phase with respect to mode number.

Let us assume a beam described by the beam parameter $q_1$ being injected
into a segment described by the parameter $q_2$.  Let the optical axis of
the beam be misaligned: the coordinate system of the beam is given by ($x,
y, z$) and the beam travels along the $z$-axis.  The beam segment is
parallel to the $z'$-axis and the coordinate system ($x', y', z'$) is given
by rotating the ($x, y, z$) system around the $y$-axis by the
\emph{misalignment angle} $\gamma$.  The coupling coefficients are defined
as:
\begin{equation}
u_{n m}(q_1)\mExB{\I(\w t -k z)}=\sum_{n',m'}k_{n,m,n',m'}u_{n'
m'}(q_2)\mExB{\I(\w t -k z')},
\end{equation}
where $u_{n m}(q_1)$ are the \HG modes used to describe the injected
beam and $u_{n' m'}(q_2)$ are the `new' modes that are used to describe the
light in the beam segment.  Please note that including the plane wave phase
propagation into the definition of coupling coefficients is very important
because it results in coupling coefficients that are independent of the
position on the optical axis for which the coupling coefficients are
computed.

Using the fact that the \HG modes $u_{n m}$ are orthonormal, we can compute
the coupling coefficients by the following inner product \cite{bayer}:
\begin{eqnarray}\label{eq:tem_conv}
k_{n,m,n',m'}&=&\mEx{\I 2 k z'
\sin^2\left(\frac{\gamma}{2}\right)}\iint\!dx'dy'~
u_{n' m'}\mEx{\I k x' \sin{\gamma}}~u^*_{n m}\\
&=&\mEx{\I 2 k z'\sin^2\left(\frac{\gamma}{2}\right)} \iprod{u_{n' m'}\mEx{\I k x' \sin{\gamma}},u_{n m}}.
\end{eqnarray}
Since the Hermite-Gauss modes can be separated with respect to $x$ and $y$,
the coupling coefficents can also be split into $k_{n m n' m'}=k_{n n'}k_{m
m'}$.  These equations are very useful in the paraxial approximation as the
coupling coefficients decrease with large mode numbers. In order to be
described as paraxial, the angle $\gamma$ must not be larger than the
diffraction angle. In addition, to obtain correct results with a finite
number of modes the beam parameters $q_1$ and $q_2$ must not differ too
much, see \Sec{sec:limits_paraxial}.

The convolution given in \eq{eq:tem_conv} can be directly computed using
numerical integration. This is computationally very expensive.  In
\cite{bayer} the above projection integral is partly solved and the coupling
coefficients are given by simple sums as functions of $\gamma$ and the mode
mismatch parameter $K$, which are defined by:
\begin{equation}
K=\frac12(K_0+\I K_2),
\end{equation}
where $K_0=(z_R-z_R')/z_R'$ and $K_2=((z-z_0)-(z'-z_0'))/z_R'$.
This can be also written as (using $q=\I\zr +z-z_0$):
\begin{equation}
K=\frac{\I (q-q')^*}{2 \Im(q')}.
\end{equation}

The coupling coefficients for misalignment and mismatch (but no lateral
displacement) can be then be written as:
\begin{equation}\label{eq:ccoeff}
k_{n n'}=(-1)^{n'} E^{(x)} (n!n'!)^{1/2} (1+K_0)^{n/2+1/4} 
(1+K^*)^{-(n+n'+1)/2}\left\{S_g-S_u\right\},
\end{equation}
where:\vspace{-5mm}
\begin{equation}
{\renewcommand{\arraystretch}{2.5}\begin{array}{l}
S_g=\sum\limits_{\mu=0}^{[n/2]}\sum\limits_{\mu'=0}^{[n'/2]}
\frac{(-1)^\mu \bar{X}^{n-2\mu}X^{n'-2\mu'}}{(n-2\mu)!(n'-2\mu')!}
\sum\limits_{\sigma=0}^{\min(\mu,\mu')}\frac{(-1)^\sigma
\bar{F}^{\mu-\sigma} F^{\mu'-\sigma}}
{(2\sigma)! (\mu-\sigma)! (\mu'-\sigma)!},\\
S_u=\sum\limits_{\mu=0}^{[(n-1)/2]}\sum\limits_{\mu'=0}^{[(n'-1)/2]}
\frac{(-1)^\mu \bar{X}^{n-2\mu-1}X^{n'-2\mu'-1}}{(n-2\mu-1)!(n'-2\mu'-1)!}
\sum\limits_{\sigma=0}^{\min(\mu,\mu')}\frac{(-1)^\sigma
\bar{F}^{\mu-\sigma} F^{\mu'-\sigma}}
{(2\sigma+1)! (\mu-\sigma)! (\mu'-\sigma)!}.
\end{array}}
\end{equation}
$S_u$ does not exist for $n=n'=0$, due to negative factorials in the denominator. The respective formula for $k_{m m'}$ can be obtained by replacing the
following parameters:
$n\rightarrow m$, $n'\rightarrow m'$, $X,\bar{X}\rightarrow 0$ 
and $E^{(x)}\rightarrow 1$ (see below).  The notation $[n/2]$ means:
\begin{equation}
\left[\frac{m}{2}\right]=\left\{
\begin{array}{ll}
m/2 &\mbox{if $m$ is even,}\\
(m-1)/2&\mbox{if $m$ is odd.}
\end{array}\right.
\end{equation}
The other abbreviations used in the above definition are:
\begin{equation}
{\renewcommand{\arraystretch}{1.5}
\begin{array}{l}
\bar{X}={(\I \zr'-z')\sin{(\gamma)}}/({\sqrt{1+K^*}w_0}),\\
X={(\I \zr+z')\sin{(\gamma)}}/({\sqrt{1+K^*}w_0}),\\
F={K}/({2(1+K_0)}),\\
\bar{F}={K^*}/{2},\\
E^{(x)}=\mEx{-\frac{X\bar{X}}{2}}.
\end{array}}
\end{equation}


In general, the Gaussian beam parameter might be different for the sagittal
and tangential planes and a misalignment can be given for both possible axes
(around the $y$-axis and around the $x$-axis), in this case the coupling
coefficients are given by:
\begin{equation}
k_{nm m'n'}=k_{n n'} k_{m m'},
\end{equation}
where $k_{n n'}$ is given above with
\begin{equation}
{\renewcommand{\arraystretch}{1}
\begin{array}{l}
q \rightarrow q_t\\
\text{and}\\
w_0\rightarrow w_{t,0} \text{, etc.},
\end{array}}
\end{equation}
and $\gamma \rightarrow \gamma_y$ is a rotation about the $y$-axis.
The $k_{m m'}$ can be obtained with the same formula, replacing:
\begin{equation}
{\renewcommand{\arraystretch}{1}
\begin{array}{l}
n \rightarrow m,\\
n' \rightarrow m', \\
q \rightarrow q_s,\\
\text{thus}\\
w_0\rightarrow w_{s,0} \text{, etc.},
\end{array}}
\end{equation}
and $\gamma \rightarrow \gamma_x$ is a rotation about the $x$-axis.

\begin{figure}[htb]
\begin{center}
\IG [viewport= 0 0 380 120,scale=0.5] {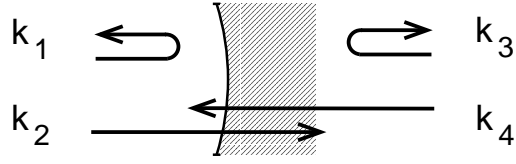} \\
\end{center}
\caption[Coupling coefficients for TEM modes]
{Coupling coefficients for \HG modes: for each optical element and each
direction of propagation complex coefficients $k$ for transmission and
reflection have to be computed. In this figure $k_1$, $k_2$, $k_3$, $k_4$
each represents a matrix of coefficients $k_{n m n' m'}$ describing the
coupling of \M{n,m} into \M{n',m'}.}
\label{fig:knm_mi}
\end{figure}

At each component a matrix of coupling coefficients has to be computed for
transmission and reflection; see \fig{fig:knm_mi}. 


\subsection{Alignment transfer function}

An alignment transfer function in this context is the ratio between any
interferometer output signal and a periodic change of the alignment angle of
a mirror or \bs.  Such transfer functions are needed for calculating the
coupling of alignment noise into longitudinal interferometer signals and for
the design of auto-alignment control systems.

As in the plane-wave approximation the source signal is injected into the
interferometer via the command {\Co fsig}. A modulation of the alignment
of a mirror or \bs creates sidebands at the modulation frequency.  We now
derive the amplitude and phase of these sidebands (with respect to the
impinging light).

First, the amplitude coefficients $a_{jnm}$ of the incoming light field are
stored, with $j$ as the frequency index and $n,m$ as the TEM mode indices.
Then the static couplings between TEM modes due to a possible static
misalignment or mode mismatch are computed, so that the amplitude
coefficients of the reflected field can be computed:
\begin{equation}
a'_{jn'm'}= \sum\limits_{n,m}k_{n'm'nm} a_{jnm}\mEx{\I \varphi_{\rm sb}},
\end{equation}
where $\varphi_{\rm sb}$ is the plane wave phase shift acquired through a
possible detuning of the component from the reference plane as given by
\eq{eq:plane-phisb}.

The beam, now given by $a'_{jn'm'}$ is again misaligned by an angle of
$\gamma=\epsilon\sin{(\omega_m t)}$.  The coupling coefficients simplify
if only misalignment is considered:
\begin{equation}
k_{n n'}=(-1)^{n'} E^{(x)} (n!n'!)^{1/2}\sum\limits_{\mu=0}^{\min{(n,n')}}
\frac{(-1)^\mu \bar{X}^{n-\mu} X^{n'-\mu}}{\mu!(n-\mu)!(n'-\mu)!},
\end{equation}
where
\begin{equation}
\bar{X}=X^*=\frac{-(z'-\I \zr')\sin{\gamma}}{w'_0},
\end{equation}
For the transfer function, only the terms linear in $\gamma$ and thus linear
in $X$ have to be considered. Only one addend of the above sum can be linear
in $X$ and only if $|n-n'|=1$.  In edition the exponential function
can be neglected $E^{(y)}=1, E^{(x)}=1+O(\gamma^2)$. Thus the sum reduces to:
\begin{equation}
\sum\limits_{\mu=0}^{\min{(n,n')}}
\frac{(-1)^\mu \bar{X}^{n-\mu} X^{n'-\mu}}{\mu!(n-\mu)!(n'-\mu)!}=
O(X^2)+\left\{\begin{array}{ll}
\frac{(-1)^{n'}\bar{X}}{n'!}&\mbox{for}\quad  n=n'+1\\
\frac{(-1)^{n}X}{n!}&\mbox{for}\quad  n'=n+1\\
1&\mbox{for}\quad n'=n\\
0&\mbox{otherwise}
\end{array}\right.
\end{equation}
The coupling coefficients can now be written
as:
\begin{equation}
k_{nn'}=\left\{\begin{array}{cl}
\sqrt{n}\bar{X}&\mbox{for}\quad  n=n'+1,\\
-\sqrt{n'}\bar{X}^*&\mbox{for}\quad  n'=n+1,\\
1&\mbox{for}\quad n'=n,\\
0&\mbox{otherwise}.
\end{array}\right.
\end{equation}

\section{Mirror surface maps}\label{sec:mirrormaps}
The term \emph{mirror map} often refers to a scan of a manufactured mirror obtained by
interferometric means, the resulting map contains, for example, measurements of the 
surface height or substrate transmission measured at the nodes of an $x,y$ grid. 
More generally we can think of a \emph{surface map} 
as a two dimensional data array describing properties of the reflection or transmission
of an optical surface as a function of the position on the surface. 

Typically a mirror map is
subject to pre-processing to remove average effects, like a tilt of the entire surface
arising from the measurement process. 
Often, further processing is done to remove the expected optical profile of the 
mirror in order to measure only the residual deviations, for example the surface
roughness.

Phase maps, for example, are of great importance for the purpose of estimating 
noises introduced by mirror surface aberrations. In the initial design of an 
interferometer we assume the mirrors to have perfectly 
smooth surfaces, yet the manufacturing process introduces various surface distortions,
for example, a surface roughness on the order of a nanometre. By providing an accurate 
simulation including surface distortions, potential design problems can be identified before 
the experimental apparatus is being build.

Three types of surface maps can be applied to \cmd{mirror} components
in \Finesse: \emph{phase maps}, \emph{absorption maps} and \emph{reflectivity maps}.
`phase maps' and `absorption maps' can be set to affect only the reflected
light, only the transmitted light or affecting both light fields.
`Reflectivity maps' always change the reflected and transmitted light.
Please note that the name `phase map' refers to the effect the map 
has on the impinging light field. However, the
numerical values stored in the map files are not phases but
surface distortions in meters. The values for higher parts of the surface are defined as positive
whereas 'holes' are indicated by negative values. 
The numerical data in `absorption maps' represent power loss coefficients (values between 0 and 1)
and `reflectivity maps' are composed of power reflectivity coefficients 
(values between 0 and 1).

If the amplitude reflectivity $r$ of a mirror is constant over its entire surface
but the mirror surface is not perfectly smooth and the surface data is available, the
reflection of a field from a mirror can be described by a phase map.
Alternatively we can imagine mirrors which are considered perfectly smooth but
are subject to absorption that varies with the exact position on the
surface. This effect can be studied using absorption maps.
Real mirrors are never perfectly smooth, nor do they
feature a perfectly homogeneous reflectance or transmittance. Therefore, in general a 
mirror surface could be best described with a map storing phase and 
amplitude information together. However, the implementation of surface maps
in \Finesse uses separate maps, giving information either on the effects
on the light phase, the absorption or the reflectivity. 
Yet, several maps of different types can be applied to the same surface 
simultaneously.

\subsection{Phase maps}
The effect of a phase map on the reflected field can be described mathematically as:
\begin{equation}
E_{\rm refl}~=~r~\mEx{\I\OPh(x,y)}~E_{\rm in}
\end{equation}
with
\begin{equation}
\OPh(x,y)=2\frac{n_1 \w}{\lambda\, \w_0}\Tun_p(x,y)
\end{equation}
and $n_1$ the index of refraction of the medium we are in and 
$\Tun_p$ the measured surface map of the mirror surface, given in metres. 
Positive values refer to higher parts of the surface and negative to lower.

The frequencies are defined as usual, $\omega$ is the frequency of the 
light field impinging on the surface and $\omega_0$ the frequency 
corresponding to the default wavelength $\lambda_0$.
For simplicity we assume that $\lambda$ can be replaced by $\lambda_0$
and the factor $\w/\w_0$ can be approximated as $1$. We then obtain:
\begin{equation}
E_{\rm refl}~=~r~\mEx{2\I\, n_1\,\Tun_p(x,y)/\lambda_0}~E_{\rm in}
\end{equation}
The transmission can be written as
\begin{equation}
E_{\rm trans}~=\I~t~\mEx{\I\,(n_1-n_2)\,\Tun_p(x,y)/\lambda_0}~E_{\rm in}
\end{equation}

\subsection{Absorption maps}
The reflected field subject to an absorption map can be written as
\begin{equation}
E_{\rm refl}~=~r~\sqrt{1-L(x,y)}~E_{\rm in}
\end{equation}
with $L(x,zy)$ the measured absorption map of the mirror surface, given in 
power coefficients. The transmission can be written as
\begin{equation}
E_{\rm trans}~=\I~t~\sqrt{1-L(x,y)}~E_{\rm in}
\end{equation}

\subsection{Reflectivity maps}
Reflectivity maps are implemented slightly differently because they replace the
mirror parameters $r$ and $t$ given in the {\Co mirror} command. These values
are used to compute a loss factor $L=1-r^2-t^2$ and then are set internally to
$1$. 
The reflected field can then be written as
\begin{equation}
E_{\rm refl}~=~\sqrt{R(x,y)}~\sqrt{1-L}~E_{\rm in}
\end{equation}
with $R(x,zy)$ the measured reflectivity map of the mirror surface, given in 
power coefficients. The transmitted field can be written as
\begin{equation}
E_{\rm trans}~=\I~\sqrt{1-R(x,y)}~\sqrt{1-L}~E_{\rm in}
\end{equation}

\subsection{Coupling coefficients from mirror maps}
In \Finesse the shape of a field is not given as a function of $x$,$y$ coordinates
but by a sum of Hermite-Gauss modes of different orders.
Therefore the effect of a mirror map needs to be described as scattering into 
higher order modes. The coupling coefficients can be computed by the usual integral. 
In the case of the reflection we obtain, for example:
\begin{equation}
k_{n,m,n',m'}=\int\int dx'\,dy'~ u_{n',m'}\,\exp(2\I\,n_1 \,\Tun_p(x',y')/\lambda_0)\,u^*_{n,m} 
\end{equation}
which can be computed directly using a numerical integration routine or in some specific cases analytically.

In order to be compatible with already existing coupling coefficients the following 
approach has been chosen: The coupling coefficients due to a mirror map are computed 
independently of any misalignment or change in Gaussian beam parameter that
occurs at the given mirror. Therefore for any given map function 
$A(x,y)$ the coefficients with respect to a field impinging on the front face of
the mirror are computed as:
\begin{equation}\label{eq:mapint}
k^{\rm map}_{n,m,n',m'}=\int\int dx'\,dy'~ u(n',m',q_1,n_1)\,A(x',y')\,u^*(n,m,q_1,n_1) 
\end{equation}
with $q_1$ and $n_1$ being the Gaussian beam parameter and the index of refraction
of the node in front of the mirror, respectively. This equation is true for reflection as well
as transmission (the map function $A$ would be different between those cases
of course). The coefficients are then merged with the other coupling
coefficients, see Appendix~\ref{sec:mapknm}.
This approach allows to seamlessly use together multiple maps as well as 
additional attributes to mirrors such as radius of curvature and 
alignment angle. However, the sperate computation and subsequent
merging of coupling coefficients is an approximation when only
a finite set of modes is used. In consequence, one has to be very
careful in setting up a model using maps; there are a few
configuration commands which can be used to select the best
method for computing and merging the coefficients, see \Sec{sec:merging_maps}.

The following table\footnote{Please note in the different entries of the table 
$B((x,y)$ represents different data types of different dimension.} gives an overview of the map functions in the different cases.
$B(x,y)$ shall be a matrix of real numbers representing the data 
stored in the map file.

\begin{tabular}{|l|l|l|l|l|}
\hline
{\it type of map} & {\it fields affected}  & $A(x,y)${\it  (refl.)}& $A(x,y)$ {\it (trans.)}\\
\hline
\hline
phase& reflection & $\exp(\I 2 k\, n_1 B(x,y))$ & 1 \\
\hline
phase & transmisson & 1 & $\exp(-\I k B(x,y))$ \\
\hline
phase& both & $\exp(\I 2 k\, n_1 B(x,y))$ & $\exp(-\I k\, (n_1-n_2) B(x,y))$ \\
\hline
absorption & reflection & $\sqrt{1-B(x,y)}$ & 1 \\
\hline
absorption & transmission & 1 & $\sqrt{1-B(x,y)}$ \\
\hline
absorption & both & $\sqrt{1-B(x,y)}$ & $\sqrt{1-B(x,y)}$ \\
\hline
reflectivity & both & $\sqrt{1-L}\sqrt{B(x,y)}$ & $\sqrt{1-L}\sqrt{1-B(x,y)}$ \\
\hline
\end{tabular}

\noindent
with $n_1$, $n_2$ the indices of refraction of the medium before and after the
surface respectively.

\subsection{The map file format}
A mirror map file can contains the mirror map as a grid $B(x,y)$. The data grid $B(x,y)$ must be stored as follows. 
The data is preceded by a header consisting of seven lines, for example:
\begin{finesse}
\end{finesse}
The first line indicates that a map of grid data follows, the second line specifies the name
of the map and the third line the type of the map. Possible types are:
\begin{itemize}
\item phase transmission
\item phase reflection
\item phase both
\item absorption transmission
\item absorption reflection
\item absorption both
\item reflectivity both
\end{itemize}
The fourth line states the number of rows and columns of the map data, line five gives the
optical center (in real numbers referring to grid indices, starting at zero), typically the center
is at (cols+1)/2,(rows+1)/2.

Line number six gives the physical length (in meters) of one grid elements in the x and y directions of 
one grid element. The overall size of the grid in relation to the size of the
light field must be chosen very carefully to avoid numerical errors. 
\Finesse computes a typical size of the light field as follows. The maximum diameter
in the horizontal direction is given as:
\begin{equation}
d_{\rm beam, x}=2~w_x(z)~\sqrt{{\rm maxtem} +0.5} 
\end{equation}
The maximum beam diameter is then computed as 
\begin{equation}
d_{\rm beam}=max(d_{\rm beam, x},d_{\rm beam, y})
\end{equation}
\Finesse then computes an effective size of the grid as two times the smallest distance 
from the optical center to the edge. \Finesse issues a warning if the
such computed grid size is smaller than four times the maximum beam size.

The last line of the header defines a scaling factor to be applied to
the data that follows.

This header is then followed by the grid data stored in columns and rows as given
by the `Size' in the header.
The grid can contain four different kinds of information
specified by type of the map (see list above). Phase maps
store information related to optical path length, given in meters, amplitude related maps
store power coefficients between 0 and 1. The grid data is then
used inside \Finesse to compute coupling coefficients as given
by Equation~\ref{eq:mapint}. Note that in all cases the
Gaussian beam parameter used to compute $u(n',m')$ and $u^*(n,m)$ are
identical.

\subsection{How to apply a map to a component}

Applying a map to a component means you have one of the maps types
discussed above already at your disposal. Take the script snippet
below, we have a mirror which we want to apply one or several maps to,
this is simply done using the \verb|map| command.
\begin{finesse}
m m1 0.99 0.01 0 n2 n3

# map command usage: map [component_name] [map_filename]
map m1 aperture_map.txt
map m1 surface_roughness_map.txt
map m1 reflectivity_map.txt
\end{finesse}

Above we have applied 3 maps to our mirror - the maximum is 30 though
using that many is unlikely - all three maps must have exactly the
same physical sizes and discretisations. In previous versions of
Finesse each map would have a separate matrix computed of coupling
coefficients, which were later matrix multiplied together. This
provides an interesting problem as matrix multiplication is not
commutative, so the result will differ depending on the order in which
the maps are specified in the kat file. To get around this issue the
maps are merged together to form one single merged-map. This is done
by representing the maps in complex exponential form and each of the
maps multiplied together. This merged-map is what is used to compute
the coupling coefficients.

\subsection{Accelerating calculations by saving coupling coefficients}
The process of calculation the coupling coefficient integral can be
incredibly slow by nature. If you have a simulation which requires
some coupling coefficients and you need to run the simulation multiple
times you can save the coupling coefficients to a file. This is done
using the \verb|knm| command:
\begin{finesse}
m m1 0.99 0.01 0 n2 n3
map m1 aperture_map.txt
map m1 reflectivity_map.txt

# knm usage: knm [component_name] [filename_prefix]
knm m1 test1_m1
\end{finesse}

The above will save the coupling coefficients generated by the various
maps you apply to a file called \verb|test1_m1.knm|. It will also save
4 other files which save the merged-map in amplitude and phase
components for reflection and transmission called \verb|.map|
files. It is important that these 5 files be kept together if you want
to reuse the cofficients.

The \verb|knm| command not only saves the cofficients but also tells
Finesse to load the files aswell, the \verb|filename_prefix| argument
just needs to be the same as what you saved with, without the filename
extension \verb|.knm| though.

Finesse will then try to load the 5 files. The conditions in which the
saved coefficients were calculated are stored in the \verb|.knm|
file. If the saved conditions are different to what the simulation is
now trying to run, the saved files will be ignored and a new set of
coefficients will be generated and saved. In the \verb|.knm| file you
will also notice lines like 
\verb|map0 : mymap_aperture.txt PWPU75F7XZIGxURwdEiJIA==|, 
which list the maps that have been applied
to the mirror. The random looking string is infact a hash that is
uniquely generated to the contents of map file
\verb|mymap_aperture.txt|. If anything in that file changes the
calculation will be redone. Also it is important to note that
reordering the maps in the kat file will also cause the computations
to be recalculated. The hash is an MD5 hash that when written to the
files is stored using a BASE64 conversion.

The 5 files however are not hash protected, thus if you change the
files by hand in a text editor you can load the files using the
\verb|knm| command and the computation will be done with any changes
you have made - this functionality may change in later versions as of
\verb|0.99.9|.

The integration routine can also be sped up by using the symmetric 
nature of the coupling coefficient matrix in certain conditions.
This is detailed further in appendix \ref{sec:knm_transpose}. This
speeds up the computation by $\approx \times 2$ as only the upper half
and diagonal elements of the matrix need computing, the lower half can
be inferred from the upper. This can be switched on or off from the \verb|kat.ini|
file by using the option \verb|calc_knm_tranpose| \verb|0 (off) or 1 (on)|. This is 
switched on by default.

\subsection{Coupling cofficient data files - ASCII vs binary formats}

When using particularly large maps or a large \verb|maxtem| value you
will quickly find that reading and writing the 5 map files becomes
painfully slow. To combat this we added the ability to save the files
in either ASCII (Normal readable text) or binary (unreadable)
format. Changing which format is used is done using the \verb|conf|
command per component:
\begin{finesse}
conf [component name] save_knm_binary [0 (ASCII) or 1 (Binary)]
\end{finesse}

If you try to load a binary or ASCII file when Finesse has been told to use the other, an error will occur.

\subsection{Integration and interpolation methods}\label{sec:int_routines}

The integral is performed numerically using an integrating library
named Cuba or by a simplistic Riemann sum. The Riemann sum works by
summing over all grid elements, the x, y coordinates used in the
Hermite-Gauss functions $u_{nm}$ are computed as follows:
\begin{equation}
\begin{array}{l}
    x(i=1:cols)=(i-x0)*xstep\\
    y(j=1:rows)=(j-y0)*ystep
\end{array}
\end{equation}

with $cols, rows$ the number of columns and rows in the data grid, $x0$ and $y0$ the indices
of the optical center as given in the header and $xstep, ystep$ the
lengths of one grid element. If a rotation angle is given in the \cmd{map} command, x and y are 
further rotated by minus the given angle (If no angle is given and no angle has been found in 
the file, an angle of zero degrees is assumed).

The Cuba integration routines are much more sophisticated in there
approach in solving the coupling coefficient integrals. Essentially
the routines sample the mirror map in a sparse fashion in an attempt
to calculate the integral. The integration of the map is split into
several domains, each domain has the integral calculated and an error
estimated, if the error is larger than the acceptable values the
domain is further split and more samples are used. This way the
routines should concentrate on areas of difficulty with more
evaluations, whereas simple domains only require a few and are much
more efficient.

As the routines sample the discretised map in a continuous manner,
interpolation is needed to sample within the maps data points. The
interpolation is provided by the GSL library and provides 3
options. The fastest method is Nearest-Neighbour, ideal if the map is
sampled highly or you just want a quick integration. Linear
interpolation, and the slowest, cubic interpolation is also
offered. The choice of interpolation routine will depend on what data
your map has been generated from. If the underlying data was
flat/linear then linear interpolation would be ideal, however if
original surface or data was smoothed, like a mirror surface, cubic
might give a better representation of the real surface. It is
important to remember that cubic interpolation can produce artefacts
that may not exist due to the polynomials it forces to fit the points
of the map, which is problematic for 'rough' maps. It is therefore
recommended that you use linear interpolation for most purposes. If
you are using a binary map, i.e. a mask of 1's and 0's, you must use
nearest-neighbour to avoid varying values at points inbetween a 1 and
0.

There are several commands used to set the various integration and
interpolation routines depending on your required use. First, you can
specify the default integration and interpolation methods in the
`kat.ini' file so as to apply to multiple kat files. This is done by
adding lines:
\begin{finesse}
mapintmethod 1
mapinterpmethod 1
\end{finesse}

The various methods are chosen by a number 1, 2 or 3, for integration methods:
\begin{itemize}
\item Riemann Sum - 1
\item Cuba Serial - 2
\item Cuba Parallel -3
\end{itemize}

and for interpolation methods:
\begin{itemize}
\item Nearest Neighbour - 1
\item Linear - 2
\item Cubic -3
\end{itemize}

To set the default integration method for all components in a kat file
you can use the \verb|intmethod [integration_method]| command 
in your \Finesse input file.

You can also specify integration and interpolation on a component
by component basis. This is in the case where one map might have a
rough surface which requires Cuba integration but another very smooth
and linear which could be done with a Riemann sum. This is done using
the \verb|conf| (configure) command, for as example as shown below:
\begin{finesse}
map m2 mymap_aperture.txt # Here we apply our map to the mirror
conf m2 interpolation_method 1 # This sets the interpolation to nearest neighbour
conf m2 integration_method 1 # This sets the integration method to the Riemann sum
\end{finesse}

\subsection{Map example: a focusing surface in transmission}\label{sec:maptest1}
This and the following example of using a map compare the results
achieved with a surface map to an equivalent \Finesse result without
a map. In this case we compare the focussing of beam by a lens (the
built-in \Finesse command) to the focussing by a mirror map
representing the same focal length. 

The transmission map representing a lens has been created using 
a set of SimTools functions~\cite{simtools}, see \Sec{sec:simtools}, 
the MATLAB file is:
\begin{finesse}
focallength = 200;
L1=10;
L2=150;
lambda = 1064e-9;
realsize = 0.7;
finesse_map_filename='lens_map.txt';
gridsize = 200;
map = FT_create_lens_map(gridsize,realsize,realsize,focallength);
FT_write_surface_map(finesse_map_filename,map);
\end{finesse} 

To run a \Finesse simulation using this map, the following input file can be used:
\loadkat{map_lens1.kat}

\begin{figure}[htb]
    \centerline{\IG[viewport= 95 400 535 715, width=0.8\textwidth]{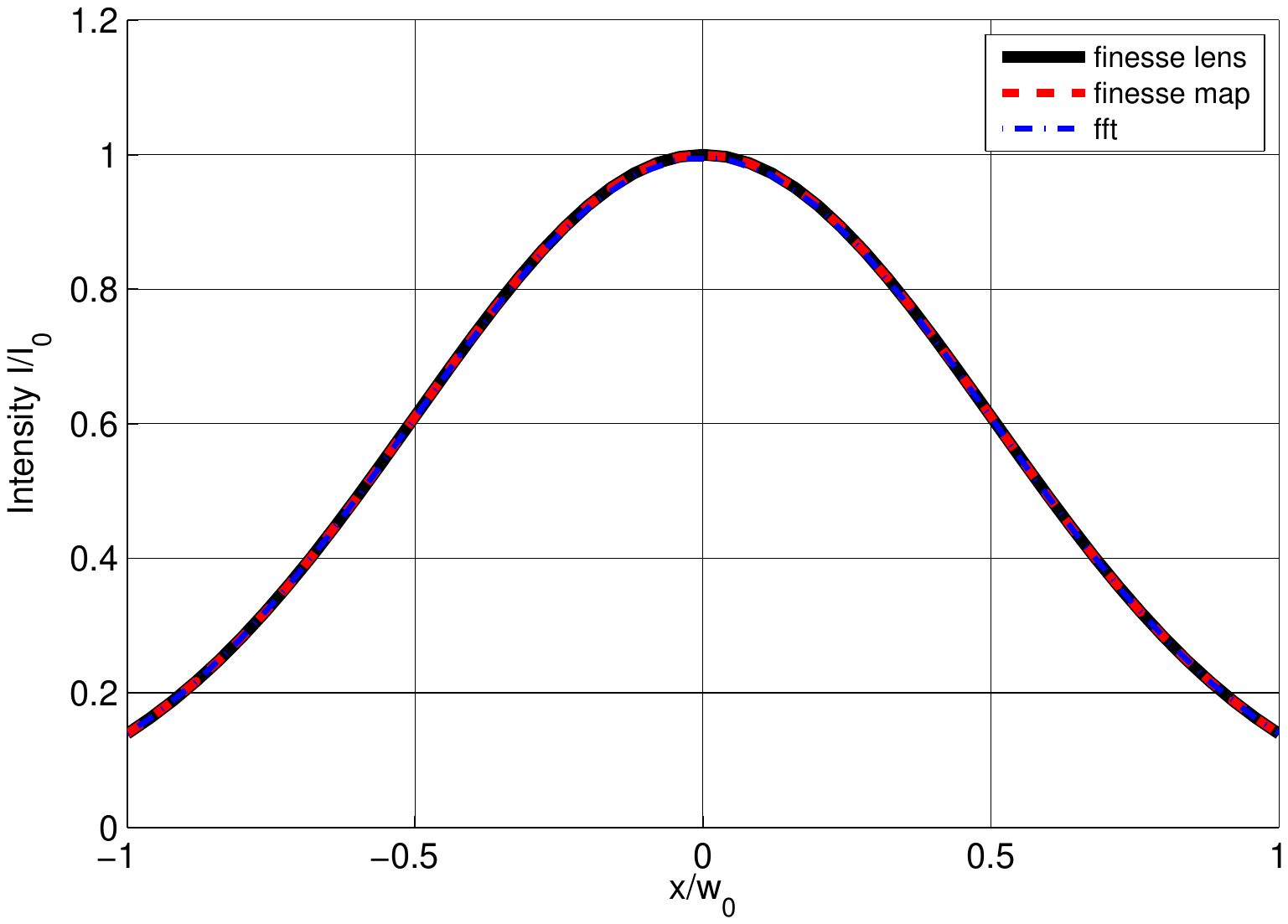}}
    \caption{Beam shape comparison between setup with a lens component and a mirror surface map
     representing the same focal length. Also shown is a result from
     an FFT propagation code using the same transmission map}\label{fig:maptest1}
\end{figure}
To check this result we can run a different file which uses a 
lens component instead of the mirror with the surface map:
\loadkat{map_lens2.kat}

Both results are shown in figure~\ref{fig:maptest1}.

\begin{figure}[htb]
    \centerline{\IG[width=8cm]{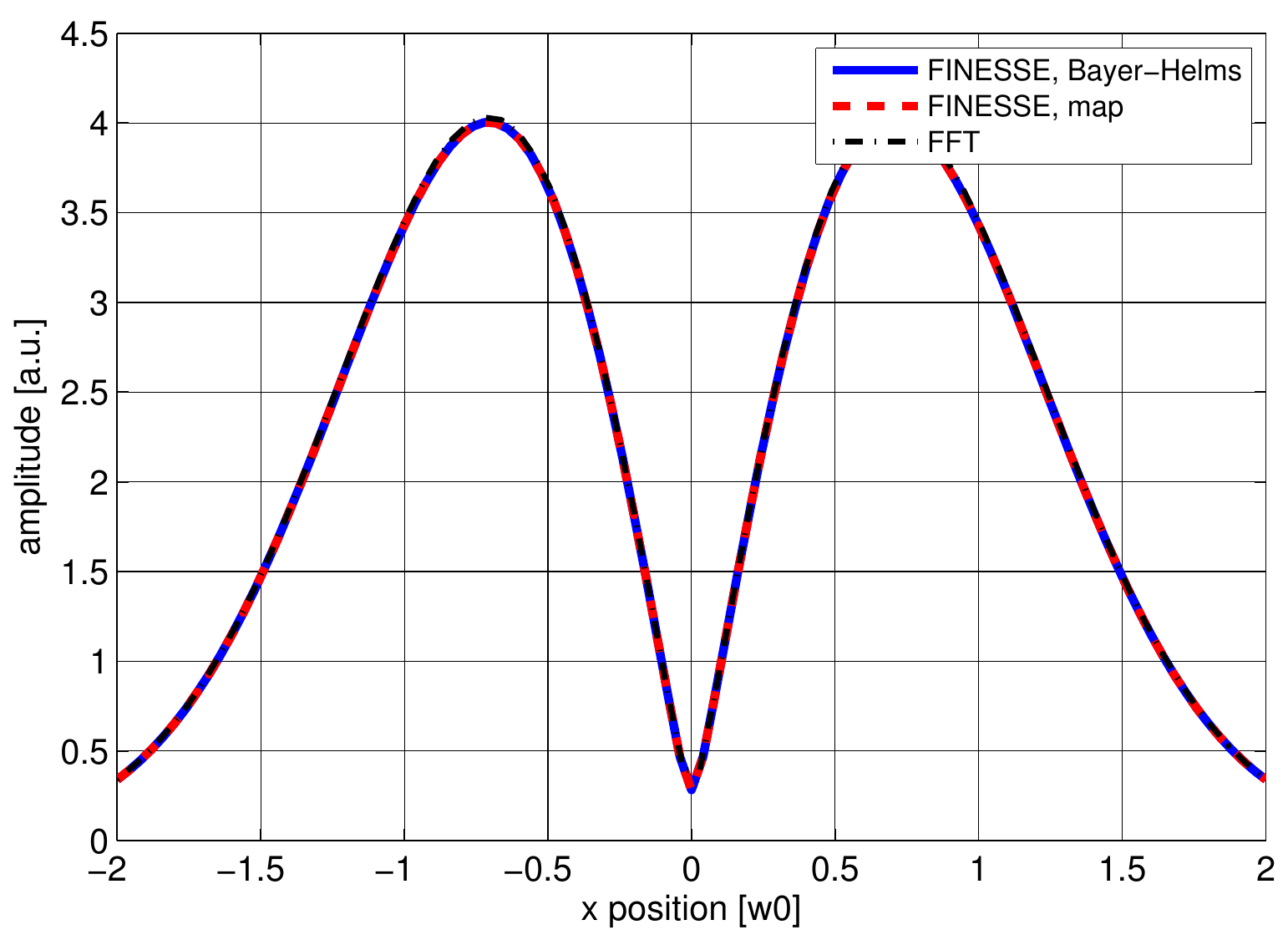}\IG[width=8cm]{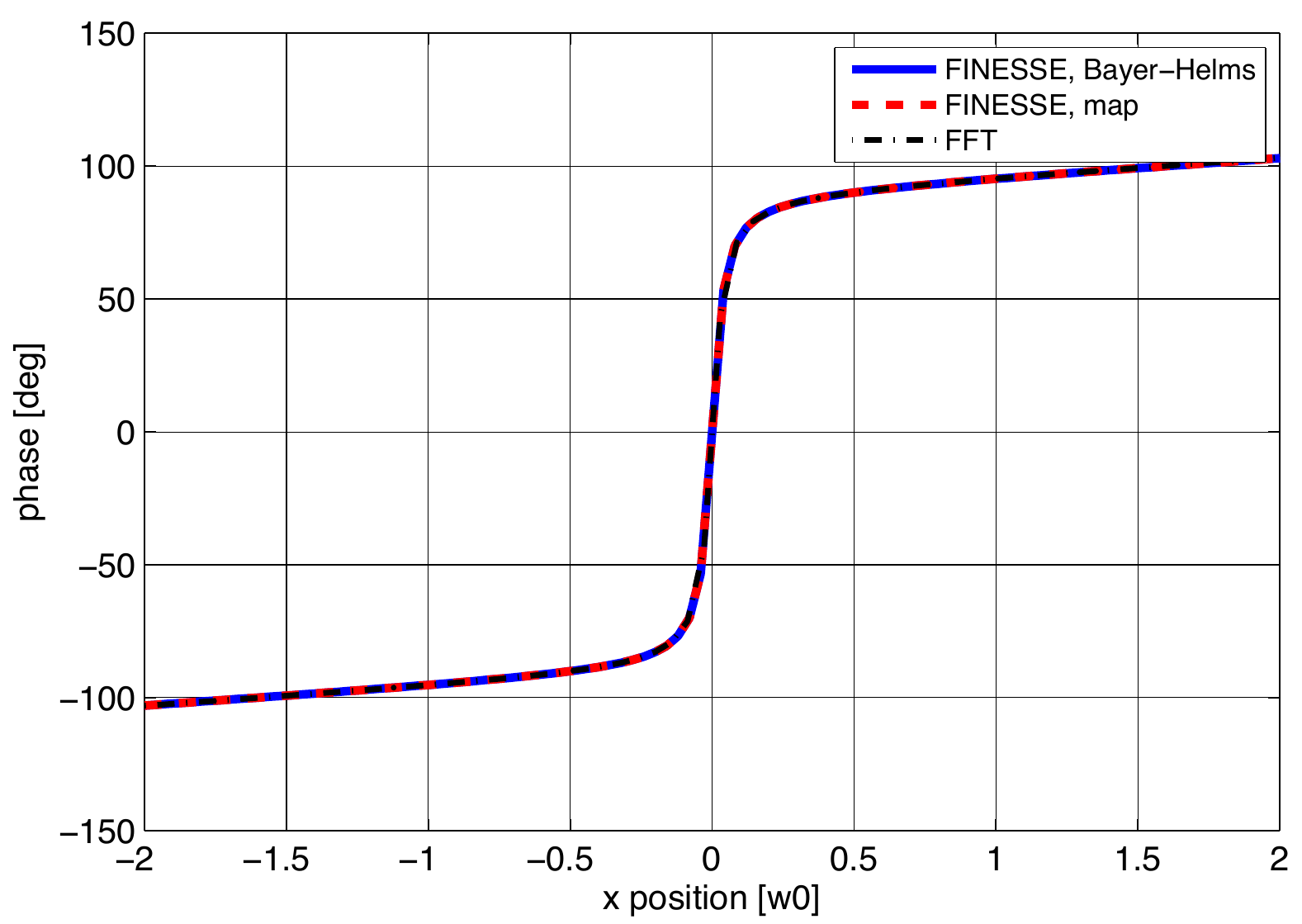}}
    \caption{Comparison of different simulations of the light
      reflected by a tilted mirror: a TEM$_{00}$ mode is reflected by 
a mirror which is tilted by 1\,$\mu$rad around the vertical axis.
In order to highlight the effect of the tilt the TEM$_{00}$ part of
the reflected beam is removed so that only the TEM$_{10}$ part
remains. The above plots show a very good agreement between
an FFT propagation with \Finesse results for using a mirror map
or the built-in \code{xbeta} command.}\label{fig:maptest2}
\end{figure}

\subsection{Surface map example: a tilted mirror in reflection}\label{sec:tilt_ex}
Another simple example is the tilted mirror: a beam in a \M{00} mode
is reflected by a mirror. This mirror can be tilted, either using the
\code{xbeta} command or by applying a tilted phase map to the mirror.
The results are for both types of simulation and an equivalent FFT
propagation are shown in Figure~\ref{fig:maptest2}. In order to
highlight the effect of the tilt, the \M{00} part of the beam has been
removed so that only the \M{10} field shows. Both the amplitude
and the phase distribution match precisely for all three methods.

The MATLAB script to generate the tile map is again based on SimTools
and look as follows:
\begin{finesse}
gridsize = 256;
R = 10;
realsize = 0.15;
xbeta = 1e-6;
ybeta = 0;
map = FT_create_tilted_map(xbeta,ybeta,R,realsize,gridsize);
FT_write_surface_map('tilted_map.txt',map)
\end{finesse}

The \Finesse input file is:
\loadkat{map_tilt1.kat}
\vspace{2cm}

\subsection{Realistic map example: thermal distortions}
A more complicated example and one which takes care to simulate 
correctly is a cavity including thermal distortions of the mirrors.
The high circulating powers in advanced gravitational wave detectors
will lead to relatively large distortions of the mirrors making up the
arm cavities.  Simulation of such a setup requires careful calculation
and preparation of the mirror maps describing the thermal distortions,
taking care to remove any curvature and offset present in the mirrors
and only including such effects in the \Finesse file.  Simulations such as
these are crucial for the commissioning of second generation gravitational
wave detectors and, as such, are a good test of \Finesse as a robust and
useful tool.  An investigation into the round-trip losses incurred in a cavity
with thermal distortions is described in section~\ref{sec:thermal_ex}, where 
we show that the correct result can be achieved with a high enough
\verb|maxtem|.

\subsection{Couling coefficients for multiple effects}\label{sec:merging_maps}
The separate computation of coupling coefficients of different effects 
at the same surface requires some care in setting up the model, in particular one
must take care configuring the Gaussian beam parameters
correctly.  

\subsubsection{Order of calculation}
\Finesse can compute three different sets of coupling coefficients:
\begin{itemize}
\item analytic coefficients using the Bayer-Helms (BH) equations, which
  includes the effects from all parameters set with the \code{attr}
command, such as misalignment and curvature.
\item coefficients for mirror maps, computed through numerical integration
\item coefficients for apertures (Currently done using numerical integration)
\end{itemize}

In principle the order in which these are computed should not matter,
however, with a finite number of modes the merging of these
coefficients remains an approximation, and thus the order in which
the coefficient matrices are merged can change the result. The
magnitude of this change might serve as an indicator of the overall
error due to the approximation of this approach. See appendix \ref{sec:mapknm}
for more mathematical details on the separation of the coupling coefficients.

The setup of a \Finesse model requires care only if a mirror map
contains some sort of curvature or astigmatism that represented in
the map instead of using the \code{attr} command. The beam
tracing algorithm in \Finesse does not know about the maps and thus
choses sub-optimal beam parameters. Therefore we strongly recommend
to remove any curvature and astigmatism from a map (using SimTools,
see \Sec{sec:simtools}) and apply these instead using the \code{attr}
command.

However, there might be special cases in which a curvature cannot be
removed from a map. In this case we need to understand how \Finesse
separates the coupling coefficient calculation into multiple matrices
to speed up the calculation of static and dynamic effects, e.g. surface
maps and tilts using the \code{xbeta} attribute (See appendix \ref{sec:mapknm}).
The coupling coefficient that includes both the effects of the surface map 
calculated by numerical integration and misalignments and mode-mismatches analytically
with Bayer-Helms can be broken down into a matrix multiplication. Matrix $K_A$ and $K_B$
can represent either the results of numerical integration or Bayer-Helms. Here $N$
and $M$ represent the incoming and outgoing mode $n'm'$ and $nm$,
\begin{eqnarray}
K_{NM} &=& [K_A K_B]_{NM}\\
&=&\sum^{\infty}_{L}  \underbrace{\iprod{U_N(q'_1) A(x,y), U_{L}(q_L)}}_{\text{\normalfont A Solver}} \underbrace{\iprod{U_{L}(q_L), B^{\ast}(x,y)U_M(q_2)}}_{\text{\normalfont B Solver}}.
\end{eqnarray}
The decision which needs to be made is what the value of $q_L$ is. This
can mathematically be any value - so long as it means the function $U$ is
defined - but we limit the choice between the incoming beam parameter $q'_1$
and the outgoing beam parameter $q_2$. It should be chosen so that the solver
that contains the mode-mismatch (Whether that be a curved map or Bayer-Helms with \code{attr})
has the incoming and outgoing $q$ values in its inner product. This can be set
using the \code{conf} command
\begin{finesse}
conf [component name] knm_change_q [1 (for q'_1) or 2 (for q_2)]
\end{finesse}
To choose whether $K_A$ or $K_B$ represents the map integration or Bayer-Helms
solver we use the command
\begin{finesse}
conf [component name] knm_order [21 (K_A = Map, K_B = Bayer-Helms) or 12]
\end{finesse}
The default values are
\begin{finesse}
knm_change_q 1
knm_order 21
\end{finesse}
The ordering is not overly important, from testing we have found the differences
are minor as the number of modes is increased. However for ppm level or computation
it might be of interest to swap the ordering to see if any commutation errors
exist between the matrices. Depending on the order the different solvers will see
different $q$ values depending on the choice of $q_L$.

\begin{figure}[htb]
    \centerline{\IG[viewport= 95 400 535 715, width=0.8\textwidth]{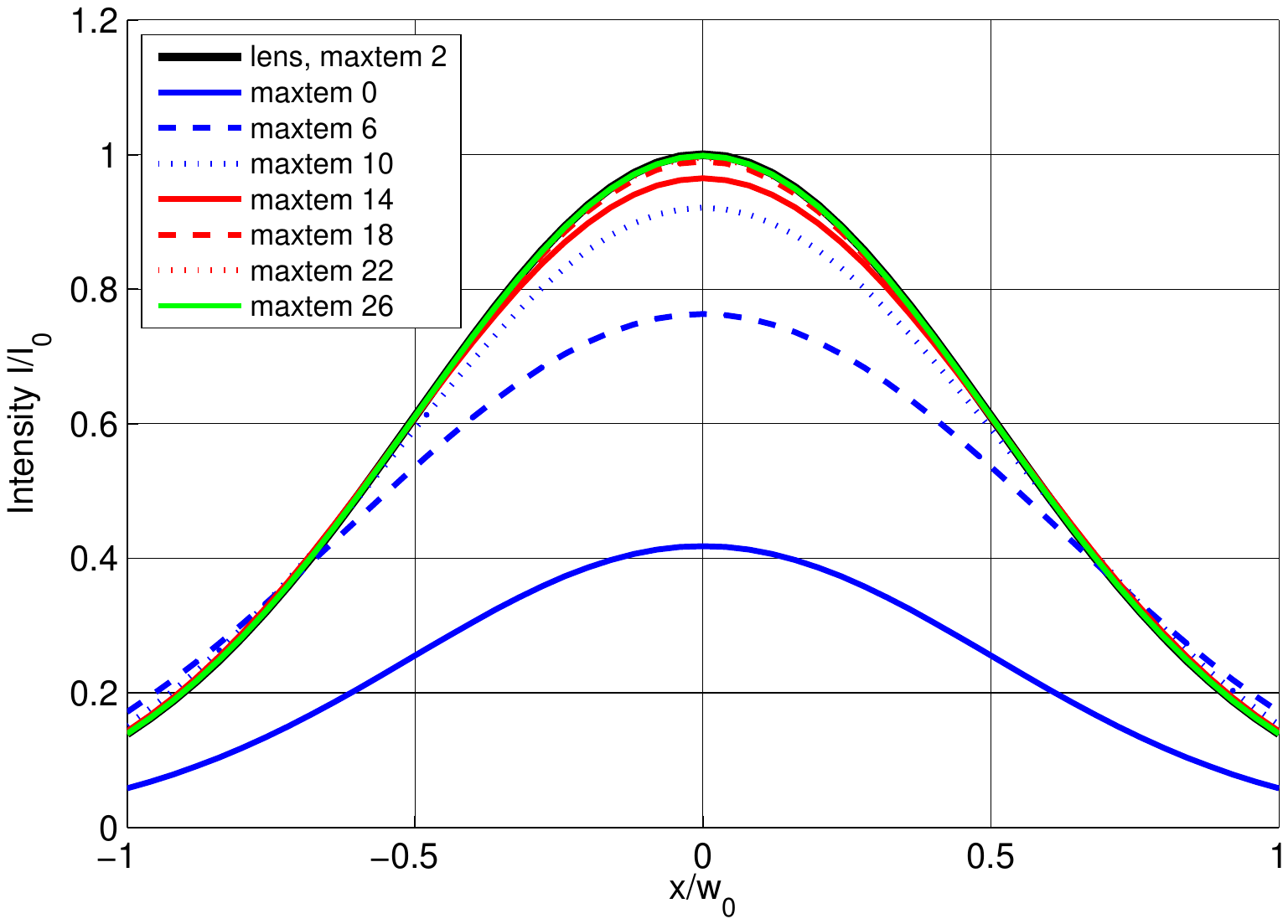}}
    \caption{Beam shape comparison between setup with a lens component and a mirror surface map
     representing the same focal length. The solid black trace shows
     the correct result obtained with the lens command. The other
     traces show the results for a mirror map for different maxtem,
     using a non-optimal configuration for the Gaussian beam
     parameter. Compare this to \mFig{fig:maptest1}, which uses a
     better configuration.}\label{fig:maptest3}
\end{figure}
\subsubsection{Choice of Gaussian beam parameter for multiple effects}
For example, if we repeat the example shown in \Sec{sec:maptest1} but
applying using a non-optimal configuration for the 
Gaussian beam parameters that are used by the numerical integration
routine, we get a substantially different result. The \Finesse input file is:
\loadkat{map_lens3.kat}

This is the same as before except for the line
\begin{finesse}
conf m1 knm_change_q 1
\end{finesse}
The results are shown in \mFig{fig:maptest3}. Whereas the previous
example gave correct results already at \code{maxtem 2}, in this
example the correct result can only be achieved using 
\code{maxtem}>20! The reason for this is the mode
mismatch at the lens, created by using an non-optimal selection
of Gaussian parameters: The mirror map acting as a lens will 
transform the beam such that it the outgoing fundamental mode is described
by a different Gaussian parameters than the incoming beam. 
By setting \code{conf m1 knm\_change\_q 1} however, the 
map coefficient calculation is forced to use the beam parameter
of the incoming beam also for the outgoing field. The difference
is substantial, the appropriate waist size for the outgoing beam
would be $w_0=5.6961815$\,mm but here the coefficients are computed
based on $w_0=10\,$mm. In Section~\ref{sec:limits_paraxial} we discuss
the limits of the paraxial approximation on which the \Finesse
algorithms are based, which includes the statement that the beam
waist sizes should not differ by more than a factor of 3. While this
example is therefore still within the limit of the paraxial
approximation, it requires a relatively high
number of modes to compute correct results.

You must make sure that if a map changes the beam parameter (i.e.
it represents a curved surface), this beam parameter change
is reflected in the Gaussian beam parameters provided to the
integration routine computing the coupling coefficient; the 
\code{conf  knm\_change\_q} command can be used for this
purpose. However, we recommend to pre-condition maps 
such that all curvature are removed and are instead entered
in the model via the radius of curvature parameters of the mirror
element. By default this will create correct results with the
least number of modes.

\section{Detection of \HG modes}

The \HG modes affect all detector types. The following sections describe
the changes with respect to the plane-wave mode.

\subsection{Amplitude detectors}
\label{sec:hgm-ad}

The amplitude detectors in the \HG mode can be specified as
\begin{finesse}
ad name n m f node
\end{finesse}
with $n$ and $m$ as the mode indices. Such a detector will plot the complex
amplitude of each field with frequency $f$ and mode indices $n, m$.

If the amplitude detector is used without specifying $n$ and $m$, the
detector tries to measure something like the phase front on the optical
axis.  To do so it computes the average of the powers in all modes at
frequency $f$ and computes the phase of the sum of these fields.  The result
can be written as follows:
\begin{equation}
S=a \exp{\left(\I \Phi\right)},
\end{equation}
where
\begin{align}
a&=\sqrt{\sum\limits_{n,m}|a_{n,m}|^2},\\
\Phi&={\rm phase}\left(\sum\limits_{n,m}a_{n,m}\right).
\end{align}
This feature has been tested to give similar values as a FFT propagation
code on the optical axis. However, this feature should be considered as
experimental.

\subsection{Photodetectors}

Since the \HG modes (as they are used here) are orthonormal, the
photocurrent upon detection on a single-element photodiode (for simplicity
shown here for one frequency component only) is proportional to:
\begin{equation}
S=\sum_{n,m}a_{nm}a_{nm}^*.
\end{equation}
(as usual the fields referring to sidebands created by \cmd{fsig} 
commands are ignored for the computation of the light power).

More interesting for the \HG modes are different detector types that are
sensitive to the shape of the beam, for instance: split photodetectors.
\Finesse can be used with such detectors of arbitrary design by defining the
\emph{beat coefficients}.  For an arbitrary split detector the photocurrent
is computed as:
\begin{equation}\label{eq:beat_coeffs}
S=\sum_{n,m}\sum_{n',m'} b_{nmn'm'} a_{nm}a^*_{n'm'},
\end{equation}
where $b$ is the beat coefficient matrix. 
The beat coefficients in the equation \ref{eq:beat_coeffs} can be
specified by the user through the \texttt{pdtype} command and the 
corresponding definition in the `kat.ini' file. 
The subsequent demodulation of the signal is performed exactly as in
plane-wave mode. 
In the following section we show
an exemplary calculation of these coefficients for a split photodetector.

\subsection{Split photodetector}\label{sec:split_pd}
A split photodetector is defined as a infinitely large plane
perpendicular to the beam axis, split into two halves along one axis. 
The output signal is computed as the difference between the signals generated from each of the two sides. 
Here we consider the $x$-split photodetector, which is divided along the $y$-axis. 
To compute the coefficients $b_{nmn'm'}$ we assume a beam consisting of two
different \HG modes is impinging on the detector.
The separability of \HG modes in $x$ and $y$ allows us to separate the beat coefficients:
\begin{equation}
b_{nmn'm'}=b_{nn'}b_{mm'}.
\end{equation}
For a detector split in the $x$-direction it is straightforward to find the beat coefficient factor 
$c_{mm'}$ corresponding to the the vertical mode indices. 
The orthonormality of the \HG functions means that:
\begin{equation}
b_{mm'}=\left\{
\begin{array}{l}
1\,\,  \mbox{when}\,\,m=m',\\
0\,\,  \mbox{when}\,\,m\neq m',
\end{array}
\right.
\end{equation}
leading to a simplification of the full beat coefficients:
\begin{equation}
b_{nmn'm'}=\left\{
\begin{array}{l}
b_{nn'}\,\,  \mbox{when}\,\,m=m',\\
0\,\,  \mbox{when}\,\,m\neq m'.
\end{array}
\right.
\end{equation}

The problem is now reduced to finding $b_{nn'}$, and as a result it is only necessary to consider the fields 
in the $x$-direction. The impinging field can now be written in one dimension as:
\begin{equation}
E(x)=a_n u_n(x) + a_{n'} u_{n'}(x).
\end{equation}
The intensity becomes
\begin{equation}
I(x)=|E|^2=a_nu_na_{n'}^*u_{n'}^* + a_n^*u_n^*a_{n'}u_{n'} +
|a_nu_n|^2+|a_{n'}u_{n'}|^2= I_1 + I_0
\end{equation}
with $I_0= |a_nu_n|^2+|a_{n'}u_{n'}|^2$.

The signal of the $x$-split detector is given by:
\begin{align}
S&=\int_0^\infty dx I(x)- \int_{-\infty}^0 dx I(x)=\int_0^\infty dx I(x)+
\int_0^{-\infty} dx I(x)\\
& =\int_0^\infty dx I(x)+ \int_0^{-\infty} dx I(x)=\int_0^\infty dx I_1(x)- \int_0^{\infty} dx I_1(-x).
\end{align}
The $I_0$ contributions are cancelled from the signal because $I_0(-x)=I_0(x)$. 
Now we want to have a closer look at
$I_1$. First we can show that $u_n u_{n'}^*$ is a real number. From
equation \ref{eq:HG_mode2} we get:
\begin{equation}
u_n u_{n'}^*=\sqrt{\frac{2}{\pi}}\sqrt{\frac{1}{2^{n+n'}n!n'!w^2}}e^{\I
    (n-n')\Psi} H_n\left(\frac{\sqrt{2}x}{w}\right)
  H_{n'}\left(\frac{\sqrt{2}x}{w}\right) e^{-\frac{2x^2}{w^2}}.
\end{equation}
However in \Finesse the Gouy phase is stored in the field amplitudes
$a_n$, so that with that in mind we calculate the following:
\begin{equation}
f_{nn'}=\left(u_n u_{n'}^*\right)_{|\mathrm{no~Gouy~phase}}=\sqrt{\frac{2}{\pi}}\sqrt{\frac{1}{2^{n+{n'}}n!{n'}!w^2}} H_n\left(\frac{\sqrt{2}x}{w}\right)
  H_{n'}\left(\frac{\sqrt{2}x}{w}\right) e^{-\frac{2x^2}{w^2}}
\end{equation}
which is a real number. Thus we can write:
\begin{equation}
I_1=a_nu_na_{n'}^*u_{n'}^* + a_n^*u_n^*a_{n'}u_{n'} = \left(a_na_{n'}^* + a_n^*a_{n'}\right) f_{nn'}
\end{equation}
and the split detector signal becomes
\begin{equation}
S=\left(a_na_{n'}^* + a_n^*a_{n'}\right) \left(\int_0^\infty dx f_{n{n'}}(x)-
  \int_0^{\infty} dx f_{nn'}(-x)\right) = \left(a_na_{n'}^* + a_n^*a_{n'}\right) b_{nn'}.
\end{equation}
From the definition of the Hermite polynomials in
Appendix~\ref{sec:h_poly} we can conclude that $f_{nn'}$ is an odd function when
$n+n'$ is odd and an even function when $n+n'$ is even.
Thus we get
\begin{equation}
b_{nn'}=\left\{
\begin{array}{ll}
0 &\quad \mbox{for} \quad n+n'~ \mbox{even},\\
2 \int_0^{\infty} dx f_{nn'}(x) & \quad \mbox{for} \quad n+n'~\mbox{odd}.\\
\end{array}\right.
\end{equation}
Next we would like to find an analytical solution for the integral $\int_0^{\infty} dx f_{nn'}(x)$.  
\begin{align}
\int_0^{\infty} dx \,f_{n{n'}}(x)&= \int_0^{\infty} dx\,
\sqrt{\frac{2}{\pi}}\sqrt{\frac{1}{2^{n+{n'}}n!{n'}!w^2}}
H_n\left(\frac{\sqrt{2}x}{w}\right)
H_{n'}\left(\frac{\sqrt{2}x}{w}\right) e^{-\frac{2x^2}{w^2}}\\
 &=\sqrt{\frac{2}{\pi}}\sqrt{\frac{1}{2^{n+{n'}}n!{n'}!w^2}}
 \int_0^{\infty} dx\,
H_n\left(\frac{\sqrt{2}x}{w}\right)
H_{n'}\left(\frac{\sqrt{2}x}{w}\right) e^{-\frac{2x^2}{w^2}}\\
 &=\sqrt{\frac{2}{\pi}}\sqrt{\frac{1}{2^{n+{n'}}n!{n'}!w^2}} ~B_1
\end{align}
The integral $B_1$ can be simplified by applying a variable
substitution:
\begin{equation}
v=\frac{\sqrt{2}x}{w}
\end{equation}
and we obtain:
\begin{equation}
B_1=\frac{w}{\sqrt{2}} \int_0^{\infty} dv\,
H_n\left(v\right)H_{n'}\left(v\right)
e^{-v^2}.
\end{equation}
To solve this we require the following two useful identities. For
$n$ being an odd number we get:
\begin{equation}
\int_0^{\infty} dx\, x^n e^{-x^2}= \frac{1}{2}\left(\frac{n-1}{2}\right)! 
\end{equation}
A Hermite polynomial can be written \cite{Abramowitz65} as:
\begin{equation}
H_n(x)=n!\sum_{l=0}^{\left[\frac{n}{2}\right]}(-1)^l\frac{1}{l!(n-2l)!}(2x)^{n-2l}
\end{equation}
with
\begin{equation}
\left[ \frac{n}{2}\right]=\left\{
\begin{array}{ll}
\frac{n}{2} & \quad \mbox{for}\quad n~ \mbox{even},\\
\frac{n-1}{2} & \quad \mbox{for}\quad n~ \mbox{odd}.\\
\end{array}\right.
\end{equation}
We require $n+n'$ to be odd, thus only $n$ or $n'$ can be odd. Assuming $n$
to be even and $n'$ to be odd we can write:
\begin{align}
B_1=&\frac{w}{\sqrt{2}} \int_0^{\infty} du\,
H_n\left(u\right)H_n'\left(u\right)
e^{-u^2} \\
=&\frac{w}{\sqrt{2}}   n!n'!\sum_l^{n/2}\sum_{l'}^{(n'-1)/2} (-1)^l(-1)^{l'}\frac{1}{l!(n-2l)!}
\frac{1}{l'!(n'-2l')!} 2^{n-2l} 2^{n'-2l'}\\
 & \cdot 
\int_0^\infty dx\, x^{n+n'-2l-2l'}e^{-x^2}
\end{align}
Knowing that $(n+n'-2l-2l')$ is an odd number we can replace the integral
with the following:
\begin{equation}
\int_0^\infty dx\, x^{n+n'-2l-2l'}e^{-x^2}=\frac{1}{2}\left(\frac{n+n'-1}{2}-l-l'\right)!
\end{equation}
and thus we obtain:
\begin{align}
B_1=&\frac{w}{\sqrt{2}}n!n'!2^{n+n'-1}\sum_l^{n/2}\sum_{l'}^{(n'-1)/2}(-1)^{l+l'}2^{-2l-2l'}\frac{\left(\frac{n+n'-1}{2}-l-l'\right)!}{l!l'!(n-2l)!(n'-2l')!}
\end{align}
and as our final result:
\begin{align}
b_{nn'}=&2 \sqrt{\frac{2}{\pi}}\sqrt{\frac{1}{2^{n+n'}n!n'!w^2}} B_1\\
=& \sqrt{\frac{2^{n+n'}n!n'!}{\pi}} \sum_l^{n/2}\sum_{l'}^{(n'-1)/2}\left(-\frac{1}{4}\right)^{l+l'}\frac{\left(\frac{n+n'-1}{2}-l-l'\right)!}{l!l'!(n-2l)!(n'-2l')!}.
\end{align}
By using split detectors in \Finesse, one may calculate the control signals for automatic
alignment systems or other similar geometrical control systems.

The \verb|kat.ini| file distributed with \Finesse 1.0 contains the
beat coefficients as described above for modes up to 
\verb|maxtem 40|. 
These coefficients have been created using a SimTools script.

\subsection{Beam detectors}\label{sec:beamshape}
The beam detector has two modes. If the command is used without specifying a 
frequency it acts like a CCD camera, it plots the beam intensity as a function
of the $x$ and $y$ coordinate perpendicular to the optical axis. The
output is a real number computed as:

\begin{equation}
s(x,y)=\sum\limits_{ij} \sum_{nm}u_{nm}(x,y)u^*_{nm}(x,y) a_{inm} a_{jnm}^*\quad\mbox{with}\quad
\{i,j~|~i,j\in\{0,\dots,N\}~\wedge~\w_i=\w_j\}.
\end{equation}

If instead a frequency is specified the beam detector resembles an amplitude detector,
it outputs the amplitude and the phase of a the light field at the given frequency
as a function of the $x$ and $y$ coordinate. The light field at frequency $\w_i$
is given by a complex number ($z$), and is calculated as follows:
\begin{equation}
z(x,y)=\sum_j\sum_{nm} u_{nm}(x,y) a_{jnm} \quad \mbox{with}\quad
\{j~|~j\in\{0,\dots,N\}~\wedge~\w_j=\w_i\}.
\end{equation}

\section{Limits to the paraxial approximation}
\label{sec:limits_paraxial}

The decomposition of a laser beam into a set of Hermite-Gauss modes is
merely an approximation. Also, the coupling coefficients as given
in~\cite{bayer} are derived using additional approximations.  In order to
obtain sensible results one has to understand the limits of these
approximations. From references given within \cite{bayer} the following
simple criteria can be determined. The paraxial approximation can be
understood as a first order approximation in the parameter $\kappa$ with:
\begin{equation}\label{eq:kappa}
\kappa=\left(\frac{w_0}{2\zr}\right)^2=\left(\frac{\lambda}{2\pi
w_0}\right)^2.
\end{equation}
In general we can assume that the approximation is valid for $\kappa\ll 1$
and the error will be of the order of $\kappa^2$. In the case of coupling
one beam into another, the two characteristic parameters should be of the
same order of magnitude.

In order to calculate some limits the above criteria are translated into:
\begin{itemize}
\item{$\kappa<0.1$}
\item{$0.1<\kappa_1/\kappa_2<10$}
\end{itemize}
From the limit on $\kappa$ one can directly derive that the divergence angle
of the beam should be approximately less than $35^\circ$, which corresponds
to limits computed in \cite{siegman}.  From the limit on the relative
difference of $\kappa_1$ and $\kappa_2$ one can derive that the waist size
of the two beams should not differ by more than a factor of $\sqrt{10}$.
Also assuming that the beam size should never exceed these limits, we can
calculate that the waist position should not differ by more than three times
the (smaller) Rayleigh range.


In conclusion, we believe that the following criteria can be used as a rough
guide to judge whether the computation stays within the limits of the
relevant approximations:
\begin{itemize}
\item the diffraction angle of every beam should be less than $30^\circ$;
\item any misalignment between two beams should not be larger than their
diffraction angles;
\item the waist sizes of the beams should not differ by more than a factor
of three;
\item the distance between the waist positions of the beams should be
smaller than three times the Rayleigh ranges;
\end{itemize}
Please note that the above limits do \emph{not} imply that correct results
can be reached by using a reasonable number of modes. In practice, much
stronger limits have to be applied to reach acceptable computation times;
see below.


In summary, for a perfectly aligned and mode-matched interferometer the
results will be correct. Both misalignment and mode mismatch (or not
optimally chosen Gaussian beam parameters) result in light being transferred
into higher-order modes. In general, the number of modes that have to be
taken into account depends on the amount of the misalignment or the amount
of mode mismatch. 

\section{Mode mismatch in practice when using \Finesse}

Mode matching effects can complicate any simulation of interferometer layouts 
using higher order modes. A mode mismatch in this context refers
to any interference between two beams which are best given in 
separate base systems, i.e. parameters beam waist size
and beam waist position associated with the two beams differ.
Thus, on interference and probably for the resulting beam no
optimum base system can be defined. Consequently the phase
information of the beam is spread of a number of transversal 
modes. Mathematically this does not pose a problem, as long as
a sufficiently large number of higher modes are used to describe the beam. 
However, in practice the definition of operating points becomes
much more difficult. And much more care is required to 
assure the simulation is set up correctly. The following sections
illustrate the problem with some examples and give some advice
on how to use \Finesse in the presence of mode mismatching. 

\subsection{Phases and operating points}  
On operating point can be defined as the microscopic positions of the 
interferometer optics. More precisely it is given by the phases
of light fields at the location (optical surface) of interference.

In \Finesse (and many other numeric simulations) the parameters accessible
by the user include the microscopic positions of optical surfaces
but not the phases of light fields. The latter describe an output
of the simulation rather than an input. Thus it is up to the user
to define the microscopic position of optical surfaces such that
the light fields feature the correct phase upon interference.

\Finesse tries to ease this task by several measures, some of which
are optional. The following features reflect design choices which apply
to both modes (plane wave and Hermite-Gauss):
\begin{itemize}
\item{The length of space components is defined as a inter multiple
of the default wavelength $\lambda$. Without Hermite-Gauss modes, i.e. when
the Gouy phase is not considered, this ensures that a space is 'resonant' to the
carrier field, i.e the phase of the field leaving the space is the same 
as on entering it.}
\item{Microscopic positions are given as \emph{tunings} which provides an intuitive
user input as often operating points can be set with tunings of 0, 45 or 90 degrees.}
\end{itemize}

In the Hermite-Gauss mode the following two simplifications can be used:
\begin{itemize}
\item{The {\tt cav} command and the automatic beam trace routine allows to
use cavity eigenmodes wherever possible.}
\item{The {\tt phase} command can be used to zero the Gouy of the \M{00} mode and
of the coupling coefficients for the \M{00} mode, see below.}
\end{itemize}

And most importantly the {\tt lock} command provides the means to 
reach the operating point accurately when the operating point
cannot easily be set by the user manually.

\begin{figure}[htb]
\begin{minipage}{\textwidth}
\IG [scale=0.7, viewport=0 0 295 250] {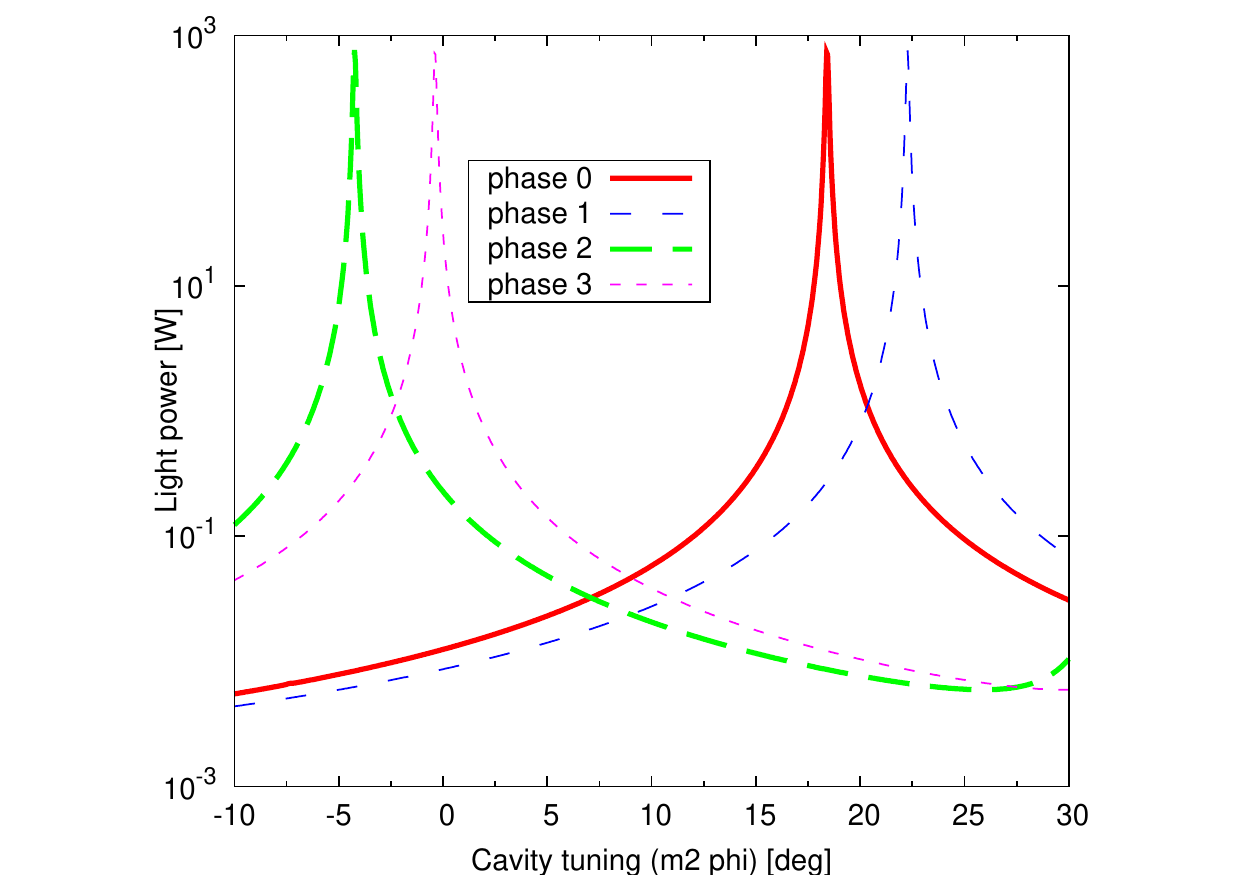}
\IG [scale=0.7, viewport= 0 0 295 250] {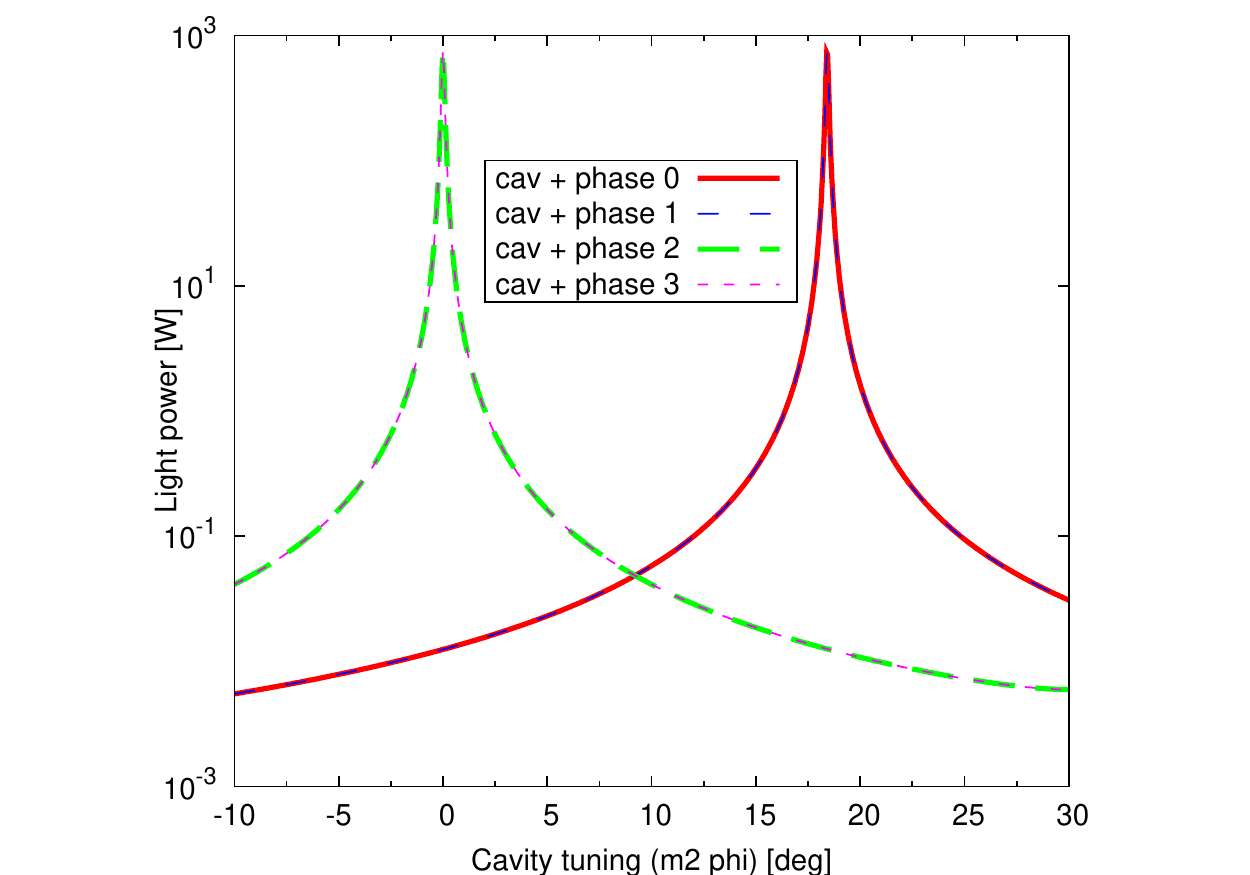}
\end{minipage}
\caption[Effects of the phase command]
{\label{fig:phaseeffect}The light power in the cavity while one mirror in tuned around the
cavity resonance. The left plot was created not using the \cmd{cav} command whereas the right
plot shows the results obtained using \cmd{cav} and thus the cavity's eigenmodes. 
The different simplifications set by the {\tt phase} command change the
mirror tuning at which resonance is achieved. The respective offset values are listed
in \tab{tab:phaseeffect}.}
\end{figure}
\subsection{The {\tt phase} command and its effects}\label{sec:phaseeffect} 
The {\tt phase} command can be used to switch on/off some simplification
with respect to the phase of the light field. The syntax is as follows: 
\begin{itemize}
\item{phase 0:  No simplification. This means for example that the Gouy phase of 
a \M{00} is \emph{not} zero and thus a space of arbitrary length is in general not resonant
to the carrier field} 
\item{phase 1: The phase of coupling coefficients is shifted so that the phase of $k_{00}$ is 0.
The phases of all coupling coefficients $k_{nm}$ for one field coupling, for example, a reflection at
one side of a mirror, are changed by the same amount. To some effect that resembles the 
movement of the optical surface. However, since this is independently applied to all 
coupling coefficients of the surface (e.g. two reflections and two transmissions), this
does not describe a possible real situation. In fact, it might validate energy
conservation or produce other weird effects.}
\item{phase 2:   The phase accumulated in a 'space' components is adjusted to that the phase of \M{00} 
set to 0. This simply removed the effect from the Gouy phase on the 'resonance' of the space
components, i.e. the length of a space is made to be \emph{not} anymore a integer multiple
of the wavelength in order to produce the desired effect of 'resonance' for the \M{00} mode.}
\item{phase 3:  Both of the above (1+2)}
\end{itemize}

Currently the default setting in \Finesse is {\tt phase 3}. The motivation
for this has been to provide the beginner with default settings that
yield intuitive results straight away. Experienced users should
check whether they can develop the habit of using {\tt phase 2} instead
which can be a bit laborious to use but always produces physically correct results.

\begin{table}[h]
\begin{center}
\begin{tabular}{|c|c|c|c|c|c|c|c|c|}
\hline
phase & 0 & 1 & 2 & 3 & 0 & 1 & 2 & 3 \\
cav  & no & no & no & no & yes & yes & yes & yes \\
offset & 18.4 & 22.3 & -4.2 & -0.4 & 18.4 & 18.4 & 0.0 & 0.0\\ 
\hline
\end{tabular}
\end{center}
\caption[Operating points depending on the phase command]
{\label{tab:phaseeffect}Tuning offsets for different use of the {\tt cav} and {\tt phase} command. 
The numbers shown here correspond to the graphs in \fig{fig:phaseeffect}.}
\end{table}

The effects of the {\tt phase} command on the operating point of a simple
Fabry-Perot cavity is shown in \fig{fig:phaseeffect}. In this and the
following examples the cavity parameters are set to:\\
\begin{center}
\begin{tabular}{lllllll}
cavity length & Rc m1 & Rc m2 & T m1 & L m1 & T m2 & L m2\\
\hline
3995 m & -2076 m & 2076 m & $5e^{-3}$&$4.5e^{-6}$&$10e^{-6}$&$50e^{-6}$\\
\end{tabular}
\end{center}
With m1 being the input mirror and m2 the end mirror of the cavity.
These parameters represent a cavity similar to a LIGO/AdLIGO arm cavity.

\fig{fig:phaseeffect} shows that the operating point is strongly dependent on the
use of the {\tt phase} and {\tt cav} commands. The offsets represented as tunings
of the end mirror are listed in \tab{tab:phaseeffect}. The numbers present
the tuning which need to be set by the user to set the cavity on resoance (assuming
a tuning of 0 degrees for the input mirror).

\begin{figure}[htb]
\begin{minipage}{\textwidth}
\IG [scale=0.7, viewport=0 0 295 250] {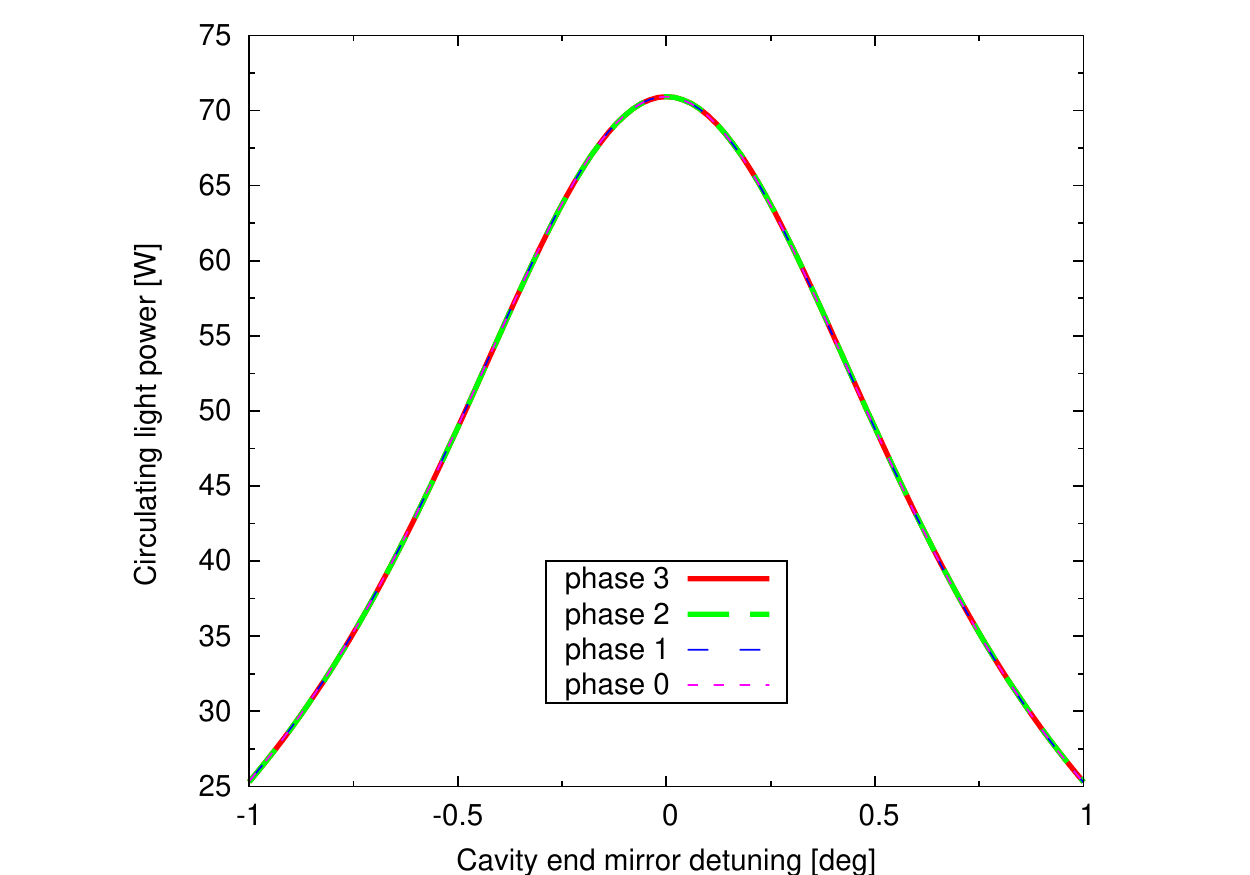}
\IG [scale=0.7, viewport= 0 0 295 250] {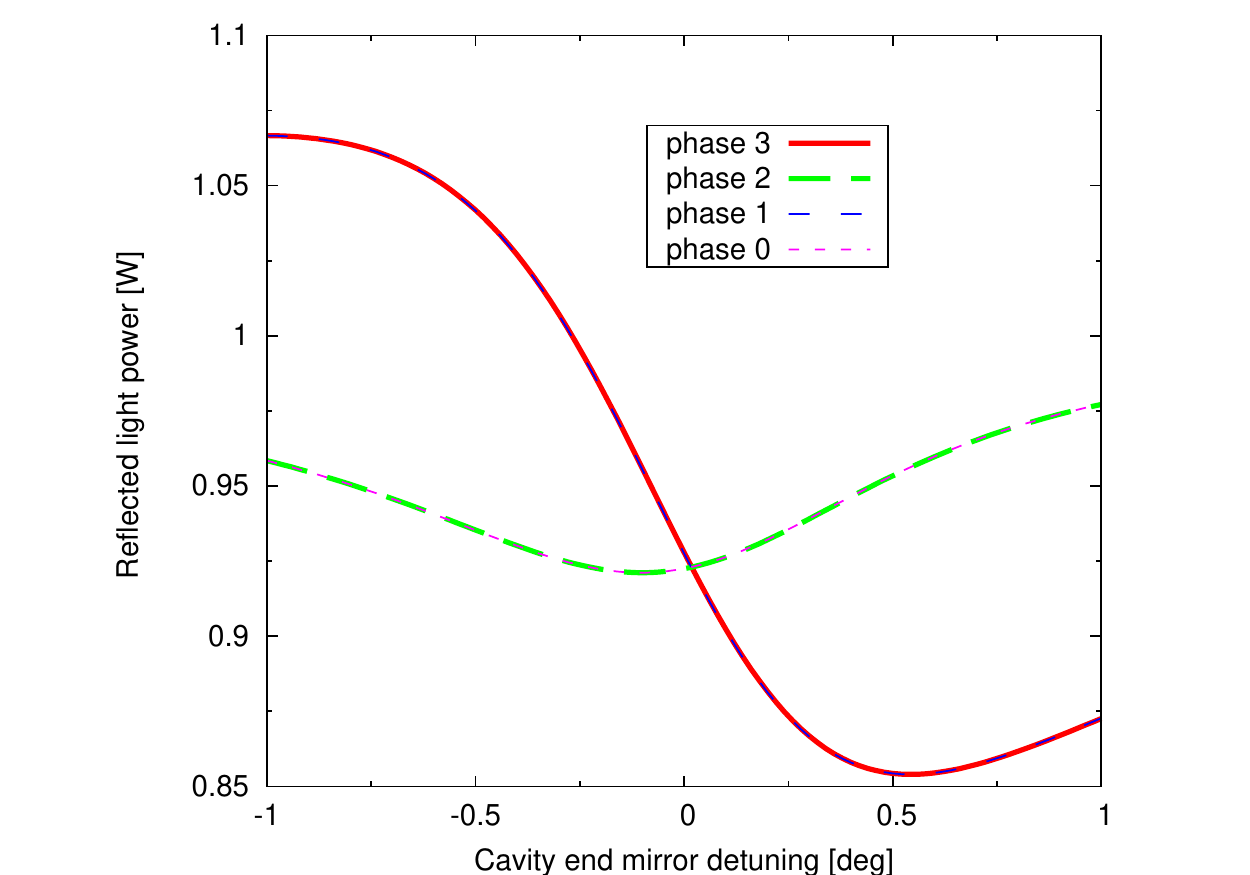}
\end{minipage}
\caption[Failure of energy conservation]
{\label{fig:phaseeffect1}
The two plots show the circulating power (left) and the reflected power (right)
of a Fabry-Perot cavity for different {\tt phase} settings. The input power is set to 1\,W. 
The two phase settings which reset the phase of
the coupling coefficients ( {\tt phase 3} and {\tt phase 1})
can `create' energy as the reflected light can
be larger than the injected light. This illustrates that they can produce
non-physical results whereas {\tt phase 2} and {\tt phase 0} do not show this
problem.  In these examples the cavity is set to use eigenmodes and the operating point has been
adjusted manually. The input beam is not matched to the cavity modes and {\tt maxtem 2}
has been used. Please note, the parameters of \tab{tab:phaseeffect} were \emph{not}
used to generate these plots, arbitrary parameters have been chosen instead to demonstrate the effect.}
\end{figure}

\fig{fig:phaseeffect1} demonstrates another curious effect of the phase command.
In the case of a simple two mirror cavity one can show that {\tt phase 3} and 
{\tt phase 1} can violate energy conservation. In this example the input beam is not 
mode-matched to the cavity, the calculation is performed using cavity eigenmodes and
the operating point has been adjusted manually. Even though the effect is small
and would probably not be noticeable in many simulation results, it shows that the
{\tt phase} command must be used with care.

In summary, a simple intuitive number to
be set manually by the user can only be achieved when using eigenmodes and {\tt phase 2/3}
(it should be noted that in the presence of a mode mismatch at a beam splitter only
{\tt phase 3} will provide an intuitive tuning for the operating point). However, physically
correct results are only guaranteed with {\tt phase 2/0}.

\subsection{Mode mismatch effects on the cavity phase}
In high-finesse cavities a small change of the light phase can quickly
detune the cavity. For example, a mathematical mode-mismatch \emph{inside} the
cavity, which occurs when the beam parameters used in the calculation are not
exactly equal those of the circulating beam, can easily lead to wrong results and has to
be treated with care.

\begin{figure}[htb]
\begin{minipage}{\textwidth}
\IG [scale=0.7, viewport=0 0 295 250] {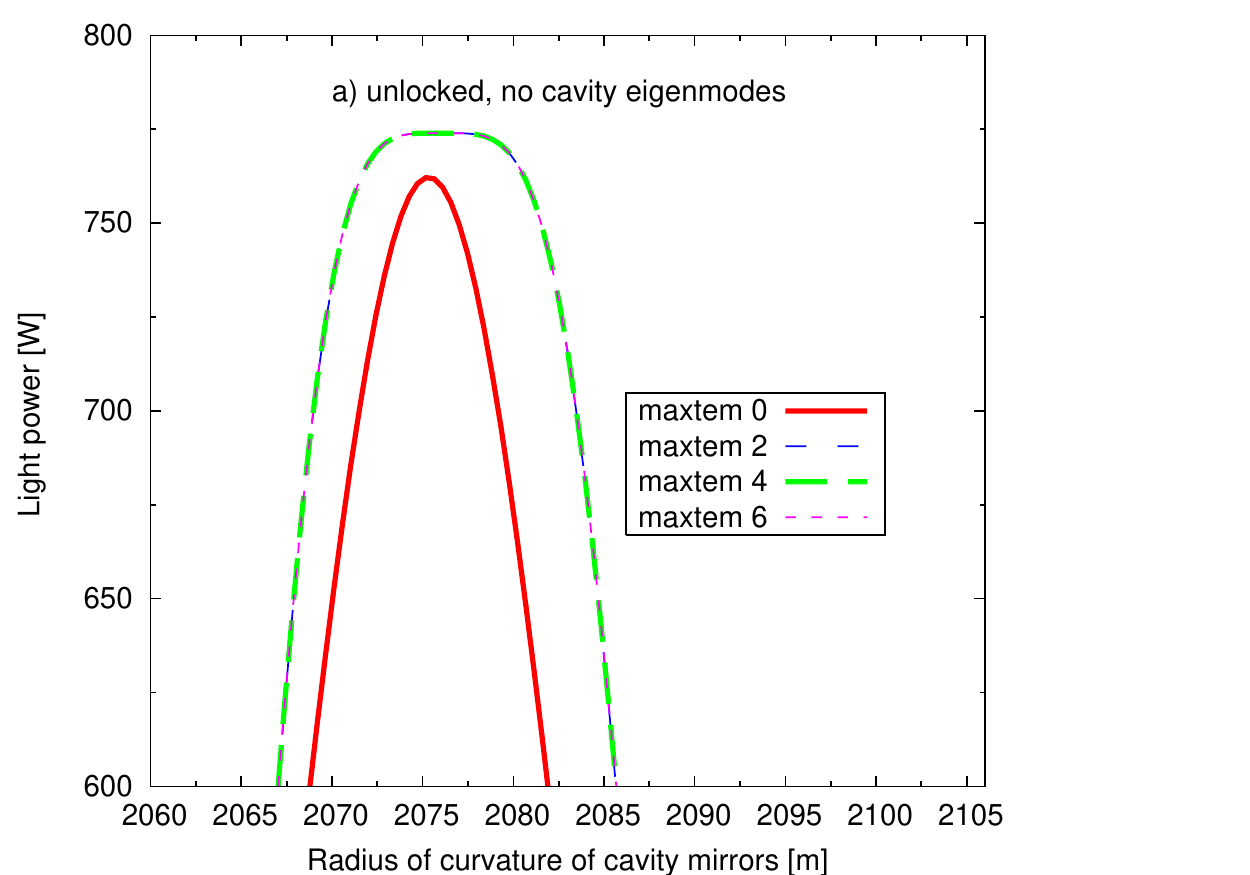}
\IG [scale=0.7, viewport= 0 0 295 250] {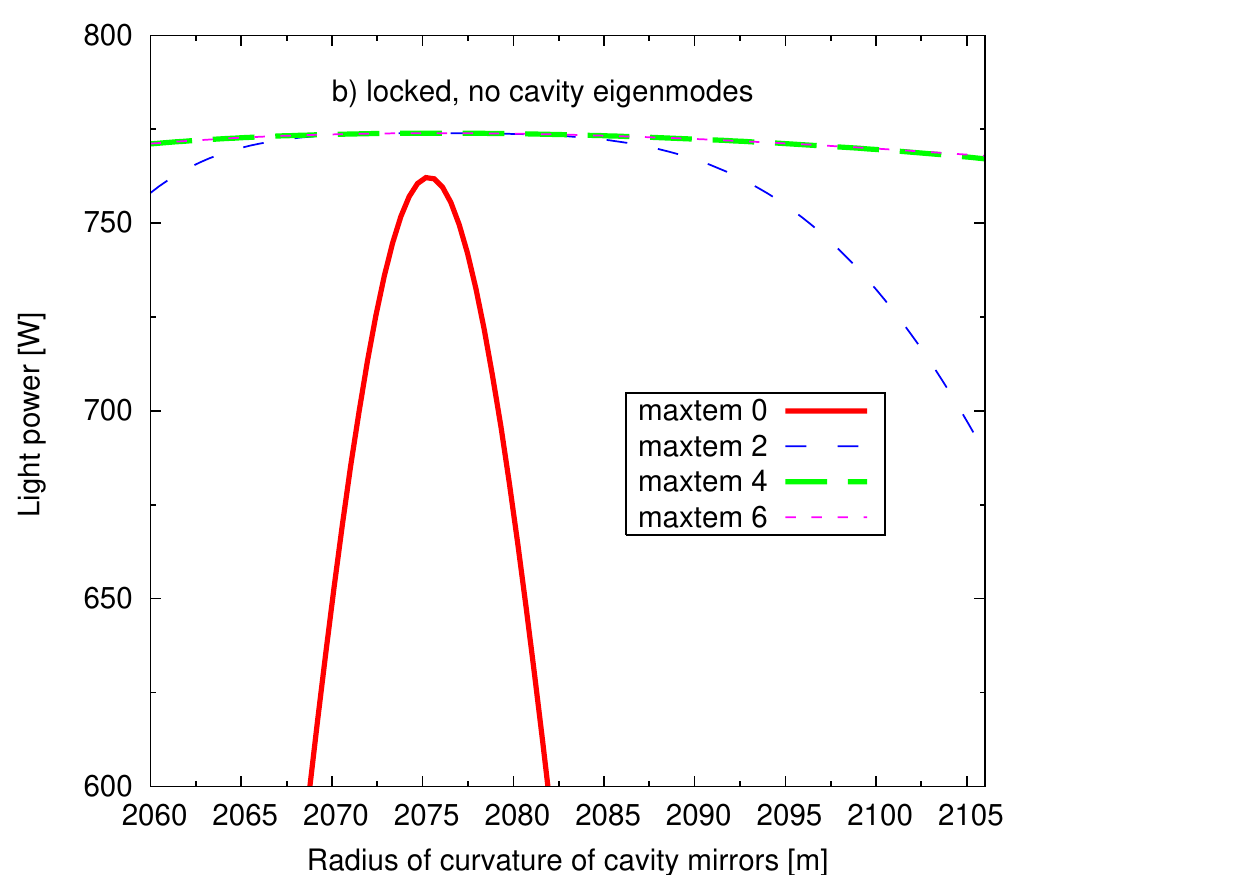}\\
\IG [scale=0.7, viewport= 0 0 295 250] {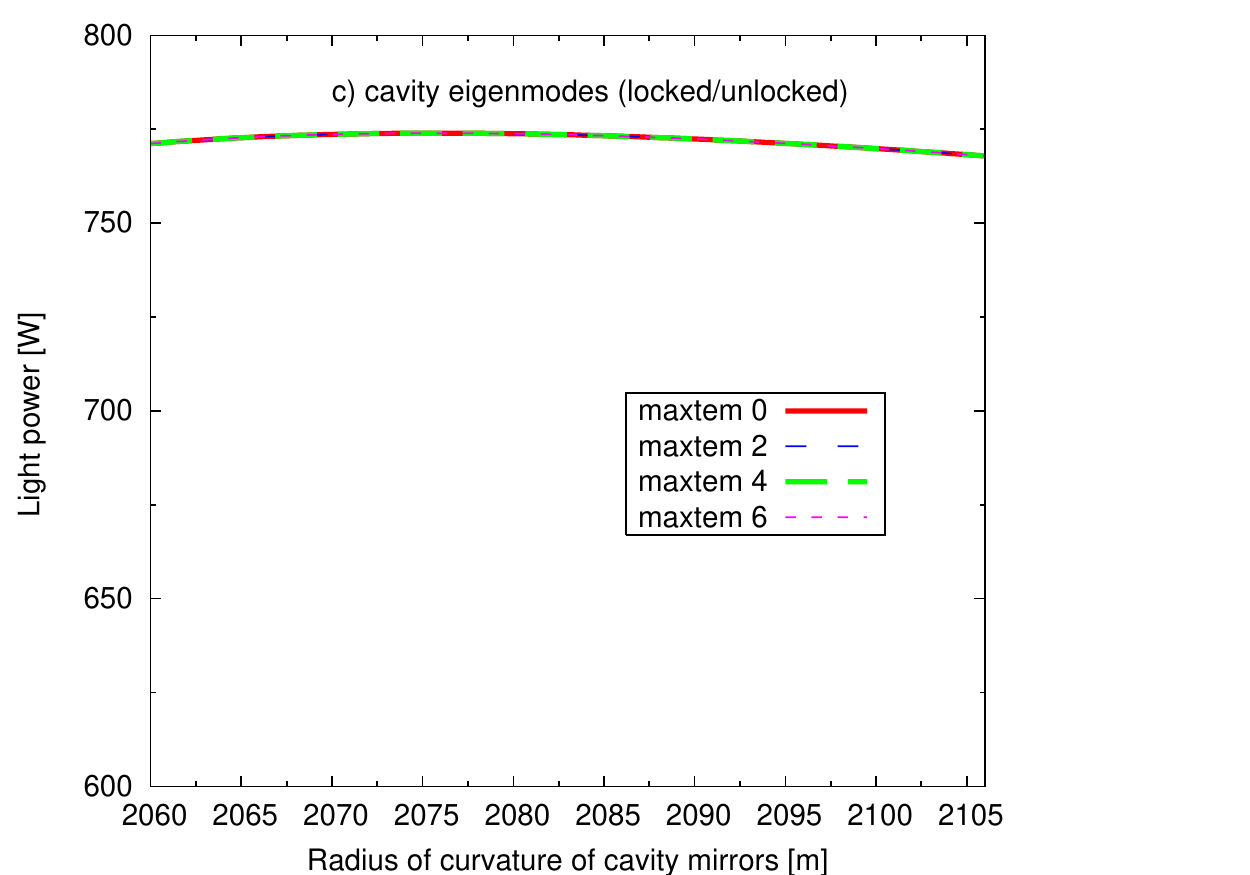}
\end{minipage}
\caption[Cavity Eigenmodes or locking]
{\label{fig:phaseeffect2}The three graphs compare the cavity power as a function
of the radii of curvature of the cavity mirrors (the radii of curvature are 
changes symmetrically, i.e. $Rc_{1}=-Rc_2$). The results do not change for 
different usage of the {\tt phase} command. The input beam has been
set to be matched to the eigenmodes of the cavity for $Rc=2076$\,m. 
Graph a) shows the situation when the {\tt cav} command is not used and
thus the system in analysed with a non-opimal base. A very rapid drop of the cavity power 
can be observed when the radius of curvature deviates from 2076\,m. This is due to a change in the
operating point of the cavity. This can be easily shown by redoing the simulation 
using a {\tt lock} command to keep the cavity on resonance, as shown in graph b).
In this case, using the cavity eigenmodes produces always the correct
result without the need for using a lock or higher order modes, as shown in graph c). 
Of course 
a mode mismatch at a beam splitter does not have a proper base system
like a single cavity. In that case a lock is often the only way to ensure
correct results.}
\end{figure}

\fig{fig:phaseeffect2} shows the power inside a linear cavity as a function of the radii of
curvature of the cavity mirrors (the cavity is symmetric). The importance of using
cavity eigenmodes is demonstrated by the fact that the correct results  
(in this example) are only achieved by either using many higher-order modes, 
preferably with a \cmd{lock} command, or by 
using the cavity eigenmodes. Note that in this case the results do not depend on the setting of the \cmd{phase} command.

\newpage

\section{Misalignment angles at a beam splitter}
\label{sec:a_bs}

The coupling of \HG modes in a misaligned setup as described above is
defined by a misalignment angle.  However, in the case of a \bs under
arbitrary incidence the analysis of the geometry is complicated because it
is commonly described in three different coordinate systems. The purpose of this section
is to derive a precise description of the problem. \Finesse uses an approximation
and the calculations below can be used to estimate the (very small) error of that
approach. 

Our discussion
will be limited to the following setup: a beam travelling along the $z$-axis
(towards positive numbers) and a beam splitter (surface) located at $z=0$,
which may be rotated around the $y$-axis by an angle $\alpha$ ($|\alpha|=$
angle of incidence). This shall be the `aligned' setup. 

To describe a misalignment of the beam splitter, one usually refers to a
coordinate system attached to the beam splitter. This coordinate system is
called $(x',y',z')$ in the following and can be derived---in this case---by
rotating the initial coordinate system by $\alpha$ around the $y$-axis. The
misalignment can be quantified by two angles $\beta_x$, $\beta_y$ that
describe the rotation of the beam splitter around the $x'$-axis and the
$y'$-axis, respectively. Rotation around the $x'$-axis is often called
\emph{tilt}, and rotation around the $y'$-axis simply \emph{rotation}.
Whereas the initial rotation $\alpha$ may be large, the misalignment angles
$\beta_x$ and $\beta_y$ are usually small.  In fact, most models describing
the effects of misalignment use approximations for small perturbations.

Here we are interested in the exact direction of the reflected beam. The
reflected beam, though, may be characterised in yet another coordinate
system ($x'',y'',z''$) with the $z''$-axis being parallel to the reflected
beam. This coordinate system  can be derived from ($x,y,z$) by a rotation of
$2\alpha$ around the $y$-axis.  A misalignment of the beam splitter will
cause the beam to also be misaligned.  The misalignment of the beam is given
by the two angles $\gamma_x, \gamma_y$ that describe the rotation around the
$x''$-axis and the $y''$-axis, respectively.

It can easily be shown that for $\beta_x=0$, the misalignment of the beam is
$\gamma_x=0$ and $\gamma_y=2\beta_y$. For normal incidence ($\alpha=0$) we
get a similar result for $\beta_y=0$: $\gamma_y=0$ and $\gamma_x=2\beta_x$.
For arbitrary incidence, the geometry is more complex. In order to compute
the effect caused by a tilt of the beam splitter we need basic vector
algebra. Please note that the following vectors are given in the initial
coordinate system ($x,y,z$). First, we have to compute the unit vector of
the beam splitter surface $\vec{e}_{\rm bs}$. This vector is rooted at
(0,0,0), perpendicular to the surface of the beam splitter and pointing
towards the negative $z$-axis for $\alpha=0$.

For $\alpha=0$ this vector is $\vec{e}_{\rm bs}=-\vec{e}_z$. Turning the
beam splitter around the $y$-axis gives:
\begin{equation}
\vec{e}_{\rm bs}=\left(\sin(\alpha),0,-\cos(\alpha)\right).
\end{equation}
Next, the beam splitter is tilted by the angle $\beta_x$ around the
$x'$-axis.  Thus, the surface vector becomes:
\begin{equation}
\vec{e}_{\rm
bs}=\left(\sin(\alpha)\cos{\beta_x}, -\sin(\beta_x),
-\cos(\alpha)\cos{\beta_x}\right).
\end{equation}

\begin{figure}[htb]
\begin{center}
\IG [viewport= 0 0 385 400, clip,scale=.4] {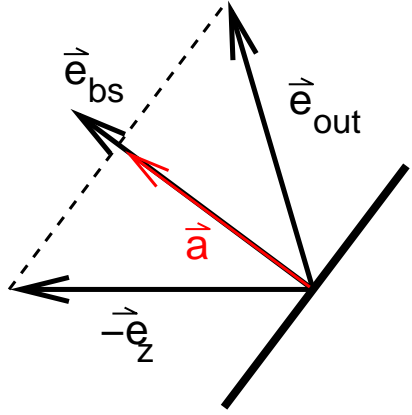} 
\end{center}
\caption[Mirroring a propagation vector at a \bs surface]
{Mirroring of vector $-\vec{e}_z$ at the unit vector of the beam splitter
surface $\vec{e}_{bs}$.}
\label{fig:a_bs_geom}
\end{figure}

In order to calculate the unit vector parallel to the reflected beam, we have
to `mirror' the unit vector parallel to the incoming beam $-\vec{e}_z$ at
the unit vector perpendicular to the beam splitter.  As an intermediate
step, we compute the projection of $-\vec{e}_z$ onto $\vec{e}_{\rm bs}$ (see
\fig{fig:a_bs_geom}):
\begin{equation}
\begin{array}{rcl}
\vec{a}&=&-(\vec{e}_z\cdot\vec{e}_{\rm bs})~\vec{e}_{\rm
bs}=\cos(\beta_x)\cos(\alpha)\vec{e}_{\rm bs}.
\end{array}
\end{equation}
The reflected beam ($\vec{e}_{\rm out}$) is then computed as:
\begin{eqnarray}
\vec{e}_{\rm out}&=&-\vec{e}_z+2(\vec{a}+\vec{e}_z)=2\vec{a}+\vec{e}_z\\
&=&(2\cos^2(\beta_x)\cos(\alpha)\sin(\alpha),
-2\cos(\beta_x)\cos(\alpha)\sin(\beta_x),
-2\cos^2(\beta_x)\cos^2(\alpha)+1)\nonumber\\
&=:&(x_o,y_o,z_o).\nonumber
\end{eqnarray}

To evaluate the change of direction of the outgoing beam caused by the tilt
of the beam splitter $\beta_x$, we have to compare the general output vector
$\vec{e}_{\rm out}$ with the output vector for no tilt $\vec{e}_{\rm
out}\bigl|_{\beta_x=0}$. Indeed, we want to know two angles: the angle
between the two vectors \emph{in} the $x$-$z$ plane ($\gamma_y$), and the
angle between $\vec{e}_{\rm out}$ and the $x$-$z$ plane ($\gamma_x$).  The
latter is simply:
\begin{equation}\label{eq:sindelta}
\sin(\gamma_x)=2\cos(\beta_x)\cos(\alpha)\sin(\beta_x) =
\cos(\alpha)\sin(2\beta_x).
\end{equation}
For small misalignment angles ($\sin(\beta_x)\approx\beta_x$ and
$\sin(\gamma_x)\approx\gamma_x$), \eq{eq:sindelta} can be simplified to:
\begin{equation}
\gamma_x\approx2\beta_x\cos(\alpha).
\end{equation}
One can see that the beam is tilted less for an arbitrary angle of
incidence than at normal incidence. An angle of $45^\circ$, which is quite
common, yields $\gamma_x=\sqrt{2}\beta_x$.

\begin{figure}[htb]
\begin{center}
\IG [viewport=50 50 410 220 ,scale=1.2] {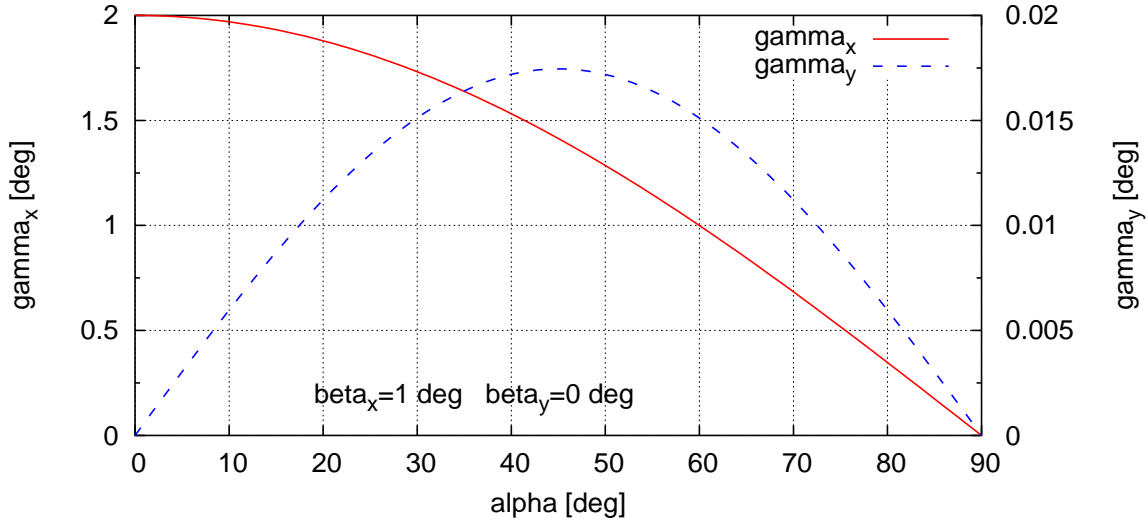} 
\end{center}
\caption[Misalignment angles as functions of the angle of incidence]
{Misalignment angles of a beam reflected by a \bs as functions of the angle
of incidence $\alpha$.  The \bs is misaligned by $\beta_x=1^\circ$ and
$\beta_y=0$.  Note that the values for $\beta$ are very large, see \Sec{sec:limits_paraxial}, and 
have been chosen to artificially enlarge the coupling between vertical and horizontal misalignment
for this demonstration.}
\label{fig:a_bs_angle}
\end{figure}

In order to calculate $\gamma_y$, we have to evaluate the following scalar
product:
\begin{equation}
\begin{array}{l}
\vec{e}_{\rm out}\bigl|_{y_o=0}\,\cdot\,\vec{e}_{\rm
out}\bigl|_{\beta_x=0}\,=\,\sqrt{x_o^2+z_o^2}\cos(\gamma_y),\\
 \\
\Rightarrow\qquad\cos(\gamma_y)=\frac{-1}{\sqrt{x_o^2+z_o^2}} 
\left(x_o\sin(2\alpha)+z_o\cos(2\alpha)\right).
\end{array}
\end{equation}
This shows that a pure tilt of the beam splitter also induces a rotation of
the beam. The amount is very small and proportional to $\beta_x^2$.  For
example, with $\alpha=45^\circ$ and $\beta_x=1$\,mrad, the rotation of the
beam is $\gamma_y=60~{\rm\mu rad}$.  \fig{fig:a_bs_angle} shows the
angles $\gamma_x$ and $\gamma_y$ as functions of $\alpha$ for
$\beta_x=1^\circ$.

\noindent In the case of $\beta_x\neq0$ and $\beta_y\neq0$, the above
analysis can be used by changing $\alpha$ to $\alpha'=\alpha+\beta_y$. 

In \Finesse the coupling between $\beta_x$ and $\beta_y$ is ignored. In other words,
the effect shown by the red (solid) trace in \fig{fig:a_bs_angle} is included in \Finesse
whereas the effect illustrated by the blue (dashed) trace is not.

\begin{figure}[htb]
\begin{center}
\IG [viewport=50 50 400 230,scale=1.1] {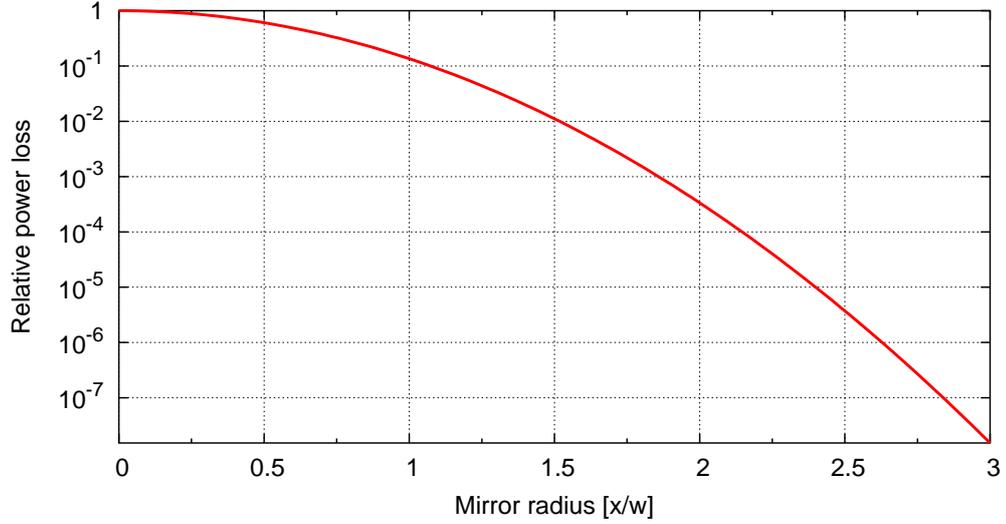} \\
\end{center}
\caption[Power loss on a small mirror]
{The plot shows a lower limit of the power loss experienced by a beam of
size $w$ on reflection at a mirror of diameter $2x$.}
\label{fig:power_loss}
\end{figure}
\section{Aperture effects and diffraction losses}\label{sec:aperture_effects}
By default \Finesse assumes that all optical components have an
infinite size transverse to the optical axis, however you
are can specify the radius of mirrors using the command:

\begin{verb}
attr mirror_name r_ap value
\end{verb}

where \verb|value| is the aperture radius in metres. It is also
possible to vary the aperture radius with the \verb|xaxis| command by
specifying a mirror and using the \verb|r_ap| attribute. The effect of
an aperture is calculated using higher order modes, therefore when an
aperture is defined a coupling coefficient matrix is computed. This
requires using the \verb|maxtem| command to choose the maximum mode
order to use in the calculations. Below we see an example on how
choosing \verb|maxtem| value affects the power loss.

Note that currently it is only possible to calculate the
coupling coefficients by computing the integral from equation~\ref{eq:mapint} numerically. 

From the intensity profile given in \eq{eq:gauss_profile} we can compute
the amount of power with respect to a given area.  The power inside a
circular disk with the radius $x$ (with the centre on the optical
axis) can be computed as:
\begin{equation}
{\renewcommand{\arraystretch}{1.5}
\begin{array}{lll}
P_{\rm disk}&=&\int_{\rm disk} I(r)=\int_0^{2\pi}d\phi\int_0^xdr\, r I(r)\\
&=&\frac{4P}{w}\int_0^xdr\, r
e^{-2r^2/w^2}\\
&=&-P\int_0^xdr \partial_r 
e^{-2r^2/w^2}=-P\left[e^{-2r^2/w^2}\right]_0^x\\
&=&P\left(1-e^{-2x^2/w^2}\right)\label{eq:r_ap_analy}.
\end{array}}
\end{equation}

\begin{figure}[htb]
\begin{center}
\IG[viewport=55 200 540 590, width=0.9\textwidth]{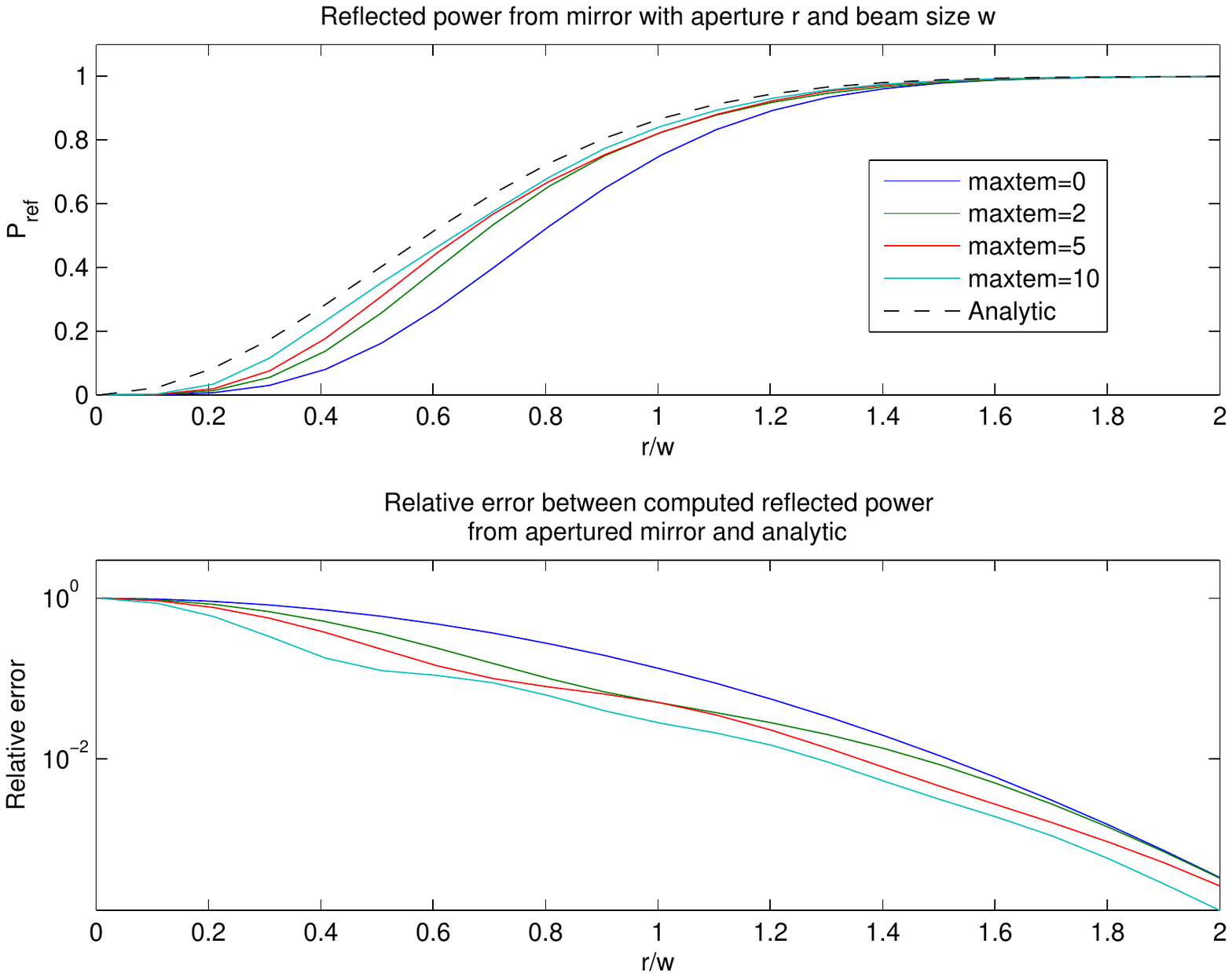}
\end{center}
\caption[Power reflected from apertured mirror of different sizes]
{ Shown in the top plot is the power reflected from an apertured
  mirror with different maxtem options. Analytic refers to equation
  \ref{eq:r_ap_analy}. The bottom plot shows the difference between
  the maxtem results and the analytic result.}
\label{fig:power_loss_r_ap}
\end{figure}
 A beam that is reflected at a mirror with diameter $2x$ will thus experience
a power loss of at least:
\begin{equation}
P_{\rm loss}~=~P\,e^{-2x^2/w^2}.
\end{equation}
This is almost the same distribution as the intensity itself. With respect
to losses we are interested in small deviations from $P$ and look at the
distribution in a different way. \fig{fig:power_loss} shows the amount of
power lost as a function of $x/w$ (the mirror \emph{radius} with respect to
the beam radius).  In modern high finesse cavities where losses due to
surface and coating imperfections can be in the range of a few ppm, the
mirror's diameter should at least be 2.5 times the beam diameter.  Typically
mirrors are designed to be at least three times the nominal beam diameter
including a safety margin allowing for imperfect alignment and some changes
in the beam diameter.

This calculation is meant for deriving limits only. In general, the effects
of apertures have to be analysed taking into account the effects of
diffraction and higher order modes.

Modelling a sharp aperture, such as a finite sized mirror, with modes
or any paraxial method, is not ideal. For example, representing a perfectly sharp
cut-off would require an infinite number of modes. It is therefore
advisable to keep in mind the amount of power that is lost due to the
aperture and also the fact a finite number of modes are used. Below a
simple test was run that looked at the power reflected from a mirror,
we then vary the size of the aperture and plot the reflected power for
increasing maxtem. 

\begin{finesse}
l i1 1 0 n1          
s s1 1 n1 n2
m m1 1 0 0 n2 dump

pd0 Pref n2
gauss g1 m1 n2 1e-3 0
conf m1 knm_flags 8 # need this to force aperture integration
conf m1 integration_method 3 # use cuba parallel integrator
xaxis m1 Rap lin 1e-5 2e-3 40
# can also be applied using attr for a single value
# attr m1 Rap 1e-4
maxtem 5
\end{finesse}

The results are plotted in figure \ref{fig:power_loss_r_ap} which
demonstrates that even a high maxtem of 20 struggled to fully
represent the beam. Using such a high maxtem causes the simulation
computation to become very slow and it is not generally recommended; a
maxtem of 10 appears to offer a good level of accuracy and is still
relatively quick to compute compared to using 20. Using the bottom
plot in figure \ref{fig:power_loss_r_ap} we can estimate the amount of
extra power we are missing compared to what the analytic equation
predicts. It would be important to consider such losses if you were
computing loses down to ppm levels.

\begin{figure}
\begin{center}
\IG[width=1\textwidth]{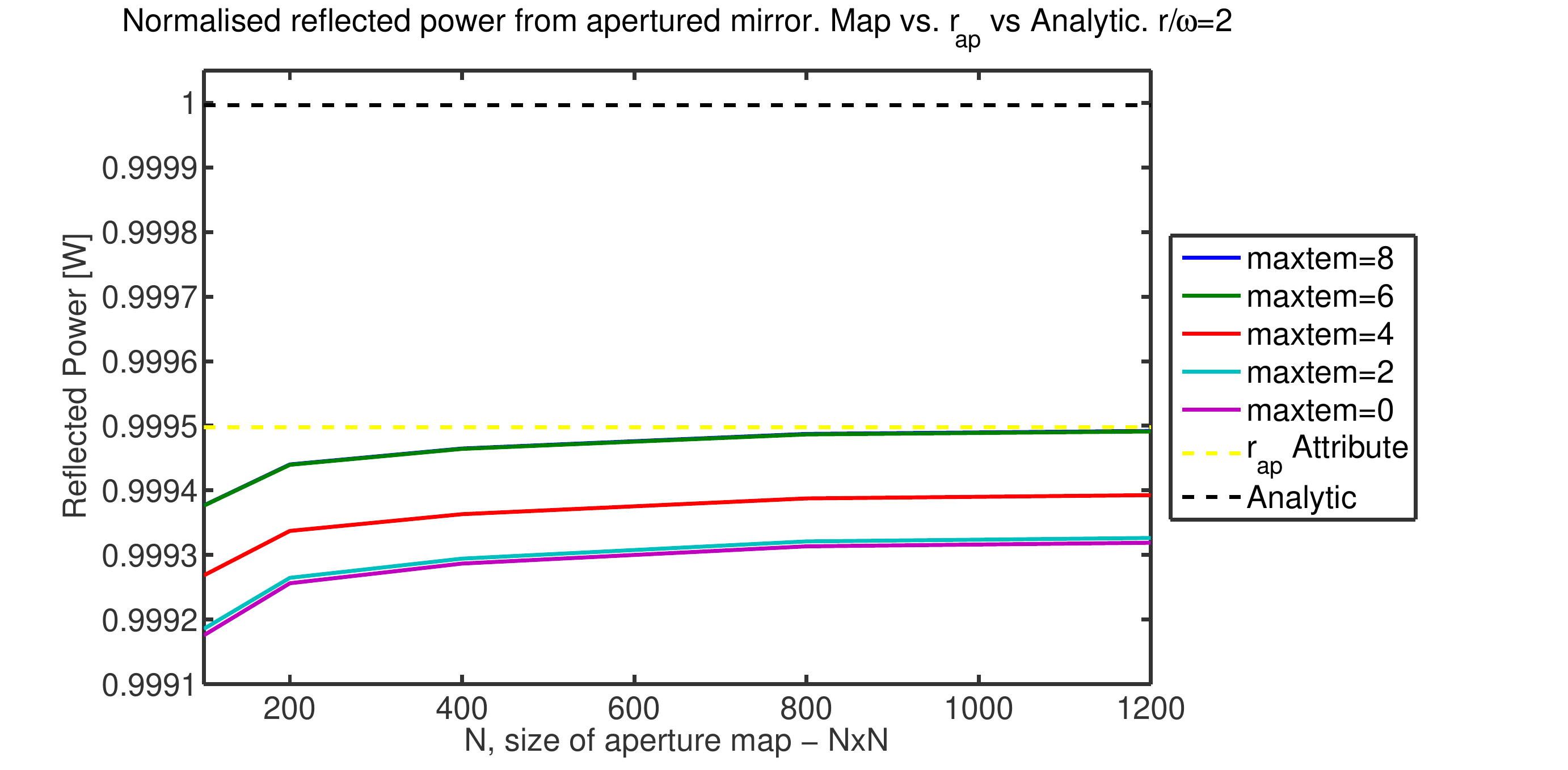}
\end{center}
\caption[Power reflected from aperture mirror map]
{Here the performance of an aperture mirror map is tested against the analytic
solution and using the $r_{ap}$ attribute of the mirror. The $r_{ap}$ attribute
was computed using maxtem 8. Here we can see
a high resolution map is required to reach the same levels of accuracy. Both do 
not reach the analytic value though.}
\label{fig:aperture_N_vs_Pref}
\end{figure}

\begin{figure}
\begin{center}
\IG[width=1\textwidth]{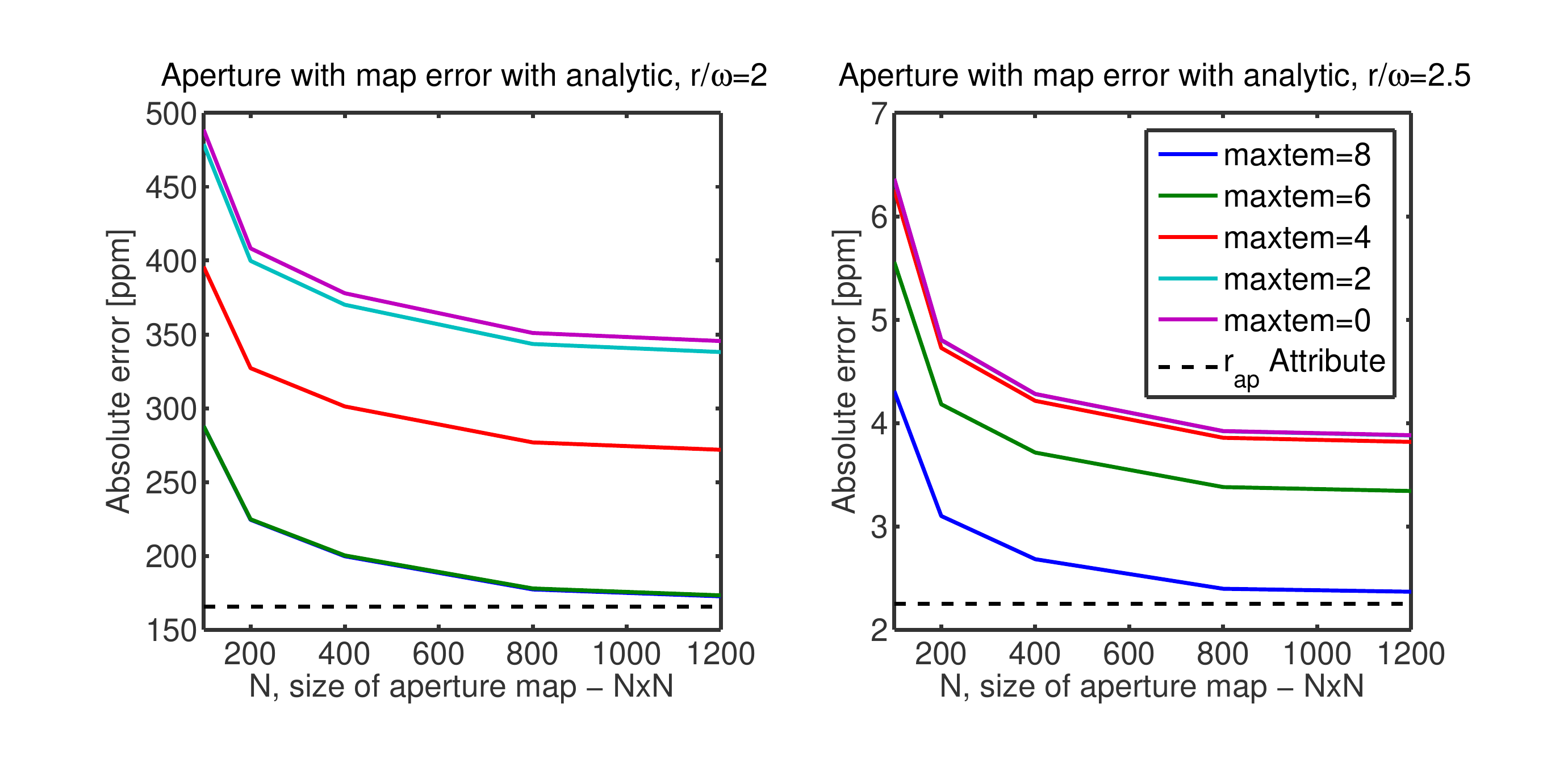}
\end{center}
\caption{The level of accuracy possible is limited by the ratio between the aperture
size $r$ and the beam size $\omega$. Unless a high resolution aperture map is 
used results differ significantly. $r_{ap}$ line is using maxtem 8.}
\label{fig:aperture_N_vs_abs_err}
\end{figure}

It should also be noted that aperture effects can also be computed
by using absorption mirror maps, which have a resolution $N\times N$ with width 
and height of the map of $2 r_{ap}$ (Thus a circular aperture fits perfectly 
within the map). However as shown in Figures 
\ref{fig:aperture_N_vs_Pref} and \ref{fig:aperture_N_vs_abs_err} the
map must be of a high enough resolution to provide accurate answers
down at ppm levels, due to circlular aperture not being perfectly represented.
 Using large maps also has computational performance
issues whereas using the $r_{ap}$ attribute can be comparitviely quick to
compute. Both $r_{ap}$ and the maps are computed using the same routines,
the former however does so in a more efficient manner, thus we recommend
when possible using the $r_{ap}$ attribute.

%% file: advanced_usage.tex
\chapter{Advanced Usage}

This chapter collects some thoughts and examples which 
might be of interest to more advanced users.

\section[\Finesse and Octave/Matlab]{F{\small INESSE} and MATLAB 
(Octave\footnote{Octave is a GNU software package similar to MATLAB. The
examples shown here can often be used with both programs, maybe after some small
changes. I have not tested any of the files for such compatibility though.})}\label{sec:octave}

MATLAB has become a quasi standard tool for solving numerical analyses in many 
areas of science, for example, the interferometer design and commissioning
of gravitational wave detectors utilises MATLAB in various ways. 
For convenience and consistency it is helpful to provide interfaces 
between \Finesse and MATLAB.

There are the following three main ways to use \Finesse with MATLAB
(or Octave):
\begin{itemize}
\item plotting via the automatically generated MATLAB function: running a \Finesse simulation
provides a number of output files, one of which is a MATLAB *.m file containing a function.
This function can be called from MATLAB to automatically plot and/or load the simulation output.
See Section~\ref{sec:mfiles} for more details on the usage of these files.
\item running \Finesse simulations from the MATLAB command window: A set of MATLAB functions,
called \emph{SimTools} is available from the \Finesse download page. The functions should
enable you to read, change and write \Finesse input files from within MATLAB, as well as to
start a simulation and read in the output data, see section~\ref{sec:simtools}.
\item Communicating directly with a running \Finesse process from within MATLAB: 
\Finesse can be used in a client-server mode, in which a
MATLAB client can talk via an internet (TCP/IP) connection to the \Finesse
process, see section~\ref{sec:servermode}. This is the most powerful method for using \Finesse with MATLAB
as it is not restricted to the usual `xaxis' tuning style but can be
used for entirely different types of simulation tasks, such as \emph{tolerancing}
which is part of many commercial packages.
\end{itemize}

\subsection{SimTools}\label{sec:simtools}
Triggered by similar work by Seiji Kawamura and Osamu Miyakawa, I started to
use Octave to post-process the output data of \Finesse. In the course of the Virgo
commissioning activity, Gabriele Vajente then developed
a set of simple scripts that automate certain computation tasks nicely.
During the design process for second generation gravitational wave detectors the
simulation tasks became more complex still and I needed to improve the
automation of tasks further. Therefore, I have started to provide
a consistent set of MATLAB functions, called \emph{SimTools}~\cite{simtools},
that can be used to read, write, edit and execute \Finesse input
files. By now the SimTools includes a great number of utility
functions to read, write and parse any text based simulation input
file but also optics function, from FFT propagation to ABCD matrix
computations\footnote{It should be notes that the SimTools package 
is far from an elegant solution, it has grown historically from 
a set of utility scripts and in particular the parsing of text is
a workaround. It has proven to be very useful but is
difficult to maintain. Eventually I would like to rewrite this
as a Python package, using a more consistent approach.}. 
We have made extensive use of SimTools for the
processing and use of mirror surface maps.

The basic idea behind SimTools is to separate \Finesse input files
logically into smaller parts which can be handled separately. Many of the 
SimTools functions deal with reading and parsing of \Finesse input files.
The two main elements of the text parsing are \emph{text lines} and \emph{text blocks}.
The latter are identified by a special comment in the input file, for example,
the following creates a block with the name `cavity':
\begin{finesse}
m m1 0.9 0.1 0 n1 n2               # mirror m1 with R=0.9, T=0.1, phi=0
s s1 1200 n2 n3                    # space  s1 with L=1200
m m2 0.8 0.2 0 n3 n4               # mirror m2 with R=0.8, T=0.2, phi=0
\end{finesse}
The SimTools function can be used to recognize, read and edit such blocks.
The following example code should give you a first idea on how this can be used:
\begin{finesse}
inname='testconsts.kat';

block=FT_read_blocks_from_file(inname);
myblock=FT_copy_block(block,'constants');

r1=FT_read_constant(myblock,'Rm1');
r2=FT_read_constant(myblock,'Rm2');
disp(sprintf('Reflectivities of m1 and m2: 

myblock=FT_write_constant(myblock,'Rm1',0.7);
\end{finesse}

SimTools are developed independently from \Finesse itself and therefore
will not be described in detail in this manual. Please download the SimTools package
from the \Finesse download page for more information. 

\subsection{Client-Server mode of F{\small INESSE}}\label{sec:servermode}
The Linux and Mac OS X versions of the \Finesse binary can be started in a so-called 
\emph{servermode} by calling the program with:
\begin{finesse}
kat --server <portnumber (11000 to 11010)> [options] inputfile 
\end{finesse}
The port number can be chosen by the user, the other options may be any of the usual options
for calling \Finesse. Also, the input file can be any unchanged input file. For example, we might
load the file \cmd{bessel.kat} with
\begin{finesse}
kat --server 11000 bessel.kat 
\end{finesse}
The input file will then be read and pre-processed as usual but instead of actually performing the 
simulation task (i.e. running along the xaxis) \Finesse will become idle and listen to incoming TCP/IP 
connection via the user-defined port (11000 in this example). 

\begin{figure}[htb]
\begin{minipage}{\textwidth}
\IG [scale=0.715] {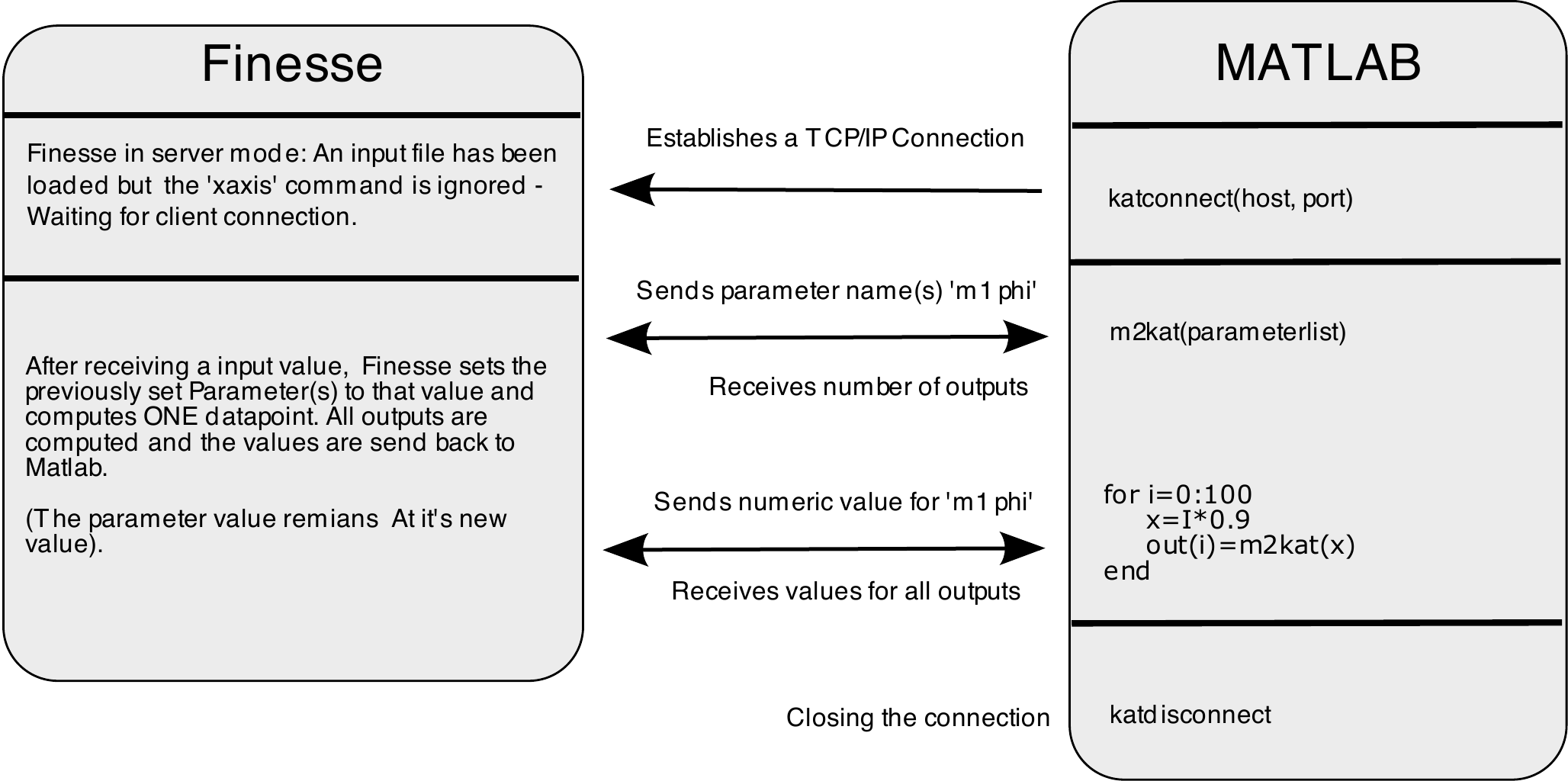}
\end{minipage}
\caption[MATLAB-Finesse communication over TCP/IP]
{\label{fig:tcpip}
This sketch illustrates how the communication between 
Finesse and MATLAB/Octave would work in a simple example. \Finesse loads an input file and then becomes idle
until a `katconnect' command from a MATLAB session opens a TCP/IP connection. Then the
`m2kat' command can be used to send or receive data over the connection. Typically 
`m2kat' is first used to specify which data is transferred (CONFIG) and then is used
again to send and receive numerical data (TUNE). After the simulation task, the command
`katdisconnect' must be used to close the TCP/IP connection again.}
\end{figure}

A MATLAB/Octave client can then send commands via TCP/IP to \Finesse, see \fig{fig:tcpip}. 
This works by three new 
MATLAB functions (from source files which need to be compiled first, see below):
\begin{itemize}
\item katconnect: establishes a connection with the \Finesse server
\begin{finesse}
usage:  socket=katconnect('server', port (11000-11010))
\end{finesse}
where `server' is the network address of the server (use `localhost' if you do all this locally
on the same computer) and `port' the port number chosen when starting the \Finesse server.
`socket' will return the index of the opened socket.
\item katdisconnect: closes a connection with the \Finesse server
\begin{finesse}
usage:   katdisconnect(socket)
\end{finesse}
with `socket' the socket number received with the `katconnect' command.
\item m2kat: performs all communications through the TCP/IP connection 
This commands can
\begin{itemize}
\item set a certain parameter to a new numeric value 
\item receive the value of a parameter
\item receive output data, for example, the photodiode outputs.
\end{itemize}
`m2kat' has three different usage modes, these are chosen automatically depending on the 
specified input and output arguments:
\begin{finesse}
 usage:
   CONFIG:
   [number_of_outputs]=m2kat(socket, number_of_parameters, 'parameter string')
   send parameter names to be tuned and get number of output data values
 or 
   TUNE:
   [output_data] = m2kat(socket, number_of_outputs, parameter_values)
   send parameter values and get output data
 or 
   INFO:
   [output_data] = m2kat(socket, number_of_parameters)
   get current values of parameters defined by a previous call of
   m2kat
\end{finesse}
The CONFIG mode is required in advance of TUNE or INFO 
commands. The CONFIG command tells the server which parameters of the
interferometer will be set or polled in the following session. With a TUNE command
one can set new numerical values to these parameters, whereas an INFO command
would return their current numerical values. A TUNE command would also return 
one output data point, i.e. one numerical value for each output specified in the 
input file. The usage of `m2kat' is a bit complex and sensitive to mistakes. It
therefore requires a careful preparation of the MATLAB client script. Please look at the provided
example for further guidance.
\end{itemize}

\paragraph{Example MATLAB client file} This example recreates a normal \Finesse simulation by tuning 
one parameter and printing the output.
\begin{finesse}

%
%
%
%
%
%
%
%
%
%
%
%
%
%
%
%

hostname='localhost';
port=11000;

noparams=2; 
parameterlist='m1 phi m2 phi'; 

N=20001;
min=-10;
max=180;
x=linspace(min,max,N);

max_calls=N;
curr_call=0;
call_range=round(N/100.0)+1;
last_print=0;
tic

socket=katconnect(hostname,port);

if (socket>0)
  nout=m2kat(socket,noparams,parameterlist);
  if (nout>0)
    out=zeros(N,nout);
    disp(sprintf(' \n'));
    disp(sprintf(' \n'));
    for i=1:N
      out(i,:)=m2kat(socket,nout,[x(i),0.0]);
      curr_call=curr_call+1;
      last_print=last_print+1;
      if (last_print>=call_range)
        disp(sprintf('\b\b\b\b\b\b\n
                                             max_calls)));
        last_print=0;
      end
    end
  end
end

disp(sprintf('\b\b\b\b\b\b\n100
toc
katdisconnect(socket);
plot(x,out);
\end{finesse}
\begin{figure}[htb]
\begin{center}
\IG [scale=0.7] {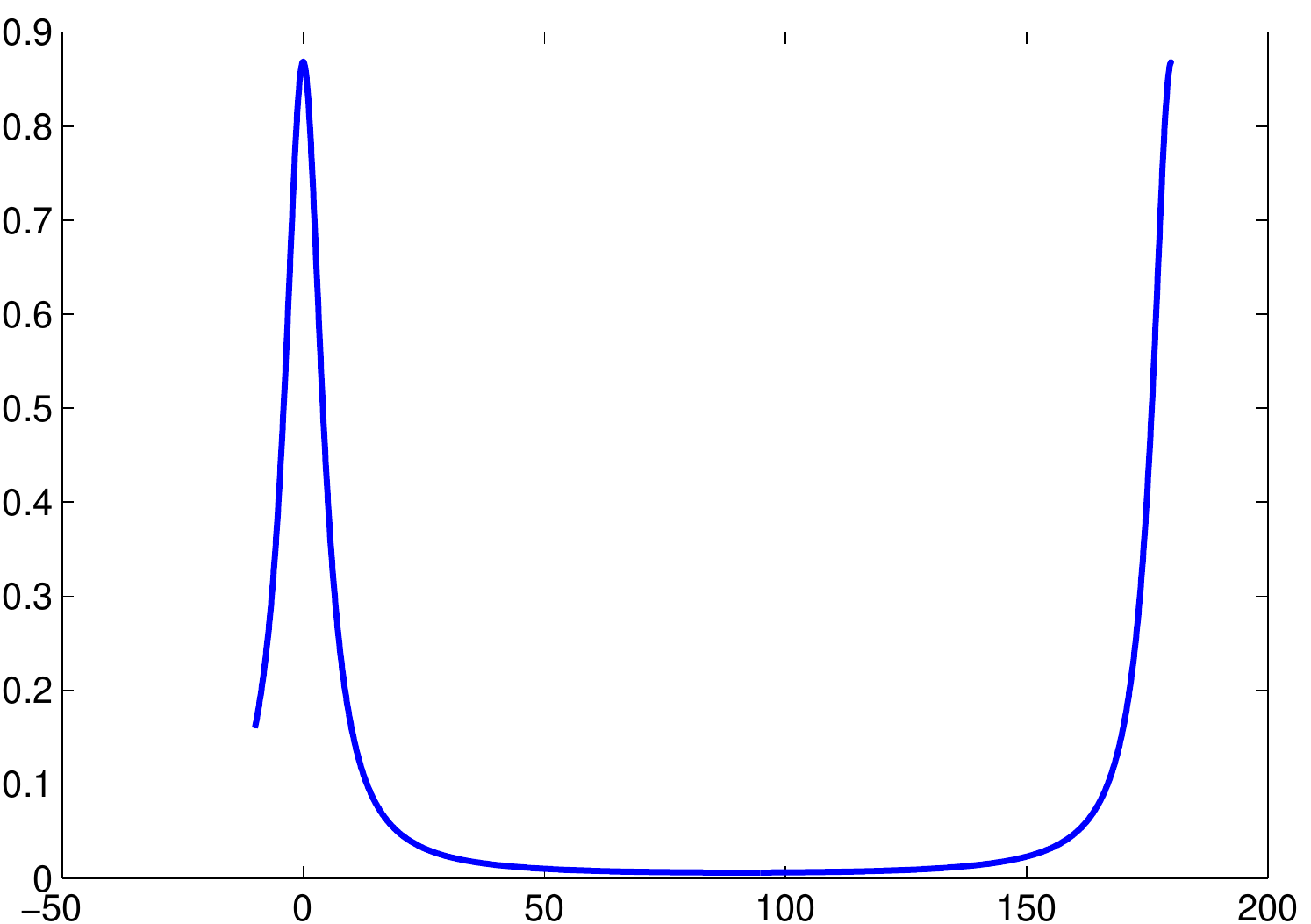}
\end{center}
\caption[MATLAB-Finesse example plot]
{\label{fig:matlab-finesse2}
Graphical output of the example script demonstrating the communication between
MATLAB and \Finesse running in server mode.
}
\end{figure}
The terminal output of \Finesse during the example above would look
similar to this:
\begin{finesse}
moon:~/work/kat/test/m2kat$ kat --server 11000 cavity1.

------------------------------------------------------------------------
                     FINESSE 0.99.7 (build 3227)
       o_.-=.        Frequency domain INterferomEter Simulation SoftwarE
      (\'".\|        03.07.2008      A. Freise  (afreise@googlemail.com)
      .>' (_--.
   _=/d   ,^\        Input file cavity1.kat,
  ~~ \)-'   '        Output file cavity1.out,
     / |             Gnuplot file cavity1.gnu 
    '  '                                        Wed Jul  9 11:33:04 2008
------------------------------------------------------------------------
*** cmd_process: processing [0]...done.
*** listen: listening on port 11000
*** cmd_process: processing [0]...done.
*** cmd_process: processing [0]...done.
*** cmd_process: processing [0]...done.

*** listen: accepted a connection from 127.0.0.1:-2491
*** listen: num connections = 1

*** receivecommands: processing CONFIG command
Number of Parameters = 2
Read parameter list: m1 phi m2 phi
*** receivecommands: received 2 parameters
*** receivecommands: processing quit command
*** listen: num connections = 0
*** cmd_process: processing [0]...done.
\end{finesse}
The first line starts \Finesse in server mode, the banner is printed and \Finesse
starts to `'listen' for incoming TCP/IP connections. It then accepts a connection
from IP number 127.0.0.1 (this means the MATLAB client I am using runs on the same computer).
The following lines acknowledge the receipt of a 'CONFIG' command. The following `TUNE'
commands do not produce any terminal output. During this example run MATLAB would
produce terminal output as follows:
\begin{finesse}
>> m2katexample
*** Creating Socket ...
*** server name is localhost
*** server internet name is localhost
*** connecting to localhost
*** Socket 9 opened!
*** CONFIG: OK

100

Elapsed time is 0.449107 seconds.

### Closing socket 9 ...   ... Done!
>> 
\end{finesse}
 The graphical output of course depends on the details of the finesse input file.
Figure~\ref{fig:matlab-finesse2} shows the output of this example.

\subsubsection{Compiling the Client programs}

The programs are scripts that run in MATLAB and Octave. For simplicity I do not provide binary versions of these,
however, they are very easy to compile.

\paragraph{Compiling in MATLAB}

The MATLAB client consists of a number of C source files which have to be compiled with the MATLAB compiler.
The files are:
\begin{finesse}
katconnect.c      : create TCP/IP connection with server
katdisconnect.c   : break TCP/IP connection
m2kat.c           : send commands to and receive data from server
m2kat.h           : include file for m2kat.c
\end{finesse}
These files can be compiled from within MATLAB with the mex command. For example, on a Macintosh the 
command (used inside the MATLAB command window)
\begin{finesse}
mex katconnect.c 
\end{finesse}
will create the binary file `katconnect.mexmaci'. Once this file exists you can call \cmd{katconnect}
or do \cmd{katconnect} just like with any MATLAB command. Please compile the three files
`katconnect', `katdisconnect' and `m2kat' and keep all files, source and binary, in 
a directory where MATLAB can find them.

\paragraph{Compiling in Octave}

The compilation from Octave works exactly as above. You can use the `mex' command from Octave with the 
same source files. The only difference is that that the mex command in Octave will create `*.o' and `*.mex' 
files storing the binary commands.

\section{Speed improvements}

The time needed for a simulation with \Finesse is of course
strongly connected with the simulation task. In most cases
\Finesse will only take seconds to produce a result in which
case a speed optimisation is not required. However, with some
more complex problems, for example, when a large number of
higher-order modes are used, the computation time
can increase dramatically. The sections below give some information
and general advise on how to avoid long computation times.

\subsection{Higher-order modes}
\begin{figure}[htb]
\begin{center}
\IG [scale=0.7] {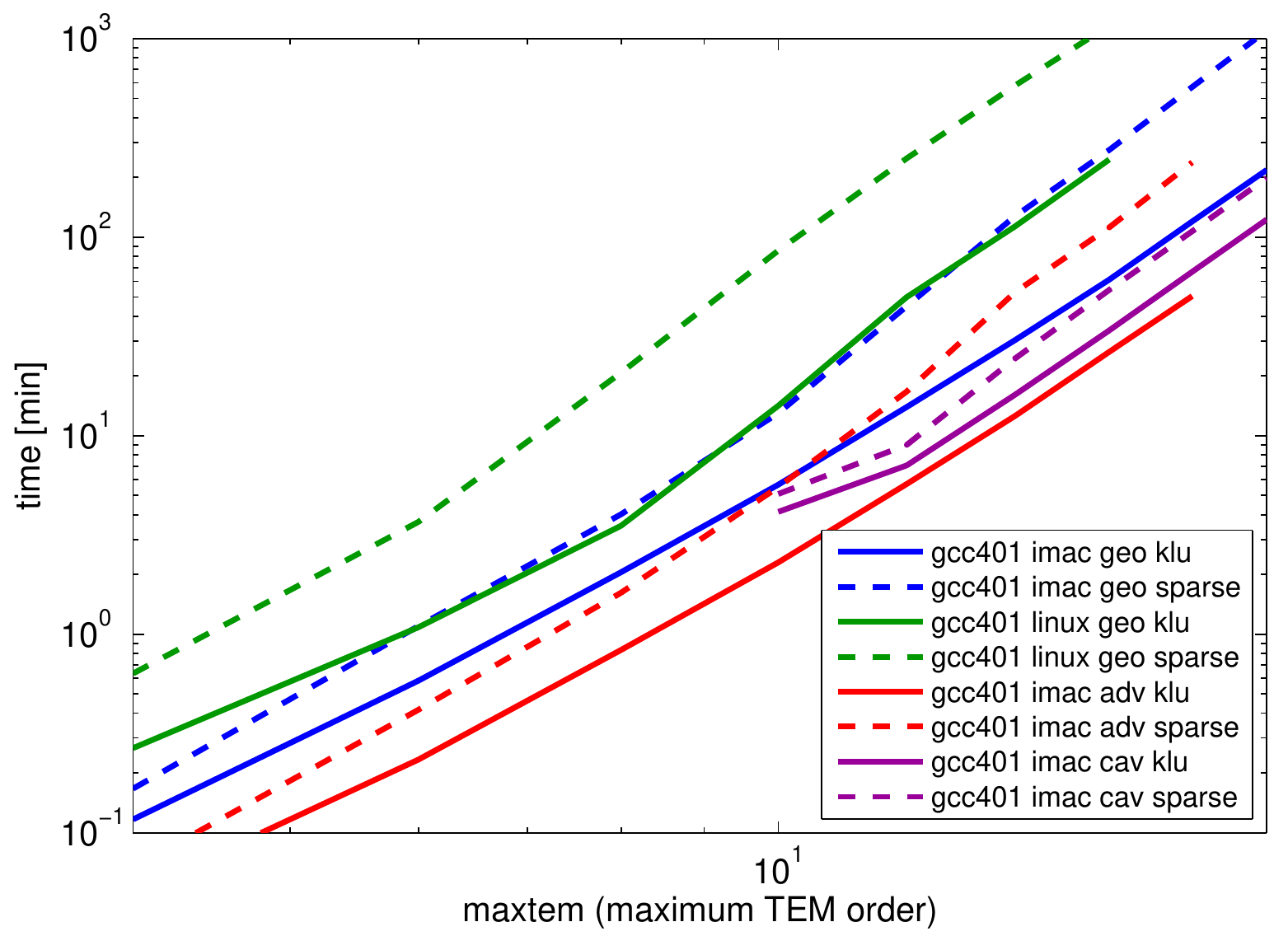}
\end{center}
\caption[Computation time for example files]
{\label{fig:speed1}The graph shows measured computation times for 
three example optical setups: `geo' represents the optical
configuration of GEO\,600, `adv' refers to Advanced Virgo and
`cav' uses a simple two-mirror cavity. All versions of \Finesse
tested here have been compiled with the gcc compiler version 4.0.1.
The test runs have been performed on an iMac or a PC running linux.
The terms `klu' and `sparse' refer the options `{-sparse}'
and `{-klu}', see text. The scaling is slightly different for 
the different files and computers. However, generally the computational
time scales like $m^{4}$ or $m^{5}$ with $m$ being the maxium TEM order
given via {\Co maxtem}.
}
\end{figure}
The most dramatic change is computation speed can be seen observed
when the number of higher order modes is changed with the {\Co maxtem}
command. \fig{fig:speed1} shows 
some measured computation times for example simulations
as a function of the value given with {\Co maxtem}.
In general, the computational
time scales like $m^{4}$ or $m^{5}$ with $m$ being the maxium TEM order.
In consequence large values for {\Co maxtem} should be used only
in special cases when a high spatial resolution of the beam is required.
\subsection{KLU versus SPARSE}
\begin{figure}[htb]
\begin{minipage}{\textwidth}
\IG [scale=0.63] {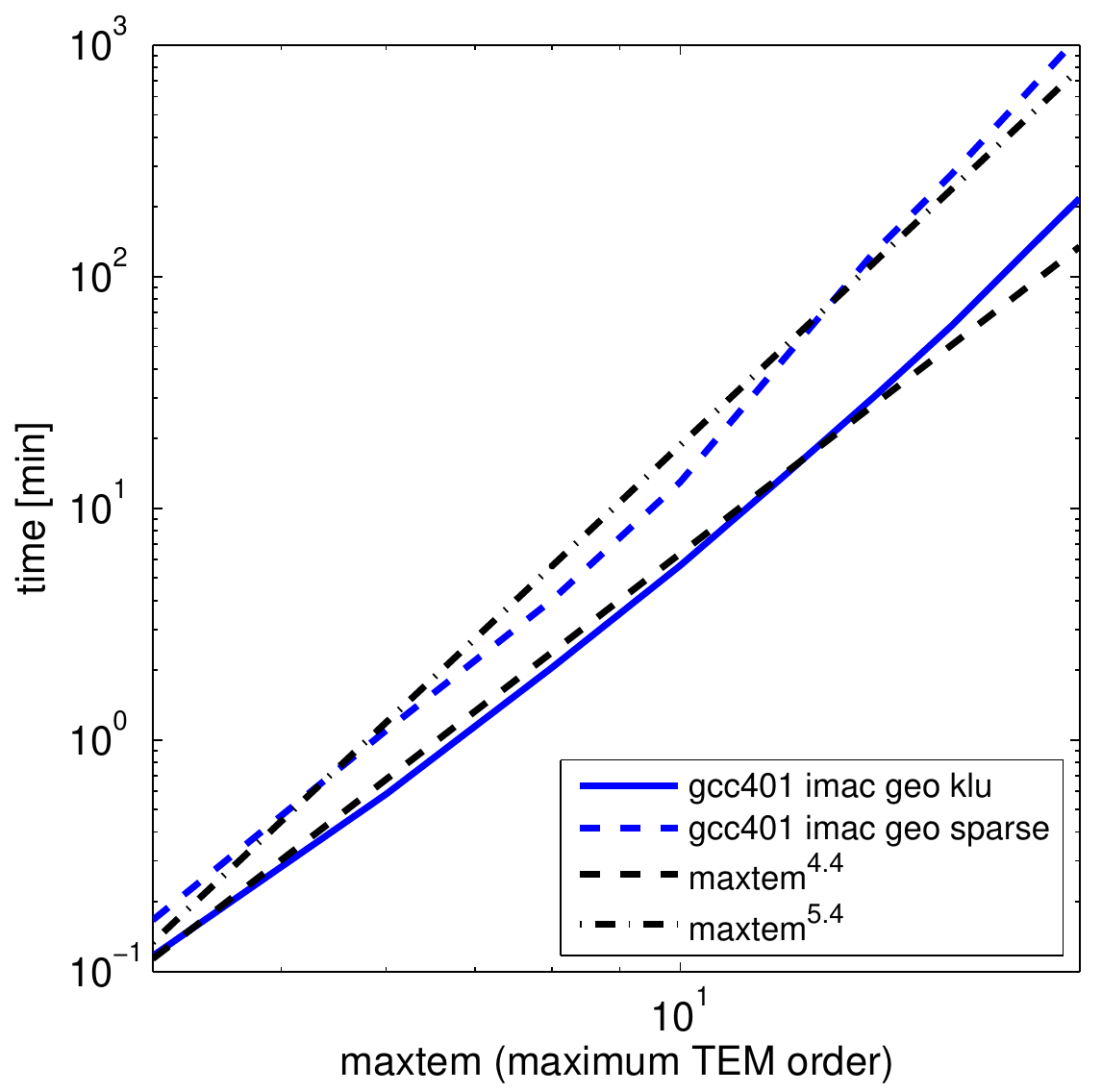} \IG [scale=0.63] {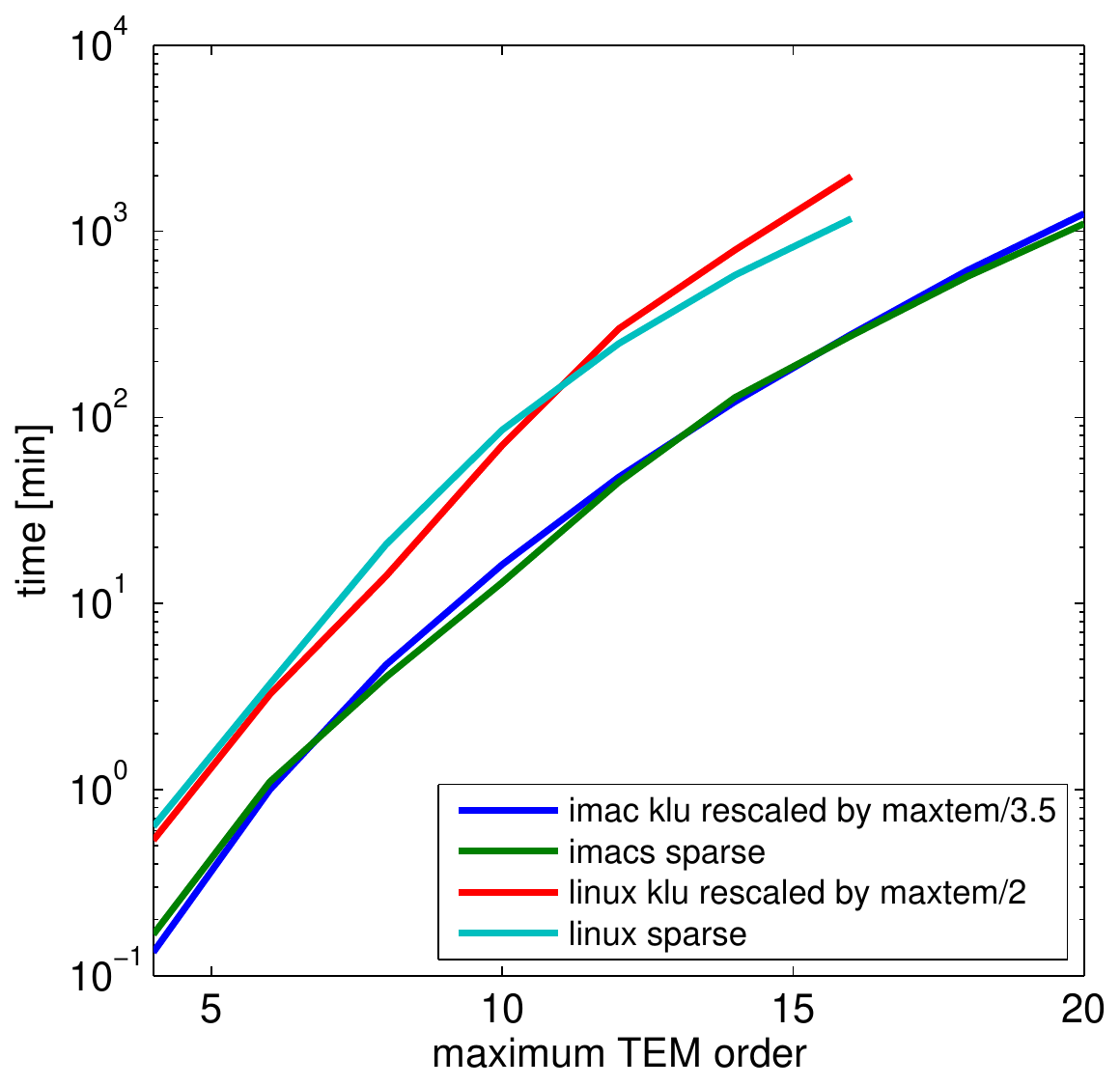}
\end{minipage}
\caption[Timing: KLU versus SPARSE]
{\label{fig:speed2}The left plot shows two traces from \fig{fig:speed1} plotted with
some power laws. The right plot compares the timing for the `klu' and the 
`sparse' options.}
\end{figure}

\Finesse currently implements two different sparse matrix solvers, the SPARSE 1.3
package~\cite{sparse} and the KLU library~\cite{DavisKLU2006}. 
In my experience the KLU package provides a better performance for very large
interferometer matrices. In particular, when many higher-order modes are used
KLU typically gives times which are lower by a factor maxtem$/3.5$, depending
on the computer as well as the input file, see, for example \fig{fig:speed2}. 
For settings of maxtem$=3$ the SPARSE package can give better results, up to a factor of two
faster. 

Therefore currently \Finesse automatically switches to KLU when {\Co maxtem} is
set to values of 4 or higher. Otherwise the SPARSE package is used by default.
This default behaviour can be changed manually with the options `{-klu}' and 
`{-sparse}' which switch to the respective package regardless of the {\Co maxtem}
value.


%% file: appendix_geo_sens.tex
\chapter{Shot-noise limited sensitivity of GEO\,600}\label{sec:geo_sens}

This section records \Finesse simulations from 2007 which were undertaken to
investigate the shot-noise limited sensitivity of the GEO\,600 detector. As it 
turns out the calculation of the sensitivity contains a fair number of factors of
2 and $\sqrt{2}$, leading to potential confusion when comparing results.
Also the GEO detector using a heterodyne read-out scheme at that time
required a slightly more complicated shot-noise computation than that given
in the usual textbooks. As a result our models of the GEO\,600 shot-noise
did not match the measured sensitivity exactly. This appendix and \Sec{sec:shotnoise}
report on the effort to validate the shot-noise model used by \Finesse and to
demonstrate how to correctly derive the shot-noise limited sensitivity.
The plots in \mFig{fig:S5sens} show that the here derived shot-noise model
correctly matches the measured data.

\section{The qshot command}
Early versions of \Finesse were able to compute only a very 
approximative shot noise level using the Schottky equation.
The current version of \Finesse features a new command \code{qshot}
which provides a more accurate approximation of the shotnoise in
a certain photo diode signal. The \code{qshot} command
can correctly compute the shotnoise level in the presence of
a number of modulation sidebands and in the absence of
squeezing and radiation pressure effects. Details of the implemented 
algorithm have been collected in a paper~\cite{Harms07}.

\section{Comparing the different methods}
The following shows how to use the \Finesse syntax correctly to compute the shot-noise
limited sensitivity of GEO\,600; it explains step by step how the apparent strain 
is computed from the transfer function end-mirror motion $\longrightarrow$ 
detector output ($T$) and the amplitude spectral density $\sqrt{S_P}$ associated
with shot-noise\footnote{Note that GEO\,600 has folded arms with the folding mirrors called
`far' mirrors. The displacement should be injected at the `central' end mirrors. If the `far' mirrors 
are used the effect is amplified by a factor two due to the double reflection which must be corrected 
manually.}.
Six extracts from \Finesse input files will be shown below, each
demonstrating a correct bit slightly different method to obtain
a shot-noise limited sensitivity. The first method is as follows:
\loadkat{geo_sens1.kat}

The method above is the most explicit and relies on several
manual conversions and corrections which are apllied via
\code{func} commands. The correction and conversion factors are:
\begin{itemize}
\item [a)] the shotnoise is first divided by $\sqrt{2}$ to simulate the effect of the mixer, see
\Sec{sec:pd}.
\item [b)] then the shotnoise is multiplied by $\sqrt{2}$. This is a upper limit for the 
noise increase due to the heterodyne type measurement.
\item [c)] The transfer function is computed by \Finesse as W/rad and must be converted
      into W/m by mulitiplication with $2\pi/\lambda$, see \Sec{sec:half_fringe}.
\item [d)] Shotnoise can then be converted into apparent displacement noise by
dividing it by the transfer function: $\sqrt{S_{\Delta L}}=\sqrt{S_P}/T$.
It is important to understand what we mean by $\Delta L$, namely the position
change of \emph{each} end mirror. Consider again the computation of the transfer 
function: We inject a signal with amplitude $x$ to both end mirrors and compute
the signal amplitude on the output photodiode; i.e. if, for example, 
any noise source would create exactly this amplitude on the diode the apparent 
displacement would correspond to a motion of \emph{each} mirror by $x$.
Thus the transfer function refers to $\Delta L$ as the arm length change
of one individual arm.
\item [e)] With this definition of $\Delta L$ we know from Martin's thesis
\cite{Hewitson04}
that the apparent strain sensitivity computes as $h= 2 \Delta L/1200$.
In the code the last two computations have been merged into
$h = 2\sqrt{S_P}/(1200 ~T)$.
\end{itemize}

The following code repeats the above in a slightly more compact form:
\loadkat{geo_sens2.kat}

\vspace{5mm}
We can also apply the signal frequencies to the spaces (this simulates GW signals)
This feature is less well tested than \code{fsig} connected to mirrors but it provides a 
good redundant check on the result above:
\loadkat{geo_sens3.kat}

\vspace{5mm}
Instead of the Schottky formula we can use the qshot detector which 
includes the effects of the RF modulation. Thus, we do not apply
an extra the 'correction factor' for modulation sidebands ($\sqrt{2}$ in the
example above) nor the $1/\sqrt{2}$ for the demodulation by the mixer:
\loadkat{geo_sens4.kat}

\vspace{5mm}
Or using qshotS:
\loadkat{geo_sens5.kat}

\vspace{5mm}
Or using displacement and qnoiseS and the scale command:
\loadkat{geo_sens6.kat}

\begin{figure}[htb]
    \centerline{\IG[viewport=90 400 530 715, width=0.9\textwidth]{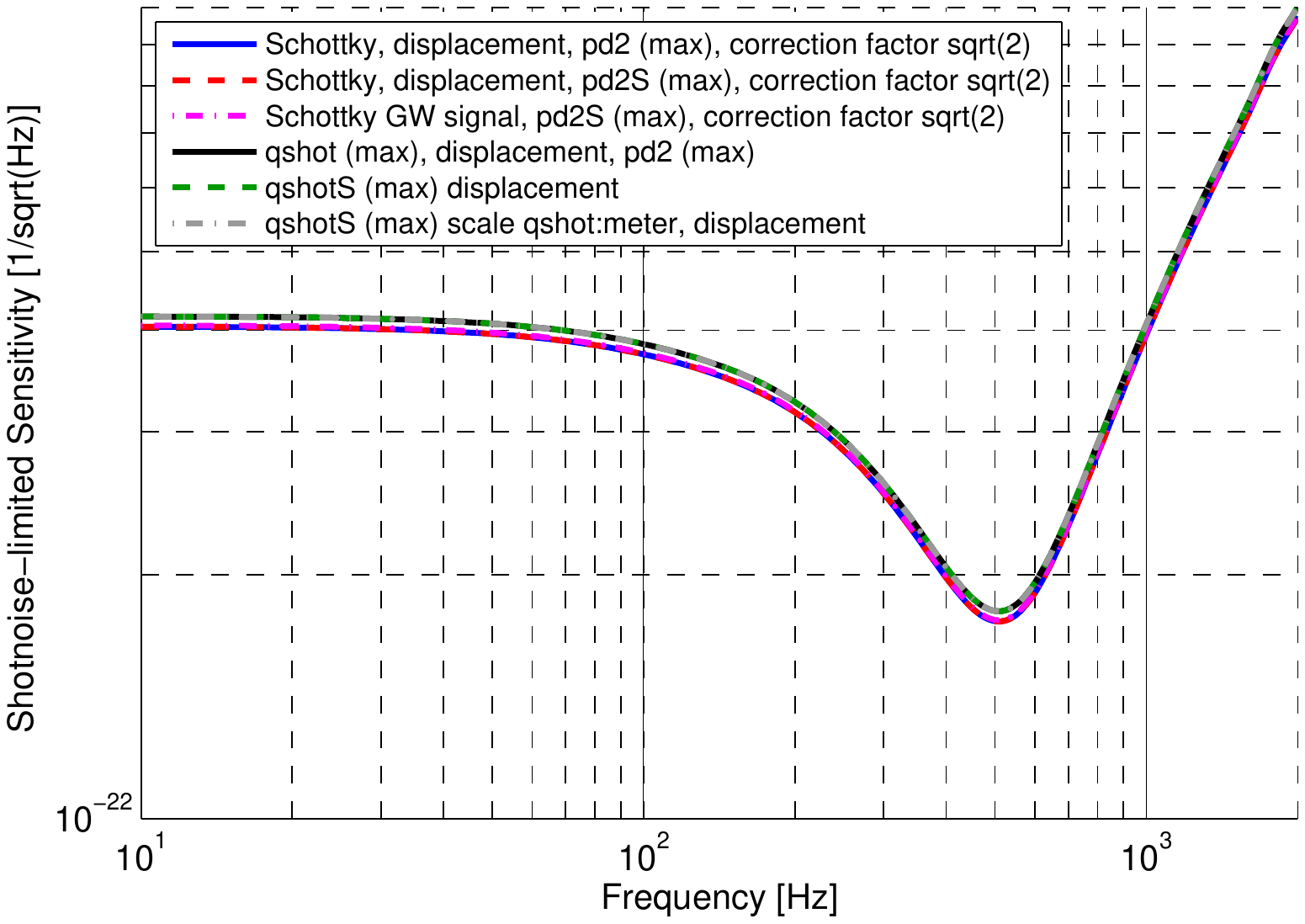}}
    \caption{Shot-noise-limited sensitivity of GEO\,600 as computed with 
    \Finesse. The plot compares the six different methods to compute
a shotnoise-limited sesnitivity mentioned in the text: {\bf Method 1 and 2}
use \code{fsig} to inject differential displacement
noise to the near end mirrors; the difference lies only in how the 
conversion into apparent strain is performed. {\bf Method 3}, however, uses
\code{fsig} to add simulated gravitational-wave signals to the
\textit{lengths} between the far and central mirrors.{\bf Method 4}
employs the new \code{qshot} detector. This detector makes use
of a new algorithm to take into account the effects of modulation 
sidebands on the shotnoise level. {\bf Method 5 and 6} also make use
of the \code{qshot} detector but include some automatic
scaling and conversion of the result. The small difference between the methods
using the \code{qshot} detector to the first three methods comes from the
fact that the applied correction of $\sqrt{2}$ in methods 1 to 3 is only approximate.}
\label{fig:sens1}
\end{figure}

\vspace{5mm}
The good news is that these methods agree with each other, see \mFig{fig:sens1}. Furthermore they 
also agree well with the measured sensitivity, see next section.

\section{Computing the shot-noise-limited sensitivity of GEO}
\begin{figure}[htb]
    \centerline{\IG[bb=50 0 705 275, width=\textwidth]{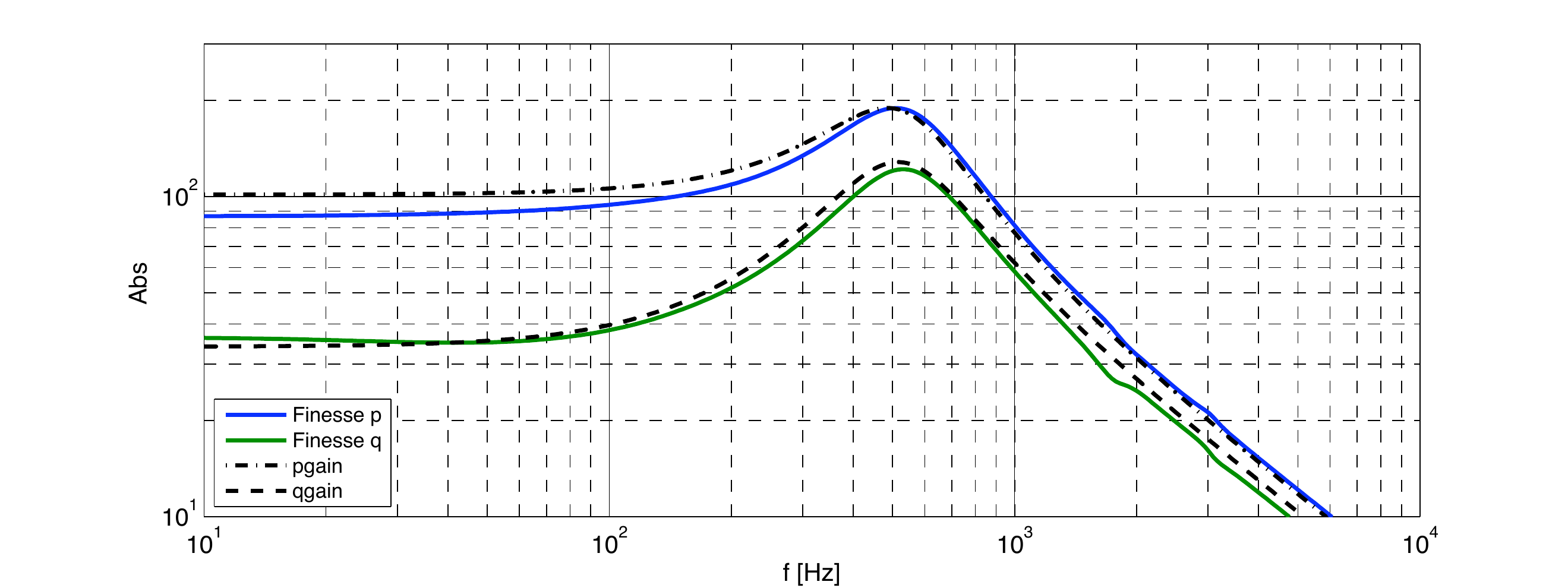}}
    \caption{comparison of the optical gains for the P and Q channel
of the dark fringe output signal. The plot shows the transfer functions
reconstructed from the parameters given on the GEO summary webpage for the 03.03.2006
and the simulated transfer
function for demodulation phases of 4 and 101 degrees.
}\label{fig:S5gain}
\end{figure}
 First we measured the light power in the south port (just after MSR) to be the same as the
measured $43$\,mW, and tune the \Finesse file accordingly
\begin{verbatim}
#l i1 3.2 0 nMU3in1            # nominal S5 corrsponds to 75deg
l i1 3 0 nMU3in1               # tuned down from 3.2 to get right
                               # power in DF                                            
\end{verbatim}

The following analysis has used the 'typical S5 sensitivity' from the GEO
sensitivity webpage (\url{http://www.geo600.uni-hannover.de/geocurves/}) as a reference.
The respective data has been taken on 03.06.2006.

We now need to tune the demodulation phase for the P and Q channel
of the dark fringe output signal. This is important to compare the 
simulation with measured data correctly. The optical gain
is modelled for the data analysis (h reconstruction) with
a simple transfer function represented by a gain, a complex pole
and a real zero. The numerical values for these parameters
on the 03.06.2006
for the P and Q channel respectively can be found
on the GEO summary pages
\url{http://www.geo600.uni-hannover.de/georeports/}.
From these parameters we can reconstruct approximately the optical gain
of the detector for the reference time. 

\begin{figure}[htb]
    \centerline{\IG[bb= 30 0 540 232, width=\textwidth]{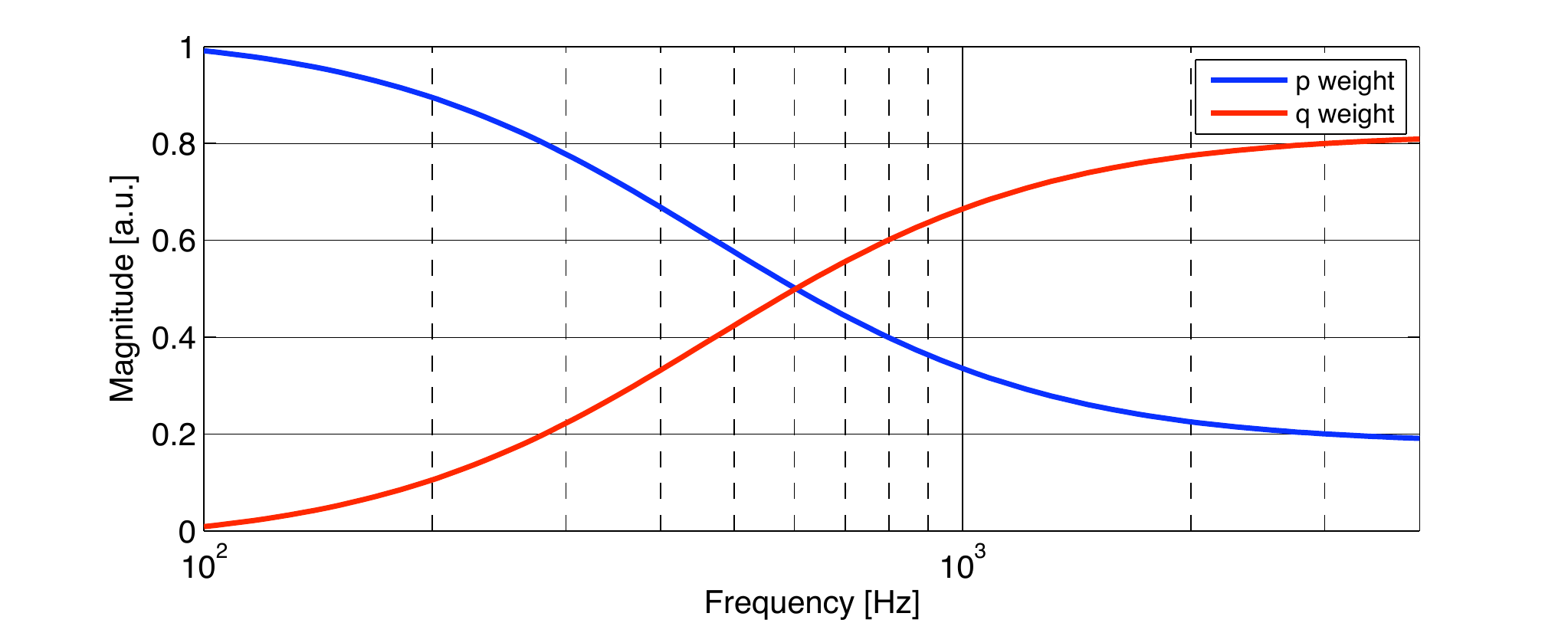}}
    \caption{Weights for P and Q quadrature.
}\label{fig:S5weight}
\end{figure}

Now, we can simulate the optical 
gain with \Finesse and tune the demodulation phase such that we obtain the
same transfer function. For one set of demodulation phases the
input file uses the following commands:
\begin{verbatim}
fsig sig1 MCN 1 0
fsig sig2 MCE 1 180

# compute transfer function  Delta_L -> DF
pd2 pdMI1 $fMI 4 1 nMSR2 
pd2 pdMI2 $fMI 101 1 nMSR2 

xaxis sig1 f log 10 10k 300
put pdMI1 f2 $x1
put pdMI2 f2 $x1
\end{verbatim}

\begin{figure}[htb]
    \centerline{\IG[width=\textwidth]{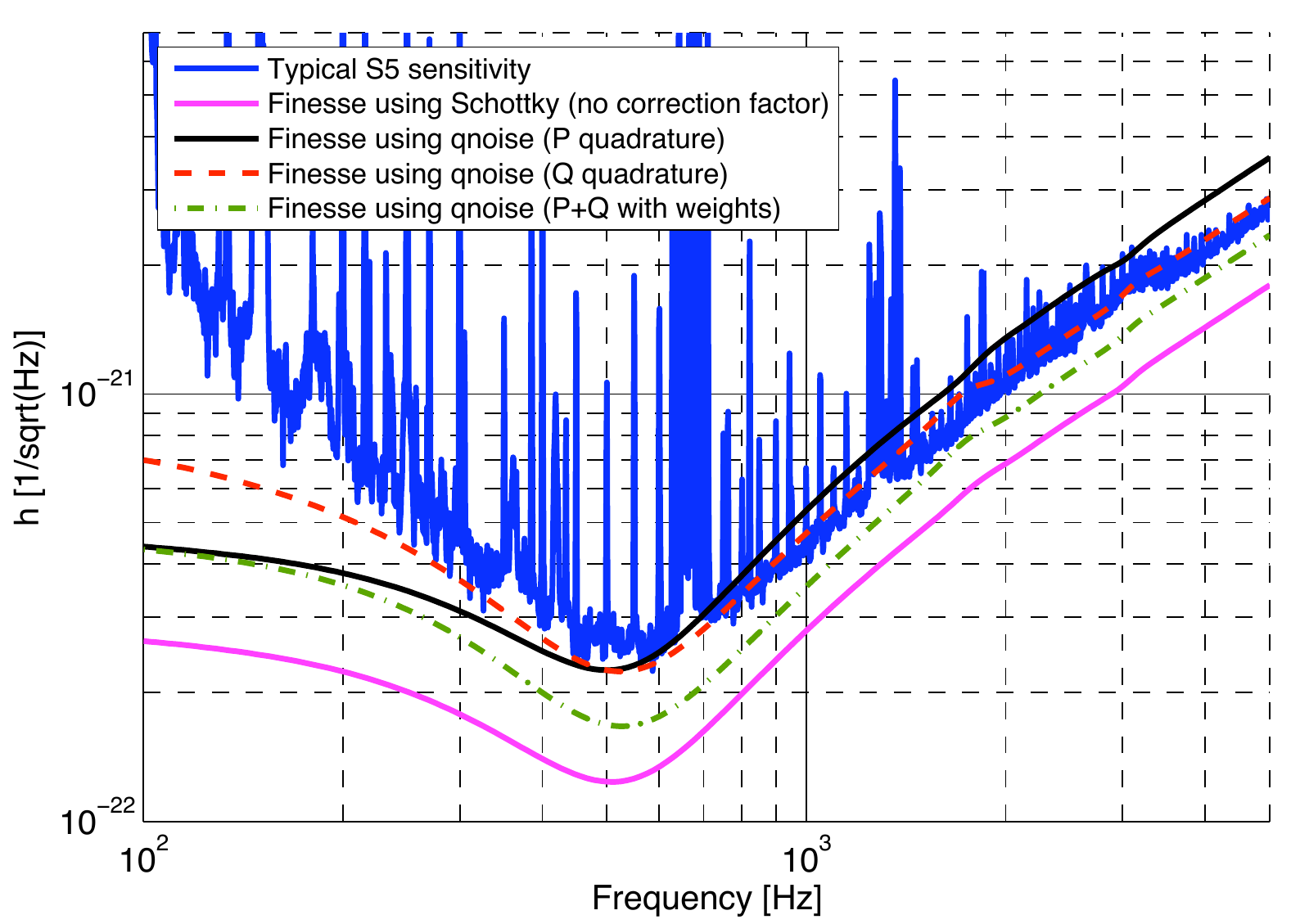}}
    \caption{Typical S5 sensitivity compared to the \Finesse computation
of the shot-noise-limited sensitivity.
}\label{fig:S5sens}
\end{figure}

\mFig{fig:S5gain} shows a comparison of the transfer function reconstructed
from the parameters given on the webpage with the simulated transfer
function for demodulation phases of 4 and 101 degrees. The resulting 
comparison of the \Finesse model with the measured GEO\,600 sensitivity
is shown in \mFig{fig:S5sens}.

%% file: appendix_thermal.tex
\chapter{Realistic thermal distortions in Advanced LIGO arm cavities}
\label{sec:thermal_ex}

In the last few years \Finesse has undergone extensive development to include the effects of mirror surface distortions.  The application of map surfaces has grown from a very basic integration routine (using a Riemann sum) to include multiple integration routines, which can be used to optimise the results for different geometric effects and drastically increase the speed of the coupling calculations (see sction~\ref{sec:int_routines}).  This involves extensive testing of the code, including internal tests of \Finesse map routines against analytical methods for calculating the effect of simple distortions (i.e. misalignment, see section~\ref{sec:tilt_ex}) and comparisons of \Finesse results with other simulation tools using significantly different simulation methods.  This has been my particular job during the development of \Finesse (Charlotte Bond).  As part of this on-going task we appealed to other simulators in the gravitational wave community for simulation examples which they believed would be a good test of \Finesse.  We were challenged by Hiro Yamamoto to simulate an Advanced LIGO arm cavity with thermally distorted mirrors and determine the loss of power during one round-trip of a light field in such a cavity.  Previous attempts to simulate this using other modal methods have failed to replicate the results achieved using Fast-Fourier-Transform (FFT) propagation methods, such as those carried out by Yamamoto~\cite{Yamamoto}.  The application of thermal effects in models of gravitational wave detectors will be crucial for the commissioning of advanced detectors, so it is vital that \Finesse can simulate these effects accurately.

The failure of previous modal models and the fact that the inclusion of thermal distortions requires delicate handling makes this setup the perfect test of \Finesse.  This is also a good example with which illustrate the specific steps required to optimise a \Finesse simulation with mirror maps and achieve accurate results with relatively low \verb|maxtem|.  In this section we present a study of the losses incurred in the arm cavities of an Advanced LIGO interferometer when the mirrors are thermally distorted.  The distortions of the mirrors in the Advanced LIGO arm cavities are expected to be relatively large, due to the high circulating powers expected in the arms.  This study is a good test for the use of mirror maps in \Finesse, as these are important simulations for commissioning and require some effort to simulate the setup correctly.

\section{Preparing mirror maps}

In order to simulate the effects of thermal distortions in \Finesse the expected distortion of the mirrors are calculated and stored in the \Finesse mirror map format.  In this case we consider the distortion after the mirror has achieved thermal equilibrium and calculate the distortions using the Hello-Vinet method~\cite{hv}.  Any absorption of the laser beam in the mirrors results in a temperature gradient in the mirror substrate, causing the material to expand and deform as well as causing a change in the refractive index.  This has two effects: 1) it distorts the surface of the mirror from an ideal curved surface; 2) it creates an effective lens in the substrate.  In this study we consider only the effect of the surface distortion.

The absorption can occur in either the mirror substrate or the coating.  For simplicity we here consider only the absorption in the coating and not the substrate.  For this example we investigate a realistic absorption level of 1\,ppm (part-per-million) per mirror.  We consider three cases:
\begin{enumerate}
\item No absorption.  The mirrors are represented as perfect spheres, the only realistic geometric effect being their finite apertures.
\item Unbalanced.  1\,ppm absorption in the end test mass (ETM) and no absorption in the input test mass (ITM).  
\item Balanced.  1\,ppm absorption in each mirror.
\end{enumerate}

\begin{figure}[t]
\centering
	\includegraphics[scale=0.42, viewport= 80 400 540 730]{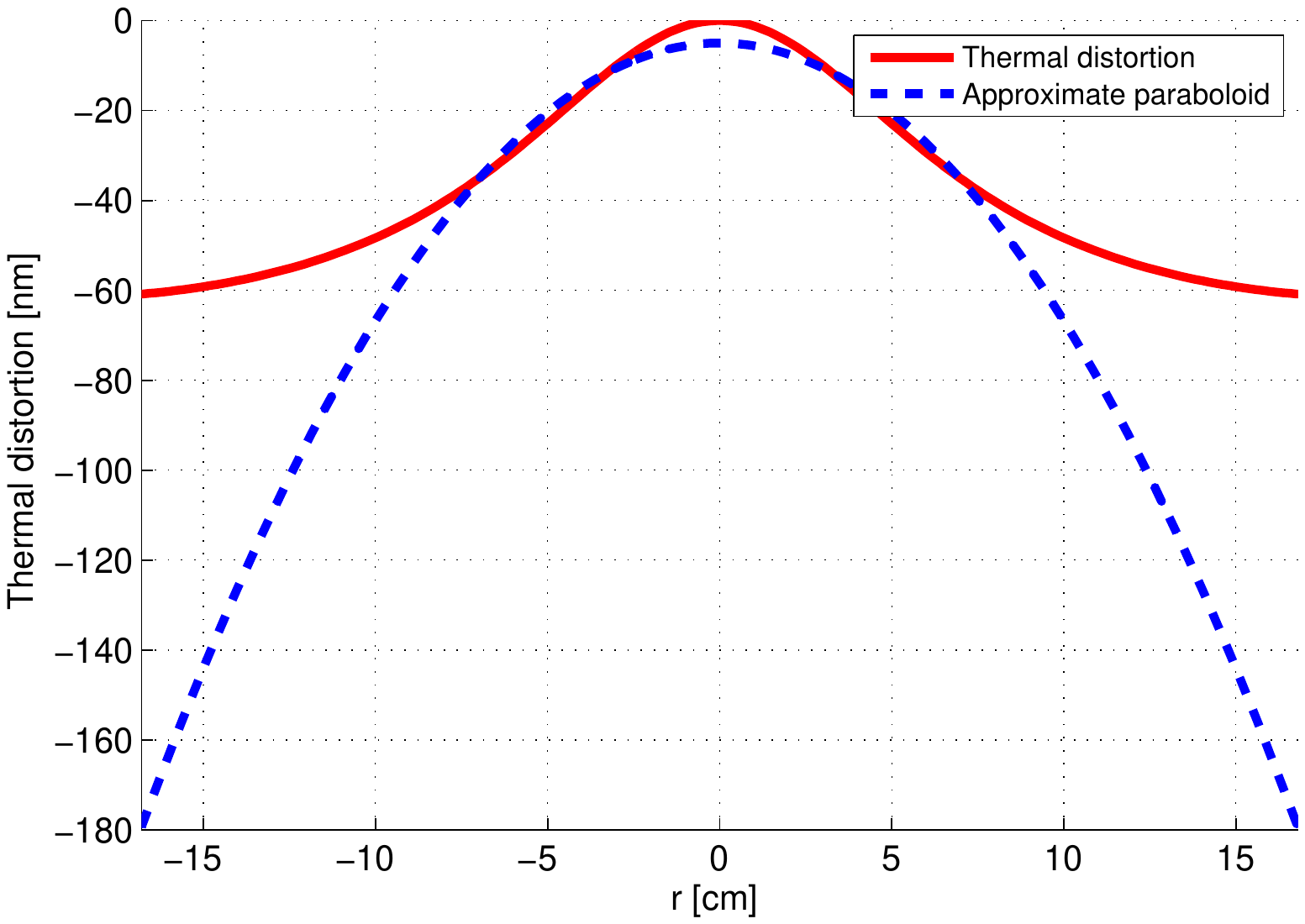}\,
	\includegraphics[scale=0.42]{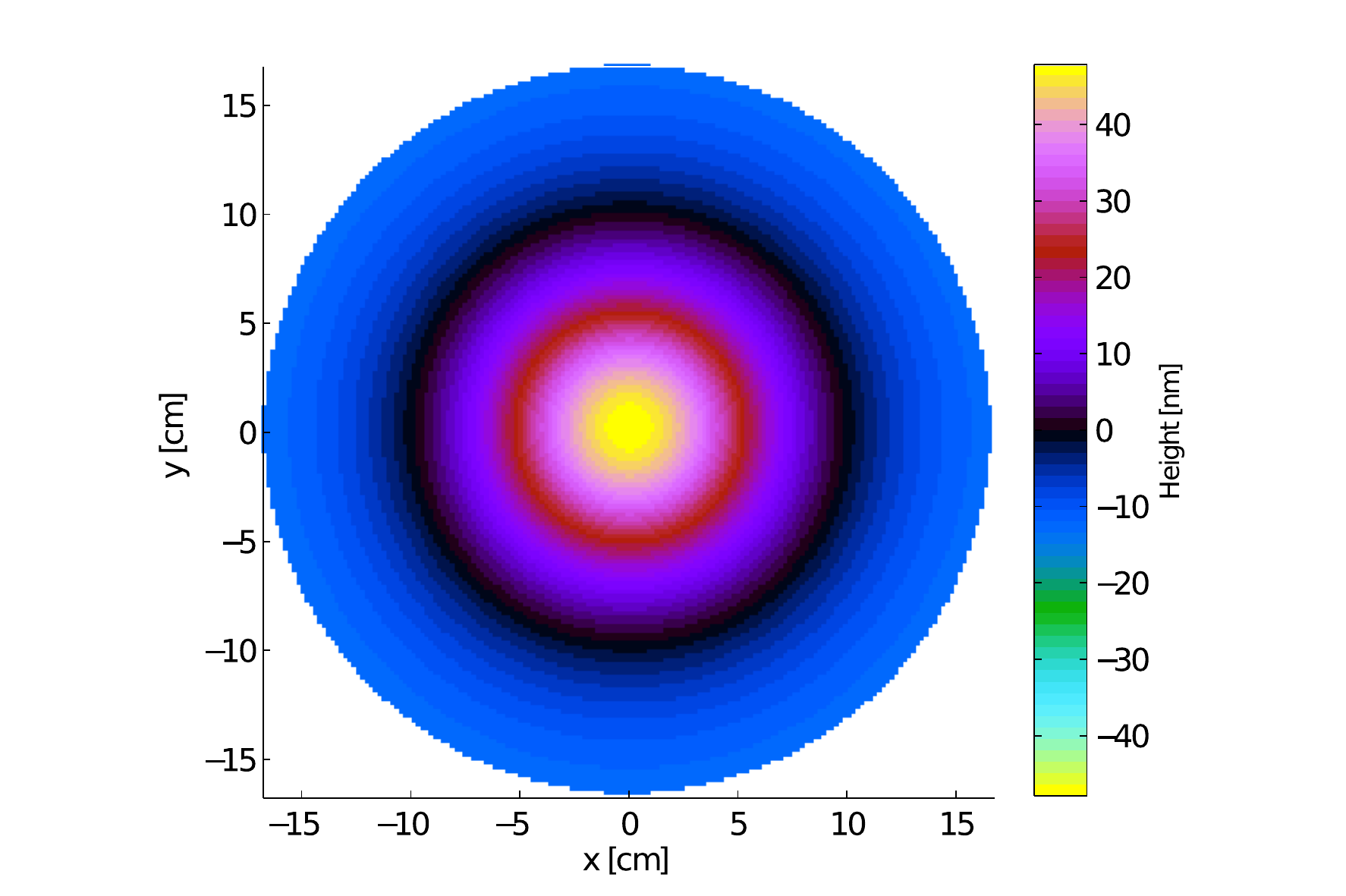}
\caption{Plots showing the expected thermal distortion of the end test mass in an Advanced LIGO arm cavity with 1\,ppm coating absorption.  The cross section (left) shows the overall distortion and an approximate paraboloid fitted to the distortion.  The equivalent mirror map is shown (right).}
\label{fig:thermal_maps}
\end{figure}

For these examples we require two mirror surface maps, one for the ITM and one for the ETM.  The distortions on each mirror will be slightly different, as although they have the same radius, thickness and material properties the beam spot sizes incident on the mirrors will be different: 5.3\,cm on the ITM and 6.2\,cm on the ETM.  The temperature gradient and resulting thermal distortion depend on the size and shape of the incident beam.  Several SimTools (see section~\ref{sec:simtools}) functions have been developed for the purposes of calculating the thermal distortions and lensing for just such a setup.  The function \verb|FT_mirror_map_from_thermal_distortion.m| was used to produce the maps for this investigation.  This function employs the Hello-Vinet formula to calculate the resulting distortion of a mirror with given dimensions and thermal properties illuminated by a gaussian beam of a given spot size.  In figure~\ref{fig:thermal_maps} plots of the thermal distortion for the end test mass are shown, both as a cross-section of the mirror surface and as a mirror surface map.  This distortion is dominated by low spatial frequencies and a substantial part can be described as a change in curvature of the mirror.

\begin{figure}[h]
\centering
	\includegraphics[scale=0.42, viewport= 80 400 540 730]{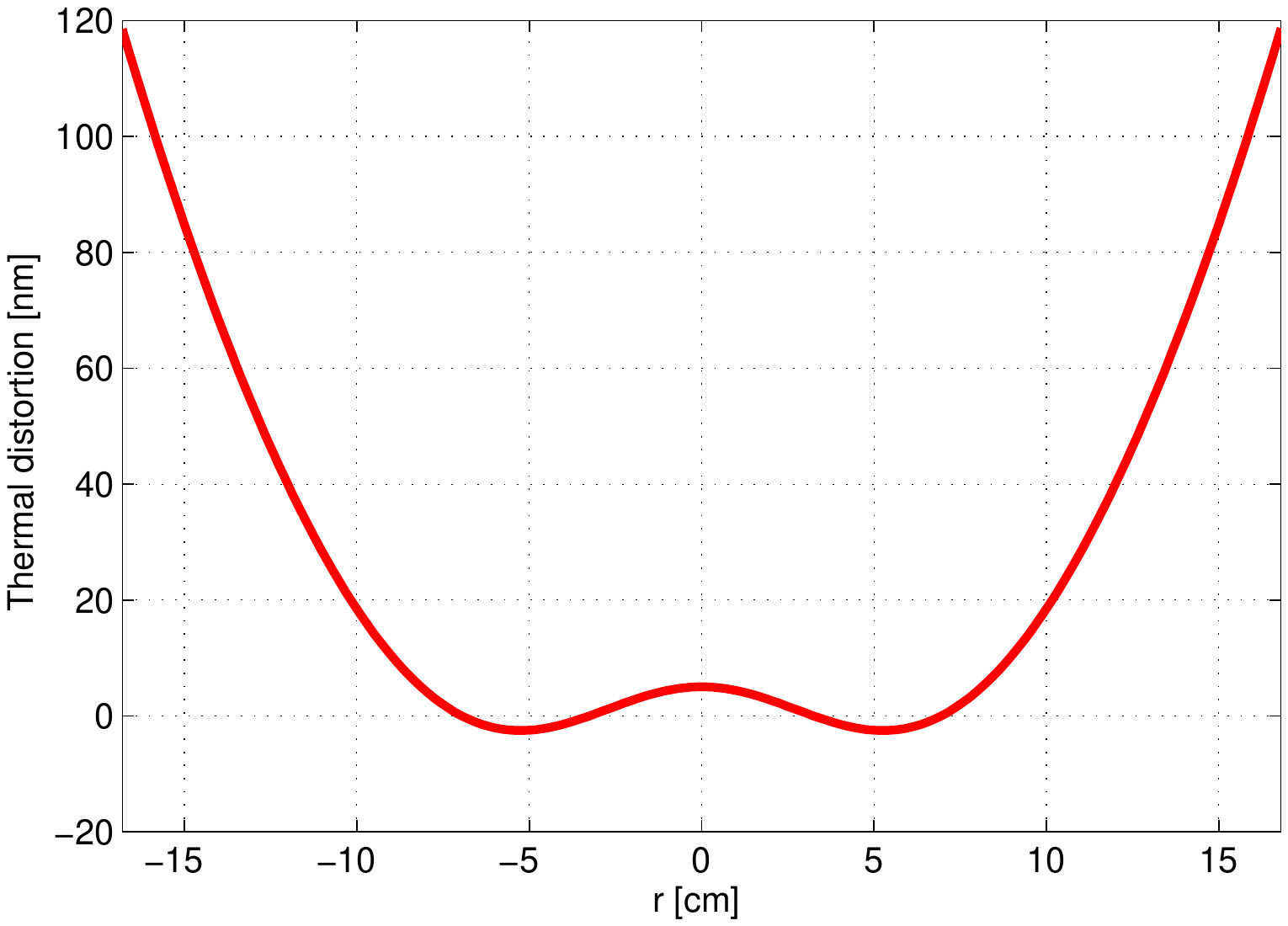}\,
	\includegraphics[scale=0.42]{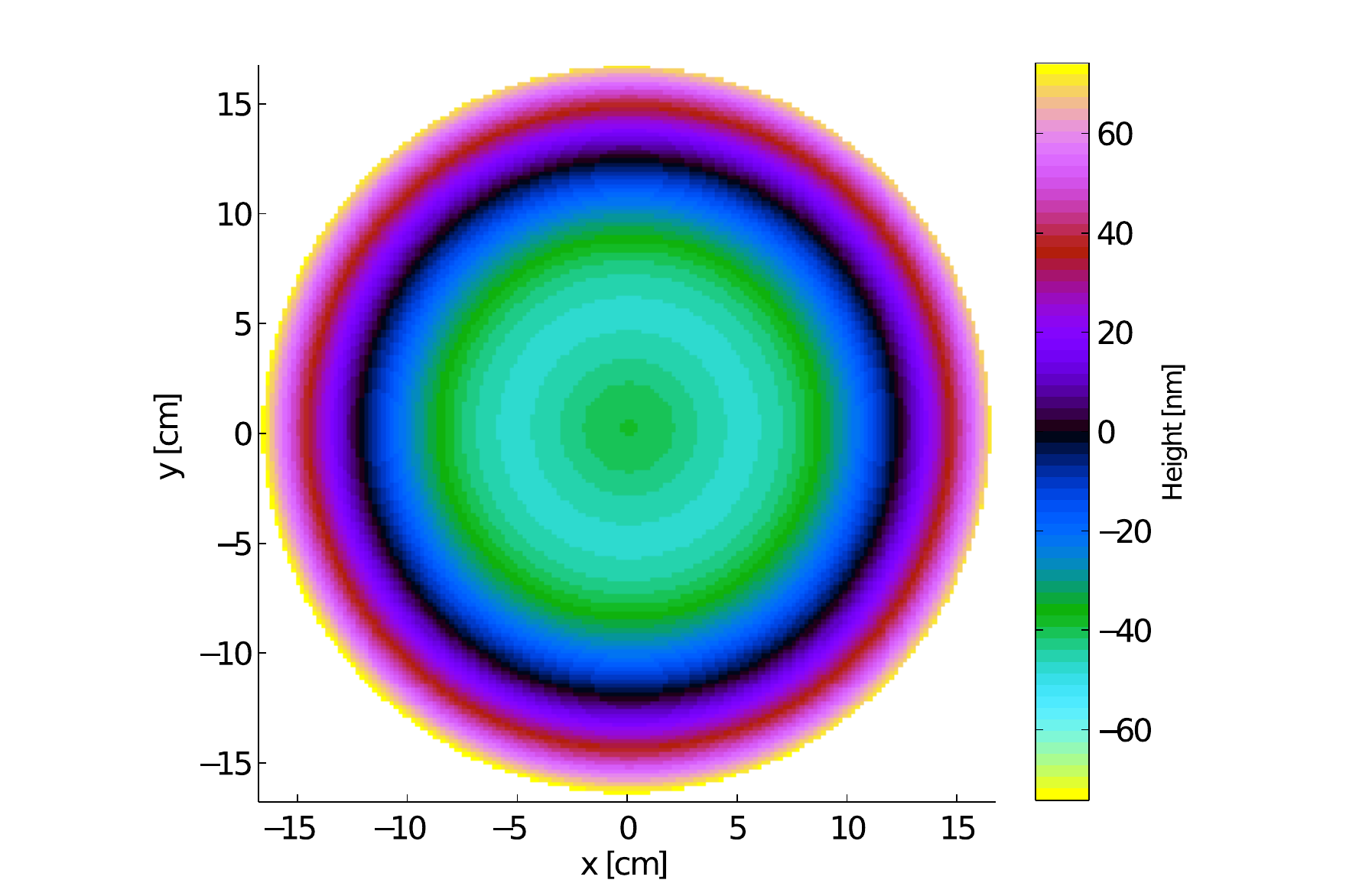}
\caption{Plots showing the cross section (left) and mirror map (right) of the expected thermal distortion of the end test mass in an Advanced LIGO arm cavity, where a Gaussian weighted curvature has been fitted and removed from the surface.}
\label{fig:thermal_res_maps}
\end{figure}

In order to successfully calculate the effects of mirror surface distortions in \Finesse the Gaussian beam parameter used to calculate the coupling coefficients must be carefully chosen.  For the case of resonant cavities the most appropriate choice of beam parameter is commonly that which matches the geometry of the cavity, in this case using a beam parameter whose curvature matches the curvature of the mirrors.  In order to apply this successfully any curvature of the mirrors should be applied in \Finesse using the \verb|attr| command, rather than being contained in a mirror map.  The \verb|cav| command can then be used to set the Gaussian parameter to be mode-matched to the cavity.  Although for small curvatures both methods should give equivalent results, a good choice of Gaussian parameter should require fewer higher order modes and a lower \verb|maxtem|.  Therefore, for this investigation the curvatures are removed from the surface maps before they are applied to the mirrors.  Again, several SimTools functions have been developed for just such a task.  Tools using Zernike polynomials have been produced for the analysis and preparation of maps, analysing the entire surface over the defined mirror disc.  However, this particular case involves fitting a curved surface to the map and then removing it (\verb|FT_remove_curvature_from_mirror_map.m|).  As the mirrors are not uniformly illuminated the curvature fitting process should be most accurate where the beam is at its most intense, i.e. the fitting process should be weighted by the appropriate gaussian beam.  This is achieved by minimising:
\begin{equation*}
x = \int_{0}^{a} \int_0^{2\pi} W(r) \left[ S_{\mathrm{map}} - S_{\mathrm{C}}  \right]^2 \pi r \mathrm{d}r
\end{equation*}    
where $S_{map}$ is the surface discribed by the map, $S_{C}$ is the fitted curved surface (described by a radius of curvature, $R_{\mathrm{c}}$) and $W(r)$ is the weighting function.  In this case $W(r)$ is the gaussian beam intensity function, with a given spot size.  For the case of mirror thermal distortions the weighted curvature (so-called \emph{approximate paraboloid}) can be calculated analytically~\cite{vpb} using the SimTools function 
\newline
\verb|FT_approximate_paraboloid_for_thermal_distortion.m|.  The cross-section of the approximate paraboloid found using this function is plotted in figure~\ref{fig:thermal_maps} for the end test mass, a curved surface with $R_{\mathrm{c}} = -80$\,km.  The curved surface is removed from the map, the residual distortion of the ETM is shown in figure~\ref{fig:thermal_res_maps}.  The distortion of the ITM was similarly calculated and a weighted curvature of $R_{\mathrm{c}}=-60$\,km was removed.  

In most cases this removed curvatures would then be included in the radius of curvature defined in the \Finesse script (using the \verb|attr| command).  However, for this investigation we assume that the Advanced LIGO thermal compensation system is working as expected, correcting for the curvature change caused by the thermal deformation.  Therefore, we assume the curvatures are corrected back to their nominal values, 2245\,m (ETM) and 1934\,m (ITM).  Generally, when using mirror maps, we would also fit and remove any tilt of the surface, effectively aligning the mirror in the simulation setup.  However, in this case the thermal distortions contain no tilt term, so we can omit this step in the map preparation.  We do, however, remove any offset from the maps using the SimTools function \verb|FT_remove_offset_from_mirror_map.m|.  Any overall offset of the mirror should be set in the \Finesse file by tuning the mirror position, not hidden in the mirror map.

\section{Simulation setup}

The simulation setup is based on the simple Fabry-Perot cavity.  The mirror parameters and cavity length are the Advanced LIGO design parameters:
\begin{verbatim}
m mITM 0.985965 0.014 0 nITM2 nITM1	
s sC 3994.5 nITM2 nETM1
m mETM 0.99996 5u 0 nETM1 dump	
\end{verbatim}
The curvature of the mirrors are defined using the \verb|attr| command and the  beam injected into the cavity is mode matched to the cavity using the \verb|cav| command:
\begin{verbatim}
attr mITM Rc 1934
attr mETM Rc 2245
cav armcav mITM nITM2 mETM nETM1
\end{verbatim}
Both mirrors are seen as concave from inside the cavity.  In \Finesse the sign of the radius of curvature is related to the order of the nodes in the mirrors definition.  As both mirrors are defined with the node inside the cavity first (\verb|nITM2| and \verb|nETM1|) the curvatures are given as positive.
For all 3 cases the finite aperture (16.8\,cm) of the mirrors must be specified using the \verb|attr| command:
\begin{verbatim}
attr mITM r_ap 0.168
attr mETM r_ap 0.168
\end{verbatim}
Finally the mirror maps must be applied to the ITM (case 3) and to the ETM (case 2 and 3).  For example, to include the map describing the thermal distortion of the ETM, the following commands are required:
\begin{verbatim}
# ETM map commands
map  mETM etm_thermal_res_map.txt
knm  mETM etm_map_coupling
conf mETM save_knm_binary 1
conf mETM interpolation_method 2
conf mETM integration_method 3
\end{verbatim}
The \verb|map| command specifies the file containing the mirror map stored in the standard \Finesse format.  A file, ``\emph{etm\_map\_coupling}'' is specified in which to save the coupling coefficients, in binary form for speed of access (\verb|save_kmn_binary|).  We also specify the linear interpolation of the mirror surface and the cuba parallel integration method for the calculation of the coupling coefficients (ref to map commands sec).  The ITM map is applied with equivalent commands.  However, care should be taken in the case of the ITM.  In \Finesse the order of the nodes specifies which way the map is applied to the mirror.  The $z$-axis of the map surface will point towards the first specified node, away from the second node.  Therefore, to orientate the ITM map with the surface facing the inside of the cavity the order of the nodes should be specified with the cavity node first.

\section{Results}
For this investigation the figure of merit we have chosen is the round-trip loss incurred for the 3 different cases.  This is a useful single number for comparison between different simulation methods.  Previous attempts with other modal based methods have failed to agree with other methods in similar investigations of the losses in thermally distorted cavities.  This makes this investigation a good test of \Finesse as a robust, accurate tool for calculating the effects of realistic optics.  

In \Finesse a single round-trip of a cavity is not simply simulated.  The round-trip loss is therefore calculated using the power circulating in the cavity:
\begin{equation}
L_{\mathrm{arm}} = T_{\mathrm{ITM}} \left( \sqrt{\frac{P_{\mathrm{FP}}(0)}{P_{\mathrm{FP}}(L_{\mathrm{arm}})}}-1 \right)
\end{equation}
where $T_{\mathrm{ITM}}$ is the transmission coefficient of the input mirror, $P_{\mathrm{FP}}(0)$ is the circulating power in the equivalent, lossless cavity and $P_{\mathrm{FP}}(L_{\mathrm{arm}})$ is the power circulating in the lossy cavity.

\begin{figure}[b]
\centering
	\includegraphics[scale=0.45, viewport= 80 400 540 730]{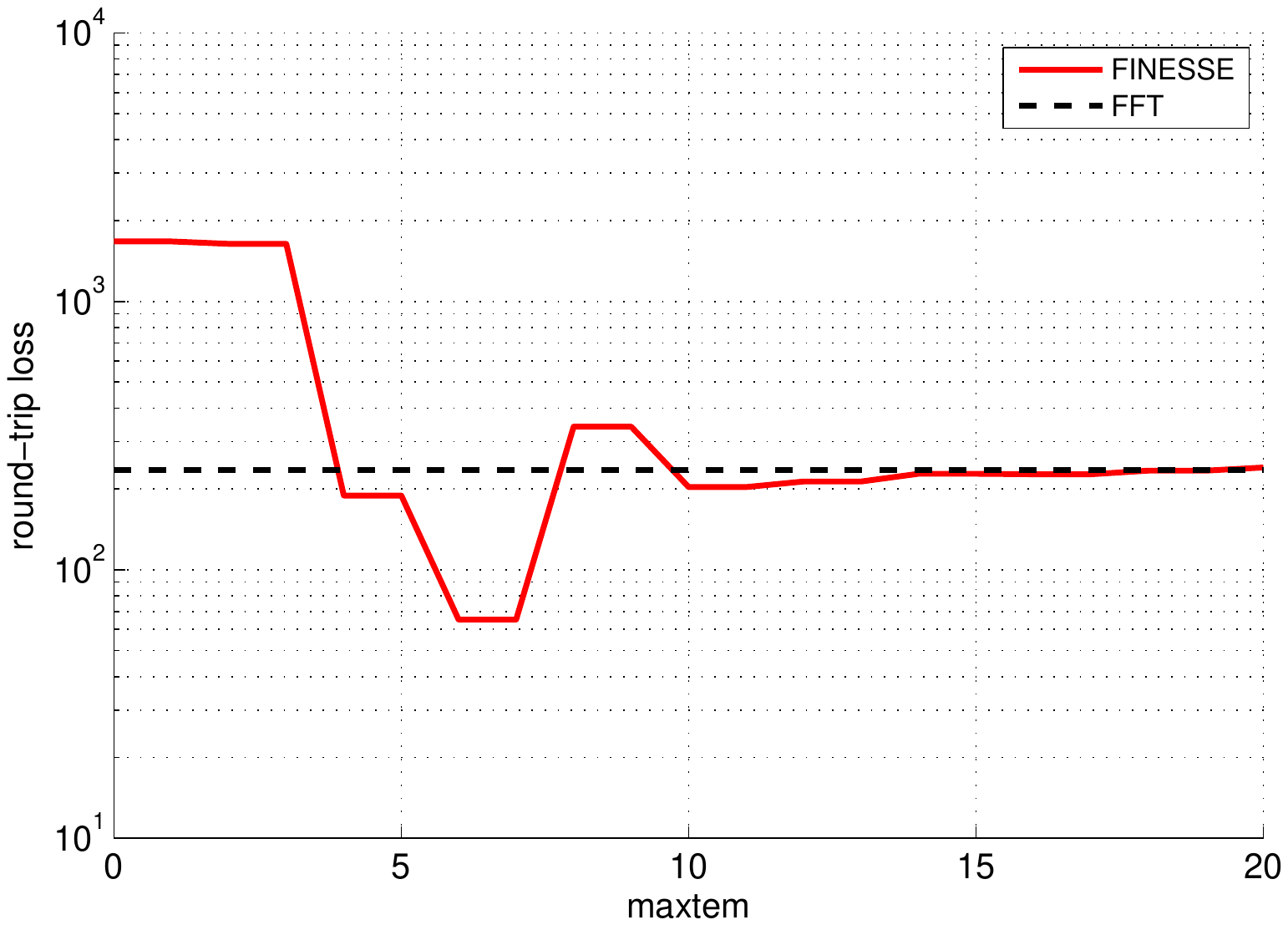}\,
	\includegraphics[scale=0.45, viewport= 80 400 540 730]{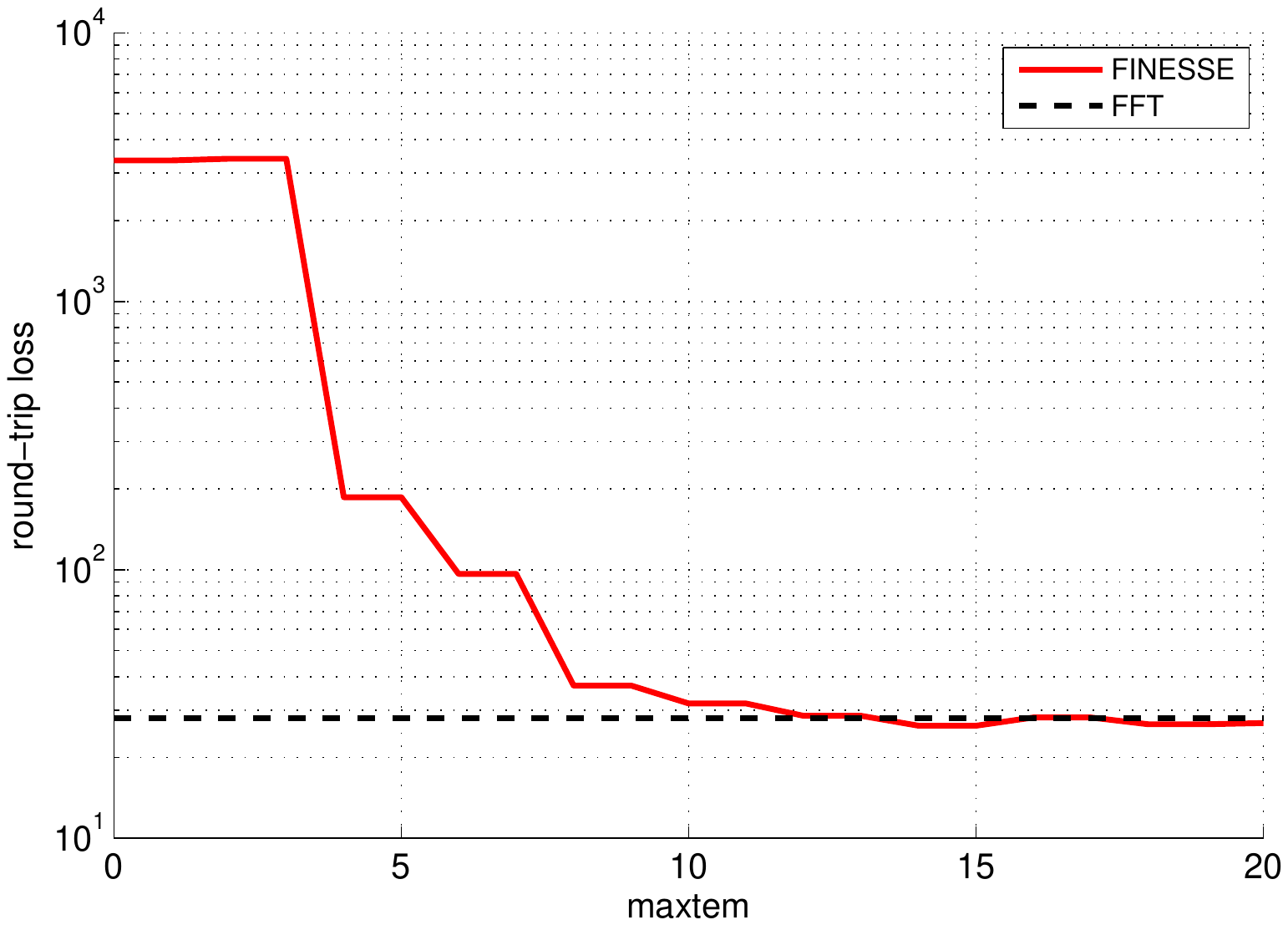}
\caption{Plots showing the round-trip loss for different values of maxtem in Advanced LIGO cavities with thermal mirror distortions simulated in \Finesse and using an FFT propagation method.  The effect of thermal distortions are added to the end mirror (left) and both mirrors (right).}
\label{fig:thermal_results}
\end{figure}

In \Finesse the setups for case 1, 2 and 3 were simulated for a range of maxtem, from 0 to 20.  Due to the relatively large nature of the thermal distortions it is expected that we will require a relatively high \verb|maxtem|.  For comparison the same simulations were also carried out using a Fast-Fourier Transform (FFT) method, based on OSCAR~\cite{OSCAR}.  The circulating power is detected when the cavity is tuned to the point where the power is at a maximum.  Locking sequences and more compex operating points are omitted in this example as these require delicate handling in the FFT code.  From the circulating power the round-trip loss is calculated for each case.  Figure~\ref{fig:thermal_results} shows plots of the round-trip losses as the \verb|maxtem| is increased for cases 2 and 3.    

These results, for the cases where thermal distortions are applied to the mirrors, demonstrates the importance of including the right number of higher-order modes in \Finesse simulations.  A lot of coupling into higher order modes occurs at lower orders, but the mid orders (6-10) still contribute significantly to the overall result.  Choosing too small a \verb|maxtem| can lead to wildly inaccurate results.  However, by including enough higher-order modes \Finesse can successfully re-create results achieved using other methods, with the added advantage that once the coupling coefficients are calculated the effects can be incorporated into the simulations without the need for re-calculation. 

The results for specific values of \verb|maxtem| for the 3 cases are summarised in table~\ref{table:loss_res}.  For case 1 the round-trip loss is incurred due to the finite size of the mirrors and the power clipped by these apertures.  The aperture is large, compared to the size of the beam, and therefore the coupling into higher order modes is small.  Therefore, the round-trip loss stabilises with low \verb|maxtem|, agreeing well with the result of the FFT propagation method.  For case 2 the distortion of the end mirror causes coupling into higher order modes, specifically modes of an even order as the distortions are symmetric, spherical aberrations.  Although the most significant coupling happens at low orders there is still a relatively large contribution from the higher orders.  At a \verb|maxtem| of around 8-10 the round-trip loss reaches the right magnitude and by \verb|maxtem| 15 the \Finesse result agrees very well with the FFT result.  Another interesting point regarding case 2 is that the addition of higher order modes does not just account for the power previously disregarded in higher-order modes.  For example if we were to simulate such a setup using only \verb|maxtem| 6 the round-trip loss would be significantly underestimated.  This is because this example deals with a cavity, where any coupling during one mirror incidence results in modes which can couple into higher order modes. 

Finally, in case 3 the round-trip loss is actually greatly reduced from the case where we consider just one distorted mirror.  As the distortions on the two mirrors are identical, as scaled by the incident beam size, the distortion of the wavefront by the end mirror will match the distorted input mirror.  This match of the distorted mirror and wavefront causes much less subsequent distortion of the beam, less coupling into higher-order modes and hence less power clipped by the finite size of the mirrors.  This result is recreated in the \Finesse simulation, an effect which previous modal models were unable to achieve, using \verb|maxtem| 10-15.

\begin{table}[htdp]
\begin{center}
\begin{tabular}{|c|c|c|c|}
\hline
Case & \Finesse (maxtem 10) & \Finesse (maxtem 15) & FFT \\
\hline
1	&	0.8		&	0.9		&	0.9 \\
2	&	203		&	228		&	234 \\
3	&	32		&	26		&	27 \\
\hline
\end{tabular}
\end{center}
\caption{Round-trip losses in simulated Advanced LIGO cavities for three different cases: 1) no thermal distortions; 2) a thermally distorted end mirror; 3) thermal distortions of the input and end mirrors.  The cavities are simulated using an FFT propagation method and \Finesse.}
\label{table:loss_res}
\end{table}%

For either case we have demonstrated that we can achieve similar results using \Finesse with a finite number of modes.  In particular, \verb|maxtem| 10 is used to achieve the correct order of magnitude, which should be good enough in this case as simulation and experiment will not match down to exact ppm losses.  However, a more accurate result can be achieved using a \verb|maxtem| of around 15.

One final consideration is the accuracy we are using in the calculation of the coupling coefficients.  This is defined in the ``\emph{kat.ini}'' file by the variable \verb|abserr|, the absolute error of the coupling coefficients.  The results summarised above were achieved using an \verb|abserr| of $1\times 10^{-6}$.  The default value set in the ``\emph{kat.ini}'' file is $1\times 10^{-3}$.  The results achieved using this default value are summarised in table~\ref{table:loss_res2}.  The do not differ by a large degree to those using $10^{-6}$

\begin{table}[htdp]
\begin{center}
\begin{tabular}{|c|c|c|}
\hline
Case & \Finesse (maxtem 10) & \Finesse (maxtem 15) \\
\hline
1	&	0.7		&	0.8              \\
2	&	202		&	225		 \\
3	&	33		&	27		\\
\hline
\end{tabular}
\end{center}
\caption{Round-trip losses in Advanced LIGO cavities for three cases: 1) no thermal distortions; 2) a thermally distorted ETM; 3) thermal distortions of the ITM and ETM.  The cavities are simulated using \Finesse and the default absolute error of $1\times 10^{-3}$.}
\label{table:loss_res2}
\end{table}%

%% file: appendix_map_knm.tex
\chapter{Maps and Coupling Coefficients}\label{sec:mapknm}


The scenario often arises where we want to apply misalignments, mode
mismatches, a surface distortion and an aperture to simulate a
realistic mirror. The coupling coefficient is more commonly refered to
as an inner product or projection of a beam
from one `basis set' into another. The basis sets we are considering
here are the orthogonal Hermite-Gaussian (HG) functions. A beam at any
point can be decomposed into a HG basis set and is described by a
vector $a_{in,N}, N\in \{0,1,2,\ldots,N_{max}\}$, where $N_{max}$ is
the highest order of HG polynomial we consider. It is also assumed
there exists some linear mapping between the HG mode number in the
sagittal and transverse directions $nm \rightarrow N$ and
$n'm'\rightarrow M$, to save writing $nm$ and $n'm'$ all the time.

The outgoing beam of mode $M$ has the beam parameter $q_{2}$, the
incoming mode $N$ is $q_1$ -- these beam parameters are known values
and are set by the user (and \Finesse's beam tracing routine). However
the incoming beam must first interact with the mirror's or
beamsplitter's known properties, i.e. its radius of curvature
represented by the application of the ABCD matrix which transforms the
incoming beam as $q_1
\rightarrow q'_1$. The projection between $q'_1$ to the basis $q_2$ is
then required. This is given by the coupling coefficient integral
(also known as an overlap integral or complex function space inner
product) between the incoming and outgoing beam basis. The coupling
between mode $N$ and $M$ is computed as
\begin{eqnarray}
k_{NM} = \left\langle U_N , U_M \right\rangle  = \iint_{-\infty}^{\infty} U_N(x,y,q'_1) U^{\ast}_N(x,y,q_2) dx dy 
\end{eqnarray}
This so far is just computing the mode-mismatch between $q'_{1}$. 
The values of $k_{NM}$ can be arranged in the form of a matrix
\begin{eqnarray}
K = 
\left[
\begin{array}{c c c c c}
\iprod{U_0,U_0} & \iprod{U_1,U_0} & \iprod{U_2,U_0} & \ldots & \iprod{U_{N_{max}},U_0} \\
\vdots & \vdots & \vdots & \ddots & \vdots \\
\iprod{U_0,U_{N_{max}}} & \iprod{U_1,U_{N_{max}}} & \iprod{U_2,U_{N_{max}}} & \ldots & \iprod{U_{N_{max}},U_{N_{max}}} \\
\end{array}
\right]
\end{eqnarray}
Which can then be applied to the mode content vector of the incoming
beam to get the outgoing beam mode content in the beam parameter $q_2$
\begin{eqnarray}
a_{out} = K a_{in}
\end{eqnarray}
The coupling coeffcient equation can also tell us if we further distort the
incoming beam, for example via a mirror surface map, how this will
project into the outgoing beam. Here, for example,  $A$
and $B$ are two separate distortions, such as a tilt of the optic surface
and some astigmatic surface curvature. Then the overlap integral becomes
\begin{eqnarray}
k_{NM} = \iint_{-\infty}^{\infty} U_N(x,y,q'_1) A(x,y) B(x,y) U^{\ast}_N(x,y,q_2) dx dy 
\end{eqnarray}
This can be represented in multiple ways using the inner product notation
\begin{eqnarray}
\iint_{-\infty}^{\infty} U_N(x,y,q'_1) A(x,y)B(x,y)U^{\ast}_N(x,y,q_2) dx dy &=& \iprod{U_N A B, U_M}\\
&=& \iprod{U_N A, B^{\ast} U_M}\\
&=& \iprod{U_N, A^{\ast} B^{\ast} U_M}
\end{eqnarray}
Assuming that the effect of $B$ is something simple and (on its own) could be
implemented using a analytic solution and $A$ is complicated and
requires a numerical integration that uses much computational
power. This arises, for example, when we have some real measurements of an optics
surface which is represented as a surface map,
and at the same time we also may want to tilt the optic by
a range of different angles to generate
alignment signa. The latter requires many
recomputations of $K$, but we posses analytic solutions in the form of
the Bayer-Helms routines. It the current form the entire matrix $K$
would need to be recomputed even though the 
mirror surface map remains unchanged.

To circumnavigate this problem we can try to separate the effects into
two sets of coupling coefficients. Mathematically this is done by "inserting unity" -- for lack of a
better term -- into the coupling coefficient equation
\begin{eqnarray}\label{eq:unity}
k_{NM} = \sum^{\infty}_{L}\iprod{U_N A, U_L}\iprod{U_L, B^{\ast} U_M}.
\end{eqnarray}
This can be understood better perhaps in terms of dot products of
vectors, which is a more specialised inner product. Take the vectors
$a$ and $b$ of equal length which are then projected onto the infinite
basis vector set $c$
\begin{eqnarray}
a &=& \sum_{n} A_n c_n \\
b &=& \sum_{n} B_n c_n \\
\iprod{a, b} &=& \sum_{n} a_n \cdot b_n \\
&=& \sum_{l}^{\infty}(a \cdot c_l)(c_{l} \cdot b)
\end{eqnarray}

Using this trick for the HG coupling coefficients we create
2 inner products: one that can be solved analytically and
another which requires numerical integration; 
we essentially need to compute 2 separate $K$
matrices.
\begin{eqnarray}
K_{NM} &=& \sum^{\infty}_{L}\iprod{U_N A, U_L}\iprod{U_L, B^{\ast} U_M} = [K_A K_B]_{NM} \\
 &=& \sum^{\infty}_{L} \iint_{-\infty}^{\infty} U_N(x,y,q'_1) A(x,y) U^{\ast}_L(x,y,q_L) dx dy \\
 && \ldots \iint_{-\infty}^{\infty} U_L(x,y,q_L) B(x,y) U^{\ast}_M(x,y,q_2) dx dy \label{eq:sep_int}
\end{eqnarray}
Of course our choosing the beam parameter that $U_L$ represents is
arbritary here. However, the choice $U_L$ will have a strong influence
on the numerical error in practise because only a limited set of modes
is used in the simulation
whereas the equation~\ref{eq:unity} is only correct if the complete
set is used. The two choices that should provide best performance
in most cases is to use
either $q'_1$ or $q_2$. It is also worth nothing that the positioning
of $A$ and $B$ in the inner products is also completely arbritary, in
the sense we could also compute
\begin{eqnarray}
K_{NM} = \sum^{\infty}_{L}\iprod{U_N A B, U_L}\iprod{U_L, U_M}
\end{eqnarray}
however the second inner product here is just $\delta_{LM}$ (if
$q_L=q_2$), which offers no computational improvement over what we had
originally. If $q_L=q'_1$ values differ we are essentially moving the
mode-mismatch computation into a separate matrix and computing the
distortion coupling back into the same beam basis.

Our outgoing beam is now computed as
\begin{eqnarray}
a_{out} = K a_{in} = (K_A K_B) a_{in}
\end{eqnarray}
Commutations should not exist as long as enough modes are used in the
projection to $U_L$ as the ordering of $A$ and $B$ is arbitrary. An
issue may arise if the solver chosen for $K_A$ or $K_B$ is an analytic
solution will be much more accurate $\approx 10^{-15}$ than a
numerical solver which are typically $10^{-6}$, thus commutation
errors will arise from this due to these errors.

\section{Correct implementation}
From the previous analysis we can now state the optimal routine for
separating the true coupling coefficient matrix $K$. For distortions
$A$ and $B$
the coupling from mode $N$ to mode $M$ whose beam parameters are
respectively $q'_1$ and $q_2$ is
\begin{eqnarray}
K_{NM} &=& [K_A K_B]_{NM}\\
&=&\sum^{\infty}_{L}  \underbrace{\iprod{U_N(q_2) A(x,y), U_{L}(q_L)}}_{\text{\normalfont A Solver}} \underbrace{\iprod{U_{L}(q_L), B^{\ast}(x,y)U_M(q_2)}}_{\text{\normalfont B Solver}}.
\end{eqnarray}
We further defined the solver as follows:
if a distortion ($A$ or $B$) is a surface map, the respective matrix
is to be computed by numerical integration. In all other cases
the matrix is computed using the Bayer-Helms equations. 

We provide to user command to influence the ordering and the selction
of $q_L$. The order of the matrix calculation is defined by
\begin{verbatim}
conf [component] knm_order AB [A,B = 1(Map) or 2(Bayer-Helms)]
\end{verbatim}
where the argument states which solver and distortion is applied to
which coupling coefficient matrix $K_A$ and $K_B$. The value of $q_L$
also needs to be decided, here we use the command
\begin{verbatim}
conf [component] knm_change_q [1 or 2]
\end{verbatim}
where if the argument is \verb|1| then $q_L = q'_1$ else if \verb|2|
then $q_L = q_2$. The command \verb|knm_apply_ABCD| is no longer
needed.

\section{Separating more distortions}
Although not necessary now, in the future the need to compute more
than two distortions efficiently might be needed. The method for
adding more is simply a repetition of the previous steps splitting the
initial inner product into two. Inner product $A$ or $B$ must be
chosen to be split, I will choose $B$ here arbritarily to add a
distortion $C$
\begin{eqnarray}
K_{NM} &=& \iprod{U_N(q_2) A(x,y)B(x,y)C(x,y), U_{M}(q_2)}\\
&=&\sum^{\infty}_{L}  \underbrace{\iprod{U_N(q_2) A(x,y), U_{L}(q_L)}}_{\text{\normalfont A Solver}} \underbrace{\iprod{U_{L}(q_L) B(x,y)C(x,y), U_M(q_2)}}_{\text{\normalfont B Solver}}\\
&=&\sum^{\infty}_{L}  \underbrace{\iprod{U_N(q_2) A(x,y), U_{L}(q_L)}}_{\text{\normalfont A Solver}}\\
&&\ldots \sum^{\infty}_{J} \underbrace{\iprod{U_{L}(q_L)B(x,y), U_J(q_J)}}_{\text{\normalfont B Solver}} \underbrace{\iprod{U_{J}(q_J), C^{\ast}(x,y)U_M(q_2)}}_{\text{\normalfont C Solver}}\\
&=& [K_A (K_B K_C)]_{NM}
\end{eqnarray}
Here we now have another matrix $K_C$ which we need to compute, this
requires another projection onto the basis $U_J$ whose beam parameter
is $q_J$. 

\section{Coupling coefficient integration performance improvements}\label{sec:knm_transpose}

Calculating the coupling coefficients by numerical integration is a computationally expensive
process and can last from minutes to days depending on the map and incoming/outgoing beam 
parameters. As seen previously the coupling coefficient matrix can be split up into 2 separate
ones for both the numerical integration result and Bayer-Helms. We also had to make a choice
for the value of $q_L$, if this is chosen so that the numerical integral matrix has the same
incoming and outgoing beam parameter an $\approx\times 2$ speedup can be achieved. Taking the coupling coefficient
integral we get
\begin{eqnarray}
U_{nm} &=& (2^{n+m-1}n!m!\pi)^{-1/2}\frac{1}{w(z)}e^{i\psi(z)(n+m+1)}...\nonumber \\
&&... H_{n}\left(\frac{\sqrt{2}x}{w_{o,x}}\right) H_{m}\left(\frac{\sqrt{2}y}{w_{o,y}}\right) e^{-i\frac{k r^2}{2q(z)}}\\
 &=& \frac{A_{nm}}{w(z)}e^{i\psi(z)(n+m+1)} H_{n} H_{m} e^{-i\frac{k r^2}{2q(z)}}\\
k_{nmn'm'} &=& \int U_{n'm'} F(x,y) U_{nm}^* dA \\
k_{nmn'm'} &=& \frac{A_{nm}A_{n'm'}}{w_{out}(z)w_{in}(z)} e^{i\psi_{out}(n'+m'+1)}e^{-i\psi_{in}(n+m+1)}...\nonumber \\
&&...\int F(x,y) H_{n} H_{m}H_{n'} H_{m'} e^{-i\frac{k r^2}{2}\left(\frac{1}{q_{out}(z)} - \frac{1}{q_{in}^*(z)}\right)}  dA \label{eq:knmnm_gen}
\end{eqnarray}
The interesting case for us here is when $q_{in}=q_{out}$, if this is true then the above simplifies,
\begin{eqnarray}
k_{nmn'm'} &=& \frac{A_{nm}A_{n'm'}}{w^2(z)} e^{i\psi(\Delta n+\Delta m)}\int F(x,y) H_{n} H_{m}H_{n'} H_{m'} e^{-\frac{x^2+y^2}{w^2(z)}}  dA, \\
\Delta n &=& n' - n, \\
\Delta m &=& m' - m. \\
\end{eqnarray}
We then find that the transpose elements are nearly identical, i.e. $00 \rightarrow 10$ and $10 \rightarrow 00$, expect for the an opposite sign in the Gouy phase
\begin{eqnarray}
k_{n'm'nm} &=& \frac{A_{nm}A_{n'm'}}{w^2(z)} e^{-i\psi(\Delta n+\Delta m)}\int F(x,y) H_{n} H_{m}H_{n'}H_{m'} e^{-\frac{x^2+y^2}{w^2(z)}}  dA.
\end{eqnarray}
Dividing one by the other leaves us with the final relationship between the elements in the matrix
\begin{eqnarray}
k_{n'm'nm} &=& k_{nmn'm'}e^{-i2\psi(\Delta n+\Delta m)}, \\
|k_{n'm'nm}| &=& |k_{nmn'm'}|.
\end{eqnarray}
So we see that the matrix is symmetric if Gouy phase is $0$, as it is at the beam waist. This
 is a useful relationship especially for calculating the coupling matrices as now we need only
 calculate one half of the matrix. The transpose elements can easily be found by just
 multiplying by the phase factor. The potential computational reduction increases when using more modes if mode matched by
a factor $(N - 1)/2N$, where $N$ is the number of modes used.

This performance increase is on by default and switches off in the code if not mode-matched. It
can be manually controlled from the \verb|kat.ini| file with the option \\ \verb|calc_knm_transpose [0 or 1]|.

%% file: numerical_math.tex

\chapter{Some mathematics}

This appendix gives some details about some of the formulae and
algorithms used in \Finesse.
 
\section{Hermite polynomials}
\label{sec:h_poly} 

The first few Hermite polynomials in their unnormalized form can be
given as:
\begin{equation}\label{eq:h_poly}
\begin{array}{ll}
H_0(x)=1, & H_1(x)=2x,\\
H_2(x)=4x^2-2, & H_3(x)=8x^3-12x.\\
\end{array}
\end{equation}
Further polynomial orders can be computed recursively using the following
relation:
\begin{equation}\label{eq:h_poly_rec}
H_{n+1}(x)=2xH_n(x)-2nH_{n-1}(x).
\end{equation}
In \Finesse the functions up to $H_{10}$ are hard-coded and for higher
orders the recursion relation is used. 

\section{The paraxial wave equation}
\label{sec:pax}

An electromagnetic field (at one point in time, in one polarisation, and in
free space) can in general be described by the following scalar wave
equation~\cite{siegman}:
\begin{equation}\label{eq:a_hg_1}
\left[\nabla^2+k^2\right]E(x,y,z)=0.
\end{equation}
Two well-known exact solutions for this equation are the plane wave:
\begin{equation}
E(x,y,z)=E_0 ~\mEx{-\I kz},
\end{equation}
and the spherical wave:
\begin{equation}
E(x,y,z)=E_0 ~\frac{\mEx{-\I kr}}{r}\qquad\mbox{with}\qquad
r=\sqrt{x^2+y^2+z^2}.
\end{equation}
Both solutions yield the same phase dependence along an axis (here, for
example, the $z$-axis) of $\exp(-\I kz)$. This leads to the idea that a
solution for a beam along the $z$-axis can be found in which the phase
factor is the same while the spatial distribution is described by a
function $u(x,y,z)$ which is slowly varying with $z$:
\begin{equation}
E(x,y,z)=u(x,y,z)~\mEx{-\I kz}.
\end{equation}
Substituting this into \eq{eq:a_hg_1} yields:
\begin{equation}\label{eq:a_hg_wave2}
\left(\delta_x^2+\delta_y^2+\delta_z^2\right)u(x,y,z)-2\I k\delta_z
u(x,y,z)=0.
\end{equation}
Now we put the fact that $u(x,y,z)$ should be slowly varying with $z$ in
mathematical terms. The variation of $u(x,y,z)$ with $z$ should be small
compared to its variation with $x$ or $y$. Also the second partial
derivative in $z$ should be small. This can be expressed as:
\begin{equation}
\left|\delta_z^2
u(x,y,z)\right|\ll\left|2k\delta_zu(x,y,z)\right|,\left|\delta_x^2
u(x,y,z)\right|,\left|\delta_y^2 u(x,y,z)\right|.
\end{equation}
With this approximation, \eq{eq:a_hg_wave2} can be simplified to the
\emph{paraxial wave equation}:
\begin{equation}\label{eq:pax}
\left(\delta_x^2+\delta_y^2\right)u(x,y,z)-2\I k\delta_z u(x,y,z)=0.
\end{equation}


%% file: syntax_ref.tex

\chapter{Syntax reference}
\label{sec:syntax_reference}

In order to use the program you have to know and understand the commands for
the input files. The following paragraphs give a full explanation of the
syntax. The help screens (use `kat -h' or `kat -hh') give a short syntax
reference. See other our online syntax reference at \url{http://www.gwoptics.org/finesse/reference/}.

\section{Comments}

Two different methods are available for adding comments to the \Finesse
input files.  First, each line can be `commented out' by putting a single
comment sign at the start of the line. The comment signs are {\Co \#}, {\Co
"} and {\Co $\%$}.

Any of these signs can also be used to add a comment at the end of a line,
for example:\\ {\Co xaxis mirror phi lin -20 20 100 \# tune mirror position}

The second commenting method is the use of C-style block comments with
\texttt{/* -- -- -- */}.  This is very useful for including several
different sets of commands in one input file. In the following example some
temporarily unused photodiodes have been commented out:
\begin{finesse}
/*
pd pd1 n23
pd pd2 n24
pd pd3 n25
pd pd4 n26
*/

pd1 pd1p 1M 0 n23
\end{finesse}

{\bf Please note that these comment strings must be used in empty lines,
otherwise the (very simple) parser cannot handle them correctly.}

\section{Components}

Parameters in square brackets [ ] are optional.
{
\begin{itemize}
\vspace{1cm} \hrule

\item{
{\bf
m
}%
 : mirror\\
usage : {\tt\small m name R T phi node1 node2}\\
\begin{tabular}{ccl}
R & = & power reflectivity (0<R<1)\\
T & = & power transmittance (0<T<1)\\
phi & = &tuning in degrees
\end{tabular}

A positive tuning moves the mirror from \cmd{node2} towards \cmd{node1}.
}

\vskip\baselineskip\hrule\vskip\baselineskip

\item{
{\bf
m1
}%
 : mirror\\
usage : {\tt\small m1 name T Loss phi node1 node2}\\
\begin{tabular}{ccl}
T & = & power transmittance (0<T<1)\\
Loss & = & power loss (0<Loss<1) \\
phi & = &tuning in degrees
\end{tabular}

{\bf Note:} the values are not
stored as T and L but as R and T with 0$<$R, T$<$1. Thus, only R and
T can be tuned (e.g.~with \cmd{xaxis}).
}

\vskip\baselineskip\hrule\vskip\baselineskip

\item{
{\bf
m2
}%
 : mirror\\
usage : {\tt\small m2 name R Loss phi node1 node2}\\
\begin{tabular}{ccl}
R & = & power reflectivity (0<R<1)\\
Loss & = & power loss (0<Loss<1)\\
phi & = &tuning in degrees
\end{tabular}

{\bf Note:} the values are not
stored as T and L but as R and T with 0$<$R, T$<$1. Thus, only R and
T can be tuned (e.g.~with \cmd{xaxis}).
}

\vskip\baselineskip\hrule\vskip\baselineskip

\item{
{\bf
s 
}%
 : space\\
usage : {\tt\small s name L [n] node1 node2}\\
\begin{tabular}{ccl}
L & = & length in metres  \\
n & = & index of refraction  \\
&& (default is 1.0 or specified with $n_0$ in `kat.ini')\\ 
\end{tabular}

}

\vskip\baselineskip\hrule\vskip\baselineskip

\item{
{\bf 
bs
}%
 : beam splitter \\
usage : {\tt\small  bs name R T phi alpha node1 node2 node3 node4}\\
\begin{tabular}{ccl}
R &=  & power reflectivity (0<R<1)\\
T &=  & power transmittance (0<T<1)\\
phi &=  &  tuning in degrees\\
alpha &= & angle of incidence in degrees          
\end{tabular}

A positive tuning moves the beam splitter along from \cmd{node3}/\cmd{node4}
towards \cmd{node1}/\cmd{node2} along a vector perpendicular to the beam
splitter surface (i.e.~the direction depends upon `alpha').

}

\vskip\baselineskip\hrule\vskip\baselineskip

\item{
{\bf 
bs1
}%
 : beam splitter \\
usage : {\tt\small  bs1 name T Loss phi alpha node1 node2 node3 node4}\\
\begin{tabular}{ccl}
T &=  & power transmittance (0<T<1)\\
Loss &= & power loss (0<Loss<1)\\
phi &=  &  tuning in degrees\\
alpha &= & angle of incidence in degrees          
\end{tabular}

{\bf Note:} the values are not
stored as T and L but as R and T with 0$<$R, T$<$1. Thus only R and T
can be tuned (e.g.~with \cmd{xaxis}).
}

\vskip\baselineskip\hrule\vskip\baselineskip

\item{
{\bf 
bs2
}%
 : beam splitter \\
usage : {\tt\small  bs2 name R Loss phi alpha node1 node2 node3 node4}\\
\begin{tabular}{ccl}
R &=  & power reflectivity (0<R<1)\\
Loss &= & power loss (0<Loss<1)\\
phi &=  &  tuning in degrees\\
alpha &= & angle of incidence in degrees          
\end{tabular}

{\bf Note:} the values are not
stored as T and L but as R and T with 0$<$R, T$<$1. Thus only R and T
can be tuned (e.g.~with \cmd{xaxis}).
}

\vskip\baselineskip\hrule\vskip\baselineskip

\item{
{\bf 
gr
}%
 : grating \\
usage : {\tt\small  gr[n] name d node1 node2 [node3 [node4]]}\\
\begin{tabular}{ccl}
d &=  & grating period in [nm]\\
\end{tabular}

Other parameters of the grating (some {\bf must} be set) can be 
set via the \cmd{attr} command; these are:
\begin{itemize}
\item{power coupling efficiencies: eta\_0, eta\_1, eta\_2, eta\_3, rho\_0}
\item{angle of incidence: alpha}
\item{radius of curvature: Rcx, Rcy, Rc (not yet implemented)}
\end{itemize}

Grating types are defined via their number of ports:
\begin{itemize}
\item[2]{ 1st order Littrow (eta\_0, eta\_1)}
\item[3]{ 2nd order Littrow (eta\_0, eta\_1, eta\_2, rho\_0)}
\item[4]{ not Littrow (eta\_0, eta\_1, eta\_2, eta\_3)}
\end{itemize}
}
\vskip\baselineskip\hrule\vskip\baselineskip

\item{
{\bf
isol
}%
 : isolator\\
usage : {\tt\small isol name S node1 node2}\\
\begin{tabular}{ccl}
S &=  & power suppression in dB\\
\end{tabular}

The light passes the isolator unchanged from \cmd{node1} to \cmd{node2} but
the power of the light going from \cmd{node2} to \cmd{node1} will be
suppressed:
\begin{equation}
a_{\rm out}~=~10^{-S/20}~a_{\rm in},\nonumber
\end{equation}
with $a$ as the field amplitude.
}
\vskip\baselineskip\hrule\vskip\baselineskip

\item{
{\bf
l
}%
 : laser (input light)\\
usage : {\tt\small l name P f [phase] node}\\
\begin{tabular}{ccl}
P &=  & light power in Watts \\
f &=  & frequency offset to the default frequency $f_0$\\
&&       (default frequency determined from `lambda' in `kat.ini')\\
phase &=  & phase
\end{tabular}

}

\vskip\baselineskip\hrule\vskip\baselineskip

\item{
{\bf
pd
}%
 : photodiode (plus one or more mixers)\\
usage : {\tt\small
pd[n]~name~[f1~[phase1~[f2~[phase2~[\dots]~]~]~]~]~node[*]}\\
\begin{tabular}{ccl}
n &=  & number of demodulation frequencies ($0\leq n\leq5$)\\
f1 &=  & demodulation frequency of the first mixer in Hz  \\
phase1 &=  & demodulation phase of the first mixer in degrees\\
f2 &=  &  demodulation frequency of the second mixer in Hz \\
phase2 &=  & demodulation phase of the second mixer in degrees\\
\dots & &
\end{tabular}

The photodetector generally computes the laser power in an interferometer
output.  With the command \cmd{scale ampere} the value can be scaled
to photocurrent.

Note: the number of frequencies (n) {\bf must} be given correctly. The
square brackets may be misleading here, since the parameter is not optional
but can be omitted only if the number of frequencies is zero.  Some likely
examples are:

\cmd{pd detector1  nout1} (or~~ \cmd{pd0 detector1 nout1}) : DC detector\\
\cmd{pd1 detector2 10M 0 nout2} : one demodulation\\
\cmd{pd2 detector3 10M 0 100 0 nout2} : two demodulations

All frequencies are with respect to the zero frequency $f_0$ set by `lambda'
in the init file `kat.ini' (see Section~\ref{sec:ini}). 

The phases are the \emph{demodulation phases} and describe the phase of
the local oscillator at the mixer. If the last phase is omitted the output
resembles a network analyser instead of a mixer. This differs from a mixer
because the resulting signal does contain phase information. A mixer with a
fixed demodulation phase is usually used for calculating error signals
whereas one often wants to know the phase of the signal for frequency
sweeps, i.e. for calculating transfer functions. 

The keyword `\cmd{max}' can be used in place of the fixed demodulation phase,
e.g.:

\cmd{pd2 detector1 10M max 200 max nout1}

This does use the optimum demodulation phase {\bf for each data point
independently}.  This can be useful when the `best' demodulation phase is
not yet known but in some circumstances it will give `strange' output
graphs.  {\bf Note that this kind of calculation does not represent any
meaningful way of handling real output signals. It was merely added for
convenience.}

Again, the optional asterisk behind the node name changes from the default
beam to the second beam present at this node (see
Section~\ref{sec:comp+nodes} for the definition of the default beam).
}

\vskip\baselineskip\hrule\vskip\baselineskip

\item{
{\bf
pdS
}%
 : shot noise limited sensitivity \\
Usage is the same as for \cmd{pd}. It calculates the shot noise in the
output using the DC photocurrent and divides it by a signal.  For example,

\cmd{pdS2 name 10M 90 100 0 n2}

would be :\\
shot noise(pd n2) / (pd2 name 10M 90 100 0 n2)

i.e.~the shot noise at node n2 divided by the signal at 100 Hz (phase$ =
0\degrees$) in the photocurrent at n2 demodulated at 10 MHz (phase$ =
90\degrees$).

Note: This detector relies on a simple approximation for the shotnoise and 
only gives the correct result when no modulation sidebands 
are present, see \Sec{sec:shotnoise}.

For this sensitivity output, all demodulation phases have to be set.

If the output is a transfer function and \cmd{fsig} was used to add
signals to mirrors or beam splitters it can be further normalised to $\rm m
/ \sqrt{\rm Hz}$ by \cmd{scale meter} (see below).  }

\vskip\baselineskip\hrule\vskip\baselineskip

\item{
{\bf
pdN
}%
 : photocurrent normalised by shot noise\\
Usage is the same as for {\tt pd} or {\tt pdS}. It calculates the inverse
of {\tt pdS} which gives the signal to shot noise ratio.

Note: This detector relies on a simple approximation for the shotnoise and 
only gives the correct result when no modulation sidebands 
are present, see \Sec{sec:shotnoise}.

}

\vskip\baselineskip\hrule\vskip\baselineskip
\item{
{\bf
ad
}%
 : amplitude detector \\
usage : {\tt\small ad name [n m] f node[*]}\\
\begin{tabular}{ccl}
n, m & = & TEM mode numbers. n refers to the $x$-direction.\\
f & = &  sideband frequency in Hz (as offset to the input light)
\end{tabular}

The optional asterisk behind the node name changes from the default beam to
the second beam present at this node (see Section~\ref{sec:comp+nodes} for
the definition of the default beam). 

The amplitude detector calculates field amplitude and phase of all light
fields at {\bf one} frequency. For correct absolute values you have to set
`epsilon\_c' correctly in `kat.ini' (see section~\ref{sec:ini}). See the
command `\cmd{yaxis}' for the various possibilities for plotting the
computed values. 

If higher order modes are used and no indices $n$, $m$ are given, the
amplitude detector tries to compute the value for the `phase front' on the
optical axis.  See \Sec{sec:hgm-ad}.
}

\vskip\baselineskip\hrule\vskip\baselineskip

\item{
{\bf
shot
}%
 : shot noise \\
usage : {\tt\small shot~name~node[*]}

It calculates the shot noise in the
output using the DC light power $P$ as
\begin{equation}
\Delta P =\sqrt{\frac{2\,h\,c~}{\lambda_0}P~}.
\end{equation}

Note: This detector relies on a simple approximation for the shotnoise and 
only gives the correct result when no modulation sidebands 
are present, see \Sec{sec:shotnoise}.

}

\vskip\baselineskip\hrule\vskip\baselineskip

\item{
{\bf
mod
}%
 : modulator\\
usage : {\tt\small mod name f midx order am/pm [phase] node1 node2}\\
\begin{tabular}{ccl}
f & = & modulation frequency in Hz           \\
midx & = &modulation index            \\
order & = & number of sidebands (or `\cmd{s}' for single sideband)           \\
am/pm & = & amplitude or phase modulation           \\
phase & = & phase of modulation 
\end{tabular}

Phase modulation :\\
`order' sets the order up to which sidebands are produced by the 
modulator. For example, given an input light with `f' equal to zero

{\tt\small
mod mo1 10k 0.3 2 pm n1 n2
}

produces sidebands at -20kHz, -10kHz, 10kHz and 20kHz. The maximum possible
order is 6. If order is set to `\cmd{s}' then the modulator produces a
single sideband.

Amplitude modulation:\\
`order' is always set to 1, and \cmd{midx} must be between 0 and 1.
}

\vskip\baselineskip\hrule\vskip\baselineskip
\item{
{\bf
fsig
}%
 : signal frequency\\
usage : {\tt\small fsig name component [type] f phase [amp]}\\
\begin{tabular}{ccl}
component  &=  & one of the previously defined component names\\
type &=& type of modulation\\
f &=  &  signal frequency\\
phase &= & signal phase \\
amp &= & signal amplitude
\end{tabular}

Used as the input signal for calculating transfer functions, see also 
Section~\ref{sec:trans+err}.
For example\\ 
{\tt\small fsig sig1 m1 10k 0}\\
shakes the previously defined component (e.g.~a mirror) `m1' at 10 kHz.
Only one signal frequency can be used in one calculation but that can be fed
to several components, e.g.

{\tt\small
fsig sig1 m1 10k 0\\
fsig sig2 m2 10k 0
}

inserts the signal at two mirrors (in phase).

For the moment the following types of signal modulation are implemented
(default type marked by *):
\begin{itemize}
\item{mirror:  phase*, amplitude, xbeta, ybeta}
\item{beam splitter: phase*, amplitude, xbeta, ybeta}
\item{space: amplitude, phase*}
\item{input: amplitude, phase, frequency*}
\item{modulator: amplitude, phase*}
\end{itemize}

To tune only the signal frequencies one has to be explicitly tuned with {\Co
xaxis} or {\Co put}, because only one signal frequency is allowed. E.g.~in
the example above, tuning {\Co sig1} will also tune {\Co sig2}.

}
\end{itemize}
}
\section{Hermite-Gauss extension}\label{sec:hg_commands}
This section gives the syntax of components and commands which 
are part of the Hermite-Gauss extension of \Finesse as described 
in \chap{sec:HGmodes}. The command \cmd{maxtem} is used to
switch between plane-waves and Hermite-Gauss beams, see below.

{
\begin{itemize}

\vspace{.5cm} \hrule \vspace{.5cm}
\item{
{\bf
tem
}%
 : distribute input power to various TEM modes\\
usage : {\tt\small  tem input n m factor phase}\\
alternative : {\tt\small  tem* input p l factor phase}\\
\begin{tabular}{ccl}
input & = & name of an input component\\
n, m & = & $\rm Hermite-Gaussian TEM_{\rm nm}$ mode numbers\\
n, m & = & $\rm Laguerre-Gaussian TEM_{\rm pl}$ mode numbers\\
factor & = & relative power factor \\
phase & = & relative phase in degrees
\end{tabular}

When an input (component `\cmd{l}') is specified, the given laser power is
assumed to be in the $\rm TEM_{00}$ mode. The command \cmd{tem} can change
that.  Each \cmd{tem} command can set a relative factor and phase for a
given TEM mode at a specified input.  Several commands for one input are
allowed. 

Please note that the \cmd{tem} command is intended to \emph{add} higher
order modes, i.e.:\\
{\tt\small 
tem input1 0 1 1.0 0.0\\
}
adds a \M{01} mode to the \M{00} mode. Both fields have the same amplitude.

In order to create a pure higher order mode the \M{00} amplitude has to
be explicitly set to zero. For example:\\
{\tt\small 
tem input1 0 0 0.0 0.0\\
tem input1 0 1 1.0 0.0\\
}
would put all power into the  $\rm TEM_{01}$ mode. 

Another example:\\
{\tt\small 
tem input1 0 0 1.0 0.0\\
tem input2 2 1 1.0 90.0\\
}
specifies that exactly the same amount of power as in the  $\rm TEM_{00}$
mode should be in $\rm TEM_{21}$, but with a phase offset of 90 degrees. 

You can also specify an input Laguerre-Gaussian beam using \verb|tem*|.
For this version of the command $p>0$ but $l$ can be a positive or negative
integer. 

Note: `\cmd{tem}' does not change the total power of the laser beam.

}

\vskip\baselineskip\hrule\vskip\baselineskip

\item{
{\bf
lens
}%
 : thin lens\\
usage : {\tt\small lens name f node1 node2}\\
\begin{tabular}{ccl}
f &=  & focal length in metres\\
\end{tabular}

A lens does not change the amplitude or phase of a passing beam. When
Hermite-Gauss modes are used, the beam parameter $q$ is changed with
respect to f.
}

\vskip\baselineskip\hrule\vskip\baselineskip

\item{
{\bf
bp
}%
 : beam parameter detector\\
usage : {\tt\small bp name x/y parameter node}\\
\begin{tabular}{ccl}
x/y &=& direction for which the parameter should be taken\\
parameter &=& a parameter derived from the Gaussian beam parameter, see
below.
\end{tabular}

This detector can plot a variety of parameters able to be derived from
the Gaussian beam parameter which is set to the respective node:\\
\begin{tabular}{lcl}
w  &:& beam radius in metres\\
w0 &:& waist radius in metres\\
z  &:& distance to waist in metres\\
zr &:& Rayleigh range in metres\\
g  &:& Gouy phase in radians\\
r  &:& Radius of curvature (of phase front) in meters\\
q  &:& Gaussian beam parameter\\
\end{tabular}

Please note that the Gouy phase as given by {\Co bp} is not the accumulated
Gouy phase up to that node but just the Gouy phase derived from the beam
parameter is:
\begin{equation}
\Psi(z)=\arctan\left(\frac{\myRe{q}}{\myIm{q}}\right)\nonumber.
\end{equation}
The accumulated Gouy phase can be plotted with the detector {\Co gouy}, see
below.
}

\vskip\baselineskip\hrule\vskip\baselineskip

\item{
{\bf
cp
}%
 : cavity parameter detector\\
usage : {\tt\small cp cavity--name x/y parameter}\\
\begin{tabular}{ccl}
cavity--name &=& name of cavity of which the parameter should be taken\\
x/y &=& direction for which the parameter should be taken\\
parameter &=& a parameter derived by the cavity trace algorithm, see below.
\end{tabular}

Please note that this detector type has not a unique name. Instead the name of
the respective cavity must be given.

This detector can plot a variety of parameters that are derived by a cavity
trace. The cavity must have been specified by the `cav' command.
Please note that these parameters are only filled with meaningful numbers
when a cavity trace is executed. You can use the command `retrace'
to force a trace for each data point. The available parameters
are the same that can be printed to the terminal using `trace 2':\\
\begin{tabular}{lcl}
w  &:& beam radius at the first cavity node in metres\\
w0 &:& waist radius at the first cavity node in metres\\
z  &:& distance to waist at the first cavity node in metres\\
r  &:& radius of curvature of beam phase front at the first cavity node in meters\\
q  &:& Gaussian beam parameter at the first cavity node\\
finesse &:& cavity finesse\\
loss  &:& round trip loss (0<=loss<=1)\\
length &:& optical path length of the cavity in meters (counting a full round-trip)\\
FSR  &:& free spectral range in Hz\\
FWHM  &:& cavity linewidth as Full Width at Half Maximum in Hz\\
pole  &:& cavity pole frequency in Hz (=0.5*FWHM)\\

\end{tabular}

Please note that the direction parameter (x/y) only applies to the parameters 
related to beam size but must always be given so that the parsing of the `cp'
command will function.

}

\vskip\baselineskip\hrule\vskip\baselineskip

\item{
{\bf
gouy
}%
 : gouy phase detector\\
usage : {\tt\small gouy name x/y space-list}\\
\begin{tabular}{ccl}
x/y &=  & direction for which the phase should be taken\\
space-list &= & a list of names of `space' components
\end{tabular}

This detector can plot the Gouy phase accumulated by a propagating beam.
For example:\\
{\Co gouy g1 x s1 s2 s3 s4 sout}\\
plots the Gouy phase that a beam accumulates on propagating through
the components {\Co s1}, {\Co s2}, {\Co s3}, {\Co s4} and {\Co sout}.
}

\vskip\baselineskip\hrule\vskip\baselineskip

\item{
\refstepcounter{dummy}
\label{sec:pdtype}
{\bf
pdtype
}%
 : defines the type of a photodetector\\
usage : {\tt\small pdtype detector-name type-name}\\

This command defines the type of a photodetector with respect to the
detection of transverse modes. The standard detector is a simple photodiode
with a surface larger than the beam width. With `pdtype' more complex
detectors can be used, like for example, split photodetectors.

In the file `kat.ini' a number of different types can be defined by giving
scaling factors for the various beat signals between the different
Hermite-Gauss modes. For example, if a photodetector will see the beat
between the \M{00} and \M{01}, then the line `0 0 0 1 1.0' (mode factor)
should be present in the description. The definitions in the `kat.ini' file
are given a name. This name can be used with the command `pdtype' in the
input files. Many different types of real detectors (like split detectors)
or (spatially) imperfect detection can be simulated using this feature.  The
syntax for the type definitions: 

\begin{finesse}
PDTYPE name
...
END
\end{finesse}

Between {\Co PDTYPE} and {\Co END} several lines of the following 
format can be given:
\begin{itemize}
\item[1.]{`\cmd{0 1 0 2 1.0}',
       beat between \M{01} and \M{02} is scaled with factor $1.0$}
\item[2.]{`\cmd{0 0 * 0 1.0}',
       '*' means `any': the beats of \M{00} with \M{00}, \M{10},
       \M{20}, \M{30},... are scaled with $1.0$}
\item[3.]{`\cmd{x y x y 1.0}',
       `\cmd{x}' or `\cmd{y}' also mean `any' but here all instances of
       `\cmd{x}' are always the same number (same for `\cmd{y}'). So in this
       example all beats of a mode with itself are scaled by $1.0$}
\end{itemize}
Please note that {\bf all beat signals which are not explicitly given are
scaled with 0.0}.  (`\cmd{debug 2}' somewhere in the input file will cause
\Finesse to print  all non-zero beat signal factors for all defined types.)
Please take care when entering a definition, because the parser is very
simple and cannot handle extra or missing spaces or extra characters.

The file `kat.ini' in the \Finesse package includes the definitions
for split photodetectors, see \Sec{sec:split_pd}.
}
\vskip\baselineskip\hrule\vskip\baselineskip
\item{
{\bf
beam
}%
 : beam shape detector \\
usage : {\tt\small beam name [f] node[*]}\\
\begin{tabular}{ccl}
f &=  & frequency of the field component in Hz
\end{tabular}

The optional asterisk behind the node name changes from the default beam to
the second beam present at this node (see Section~\ref{sec:comp+nodes} for
the definition of the default beam). 

With the beam analyser one can plot the cross-section of a beam, see 
Section~\ref{sec:beamshape}.

If no frequency is given the beam detector acts much like a CCD camera,
it computes the light intensity per unit area as a function of $x$ and $y$. 
The $x$-axis has to be set to either `x' or `y'. 
For example {\Co xaxis beam1 x lin -10 10 100} sets the $x$-axis to tune 
the position in the $x$-direction from
$-10 w_0$ to $10 w_0$ in 100 steps (for beam analyser `beam1').  A second
$x$-axis can be set to the perpendicular direction in order to plot the two
dimensional cross-section of the beam. Thus the axes of the plot are
scaled automatically by $w_0$, the waist size computed from the Gaussian
beam parameter at the beam detector. The values for the waist size are 
printed to the terminal and given in the plot labels.

If a frequency is given the beam detector outputs the amplitude and phase of the
light field at the given frequency (you can use the \cmd{yaxis} command to
define whether the amplitude, phase or both should be plotted).
}
\vskip\baselineskip\hrule\vskip\baselineskip
\item{
{\bf
mask
}%
 : mask out certain TEM modes in front of a photodetector or a beam analyser\\
usage : {\tt\small  mask detector n m factor}\\
\begin{tabular}{ccl}
detector & = & name of a photodetector or beam analyser\\
n, m & = & $\rm TEM_{\rm nm}$ mode numbers\\
factor & = & power factor [0,1]\\
\end{tabular}

Several \cmd{mask} commands can be used per detetector.

Without this command all photodetectors (for which `pdtype' is not used)
detect the power of all  TEM modes, for example :
\begin{equation}
S(f_1,f_2)=\sum_{\rm nm}2\Re\left(a_{\rm nm}(f_1)a_{\rm
nm}(f_2)\right)=\sum_{\rm nm} S_{\rm nm},
\end{equation}
where $a_{\rm nm}(f)$ is the amplitude of the $\rm TEM_{\rm nm}$ mode at
frequency $f$.  Note that other detectors, like split detectors, quadrant
cameras or bulls-eye detectors use a special geometry to detect certain
cross-terms. Setting a mask for a $\rm TEM_{\rm nm}$ will scale the detected
power by the given factor:
\begin{equation}
S_{\rm nm}(f_1,f_2)=\mbox{\rm factor}~2\Re\left(a_{\rm nm}(f_1)a_{\rm
nm}(f_2)\right).
\end{equation}
}

\vskip\baselineskip\hrule\vskip\baselineskip
\item{
{\bf
attr
}%
 : additional (optional) attributes for mirrors, beam splitters and spaces\\
usage : {\tt\small attr component M value Rc[x/y] value x/ybeta value gx/y value}\\
\begin{tabular}{ccp{10cm}}
component &=  &  the component for the attributes to be set to\\
 M &=  & mass in grammes \\
 Rc&=  & radius of curvature, in metres (zero is used for plane surface)\\
 Rcx&=  & radius of curvature in interferometer plane\\
 Rcy&=  & radius of curvature in plane perpendicular to interferometer\\
 Rap&=  & Radius of aperture (size of optic) [m] (Only for mirrors)\\
 xbeta&=  & angle of mis-alignment in the interferometer plane in radian\\
 ybeta&=  & angle of mis-alignment perpendicular to the interferometer plane in radian\\
 g &=  & Gouy phase of a space component in degrees (see \Sec{sec:gouy})\\
 gx &=  & Gouy phase of a space component (horizontal component)\\
 gy &=  & Gouy phase of a space component (vertical component)\\
 value&=  & numerical value for the specified attribute
\end{tabular}

Note, in contrast to phases {\bf the alignment angles {\tt\small xbeta} and
{\tt\small ybeta} are given in radians}.

The various attributes are optional. For example one can simply set the
radius of curvature of a mirror `m1' to 10 metres with the command:
\begin{finesse}
attr m1 Rc 10
\end{finesse}
The sign for the radius of curvature is defined as follows: if the surface
seen from the {\bf first specified node} (specified at the respective mirror
or beam splitter) is concave, then the number for the radius of curvature is
{\bf positive} (see \Sec{sec:abcd}).

Please note that when the attributes `Rc' or `g' are used you cannot tune
the parameter itself. Instead, the separate directions i.e.~`Rcx' and/or
`Rcy' and `gx' and/or `gy' must be used for further tuning, e.g.~with the
{\Co xaxis} command.
}

\vskip\baselineskip\hrule\vskip\baselineskip

\item{
{\bf
map
}%
 : loads and applies a surface map to a mirror component\\
usage : {\tt\small map component filename}\\
\begin{tabular}{ccp{10cm}}
component &=  &  mirror to which the map should be applied\\
filename &=& name of the file containing the map data
\end{tabular}

This command reads a file given by filename and searches for a surface map
given by a grid or by coupling coefficients (previously computed by
\Finesse). The data must be provided in a special structure, 
see \ref{sec:mirrormaps}. You can apply multiple maps to a mirror surface by repeating 
the command with the same component.

}

\vskip\baselineskip\hrule\vskip\baselineskip

\item{
{\bf
gauss
}%
 : setting the $q$ parameter for a Gaussian beam\\
usage : {\tt\small gauss name component node w0 z [wy0 zy]}\\
(alternative: {\tt\small gauss* name component node q [qy] })\\
(alternative: {\tt\small gauss** name component node w(z) Rc [wy(z) Rcy] })\\
\begin{tabular}{ccp{10cm}}
w0 &=  &  beam waist size in metres\\
z &=  &  distance to waist in metres\\
Rc &=  &  Radius of curvature \\
w(z) &=  &  beam spot size at node \\
q &=  &  complex beam parameter (given as `$z + \mathrm{i}\zr$',
i.e.~`distance-to-waist Rayleigh-range')
\end{tabular}

A Gaussian beam at a certain point $z'$ on the optical axis can be
characterised by two parameters. The first common method is to specify the
waist size $w_0$ and the distance to the waist $z$. In \Finesse the complex
parameter for Gaussian beams is used: $$q=z+ \I \zr,$$ with $\zr$ the
Rayleigh range and $z$ the distance to the beam waist.

The distance to the waist can be positive or negative. {\bf A positive value
means the beam has passed the waist, a negative number specifies a beam
moving towards the waist}.  It is clear that the $q$ parameter has to be set
for a certain direction of propagation (the other direction then has the
parameter $q'=-q^*$). The direction of propagation is set with `component'.
The node at which the Gauss parameter is to be set has to be connected to
the specified component. The direction of propagation is defined as: {\bf
from the component towards the node}.

In general a Gaussian beam may have two different beam parameters for the
$x$- and the $y$-direction. When two parameter sets are given with
\cmd{gauss} the first set is assumed to be valid for the $x$-direction and
the second for the $y$-direction.  If only one set is given then it is used
for both directions.
}

\vskip\baselineskip\hrule\vskip\baselineskip

\item{
{\bf
cav
}%
 : tracing a beam through a cavity and compute the $q$-eigenvalues\\
usage : {\tt\small  cav name component1 node1 component2 node2}

The components and nodes specify the start and end point of a beam path
through a possible cavity. `\cmd{node1}' has to be connected to
`\cmd{component1}' and `\cmd{node2}' to `\cmd{component2}'. There are only
two possibilities for specifying a cavity in \Finesse:
\begin{itemize}
\item a linear cavity: the start component and end component are two
different mirrors.
\item a ring cavity: the start component and end component are the same beam
splitter and the nodes are either 1 and 2 or 3 and 4, so that the beams are
connected to each other via a reflection.
\end{itemize}

When the cavity is stable (not critical or unstable) the eigenmatrix is
computed. The resulting eigenvalues for the Gaussian beam ($q$ parameters)
are then set for all cavity nodes.

Use `trace' in the input file to see what cavity nodes are found and which
$q$-values are set, see below.

}
\vskip\baselineskip\hrule\vskip\baselineskip
\item{
{\bf
trace
}%
 : set verbosity for beam tracing\\
usage : {\tt\small trace n}\\
\begin{tabular}{ccl}
n &=  & an integer which sets the verbosity level.
\end{tabular}

When the trace is set, \Finesse will print some information while tracing a
beam through the interferometer, through a cavity, or during other
computation tasks that are connected to the Hermite-Gauss extension. The
integer `n' is bit coded, i.e.~n=2 gives different information to n=4,
while n=6 will give both. 

{\vspace{\baselineskip}\centering
\begin{tabular}{l  p{12cm}}\hline
n & output\\\hline
 1 & list of TEM modes used\\
 2 & cavity eigenvalues and cavity parameters like free spectral range, 
 optical length and finesse\\
 4 & mode mismatch parameters for the initial setup (mismatch parameter as
 in \cite{bayer})\\
 8 & beam parameters for every node, nodes are listed in the order found by
 the tracing algorithm\\
 16 & Gouy phases for all spaces\\
 32 & coupling coefficients for all components\\
 64 & mode matching parameters during calculation, if they change due to a
 parameter change, for example by changing a radius of curvature\\
 128 & nodes found during the cavity tracing\\ 
\hline
\end{tabular}
}
}\label{tab:trace}

\vskip\baselineskip\hrule\vskip\baselineskip
\item{
{\bf
retrace [off]
}%
 : recomputes the Gaussian parameters at each node for every data point\\
usage : {\tt\small retrace}

\Finesse needs to trace the beam through the interferometer
in order to set the Gaussian beam parameters (see \Sec{sec:trace}
for a detailed description). This is always done once at the start of
a simulation if higher-order modes are used.
If \cmd{xaxis} or \cmd{put} are used to tune a parameter like a length 
of a space or the focal length of a lens or the radius of curvature of a mirror
the beam parameters are locally changed and the beam tracing should be
repeated. Without re-computing a proper base of Gaussian beam parameters such a tuning
introduces a virtual mode mismatch which can lead to wrong results.  

\Finesse automatically detects whether a re-tracing is required and if so
computes a new set of base parameters for each
data point.
The {\Co retrace} command can be used to over-ride the automatic behaviour:
\cmd{retrace} will force a retracing and \cmd{retrace off} will prevent
it, regardless of the \cmd{xaxis} settings. 

Please note that the re-tracing cannot
avoiding all unwanted mode-mismatches. 

}

\vskip\baselineskip\hrule\vskip\baselineskip
\item{
{\bf
startnode
}%
 : recomputes the Gaussian parameters at each node for every data point\\
usage : {\tt\small startnode node}

This command allows one to explicitly set the node at which the automatic
beam trace algorithm starts. The node must have a beam parameter associated
with it. This means the node must be inside a cavity that has been traced
with the {\Co cav} command or the parameter must be explicitly set via the
{\Co gauss} command.
}

\vskip\baselineskip\hrule\vskip\baselineskip
\item{
{\bf
maxtem
}%
 : set the maximum order for Hermite-Gauss modes\\
usage : {\tt\small maxtem order}\\
\begin{tabular}{ccl}
order &=  &  maximum order, i.e.~$n+m$ for $\rm TEM_{\rm  nm}$ modes.
\end{tabular}

maximum number : 1\\

This defines the maximum order for $\rm TEM_{\rm  nm}$ and thus the number
of light fields used in the calculation. The default `order' is 0, the
maximum value is 100. For large values the interferometer matrix becomes
very large and thus the simulation extremely slow. Please note that the \HG
mode is automatically switched on if at least one attribute or command
referring to transverse modes is entered. You can explicitly switch off the
\HG mode by using `maxtem off'.

}

\vskip\baselineskip\hrule\vskip\baselineskip
\item{
{\bf
phase  
}%
 : switches between different modes for computing the light phase\\
usage : {\tt\small phase number}\\

Four different modes are available:\\
\begin{tabular}{ccl}
 0& = & no change\\
 1& = & the phase for coupling coefficients of \M{00} is scaled to 0\\
 2& = & the Gouy phase\index{Gouy phase} for \M{00} is scaled to 0 \\
 3& = & combines modes 1 and 2 (default)
\end{tabular}

The command `phase' can be used to change the computation of light field
phases in the \HG mode. In general, with higher order modes the spaces are
not resonant any more for the \M{00} mode because of the Gouy phase.
Furthermore, the coupling coefficients $k_{nmnm}$ contribute to the phase
when there is a mode mismatch. For correct analysis these effects have to be
taken into account. On the other hand, these extra phase offsets make it
very difficult to set a resonance condition or operating point intuitively.
In most cases another phase offset would be added to all modes so that the
phase of the \M{00} becomes zero.  With the command `phase' these phase
offsets can be set for the propagation through free space, for the coupling
coefficients or both: `phase 1' ensures that phases for the coupling
coefficients $k_{0000}$ (\M{00} to \M{00}) are 0, `phase 2' ensures that all
Gouy phases for \M{00} are 0 and `phase 3' combines both effects. The phases
for all higher modes are changed accordingly, so that the relative phases
remain correct. {\bf Please note that only {\tt phase 0} and  {\tt phase 2}
guarantee always correct results, see \Sec{sec:phaseeffect} for 
more details.}
}

\vskip\baselineskip\hrule\vskip\baselineskip
\item{
{\bf

knm  
}%
 : specifies a file which the coupling coefficients for a given
   component generate and saves/loads them from the file\\
usage : {\tt\small knm \verb|component_name| \verb|filename_prefix|}\\

\textbf{A different command of the same name existed in previous versions. Usage has
  changed from version 0.99.8. The old functionality has been moved
  into the \cmd{conf knm\_flags} command.}

With `knm' the user can specify a file which the coupling coefficents for a component with a 
map applied can save and load the coefficients from. This is primarily used to save computational
time as map coefficients can be expensive to compute. The only components this can be applied to
at the moment are mirrors. \verb|filename_prefix| states the filename prefix.

The numeric integration uses a fast self-adapting routine \cite{dcuhre} and Cuba routines but
nevertheless will be very slow in comparison to the simple formula
calculation.  The numeric integration algorithm can be customised with the
following parameters in the kat.ini file:\\
\begin{tabular}{ccl}
    maxintop &:& maximum function calls of the numeric integration\\
             && algorithm (default 400000)\\
    abserr  &:& absolute error requested by integration routine\\
            &&  (default 1e-6)\\
    relerr  &:& relative error requested by integration routine\\
            &&  (default 1e-6)\\
	maxintcuba &:& Maximum integrand calculations for Cuba routines\\
			&&  (default 1e6)\\
\end{tabular}

}
\end{itemize}
}

\section{Commands}

{
\begin{itemize}

\vspace{.5cm} \hrule \vspace{.5cm}
\item{
{\bf
xaxis
}%
 : $x$-axis definition, i.e.~parameter to tune\\
usage : {\tt\small xaxis component parameter lin / log min max steps}\\
(alternative: {\tt\small xaxis* component parameter lin / log min max steps})\\
\begin{tabular}{ccl}
 component& = & one of the previously defined component names  \\
 parameter& = & a parameter of the component e.g.~`L' for a space  \\
 lin/log& = &  defines a linear or logarithmic $x$-axis \\
 min& = & start value  \\
 max& = & stop value  \\
 steps& = & number of steps from min to max \\
\end{tabular}

maximum number : 1\\

The previous definition of the interferometer yields exactly one output
value for every detector. To create a plot we have to define a parameter
that is changed (tuned/swept) along the $x$-axis of the plot. Exactly one
{\tt\small xaxis} must be defined. For example,

{\tt\small
xaxis s1 L lin 1 10 100
}

changes the length of space s1 from 1 metre to 10 metres in 100 steps.

Another useful example is to sweep the laser frequency using:

{\tt\small
xaxis i1 f lin 0 10k 500
}

When the optional asterisk is used then the previously defined value for the
parameter to tune is used as an offset. For example:

{\tt\small
s s1 L 5\\
xaxis* s1 L lin 1 10 100
}

tunes the length of space \cmd{s1} from 6 to 15 metres. 

When the axis is logarithmic the min/max values are multiplied to the
previously defined value, e.g.

{\tt\small
s s1 L 5\\
xaxis* s1 L log .1 10 100
}

tunes the length of space \cmd{s1} from 0.5 to 50 metres. Note that the
parameters used as $x$-axis in the output plot are those given in the
{\tt\small xaxis*} statement, {\bf not} the computed values which are really
used in the calculation. This feature allows one to specify the tunings of
the operating point in the interferometer description and then always tune
{\em around} that operating point by $\mathrm{min} = -\mathrm{max}$.

}

\vskip\baselineskip\hrule\vskip\baselineskip
\item{
{\bf
x2axis
}%
 : second $x$-axis definition, creates 3D plot\\
usage as for \cmd{xaxis}\\

This command defines a second $x$-axis. A 3D plot is created. 
}

\vskip\baselineskip\hrule\vskip\baselineskip
\item{
{\bf
yaxis
}%
 : $y$-axis definition (optional)\\
usage : {\tt\small yaxis [lin / log] abs:deg / db:deg / re:im / abs / db /
de}\\
\begin{tabular}{ccl}
abs:deg &=  &  amplitude and phase\\
db:deg &=  &  amplitude in dB and phase\\
re:im &=  &  real and imaginary part\\
re &=  &  real part\\
im &=  &  imaginary part\\
abs &=  &  amplitude\\
db &=  &  amplitude in dB\\
deg &=  & phase  
\end{tabular}

maximum number : 1\\

This defines the (first plus an optional second) $y$-axis.
}

\vskip\baselineskip\hrule\vskip\baselineskip
\item{
{\bf
scale
}%
 : rescaling of output amplitudes (optional)\\
usage : {\tt\small scale factor [detector]}\\
\begin{tabular}{ccl}
factor & = & scale factor \\
detector & = & output name 
\end{tabular}

All or a specified output signal is scaled by factor.  (The scaling is done
after demodulations.)

If the keyword \cmd{meter} is used instead of a number for {\tt\small
factor} the output is scaled by $2\pi/\lambda_0$ (or $\lambda_0/2\pi$ for
\cmd{pdS}). In case the output is a transfer function and the signals have
been added to mirrors or beam splitters the transfer functions are thus
normalised to ${\rm W/m}/\sqrt{\rm Hz}$ (${\rm m}/\sqrt{\rm Hz}$ for
\cmd{pdS}).

If the keyword \cmd{ampere} is used instead of a number for {\tt\small
factor} the output is scaled by $e~q_{\rm eff}~\lambda_0/(hc)$. This
converts light power (Watt) to photocurrent (Ampere).

If the keyword \cmd{deg} is used the output will be scaled by $180/\pi$.
}

\vskip\baselineskip\hrule\vskip\baselineskip
\item{
{\bf
diff
}%
 : differentiation\\
usage : {\tt\small diff component parameter}\\
\begin{tabular}{ccl}
component  & = & one of the previously defined component names \\
parameter & = &  a parameter of the component e.g. `L' for a space
\end{tabular}

maximum number : 3\\

Instead of the standard result of the calculation, partial
differentiation with respect to the specified parameter will be plotted. For
a higher order differentiation you can specify the command again with the
same parameter. The differentiation is calculated as:

{\tt\small
diff = (f(x+h/2) - f(x-h/2))/h
}

The step size \texttt{h} can be specified via the constant {\tt deriv\_h}
in `kat.ini', the default is 1e-3. You can also overwrite the 
value from the `kat.ini' file with the command {\tt deriv\_h} in an
input file. This is useful when several files which require a different
step size are located in the same directory.

Please note that if {\tt\small put} is used, the parameter specified in
{\tt\small put} is linked to the parameter from the {\tt\small xaxis}
command, so a differentiation with respect to the parameter specified
in {\tt\small put} is not possible and a differentiation with respect to the
parameter stated in {\tt\small xaxis} will automatically perform a
differentiation with respect to the connection of parameters (which was
introduced by {\tt\small put}).

}
\vskip\baselineskip\hrule\vskip\baselineskip
\item{
{\bf
const
}%
 : constant definition\\
usage : {\tt\small const name value}\\
\begin{tabular}{ccl}
name  & = & user-defined name, less than 15 characters long \\
value & = & numerical or string value
\end{tabular}

The constants are used during pre-processing of the input file.  If anywhere
in the file the command `\cmd{const name value}' is defined, every instance
of `\cmd{\$name}' in the input file is replaced by `\cmd{value}'.  
}

\vskip\baselineskip\hrule\vskip\baselineskip
\item{
{\bf
variable
}%
 : definition of a dummy variable\\
usage : {\tt\small variable name value}\\
\begin{tabular}{ccl}
name  & = & user-defined name, less than 15 characters long \\
value & = & numerical or string value
\end{tabular}

The sole purpose of this command is to provide a dummy variable that
can then be tuned by the \cmd{xaxis} command and connected to the
interferometer with \cmd{put} commands. A typical application would be the
tuning of a differential arm length as follows:
\begin{finesse}
variable deltax 1
xaxis deltax abs lin -1 1 100
put* ETMX phi $x1
put* ITMX phi $mx1
\end{finesse}

}

\vskip\baselineskip\hrule\vskip\baselineskip
\item{
{\bf
set
}%
 : variable definition\\
usage : {\tt\small set name component parameter}\\
\begin{tabular}{ccl}
name  & = & user defined name, less than 15 characters long \\
component& = & one of the previously defined component names  \\
parameter& = & a parameter of the component e.g. `L' for a space  \\
\end{tabular}

The variables are used for creating input variables that can be used with
functions, see below. 

With `set', all tunable parameters in the input file can be   accessed with
the usual syntax `\cmd{component parameter}'.  The {\Co set} command will
link the variable `\cmd{\$name}' to the parameter value of the named
component.  In addition, the output of any detector can be stored in a
variable. The syntax is:

{\Co set name detector-name re/im/abs/deg}

where \cmd{re/im/abs/deg} indicate which real number to use if the detector
output is a complex number.  (NOTE: for detectors with a real output, use
`\cmd{re}' and NOT `\cmd{abs}' since `\cmd{abs}' will remove the sign!)

The \cmd{set} commands are executed for each data point, i.e.~if the
component parameter is changed e.g.~with the {\Co xaxis command} the
variable \cmd{\$name} will change accordingly.

In addition to user defined variables, several internal
variables have been defined: `\cmd{\$x1}',
`\cmd{\$x2}' and `\cmd{\$x3}' have been pre-defined and point to the current
value of the \cmd{xaxis} (or \cmd{x2axis}, \cmd{x3axis} respectively).  
in addition, `\cmd{\$mx1}', `\cmd{\$mx2}' and `\cmd{\$mx3}' are defined
as minus the corresponding `\cmd{\$x}' variable.
These predefined names must not be used with `\cmd{set}'.

}
\vskip\baselineskip\hrule\vskip\baselineskip
\item{
{\bf
func
}%
 : function definition\\
usage : {\tt\small func name = function-string}\\
\begin{tabular}{ccl}
name  & = & user-defined name, less than 15 characters long \\
function-string & = & a mathematical expression
\end{tabular}

For example, {\Co  func y = \$Lp+2} defines the new variable `y = Lp+2',
with `Lp' being a previously defined variable.  Such previously defined
variables are entered with a `\$' sign. The new variable (i.e.~the function
result) will be plotted as a new output, like a detector output.  Any
previously defined variable via {\Co set}, {\Co func} or {\Co lock} (see
below) can be used like the function string.  The functions are
exectuted for each data point.  (Please note that if you use two similar
function names like `function' and `function1' the parser might have
problems to distinguish between the two.)

This new feature uses the mathematical expression parser Formulc 2.22 by
Harald Helfgott.  The following functions are available in the function
string: exp(), ln(), sin(), cos(), tan(), asin(), acos(), atan(), atan2(),
abs(), sqrt(), pi() (with no parameters inside the parentheses) and rnd() (a
random number between 0 and 1).

Numbers have to be given numerically, e.g.~`3.0E-9' instead of `3n'. Please
note that `3.0e-9' does not work. Multiplication with negative numbers
requires parentheses, e.g.:\\ {\Co y = (-1)*\$x1}

For a detailed description of the parser syntax, please see the
documentation of Formulc 2.22.
}

\vskip\baselineskip\hrule\vskip\baselineskip
\item{
{\bf
put[*]
}%
 : write variable into interferometer parameter\\
usage : {\tt\small put component parameter \$variable}\\
\begin{tabular}{ccl}
component& = & one of the previously defined component names  \\
parameter& = & a parameter of the component e.g.~`L' for a space  \\
variable  & = & previously declared variable
\end{tabular}

For example, {\Co put space2 L \$y} writes the content of variable $y$ into
the length of `\cmd{space2}'. ({\Co put*} always adds to the initially set
lengths of `\cmd{space2}') 

All {\Co put} commands are executed once before first data point is
computed.  If photodetector outputs are used in {\Co put} or {\Co func} they
are set to 0.0 for the first data point calculation.

}

\vskip\baselineskip\hrule\vskip\baselineskip
\item{
{\bf
lock
}%
 : control loop definition\\
usage : {\tt\small lock name \$variable gain accuracy}\\
\begin{tabular}{ccl}
name  & = & user defined name, less than 15 characters long \\
variable & = & previously defined variable (usually a photodetector output)\\
gain & = & loop gain\\ 
accuracy & = & threshold to decide whether the loop is locked
\end{tabular}

The command will read the variable given by `\cmd{\$variable}' and write it
into the new variable `\cmd{name}'. This variable will be also plotted as a
new output, like a detector output.  {\Co lock*} stops after the first point
so that only the initial lock is found and the rest is computed without
locking.  (Please note that if you chose two similar lock names like
`\cmd{mylock}' and `\cmd{mylock1}' the parser might have problems to
distinguish between the two.) 

\Finesse will perform an iterative computation for each data point on the
$x$-axis. In fact, it will compute the interferometer iteratively until the
condition 
\begin{equation}
\abs{(\${\rm variable})}<{\rm accuracy}\nonumber
\end{equation}
is fulfilled. 

In order to achieve this goal the command tries to mimic a control loop with
a simple integrator. The input `\cmd{\$variable}' serves as the error signal
and the output stored in `\cmd{\$name}' holds the feedback signal (which has
to be connected to the interferometer by the user with a {\Co put} command).
In each iterative step it perfroms the operation:\\

name = \$name + gain * \$variable   ( or name += gain * \$variable )\\

Several {\Co lock} commands can be active simultaneously and the lock output
variables can be used in {\Co func} commands located below the {\Co lock} in
the input file.  {\bf Please note: The order of the commands `\cmd{func}'
and '\cmd{lock}' in the input file determines the order of their
computation!}
   
Of course the lock fails miserably if:
\begin{itemize}
\item{the loop is not closed,}
\item{the error signal is not good,}
\item{the computation is not started at or close to a good operating point,}
\item{the gain is wrong (sign, amplitude) or,}
\item{the steps as given by the \cmd{xaxis} command are too large
(i.e.~move the interferometer out of the linear range of the error signal).}
\end{itemize}
A fine tuning of the gain is useful to minimise the computation time.

An example: 
\begin{finesse}
    to lock a cavity to a laser beam we can write:
    # laser and EOM
    l i1 1 0 n0 
    mod eo1 40k 0.3 3 pm n0 n1  
    # cavity:
    m m1 0.9 0.1 0 n1 n2
    s s1 1200 n2 n3
    m m2 .9 0.01 0 n3 n4
    # Pound-Drever-Hall signal
    pd1 pdh 40k 0 n1
    # tune 
    xaxis m2 phi lin 0 100 400

    # set the error signal to be photodiode output (`re' stands
    # for the real part of the diode output. `re' has to be used 
    # with diodes that provide a real output.
    set err pdh re
    # Perform the lock! Gain is -10 and the accuracy 10n ( = 1e-8)
    lock z $err -10 10n 
    # ... and connect the feedback to the interferometer
    put* m1 phi $z
\end{finesse}
The behaviour of the locking routine can be adjusted by setting some
paramaters in `kat.ini'. For example, the lock iteration can automatically
adjust the loop gains. The following parameters in the `kat.ini' file can be
used:
\begin{itemize}    
\item{\cmd{locksteps} (integer, >0, default 10000):
  maximum number of steps in which the iteration tries to achieve the lock. }
\item{\cmd{autogain}  (integer, 0,1,2 default 2):
  switch for the automatic gain control:  0 = Off, 1 = On, 2 = On with
  verbose output.}
\item{\cmd{autostop} (integer, 0,1 default 1):
  if autostop is swiched ON the locking algorithm will stop
  after it fails to reach the desired accuracy once.}
\item{\cmd{sequential} (integer, 0,1,5, default 5):
  this keyword determines if the feedback signals are computed 
  sequentially or in parallel. The sequential mode is slower
  but performs much better far away from the operating point
  or when `autogain' is needed. The default 5 uses the sequential
  mode for the first two data points and then switches to the faster
  parallel locking.}
\item{\cmd{lockthresholdhigh} (double, >0, default=1.5):
  whether or not a loop is probably oscillating with a too high gain is 
  determined using `lockthresholdhigh'. The criterion used is
  as follows (with y1,y2,... as successive error signal values):
  the oscillation condition  is defined as:\\
  if abs((y1+y3-2*y2)/accuracy/y3) > lockthresholdhigh,
  true=loop oscillates.}
\item{\cmd{lockthresholdlow} (double, >0, default=0.01):
  whether or not a loop gain is too low is determined
  using `lockthresholdlow'. The low-gain condition is defined as:\\
  if abs((y1+y3-2*y2)/accuracy/y3) < lockthresholdlow,
  true=loop gain too low.}
\item{\cmd{locktest1} (integer, >0, default 5) and
  locktest2 (integer, >0, default 40):
  `locktest1' and `locktest2' determine the number of steps that an 
  iteration is allowed to remain in an `oscillation' (or `low gain'). After 
  `locktest1' number of steps the loop state is checked. If for 
  `locktest2' number of checks the same error condition persists the loop gain 
  will be reduced or increased by the factor `gainfactor'.}
\item{\cmd{gainfactor} (double, >0, default 3).}
\end{itemize}
}

You can find two more examples in \Sec{sec:lock}.

\vskip\baselineskip\hrule\vskip\baselineskip

\item{
{\bf
showiterate
}%
 : define verbosity of the {\Co lock} commands\\
usage : {\tt\small showiterate steps}\\
\begin{tabular}{ccl}
steps  & = & number of iterations 
\end{tabular}

If `steps' is >0 the current state of the lock iteration is printed every
`steps' iterations. If `steps'$=-1$ the result is printed only after the first
succesful iteration (useful for knowing the values of the initial operating
point).
}

\vskip\baselineskip\hrule\vskip\baselineskip
\item{
{\bf
noplot
}%
 : suppress the plot of an output\\
usage : {\tt\small noplot output}\\
\begin{tabular}{ccl}
output  & = & previously defined output (detector, function, etc.)
\end{tabular}

Since {\Co func} and {\Co lock} create new outputs, the resulting plots
might become very cluttered. Therefore the command {\Co noplot output} has
been introduced. It suppresses the plotting of the given output
(photodetector, function, lock, ...). The data is stored in the *.out file
as before, only the plot command in the respective the *.gnu batch file is
changed.

Please note that \cmd{noplot} cannot be used to suppress all plotting.
One output must remain to be plotted. If you want to suppress all graphical output
please use \cmd{gnuterm~no}.

}

\vskip\baselineskip\hrule\vskip\baselineskip
\item{
\refstepcounter{dummy}
\label{sec:derivh}
{\bf
deriv\_h
}%
 : overwrites the value for deriv\_h given in `kat.ini'\\ 
usage : {\tt\small deriv\_h value}\\
\begin{tabular}{ccl}
value  & = & step size for numerical differentiation
\end{tabular}

This command can be used to overwrite the pre-defined vale for deriv\_h. This
can be useful especially when you want to differentiate alignment signals in which
numerical values of $10^{-9}$ are often required.
}

\vskip\baselineskip\hrule\vskip\baselineskip
\item{
\refstepcounter{dummy}
\label{sec:conf}
{\bf
conf
}%
 : Sets configuration options for various optical components \\ 
usage : {\tt\small conf component option value}\\
\begin{tabular}{ccl}
component & = & mirror or beamsplitter name \\
option & = & name of option to set \\
value  & = & input for option
\end{tabular}

This command is used to fine tune and alter the computational routines for a given
optical component. It does not represent anything physical about the optic. Currently
it is used to alter the behaviour of the coupling coefficient computation for both
mirrors and beamsplitters, it is not used for any other components at the moment.

\begin{itemize}    
\item{\cmd{integration\_method} (1 or 2 or 3)
      sets the numerical integration method. Cuba refers to a
      self-adapting routine which is faster but less robust: 1 - Riemann Sum, 2 - Cubature - Serial, 3 - Cubature - Parallel (default)
	  }
\item{\cmd{interpolation\_method} (1 or 2 or 3)
      set the interpolation method for the numerical integration of
      surface maps, the (use NN for maps with sharp edges):
      1 - Nearest Neighbour (default), 2 - Linear, 3 - Spline.
	  }
\item{\cmd{interpolation\_size integer} (odd integer > 0)
      sets the size of the interpolation kernel, must be odd and > 0}
\item{\cmd{knm\_flags} ( integer > 0 )
      Sets the knm computation flags which define if coeffs are calculated
      numerically or analytically if possible.}
\item{\cmd{show\_knm\_neval} (0/1)
      Shows the number of integrand evaluations used for the map integration.}
\item{\cmd{save\_knm\_matrices} (0/1)
      If true the knm matrices are saved to .mat files
      for distortion, merged map and the final result}
\item{\cmd{save\_knm\_binary} (0/1)
      If true the knm and merged map data is stored in a
      binary format rather than ascii. See -convert option
      for kat in -h for converting between the 2 formats}
\item{\cmd{save\_interp\_file} (0/1)
      If 1 then for each knm calculated a file is written
      to outputting each interpolated point. The output file
      will have 4 columns: x,y,A,phi. So for each integrand
      evaluation the interpolated point is plotted}
\item{\cmd{save\_integration\_points} (0/1)
      If 1 then the points used for integration are
      saved to files. Only use this with Riemann
      integrator, Cuba can use millions of points and is slow}
\item{\cmd{knm\_order} (12 or 21)
      changes order in which the coupling coefficient matrices
      are computed. 1 = Map, 2 = Bayer-Helms}
\item{\cmd{knm\_change\_q} (1 or 2)
      Decides the value of the expansion beam parameter q\_L.
      If 1 then q\_L = q'\_1 and if 2 then q\_L = q\_2.} 
\end{itemize}

}

\end{itemize}
}

\section{Auxiliary plot commands}

{
\begin{itemize}

\vskip\baselineskip\hrule\vskip\baselineskip
\item{
{\bf
gnuterm
}%
 : Gnuplot terminal (optional)\\
usage : {\tt\small gnuterm terminal [filename]}\\
\begin{tabular}{cc p{10cm}}
terminal & = & one terminal name specified in `kat.ini', default is `x11' or
`windows' respectively \\
filename & = & name for Gnuplot output file
\end{tabular}

maximum number : 20\\

If you do not want a Gnuplot batch file to be written use :
`gnuterm no'.
}
\vskip\baselineskip\hrule\vskip\baselineskip
\item{
{\bf
pause
}%
 : pauses after plotting\\
usage : {\tt\small pause}\\

maximum number : 1\\

Adds a command `pause -1' to the Gnuplot batch file after each plot into a
screen terminal.
}
\vskip\baselineskip\hrule\vskip\baselineskip
\item{
{\bf
multi
}%
 : switches from a single surface to multiple surfaces in 3D plots\\
usage : {\tt\small multi}\\

maximum number : 1\\

By default in a 3D plot only the first output is plotted even if multiple
outputs are present. If `multi' is set, Gnuplot plots multiple surfaces into
the same graph.

Please note that even without setting `multi' the data of all outputs is
present in the output data file.
}

\vskip\baselineskip\hrule\vskip\baselineskip
\item{
{\bf
GNUPLOT \ ...\ END  
}%
 : extra Gnuplot commands \\
usage (for example):\\
{\tt\small 
GNUPLOT\\
set view 70, 220, ,\\
set contour\\
END
}

All the Gnuplot commands specified between GNUPLOT and END will be written
to the Gnuplot batch file. This is especially useful for 3D plots (see
3D.kat for an example).
}

\end{itemize}}


%% file: ack.tex

\addsec*{Acknowledgements}
\addcontentsline{toc}{chapter}{Acknowledgements}
\thispagestyle{plain}

Gerhard Heinzel has been the major force behind the creation of \Finesse. He had the idea of
using the LISO routines on interferometer problems and he let me copy his
code for that purpose. Furthermore he has been very busy as my most faithful
beta tester and has helped me getting the right ideas in many discussions.
I have gotten many helpful bug reports from Guido M\"uller early on and always
enjoyed discussing interferometer configurations with him. The latter is
also true for Roland Schilling: For hours he would listen to me on the phone
while I was trying to understand my program - or interferometers and optics.
Ken Strain has been a constant source of help and support during the several
years of development. During my time at Virgo Gabriele Vajente and Maddalena Mantovani have
acted as faithful test pilots for the extension with the \cmd{lock} command.
Alexander Bunkowski has initiated and helped debugging the grating
implementation. Jerome Degallaix has often helped with suggestions, 
examples and test results based on his code OSCAR 
to further develop and test \Finesse. 

Paul Cochrane
has made a big difference with his help on transforming the source code from its
messy original form into a more professional package, including a testsuite, 
an API documentation and above all a readable source code.

During the last years (2011 to 2013) several current and former members from my research
group in Birmingham have put significant effort into the further
development, testing and use of \Finesse. Daniel Brown has become lead
programmer and has provided a large number of bug fixes and new
features navigating the tricky grounds of optics with high-order
modes. Due to his work, \Finesse has reached version 1.0 and is now
available as open source. Charlotte Bond is a specialist in using \Finesse, in
particular with mirror surface maps or strange beam shapes; she has
become the main contributor of our Simtools package and her help with
\Finesse has been invaluable for getting the physics of higher-order
modes right. 
Further, Keiko Kokeyama, Paul Fulda and Ludovico
Carbone have worked very hard to help making \Finesse do useful things for
the Advanced LIGO commissioning team.

Many people in the gravitational wave community have helped me with
feedback, bug reports and encouragement. Some of them are Seiji Kawamura,
Simon Chelkowski, Keita Kawabe, Osamu Miyakawa, Rainer K\"unnemeyer, Uta Weiland, Michaela
Malec, Oliver Jennrich, James Mason, Julien Marque,
Mirko Prijatelj, Jan Harms, Oliver Bock, Kentaro Somiya, Antonio
Chiummo, Holger Wittel, Hartmut Grote, Bryan Barr, Stefan Ballmer, Daniel
Shaddock and probably many more that I have not mentioned here.

Last but not least I would like to thank the GEO\,600 group, especially
Karsten Danzmann and Benno Willke, for the possibility to work on \Finesse
in parallel to my experimental work on the GEO site. \Finesse would not
exist without their positive and open attitude towards the young members of
the group.
